\documentclass[aps,prd,twocolumn,superscriptaddress,floatfix,nofootinbib,showpacs,amsmath,amssymb,altaffilletter]{revtex4-1}
\usepackage[colorlinks=true,pdfstartview=FitV,linkcolor=blue,citecolor=blue, urlcolor=blue]{hyperref}
\usepackage{rotate}
\usepackage{rotating}
\usepackage{graphicx}
\usepackage{multirow}
\usepackage{bm}
\usepackage{wrapfig}
\usepackage{paralist}
\usepackage{wrapfig}
\usepackage{subfigure}
\usepackage{longtable}
\usepackage{color}
\usepackage{appendix}
\usepackage{vruler}
\usepackage{hyperref}
\usepackage{setspace}
\usepackage{ifthen}
\usepackage{array}
\usepackage{etoolbox}

\usepackage[sc]{mathpazo}
\newcommand{\myrefs}[2]{\href{http://dx.doi.org/#2}{#1}}
\newcommand{\mref}[1]{\href{http://#1}{#1}}

\newcommand{\ET}[2]{\mbox{#2$\times$10$^{#1}$}}

\newcommand{\bg}{\mbox{$\beta/\gamma$}}

\newcommand{\sodium}{\mbox{$^{22}$Na}}

\newcommand{\rbthree}{\mbox{$^{83}$Rb}}
\newcommand{\krthree}{\mbox{$^{83m}$Kr}}

\newcommand{\bios}{\mbox{$^{207}$Bi}}

\newcommand{\lar}{\mbox{LAr}}

\newcommand{\kevr}{\mbox{keV}}

\newcommand{\sone}{\mbox{S1}}
\newcommand{\stwo}{\mbox{S2}}

\newcommand{\fno}{\mbox{\tt f$_{90}$}}
\newcommand{\pe}{\mbox{PE}}
\newcommand{\ser}{\mbox{SER}}

\newcommand{\wph}{\mbox{$W_{\rm ph}$(max)}}

\newcommand{\dst}{DarkSide-10}

\newcommand{\lartpc}{TPC}

\newcommand{\leff}{\mbox{$\mathcal{L}_{\rm eff,\,^{83m}Kr}$}}
\newcommand{\qy}{\mbox{$\mathcal{Q}_{\rm y}$}}
\newcommand{\Enr}{\mbox{$E_{\rm nr}$}}
\newcommand{\Edrift}{\mbox{$\mathcal{E}_{\rm d}$}}
\newcommand{\tdrift}{\mbox{$t_{\rm d}$}}

\newcommand{\sonenr}{\mbox{$S1_{\rm nr}$}}

\newcommand{\Rone}{\mbox{$R_{1}$}}
\newcommand{\Rtwo}{\mbox{$R_{2}$}}

\newcommand{\chisq}{\mbox{$\chi^{2}$}}
\newcommand{\Nex}{\mbox{$N_{\rm ex}$}}
\newcommand{\Ni}{\mbox{$N_{\rm i}$}}
\newcommand{\e}{\mbox{$\mathrm{e^-}$}}
\newcommand{\fn}{\mbox{$f_{\rm n}$}}
\newcommand{\fl}{\mbox{$f_{\rm l}$}}

\newcommand{\tpctof}{TPCtof}
\newcommand{\ntof}{Ntof}
\newcommand{\npsd}{Npsd}
\begin{document}
\title{Measurement of scintillation and ionization yield and scintillation pulse shape from nuclear recoils in liquid argon}
\newcommand{\Chicago}{Kavli Institute for Cosmological Physics, University of Chicago, Chicago, IL 60637, USA}
\newcommand{\FNAL}{Fermi National Accelerator Laboratory, Batavia, IL 60510, USA}
\newcommand{\Houston}{Department of Physics, University of Houston, Houston, TX 77204, USA}
\newcommand{\Indiana}{Physics Department, University of Notre Dame, Notre Dame, IN 46556, USA}
\newcommand{\LNGS}{INFN Laboratori Nazionali del Gran Sasso, Assergi 67010, Italy}
\newcommand{\Napoli}{Physics Department, Universit\`a degli Studi Federico II and INFN, Napoli 80126, Italy}
\newcommand{\NotreDame}{Physics Department, University of Notre Dame, Notre Dame, IN 46556, USA}
\newcommand{\Princeton}{Physics Department, Princeton University, Princeton, NJ 08544, USA}
\newcommand{\Temple}{Physics Department, Temple University, Philadelphia, PA 19122, USA}
\newcommand{\UCL}{Department of Physics and Astronomy, University College London, London WC1E 6BT, United Kingdom}
\newcommand{\UCLA}{Physics and Astronomy Department, University of California, Los Angeles, CA 90095, USA}
\newcommand{\UMass}{Physics Department, University of Massachusetts, Amherst, MA 01003, USA}
\newcommand{\KICPChicago}{Kavli Institute for Cosmological Physics, University of Chicago, Chicago, IL 60637, USA}
\newcommand{\LLNL}{Lawrence Livermore National Laboratory, 7000 East Ave., Livermore, CA 94550, USA}
\author{H.~Cao}\affiliation{\Princeton}
\author{T.~Alexander}\affiliation{\UMass}\affiliation{\FNAL}
\author{A.~Aprahamian}\affiliation{\Indiana}
\author{R.~Avetisyan}\affiliation{\Indiana}
\author{H.~O.~Back}\affiliation{\Princeton}
\author{A.~G.~Cocco}\affiliation{\Napoli}
\author{F.~DeJongh}\affiliation{\FNAL}
\author{G.~Fiorillo}\affiliation{\Napoli}
\author{C.~Galbiati}\affiliation{\Princeton}
\author{L.~Grandi}\affiliation{\Chicago}
\author{Y.~Guardincerri}\affiliation{\FNAL}
\author{C.~Kendziora}\affiliation{\FNAL}
\author{W.~H.~Lippincott}\affiliation{\FNAL}
\author{C.~Love}\affiliation{\Temple}
\author{S.~Lyons}\affiliation{\Indiana}
\author{L.~Manenti}\affiliation{\UCL}
\author{C.~J.~Martoff}\affiliation{\Temple}
\author{Y.~Meng}\affiliation{\UCLA}
\author{D.~Montanari}\affiliation{\FNAL}
\author{P.~Mosteiro}\affiliation{\Princeton}
\author{D.~Olvitt}\affiliation{\Temple}
\author{S.~Pordes}\affiliation{\FNAL}
\author{H.~Qian}\affiliation{\Princeton}
\author{B.~Rossi}\affiliation{\Napoli}\affiliation{\Princeton}
\author{R.~Saldanha}\affiliation{\Chicago}
\author{S.~Sangiorgio}\affiliation{\LLNL}
\author{K.~Siegl}\affiliation{\Indiana}
\author{S.~Y.~Strauss}\affiliation{\Indiana}
\author{W.~Tan}\affiliation{\Indiana}
\author{J.~Tatarowicz}\affiliation{\Temple}
\author{S.~Walker}\affiliation{\Temple}
\author{H.~Wang}\affiliation{\UCLA}
\author{A.~W.~Watson}\affiliation{\Temple}
\author{S.~Westerdale}\affiliation{\Princeton}
\author{J.~Yoo}\affiliation{\FNAL}
\collaboration{The SCENE Collaboration}


\begin{abstract}
We have measured the scintillation and ionization yield of recoiling nuclei in liquid argon as a function of applied electric field by exposing a dual-phase liquid argon time projection chamber (LAr-TPC) to a low energy pulsed narrow band neutron beam produced at the Notre Dame Institute for Structure and Nuclear Astrophysics.  Liquid scintillation counters were arranged to detect and identify neutrons scattered in the TPC and to select the energy of the recoiling nuclei.  
We report measurements of the scintillation yields for nuclear recoils with energies from 10.3 to 57.3\,keV and for median applied electric fields from 0 to 970\,V/cm. For the ionization yields, we report measurements from 16.9 to 57.3\,keV and for electric fields from 96.4 to 486\,V/cm.
We also report the observation of an anticorrelation between scintillation and ionization from nuclear recoils, which is similar to the anticorrelation between scintillation and ionization from electron recoils.  
Assuming that the energy loss partitions into excitons and ion pairs from \krthree\ internal conversion electrons is comparable to that from \bios\ conversion electrons, we obtained the numbers of excitons (\Nex) 
and ion pairs (\Ni) and their ratio (\Nex/\Ni) produced by nuclear recoils from 16.9 to 57.3\,keV.
Motivated by arguments suggesting direction sensitivity in LAr-TPC signals due to columnar recombination, a comparison of the light and charge yield of recoils parallel and perpendicular to the applied electric field is presented for the first time.

\end{abstract}

\maketitle

\section{Introduction}

We have used a monoenergetic neutron beam to characterize scintillation (\sone) and ionization (\stwo) signals produced by nuclear recoils between 10.3 and 57.3\,\kevr\ in a liquid argon time projection chamber (LAr-TPC) with and without an applied electric field.  The results described in this paper are relevant for the calibration and interpretation of data of LAr-TPC dark matter detectors~\cite{warp,ardm,darkside}.  They also lay the groundwork for a method that could be applied for the characterization of  a liquid xenon time projection chamber (LXe-TPC)~\cite{xenon,lux,pandax} and other dark matter detectors.

In a previous paper we introduced our method and compared it with prior methods~\cite{scene1,gastler,regenfus}.  We also discussed our initial measurements on S1 and reported the first observation of a dependence on drift field of the S1 yield.

In this paper we present the detailed set of results on the S1 and S2 measurements.  We also report the first observation in LAr of an anticorrelation between scintillation and ionization from neutron-induced nuclear recoils; this closely resembles the anticorrelation between scintillation and ionization from electrons~\cite{kubota78}, relativistic heavy ions~\cite{doke85}, $\alpha$ particles and fission fragments~\cite{hitachi87}.  
With the aid of a model describing the relationship between the number of ion pairs (\Ni) and the magnitude of S2, we extracted the numbers of excitons (\Nex) and ion pairs (\Ni) and their ratio (\Nex/\Ni) produced by nuclear recoils from 16.9 to 57.3 keV.  
Finally, we report a preliminary comparison of the S1 and S2 yields for recoils parallel and perpendicular to the applied field.

\section{Apparatus}
\subsection{Detectors and geometry}

The experiment was performed at the University of Notre Dame Institute for Structure and Nuclear Astrophysics in two runs in June and in October, 2013. As many of the experiment details were identical to those described in our previous paper, we have repeated the relevant descriptions from that paper here for the reader's convenience, adding additional information pertinent to the current results when necessary.
Protons from the Tandem accelerator~\cite{tandem} struck a 0.20\,mg/cm$^2$ thick LiF target, deposited on a 1-mm-thick aluminum backing, generating a neutron beam through the reaction $^7$Li(p,n)$^7$Be.  For the October~2013 run, a 0.1-mm-thick tantalum layer was interposed between the LiF target and aluminum backing to fully stop the protons before they reach the aluminum. This reduced the intensity of $\gamma$-ray background.  The proton beam was bunched and chopped to provide pulses 1\,ns wide, separated by 101.5\,ns, with an average of $6.3\times10^4$ protons per pulse. The accelerator pulse selector was set to allow one of every two proton pulses to strike the LiF target, giving one neutron beam pulse every 203.0\,ns.  During the S2 studies, the pulse selector setting was modified to allow one of every four, five, or eight pulses.  

\begin{figure}[t!]
\includegraphics[width=\columnwidth]{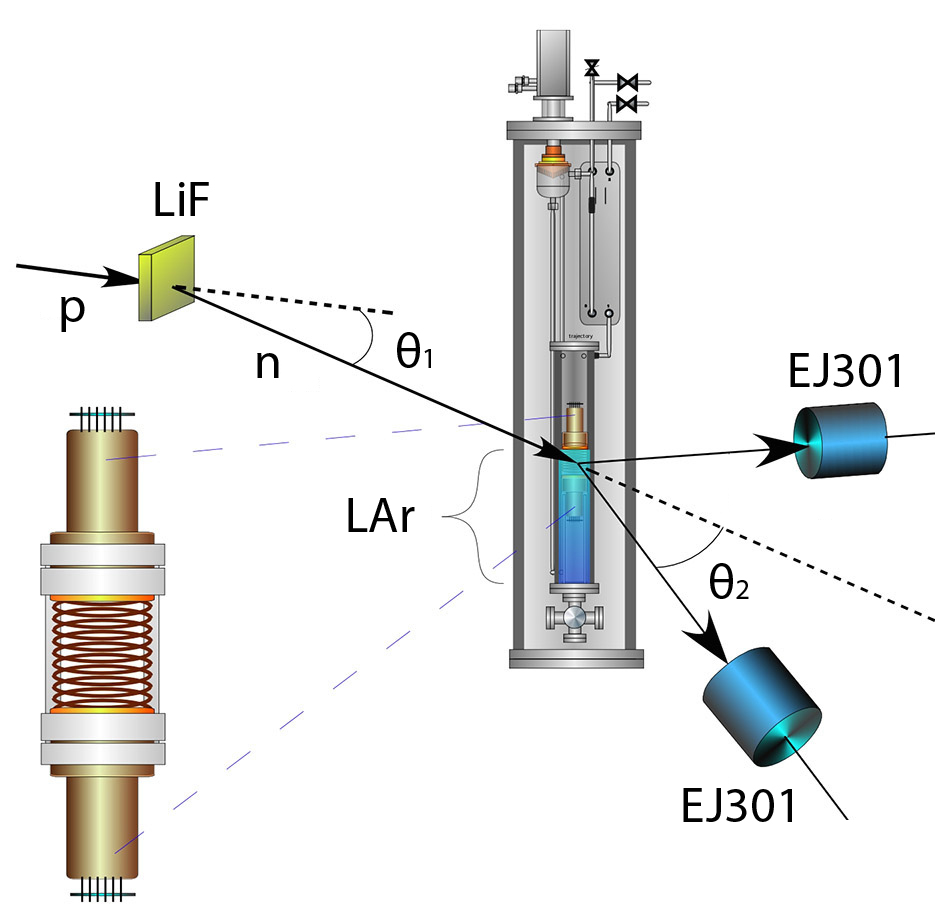}
\caption{\label{fig:schematic}A schematic of the experiment setup.  $\theta_1$ is the neutron production angle and $\theta_2$ is the scattering angle.  The inset shows a zoomed-in view of the TPC including the PMTs, field shaping rings and PTFE support structure.  It does not include the inner reflector.}
\end{figure}

The TPC was located 73.1\,cm from the LiF target in June and 82.4\,cm in October.  The average number of neutrons passing through the TPC per pulse was $\approx$\ET{-4}{3}.  Scattered neutrons were detected in three 12.7$\times$12.7\,cm cylindrical liquid scintillator neutron detectors~\cite{eljen}.  These detectors were placed on a two-angle goniometer-style stand at a distance of 71\,cm from the LAr target and at selected angles with respect to the beam direction. The angles determined both the energy of the nuclear recoils and the direction of the initial momentum of the recoils.  Figure~\ref{fig:schematic} shows a schematic of the geometry along with a zoomed-in view of the TPC, and Table~\ref{table:neutrons} lists the configurations of beam energy, detector location and the corresponding median nuclear recoil energy in the TPC.  The liquid scintillators provided timing information and pulse shape discrimination, both of which suppressed background from $\gamma$-ray interactions.  Cylinders of polyethylene (22$\times$22\,cm) shielded the neutron detectors from direct view of the LiF target for all but the 49.7\,\kevr\ data.

\begin{table*}[!ht]
\setlength\extrarowheight{4pt}
\begin{center} 
\begin{tabular}{lccccc} \hline\hline
&Proton	&Neutron	&Scattering	&Nuclear recoil  & Geometric \\
&energy	&energy	&angle		&energy & energy  \\
&[MeV]	&[MeV]	&[$^\circ$]		&[\kevr] & [\kevr] \\ \hline
\parbox[t]{3mm}{\multirow{5}{*}{\rotatebox[origin=c]{90}{Jun 2013}}} 
&2.376	&0.604	&49.9		&$10.3^{+1.5}_{-1.4}$  & 10.8 \\
&2.930	&1.168	&42.2		&$14.8^{+2.7}_{-2.6}$  & 15.2 \\
&2.930	&1.168	&49.9		&$20.5^{+3.0}_{-2.8}$  & 20.8 \\
&2.930	&1.168	&59.9		&$28.7^{+2.8}_{-2.8}$  & 29.0 \\ 
&2.930	&1.168	&82.2		&$49.7^{+3.4}_{-3.4}$  & 49.9 \\ [2pt]\hline
\parbox[t]{3mm}{\multirow{4}{*}{\rotatebox[origin=c]{90}{Oct 2013}}} 
&2.316	&0.510	&69.7		&$16.9^{+1.5}_{-1.5}$  & 16.5 \\
&3.607	&1.773	&45.0		&*$25.4^{+3.2}_{-2.9}$  & 26.1 \\
&2.930	&1.119	&69.7		&*$36.1^{+3.1}_{-3.1}$  & 36.3 \\
&3.607	&1.773	&69.7		&*$57.3^{+5.0}_{-4.9}$  & 57.6 \\ [2pt]\hline
\end{tabular}
\caption{Proton energy, neutron energy, and scattering angle settings for the two runs. Note that the neutron production angle was 25.4$^{\circ}$ in June and 35.6$^{\circ}$ in October.  To determine the nuclear recoil energy we performed a Monte Carlo (MC) simulation of neutron scattering in our apparatus taking full account of all materials and the geometry of the detectors.  The fourth column lists the median energy of the single scattering distribution obtained from the MC, and the central $68\%$ of the scatters are contained within the plus/minus band provided. For interest, we also show the recoil energy calculated directly from the scattering angle using the center of the TPC and the center of the neutron detector.  Data sets marked with an asterisk~(*) were taken with the TPC AND trigger requiring the coincidence of the two TPC PMT's, see the text for details.}
\label{table:neutrons}
\end{center}
\end{table*}

\begin{figure}[t!]
\includegraphics[width=\columnwidth]{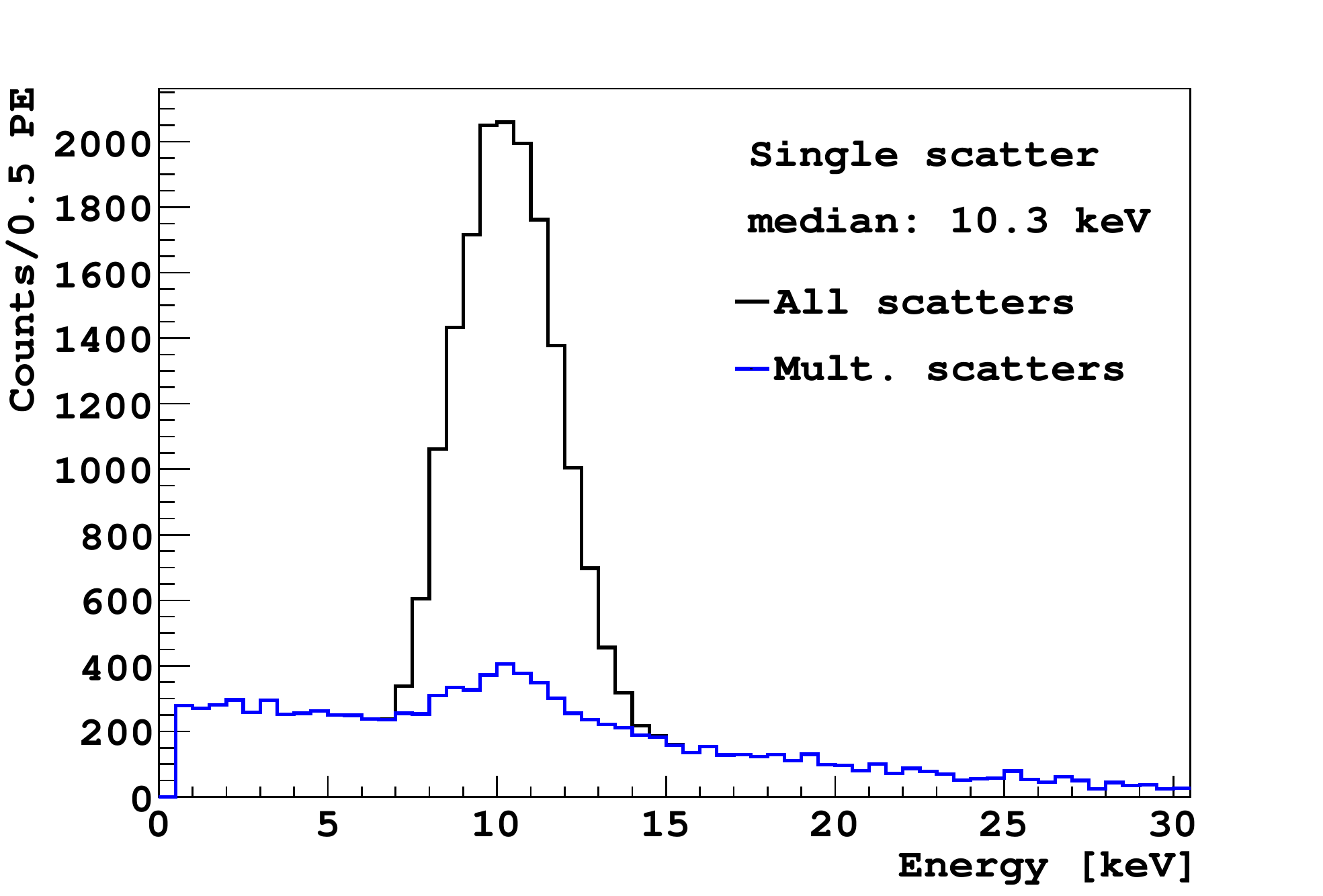}
\caption{\label{fig:mc}GEANT4-based simulation of the energy deposition in the LAr-TPC at the 10.3\,\kevr\ setting. {\bf \color{black} Black:} All scatters that produced a coincidence between the TPC and the neutron detector and survived the timing cuts discussed in the text.  {\bf \color{blue} Blue:} From neutrons scattered more than once in any part of the entire TPC apparatus before reaching the neutron detector. About $25\%$ of these events are very shallow scatters depositing minimal energy elsewhere in the apparatus. They look very much like single scatters and produce the peak in the multiple scattering distribution at 10 keV. Each setting is labeled according to the median of the simulated single scatter distribution.}
\end{figure}

The diameter and height chosen for the liquid argon target allowed the acquisition of adequate statistics with an acceptable level of contamination from multiple scattering.  Figure~\ref{fig:mc} shows energy deposition distributions from a detailed GEANT4~\cite{geant4} simulation of the detector for the 10.3\,\kevr\ setting; the multiple scattering contributes less than 32\% of the total event rate between 5 and 16\,\kevr, and the position of the single scattering peak is not affected by the background. We define each data set according to the median recoil energy of the single scattering component of the MC.

In a two-phase liquid noble gas TPC, electrons are collected by applying an electric field (the ``drift field'') to drift them to a liquid-gas interface. At the interface, they are extracted into the gas by a stronger electric field (the ``extraction field''). Once in the gas, the electrons are ultimately detected by observing the proportional scintillation light produced as they are accelerated through the gas by a ``multiplication field.''  The design of this TPC closely followed that used in \dst~\cite{darkside}.  The active volume was contained within a 68.6\,mm diameter, 76.2\,mm tall, right circular polytetrafluoroethylene (PTFE) cylinder lined with 3M Vikuiti enhanced specular reflector~\cite{vikuiti} and capped by fused silica windows.  The LAr was viewed through the windows by two 3'' Hamamatsu R11065 photomultiplier tubes (PMTs)~\cite{hamamatsu}.  The windows were coated with the transparent conductive material indium tin oxide (ITO), allowing for the application of electric field, and copper field rings embedded in the PTFE cylinder maintained field uniformity.  All internal surfaces of the detector were evaporation coated with the wavelength shifter TetraPhenylButadiene (TPB) which converted the LAr scintillation light from the vacuum UV range (128\,nm) into the blue range ($\sim$420\,nm).

A hexagonal stainless steel mesh was fixed at the top of the active LAr volume and connected to the electrical ground to provide the drift field (between the bottom window and the mesh) and the extraction and amplification fields (between the mesh and the top window).  The strips in the mesh were 50\,$\mu$m wide, and the distance between the parallel sides in each hexagon was 2\,mm.  We maintained the LAr level at 2\,mm below the mesh in June and 1\,mm above the mesh in October by keeping a constant inventory of Ar in the closed gas system at stable temperature and pressure.  We monitored the liquid level with three pairs of 10\,mm$\times$10\,mm parallel-plate capacitive level sensors, with radially symmetric positions along the circumference of the mesh.  Ar gas filled the remaining volume below the anode (the ITO coating on the top window). The gap between the mesh and the anode was 7\,mm in height.  The electric potential difference between the cathode and the mesh sets the drift field, and that between the anode and the mesh sets the electron extraction field in the liquid above the mesh and in the multiplication field in the gas region.  The cathode and anode potentials were controlled independently. This allowed us to collected data with and without the ionization signals by switching on and off the voltage applied to the anode.   

We applied nominal drift voltages of 50, 100, 200, 300, 500 and 1000 V/cm. A 3D model of the detector was implemented in GMSH~\cite{GMSH}, a finite element mesh generator, and used to calculate the electrostatic potential with ELMER~\cite{ELMER}, an open-source general-purpose finite element software package. We combine the ELMER results with the GEANT4 simulation of the location of neutron scatters to derive the neutron-weighted median field at each voltage setting.  Table~\ref{table:HV} shows those values, along with the electric field range containing the central $68\%$ of the neutron scatters.  For the remainder of the paper, each setting will be referred to by the neutron-weighted median voltage. 

\begin{table}
\begin{center} 
\begin{tabular}{c|c|c} \hline\hline
Nominal E  &  Neutron-weighted  & $68\%$ coverage  \\\ 
 field [V/cm] & median field [V/cm]	& field range \\ \hline
50 	&	49.5	& $45.5 - 53.5$ \\
100 & 96.4 & $92.5 - 108$ \\
200 &  193 & $189 - 212$ \\
300 & 293 & $285 - 322$ \\
500 &  486 & $476 - 536$ \\
1000 & 970 & $954 - 1073$ \\ \hline
 \end{tabular}
\caption{Nominal electric field values, along with the neutron-weighted median field obtained by convolving an ELMER finite element simulation of the electrostatic potential in the TPC volume and a GEANT4 simulation of neutron scattering locations. The electric field was within the range given by the last column for the central $68\%$ of the neutron scatters. For the remainder of the paper, each setting will be referred to by the neutron-weighted median field.}
\label{table:HV}
\end{center}
\end{table}

\begin{figure}[t!]
\includegraphics[width=\columnwidth]{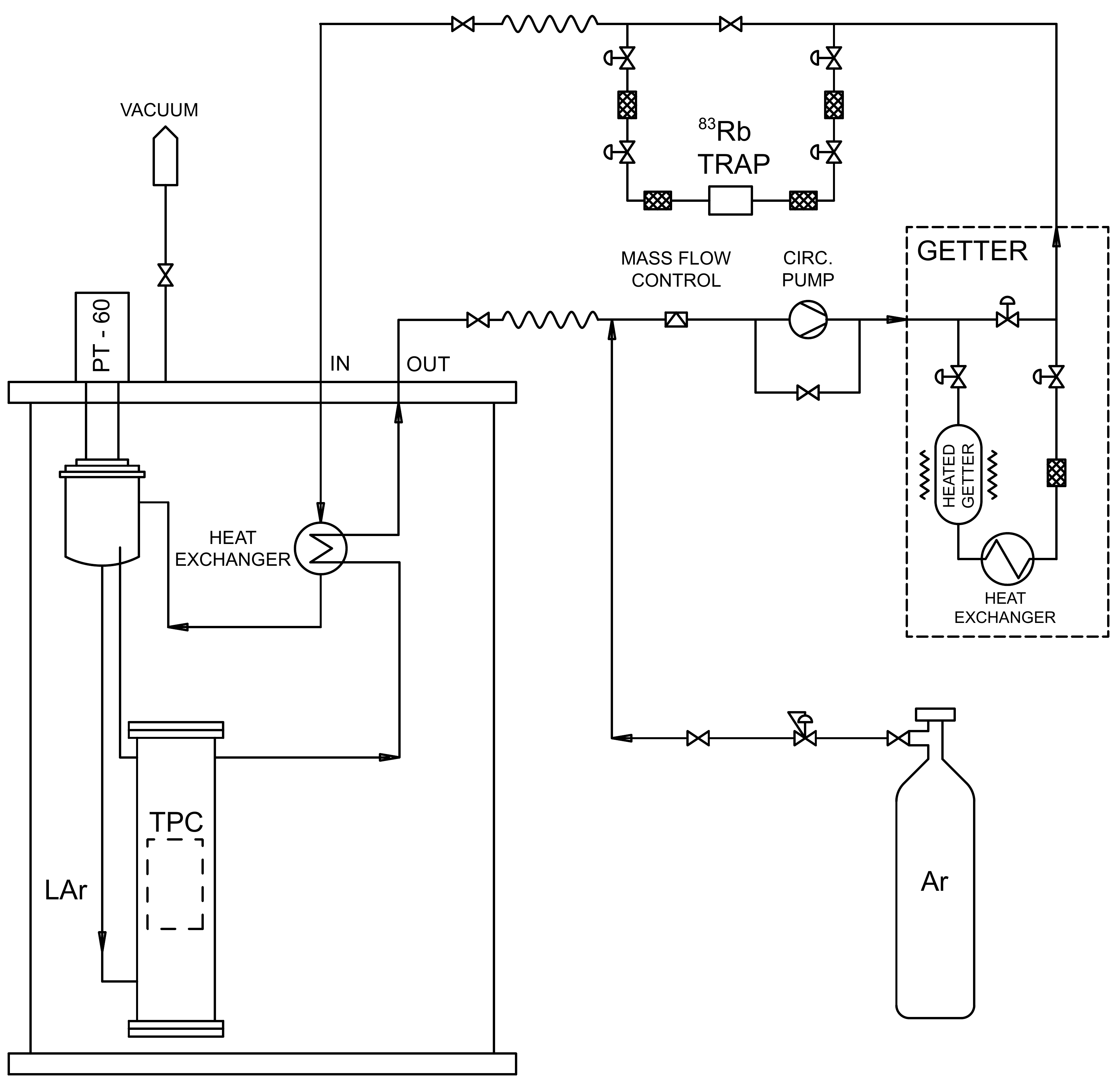}
\caption{\label{fig:pid}Ar gas system used for continuous purification of the LAr and injection of \krthree\ source.}
\end{figure}

The argon handling system is shown in Fig.~\ref{fig:pid}. The LAr detector was cooled by a Cryomech PT-60~\cite{cryomech} connected through a heater block to a condenser.  Commercial argon gas (6 9's grade~\cite{airgas}) was recirculated through a SAES MonoTorr PS4-MT3-R1 getter~\cite{saes} to remove impurities (mainly oxygen, nitrogen and water) from both the input gas and the LAr-TPC.

\subsection{Monitoring and calibration}

To monitor the scintillation yield from the LAr, \krthree\ was continuously injected by including a \rbthree\,trap~\cite{kastens,lippincott_kr,manalaysay} in the recirculation loop (see Fig.~\ref{fig:pid}).  \krthree\ has a half-life of 1.83\,hours and decays via two sequential electromagnetic transitions with energies of 9.4 and 32.1\,\kevr\ and a mean separation of 223\,ns~\cite{ensdf}. Because scintillation signals in LAr last for several microseconds~\cite{hitachi83}, we treated the two decays as a single event.  The activity of \krthree\ in the TPC was 1.2\,kBq (reduced to 0.5\,kBq in the October run). 

During the runs where ionization was measured simultaneously with scintillation, we also tracked the electron-drift lifetime, $\tau$~\cite{xe_rmp,bakale} with \krthree. This was done by measuring the correlation between the S2 pulse integral and the drift time in the TPC. The electron-drift lifetime was greater than 40\,$\mu$s at the start of October run and kept improving over the course of the data taking, reaching 120\,$\mu$s at the end of the run.  The maximum drift time in the TPC ranged from 300\,$\mu$s (drift field 50\,V/cm) to 46\,$\mu$s (500\,V/cm), as electron drift velocity increases with drift field~\cite{icarus}.  
We corrected the integral of each S2 signal for attachment of the drifting electrons by dividing by $\exp(-t_d/\tau)$, where \tdrift\ is the drift time.

\begin{figure}[t!]
\includegraphics[width=\columnwidth]{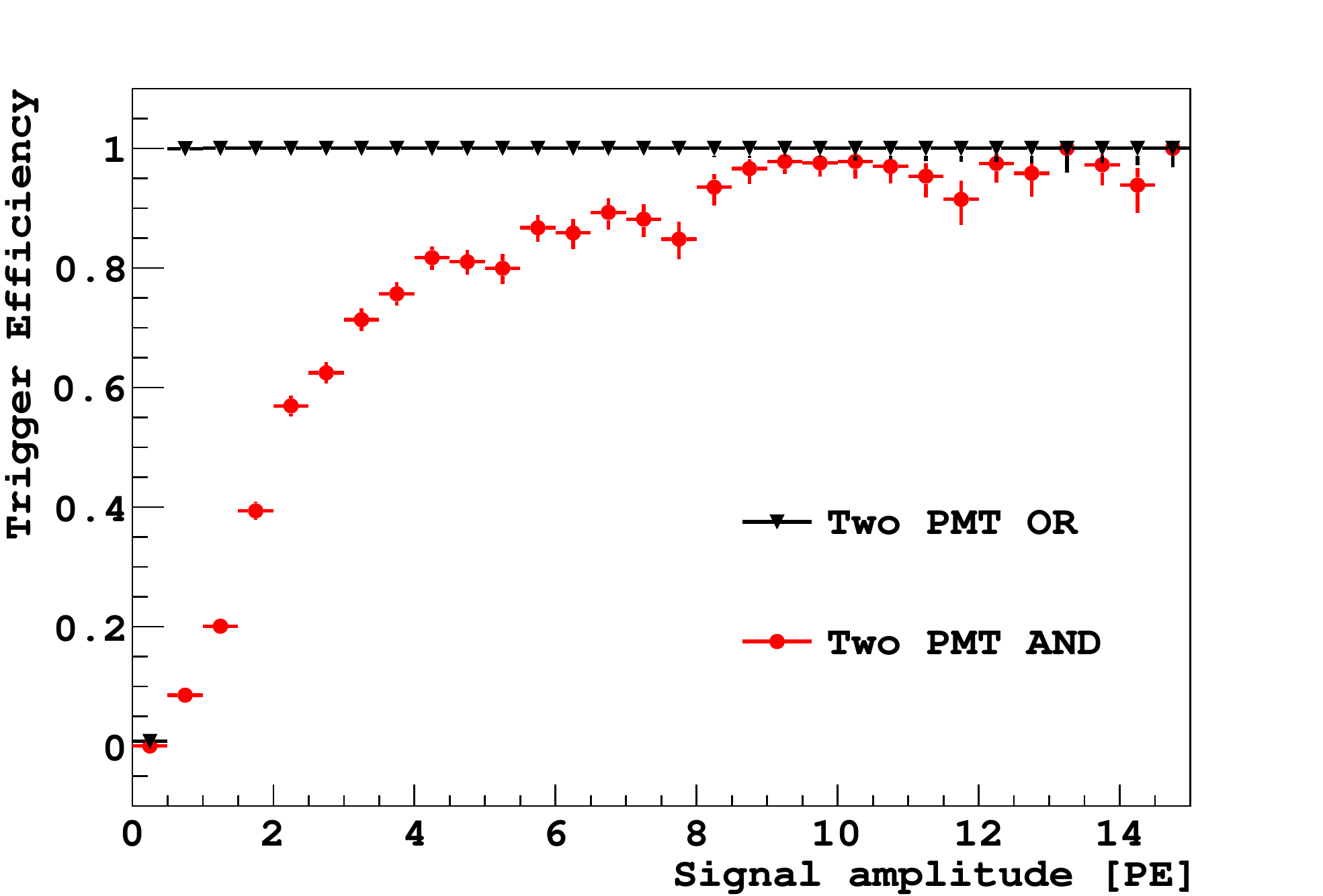}
\caption{\label{fig:trig_eff}Efficiency for the two TPC trigger conditions described in the text.  {\bf \color{black} Black:} OR of the two TPC PMT's.  {\bf \color{red} Red:} AND of the two TPC PMT's.  See text for description of the measurement of the efficiency.}
\end{figure}

The experiment trigger required a coincidence of the TPC trigger with one of the neutron detectors.  The TPC trigger was set as either the OR or the AND of the two TPC PMT's discriminator signals. The discriminator thresholds of the TPC PMTs were set to $\sim$0.2\,photoelectrons (\pe).  As shown in Fig.~\ref{fig:trig_eff}, the TPC trigger efficiency was determined to be above 90\% for pulses above 1\,\pe\ with the OR trigger (above 10\,\pe\ with the AND trigger) using positron annihilation radiation from a \sodium\ source placed between the TPC and a neutron detector, following the method described in Ref.~\cite{plante}. Use of the AND trigger was limited to the recoil energies above 25 keV (marked with * in Table~\ref{table:neutrons}). See Sec:~\ref{sec:trigger} for further details. 

In addition to the coincidence events, we recorded events triggered by the TPC alone, consisting largely of \krthree\ events, at a prescaled rate of 12\,Hz (5\,Hz in October).

The data acquisition system was based on 250\,megasample per second waveform digitizers~\cite{caen}, which recorded waveforms from the TPC, the neutron detectors and the accelerator RF signal. The data were recorded using the in-house daqman data acquisition and analysis software~\cite{daqman}.  At the times when the TPC was operated without S2 production ({\it i.e.} with zero anode voltage), the digitizer records were 16\,$\mu$s long including 5\,$\mu$s before the hardware trigger (used to establish the baseline).  At the times when the TPC was operated with S2 production, the length of the digitizer records was set to the maximum drift time plus 45\,$\mu$s.

The overall stability of the light yield was of critical importance to our measurements.  Several systems, including the wavelength shifter, the reflector, the photosensors, and the electronics, determined the light yield and its variations.

The single \pe\ response (\ser) of each PMT, determined using pulses in the tails of scintillation events, was monitored at 15-minute intervals and showed a slow decline of about 15$\%$ (26$\%$) in the top PMT and $10\%$ ($26\%$) in the bottom PMT over the course of the 6 (13) day run in June (October). The uncertainty on any given measurement of the SER is about $1\%$. 

Our data included a population of prompt events characterized by fast pulses in the TPC PMTs with times of flight slightly faster than photon-induced scintillation events [see Fig.~\ref{fig:tof-cuts}\,(a), ``Cerenkov'' events]. Data taken with no liquid in the TPC contained a similar collection of events.  The light from those events was typically concentrated in only one of the PMTs and did not exhibit the slow component characteristic of liquid argon scintillation.  We interpreted these signals as Cerenkov radiation from fast electrons passing through the fused silica windows, and therefore independent of scintillation processes in the argon. We used them to monitor for any dependence of the apparatus response on the drift field.  The spectrum of these events showed a peak at $\sim$80\,\pe\ in June which was stable within $\pm 2.5\%$ over all the electric field settings. 

During the October run, we injected light pulses of 355\,nm and $\sim$1\,ns width from a LED at a rate of 1\,Hz through an optical fiber into the TPC, and recorded the corresponding data by forcing the simultaneous trigger of the data acquisition system.  The mean pulse integral in PE on the bottom PMT drifted in a range of $\pm$4\% over the entire run (assuming perfect stability of the LED system).  We did not observe any change in the mean pulse integral immediately following the changes to the cathode voltage {\it i.e.} the drift field.

The mean response of the top PMT to LED pulses decreased by about a factor of $\sim$2 whenever ionization signals were turned on and would recover within 30 minutes when ionization signals were turned back off. The bottom PMT did not exhibit such a decrease in response, being stable to within a few percent throughout. The SER changed by as much as $10\%$ in the presence of ionization signals, with the bottom PMT more susceptible to changes than the top PMT, but as discussed above, any changes due to the SER were corrected on 15-minute time scales (and at the boundary of a given run condition). The decrease in response to the LED pulses was not apparent when the drift was on with no extraction field.  We believe this represents a reduction in efficiency at the high light levels
produced in the top PMT by the S2 signals.  Manufacturer's data shows a reduction in the maximum allowable cathode current density at reduced temperature~\cite{hamamatsu-data}.  The reduction also depended on the drift field.  Higher drift field reduced the electron-ion recombination in LAr, which increased the magnitude of S2 signals.  To correct for this variation in response, 
we divided the data into 15-minute blocks and within each block, normalized the top PMT signals to the LED response. 

The stability of the entire system was assessed throughout the data taking by monitoring the \krthree\ peak position.  At zero field, the position of the \krthree\ peak was measured to be 260 (200)\,\pe\ in June (October) and varied by less than $\pm$4\% ($\pm$4\%) over the entire run.  The reduction in light yield in October was a result of operating the TPC with the liquid level above the mesh (to allow S2 collection). The short term stability within a data set was checked with \krthree\ spectra accumulated every 15 minutes; these show negligible variations over several hours.

\section{Event Selection}
\label{sec:EventSelection}
\begin{figure}[t!]
\includegraphics[width=\columnwidth]{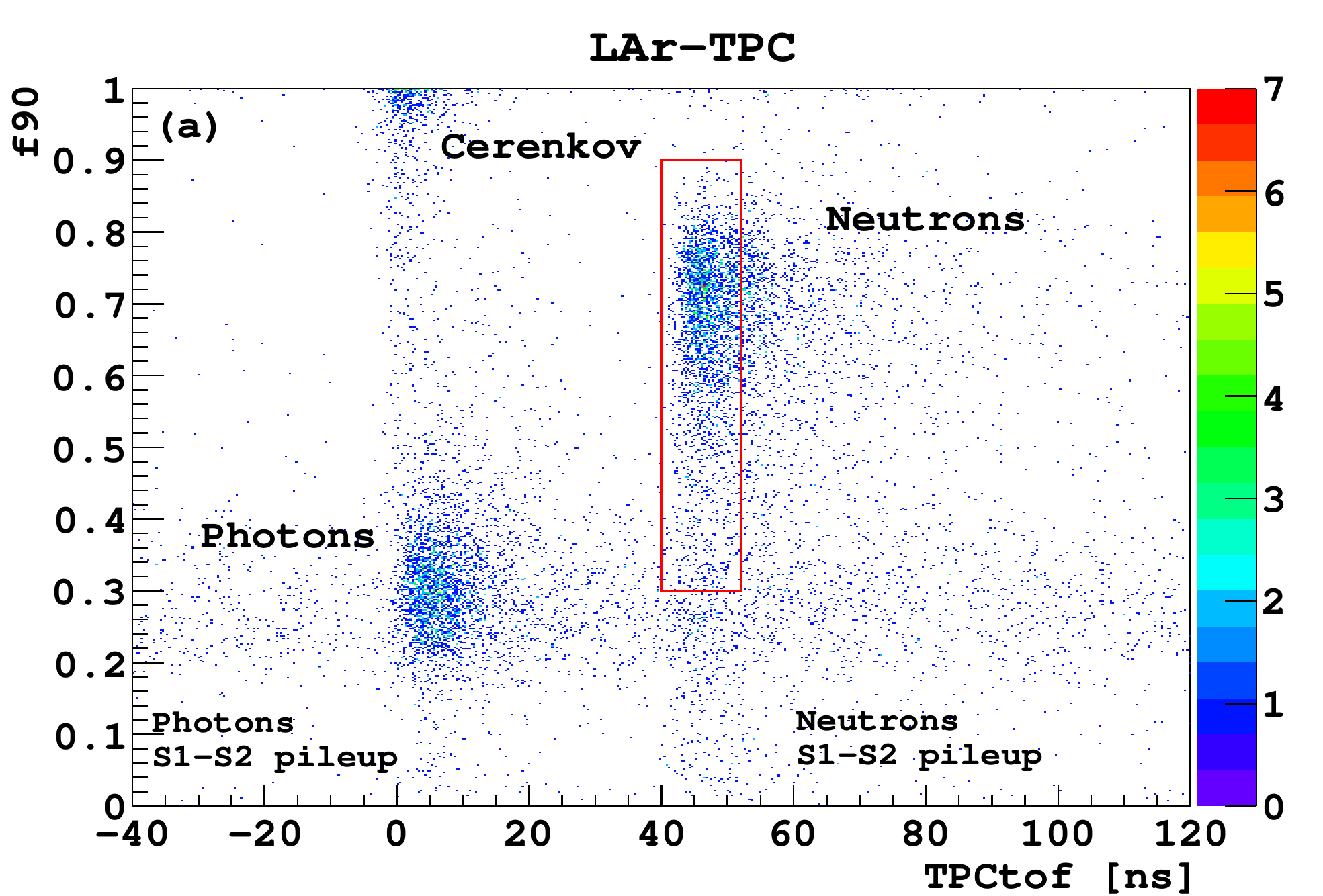}
\includegraphics[width=\columnwidth]{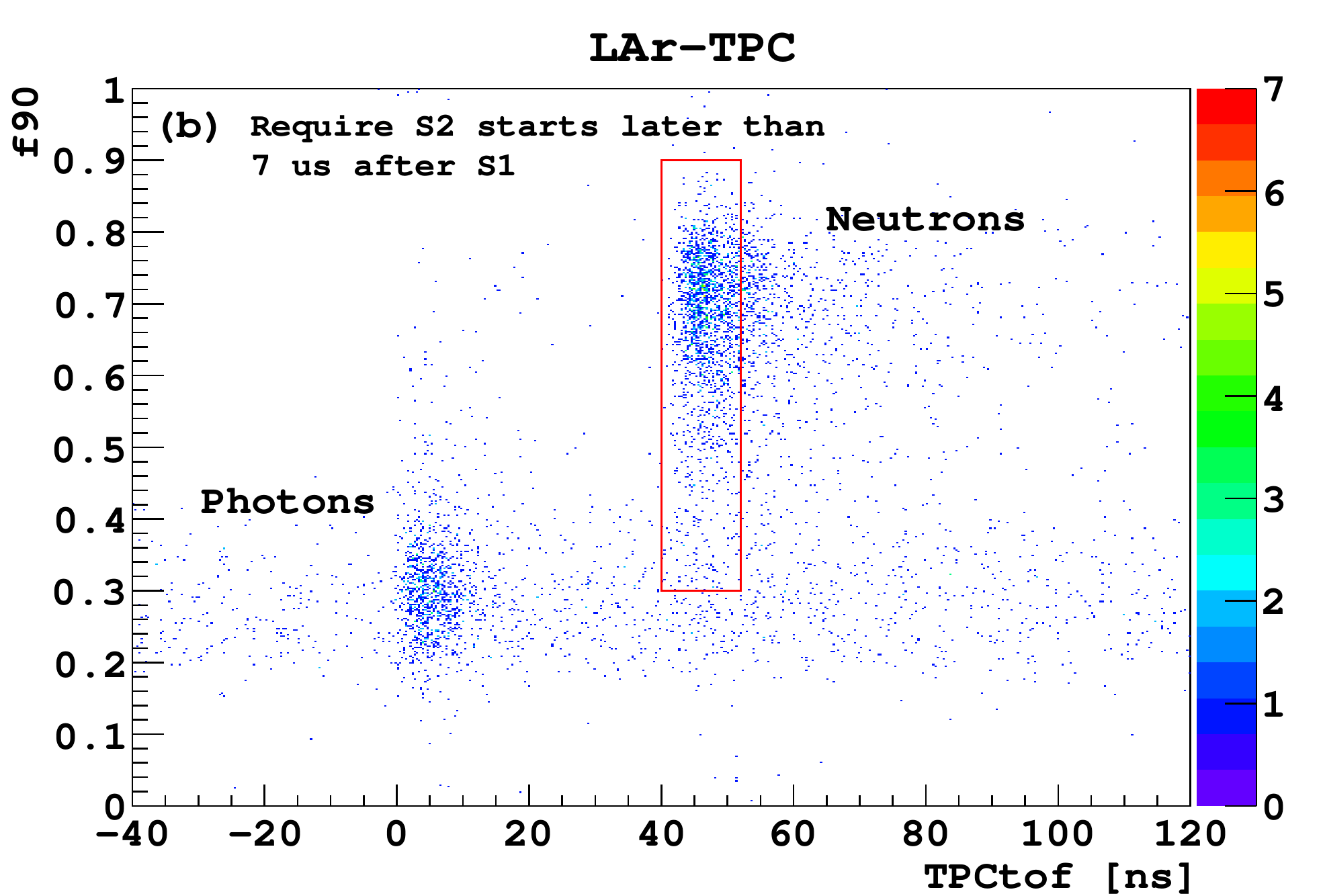}
\includegraphics[width=\columnwidth]{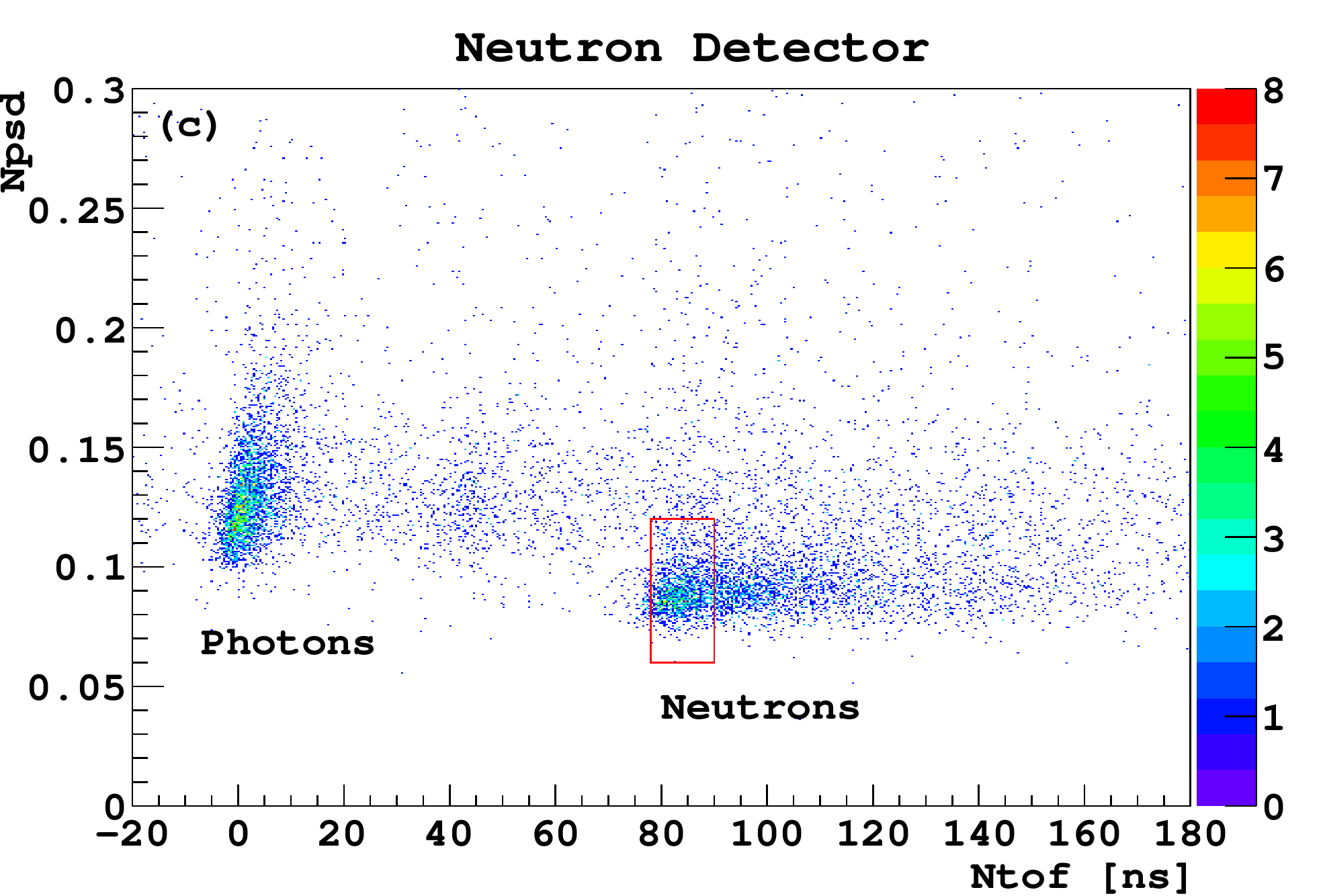}
\caption{\label{fig:tof-cuts}Distributions of pulse shape discrimination vs. time of flight for data taken in the 57.3\,\kevr\ configuration described in the text.  See the text for the definition of the variables.  Red boxes outline the regions selected by analysis cuts. Panels (a) and (b) described the TPC response and panel (c) the neutron detectors response.  In panel (a), the clusters of events with \fno$<$0.1 have S2 signals that start before the termination of S1 signals.  Panel (b) shows the distribution of events selected with the requirement that S1 and S2 signals are properly resolved.}
\end{figure}

\begin{figure*}[t!]
\includegraphics[width=\columnwidth]{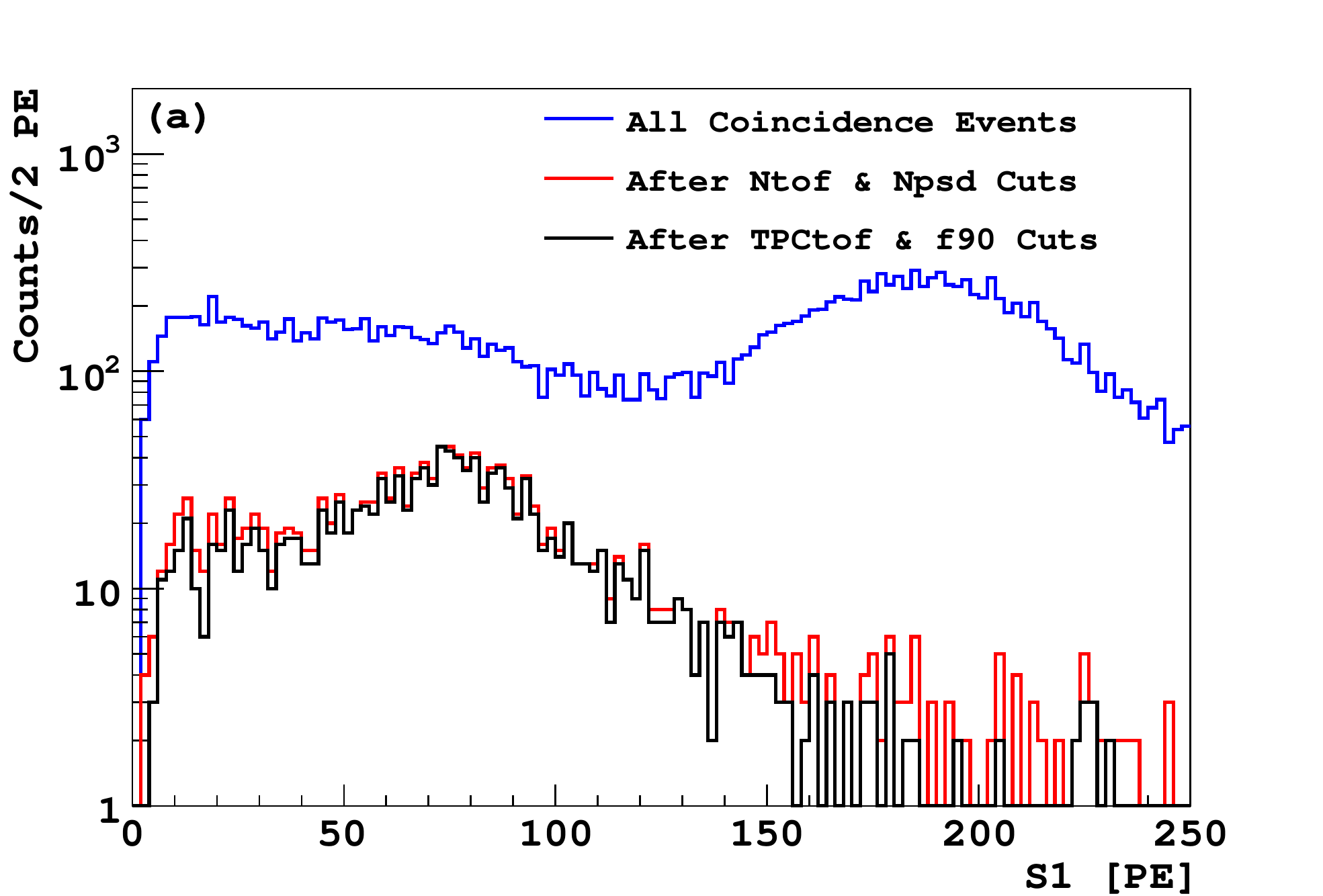}
\includegraphics[width=\columnwidth]{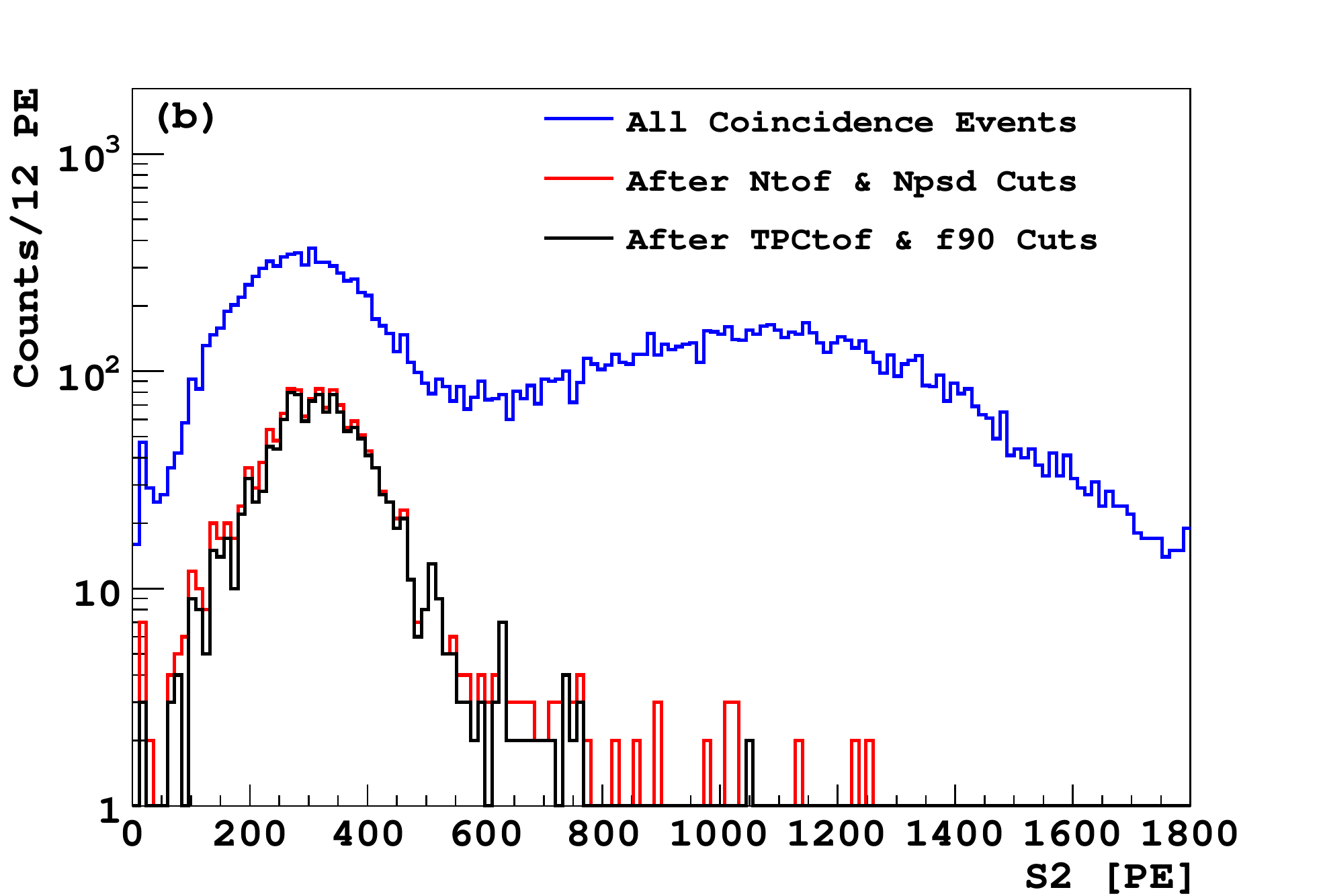}
\includegraphics[width=\columnwidth]{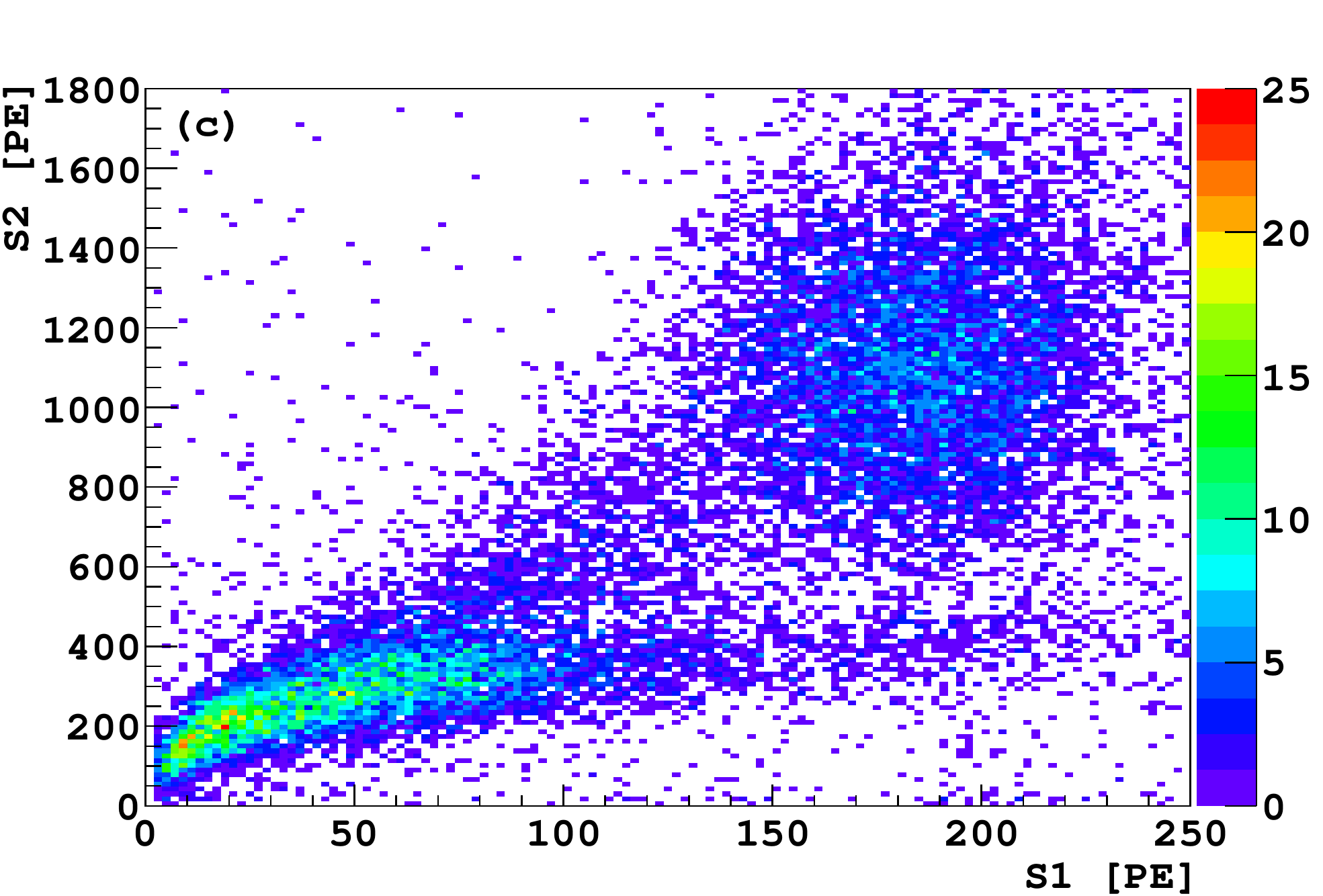}
\includegraphics[width=\columnwidth]{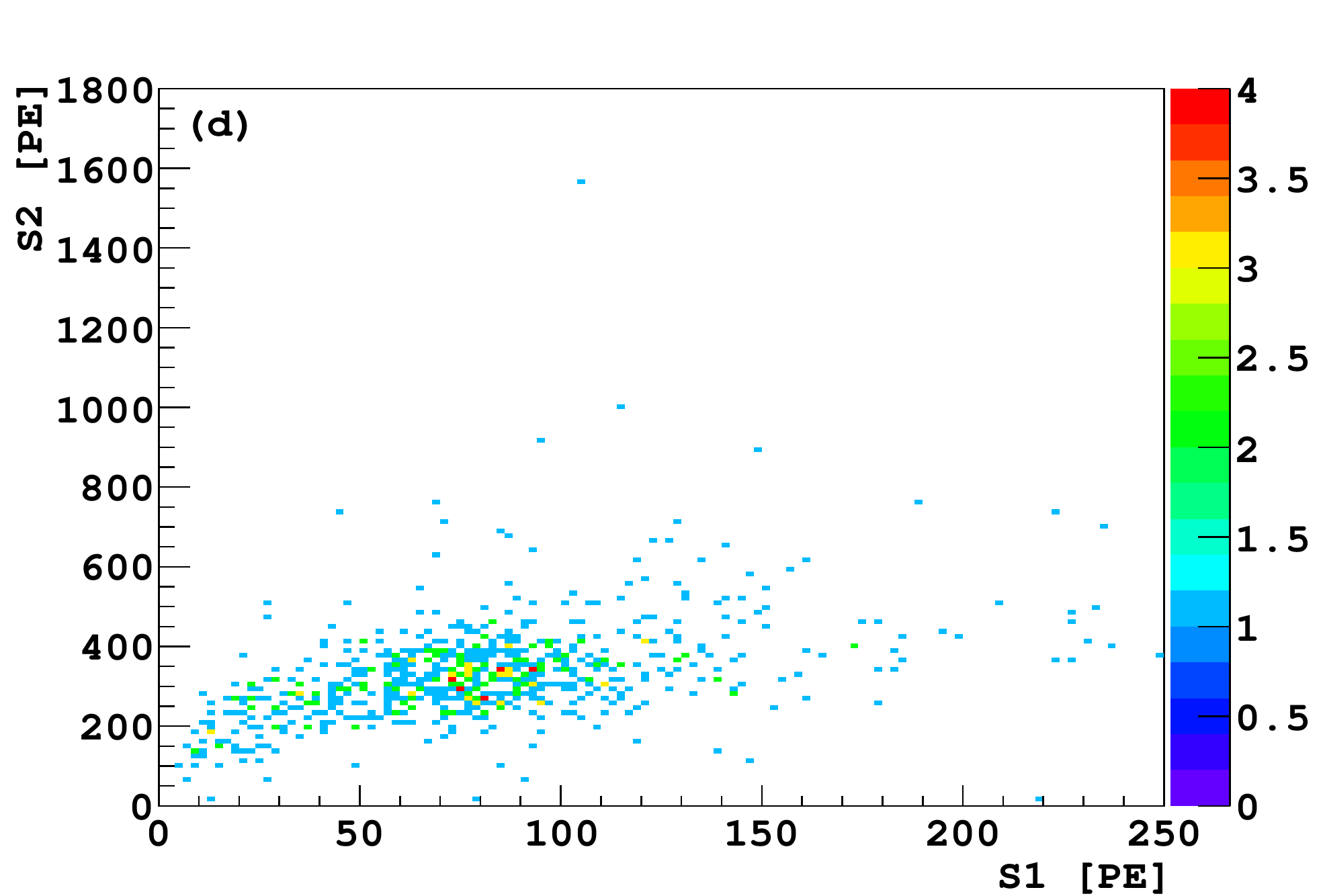}
\caption{\label{fig:s1s2}(a) Surviving primary scintillation light (\sone) distributions for 57.3\,\kevr\ nuclear recoils as the neutron selection cuts described in the text were imposed sequentially.  Data were collected with a drift field of 193\,V/cm and an extraction field of 3\,kV/cm.  The high energy peak around 187\,\pe\ is due to the \krthree\ source used for continuous monitoring of the detector.  (b) Surviving distributions of electroluminescence light from ionization (\stwo) for 57.3\,\kevr\ nuclear recoils after the same cuts. (c) \stwo\ vs.~\sone\ distribution for all events with resolved/non-overlapping \sone\ and \stwo\ before the neutron selection cuts. (d) \stwo\ vs~\sone\ distribution for the events surviving the neutron selection cuts.}
\end{figure*}

We focus on the case of events taken in a specific configuration $-$ 57.3\,\kevr\ nuclear recoils with a 193\,V/cm drift field, a 3.0\,kV/cm extraction field, and a 4.5\,kV/cm multiplication field (although the same voltage is applied, the extraction and multiplication fields have different strengths due to the change in dielectric constant between argon gas and argon liquid) $-$ to illustrate the basic criteria for event selection that were applied to the analysis of the entire set of data.  Figure~\ref{fig:tof-cuts} shows, for this data set, the relevant distributions in pulse shape discrimination parameters and time of flight (TOF), with the selection cuts marked by the red boxes.  The distributions were similar at other drift fields and recoil energies.  Figure~\ref{fig:s1s2} shows, for the same data set, the impact of the cuts based on the pulse shape discrimination parameters and time-of-flight distributions.  Again, the results of this selection were similar for all data sets within this experiment.

Figure~\ref{fig:tof-cuts}(a) shows a scatterplot of the discrimination parameter \fno~\cite{deap,lippincott}, defined as the fraction of light detected in the first 90\,ns of the S1 signal, vs. the time difference between the proton-beam-on-target and the TPC signal (\tpctof). The time of arrival of a pulse into the 250 MSPS digitizers was determined by interpolating the data to a threshold of 50\% peak amplitude. As the proton-beam-on-target signal is an RF pulse, we reference the TPC signals to the closest positive-slope zero crossing. Beam-associated events with $\gamma$-like and neutron-like \fno\ are clustered near 5 and 45\,ns respectively, as expected given the approximate 1.8\,cm/ns speed for 1.773\,MeV neutrons.  Cerenkov events are characterized by \fno\ close to 1.0 and $\gamma$-like timing.  The \krthree\ events appear with $\beta$/$\gamma$-like \fno, and are uniformly distributed in the \tpctof\ variable as expected.  For the events with vertices located a short distance from the mesh, S1 and S2 arrived too close in time to be resolved, resulting in a smaller than usual \fno\ (S1-S2 pileup). These events were removed by requiring each event to contain a second pulse that started at least 7$\mu$s after the first, with the second pulse's \fno\ less than 0.1.  Figure~\ref{fig:tof-cuts}(b) shows the same scatterplot after removal of these events.

Figure~\ref{fig:tof-cuts}(c) shows a scatterplot of the neutron pulse shape discriminant (\npsd), defined as peak amplitude divided by area in the neutron detectors, vs the time difference between the proton-beam-on-target and the neutron detector signal (\ntof).  Neutron events cluster near a \npsd\ of 0.09 and a \ntof\ of 85\,ns, while \bg\ events cluster near a \npsd\ of 0.13 and a \ntof\ of 2\,ns.  Random coincidences from environmental backgrounds are visible at intermediate times.  

We selected nuclear recoil events with \ntof\ and \tpctof\ within $\pm$6\,ns of the bin with the maximum number of events in the nuclear recoil region. For pulse shape we imposed the requirements of $0.06< \mathrm{Npsd}<0.12$ and $0.3<$ \fno $<0.9$ for all recoil energies studied.  Figure~\ref{fig:s1s2}(a) shows the 
\sone\ spectra as cuts based on the pulse shape discrimination parameters and time-of-flight distributions are imposed in sequence.  The high energy peak around 187\,\pe\ is the signal from the \krthree\ source used for continuous monitoring of the detector.  Similarly, Fig.~\ref{fig:s1s2}(b) shows the 
\stwo\ spectra as the same cuts are imposed.  The S2 peak from the \krthree\ source is located near 1100 PE.  Figures~\ref{fig:s1s2}(c) and~\ref{fig:s1s2}(d) provide a comparison of the \stwo\ vs~\sone\ distribution before and after the neutron selection cuts.  The outstanding signal to background ratio that emerges as the cuts are applied in sequence shows the power of this technique.

\subsection{Impact of trigger efficiency on S1 and S2 spectra}
\label{sec:trigger}

\begin{figure}[t!]
\includegraphics[width=\columnwidth]{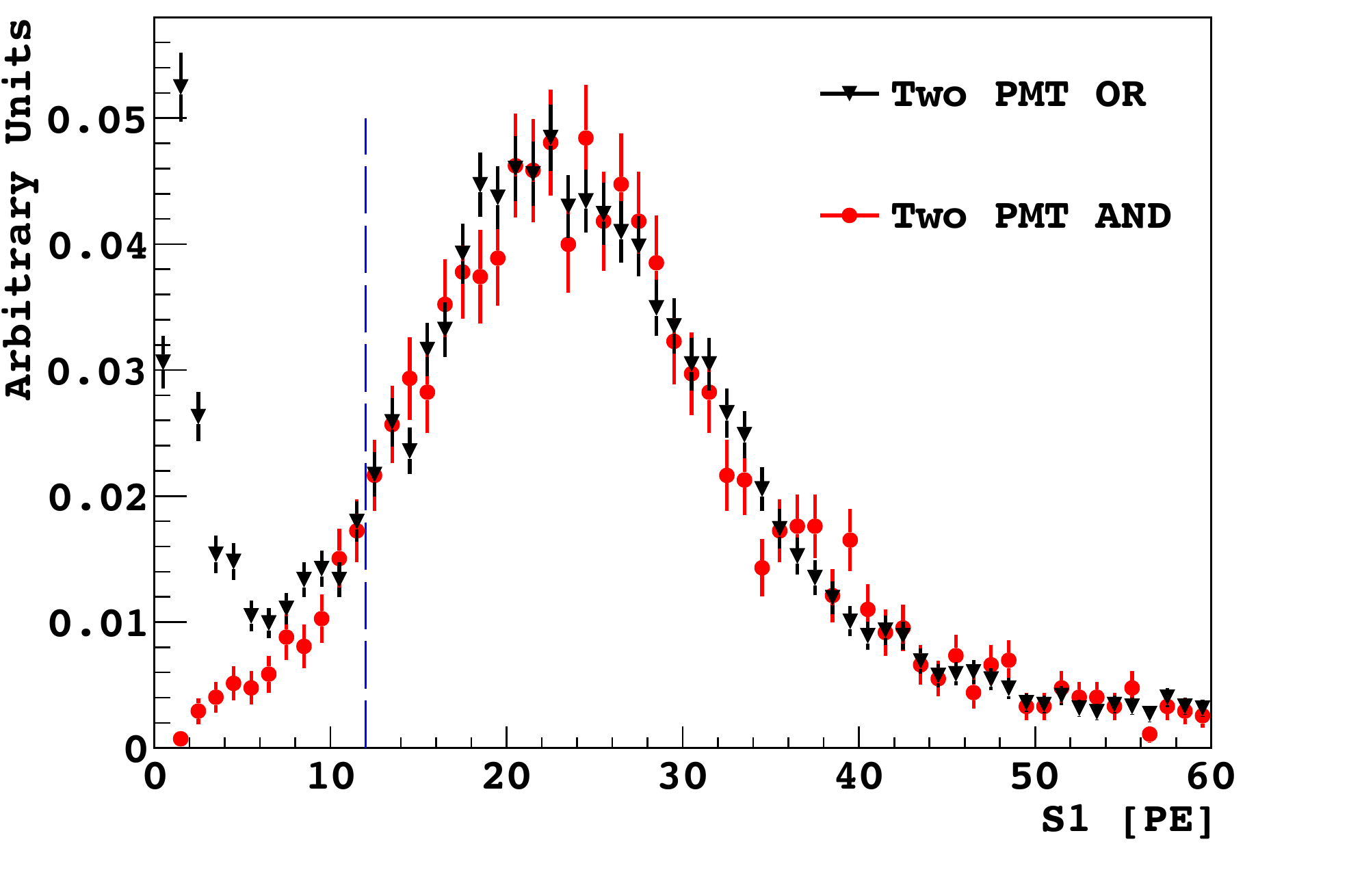}
\caption{\label{fig:trig_21keV}Comparison of recoil S1 spectra taken with the TPC PMT's OR ({\bf \color{black} black}) and AND ({\bf \color{red} red}) trigger for the 20.5 keV setting at 970 V/cm. The integral between 12 and 60\,\pe\ for each spectrum is normalized to 1. Use of the coincidence trigger had no significant effect on the spectral shape above $\sim$12\,\pe.}
\end{figure}

To assess possible distortions of the S1 spectra due to the trigger efficiency introduced by the AND trigger described earlier, we analyzed two subsets of 20.5\,\kevr\ nuclear recoils data taken with the two different TPC triggers.  As shown in Fig.~\ref{fig:trig_21keV}, the spectrum distortion induced by the choice of trigger is significant only below 10\,\pe.  This is in good agreement with the independent measurement of the trigger efficiency performed with the \sodium\ source (see Fig.~\ref{fig:trig_eff}).  The Gaussian mean of the Gaussian plus first order polynomial fit is 22.3\,$\pm$\,0.6\,PE with the OR trigger and 22.9\,$\pm$\,0.7\,PE with the AND trigger.  Hence, the fits to each provide a result that is statistically indistinguishable.  We conclude that all spectra collected with the trigger condition requiring the AND of the TPC PMT's produced undistorted spectra above 12\,\pe\ and could be used reliably, while helping reduce the amount of data written to disk by efficiently rejecting the rising background below 5\,\pe.

\begin{figure}[t!]
\includegraphics[width=\columnwidth]{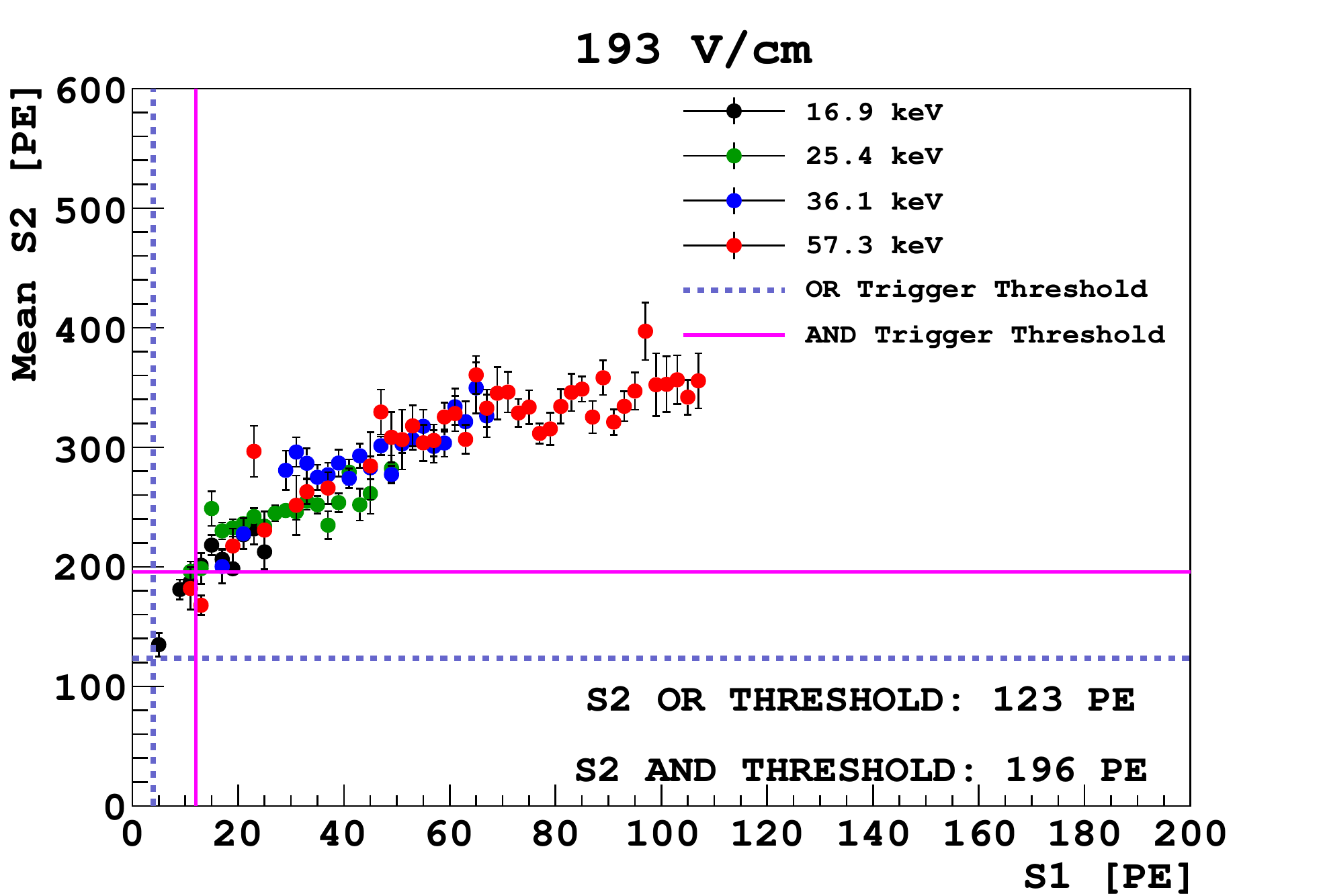}
\caption{\label{fig:S2trigger}Mean S2 signal vs. S1 for \Edrift = 193\,V/cm at all recoil energies. For this drift field, an S1 signal of 4 PE corresponds to an S2 signal of 123 PE, while an S1 signal of 12 PE corresponds to an S2 signal of 196 PE. These values form the lower bounds for the fit ranges used in the S2 analysis for data taken with the OR trigger and AND trigger respectively. }
\end{figure}

When examining the S2 response, these lower bounds on S1 correspond to much larger signals because of the large amplification in the S2 channel. To keep the same fitting bounds, we plot S2 vs. S1 for each drift voltage and find the value of S2 corresponding to an S1 of 4 PE for data taken with the OR trigger and 12 PE for data taken with the AND trigger; these S2 values form the lower fit boundaries for the S2 analysis in each trigger configuration.  Figure~\ref{fig:S2trigger} shows an example of this analysis for a drift field of 193\,V/cm, and Table~\ref{table:S2trigger} shows the corresponding lower fit boundaries for the S2 data at all drift fields. 

\begin{table}[t!]
\begin{center} 
\begin{tabular}{c|c|c} \hline\hline
Field & S2 bound & S2 bound \\
\lbrack V/cm\rbrack &(OR) [PE] & (AND) [PE]\\ \hline
49.5 & ... & 163 \\
96.4 & 104 & 174 \\
193 & 123 & 196 \\
293 & 142 & 224 \\
486 & 183 & 255 \\ \hline
\end{tabular}
\caption{Lower fit bounds for the S2 analysis at each drift field, derived from the relationship between S2 and S1, assuming a lower bound on S1 of 4 PE for the OR trigger and 12 PE for the AND trigger.}
\label{table:S2trigger}
\end{center}
\end{table}

\section{Analysis of the S1 Spectra and Determination of \leff}
\label{sec:s1fit}

We define \leff\ as the scintillation efficiency of nuclear recoils relative to that of electron recoils from \krthree\ at zero field: 
\begin{equation}
\mathcal{L}_{{\rm eff} \mbox{, } ^{83m}{\rm Kr}}\left( E_{\rm nr} \mbox{, } \mathcal{E}_{\rm d}\right) = \frac{S1_{\rm nr}\left( E_{\rm nr} \mbox{, } \mathcal{E}_{\rm d} \right) / E_{\rm nr}}{S1_{\rm Kr}\left(\mathcal{E}_{\rm d}=0 \right) / E_{\rm Kr}},
\label{eq:leff}
\end{equation}
where $E_{\rm Kr}$ is 41.5 keV, \Enr\ is the recoil energy and \Edrift\ is the drift electric field.  The measurement of \leff\ in this experiment permits the unbiased and straightforward computation of nuclear recoil scintillation yield from the measured light yield of \krthree\ in any liquid argon scintillation detector.  The first results of our experiment~\cite{scene1} demonstrated that \leff\ depends not only on \Enr\ but also on \Edrift. 

With the experiment described here we have obtained a precise determination of \leff\ in LAr.  The crucial step in the analysis of our data was the determination of the overall S1 yield, \sonenr, as a function of \Enr\ and \Edrift.
  
This was accomplished by fitting the data for each recoil angle setting (with PE as the ordinate) to Monte Carlo energy deposition spectra (with keV as the ordinate), using a single scale factor \leff\ for each experiment geometry. The ordinates for the data were computed based on a light yield 6.3$\pm 0.3$ (4.8$\pm 0.2$)\,\pe/keV measured in June (October) using \krthree. 

The simulations computed the energy deposition in the LAr, taking into account the complete kinematics and geometry of the 
LAr-TPC and the coincidence detectors, as well the TOF analysis cuts.  Before fitting, the MC distribution was convolved with a 
Gaussian energy resolution function with $\sigma_1$ parametrized as 
$\sigma_1$\,=\,$\sqrt{S1_{\rm nr}}$\,\Rone$\left(\Enr \mbox{, } \Edrift\right)$, 
where \Rone\ is a free parameter of the fit.  The fit 
procedure varies \leff\ and \Rone\ to minimize the \chisq\ defined as
\begin{equation}
\chi^{2}(\mathcal{L}_{{\rm eff} \mbox{, } ^{83m}{\rm Kr}} \mbox{, } R_{1}) = \sum\limits_{i=1}^n\frac{\left(O_i-S_i\right)^2}{S_i} ,
\end{equation}
where $n$ is the total number of bins in the chosen fit region, $O_i$ is the number of events observed in bin $i$, and $S_i$ is the number of events in bin $i$ resulting from simulations. The area of the MC spectrum was forced to match that of the data, and the fit parameters were applied to the MC before binning.

\begin{figure*}[t!]
\includegraphics[width=0.95\columnwidth]{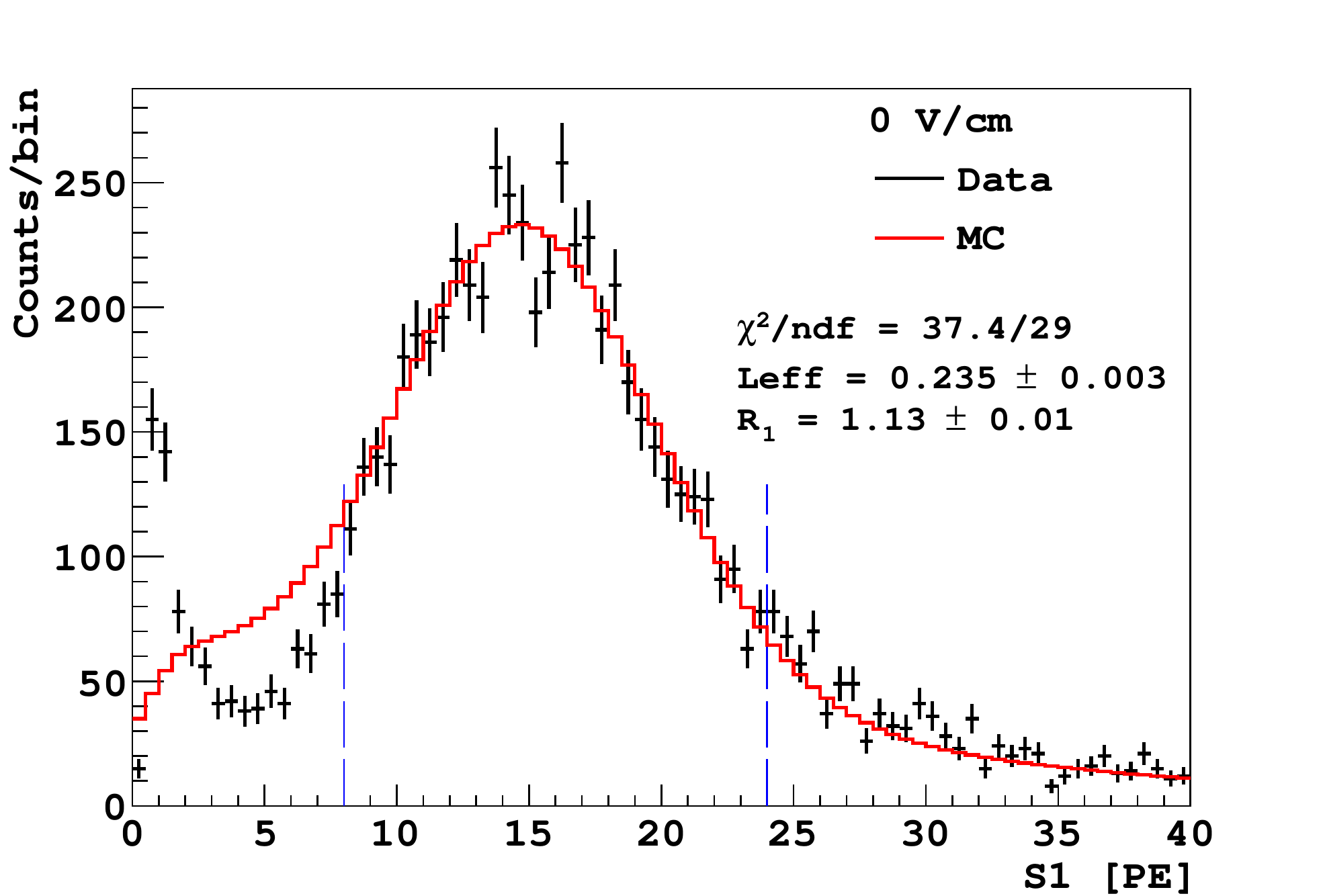}
\includegraphics[width=0.95\columnwidth]{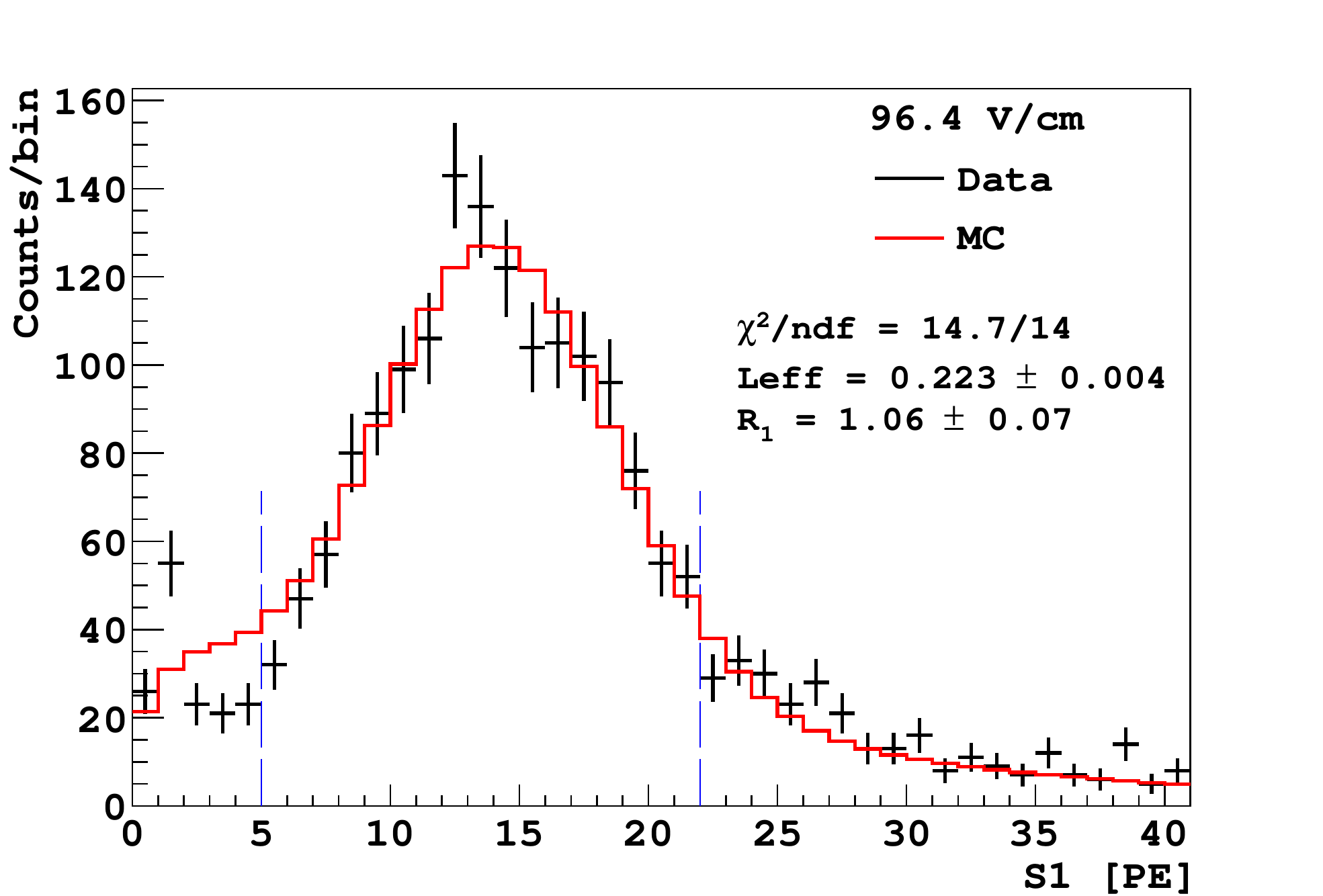}
\includegraphics[width=0.95\columnwidth]{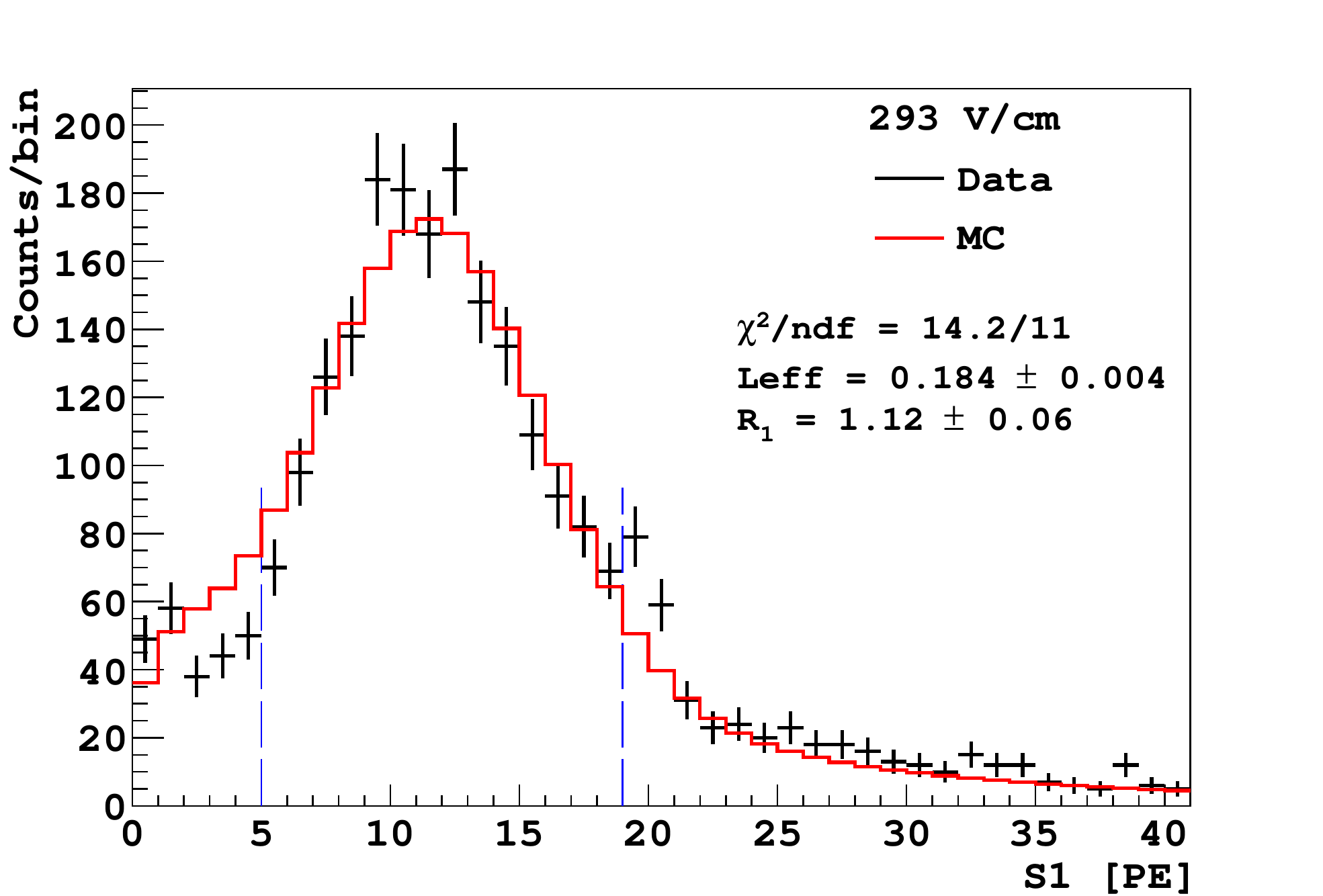}
\includegraphics[width=0.95\columnwidth]{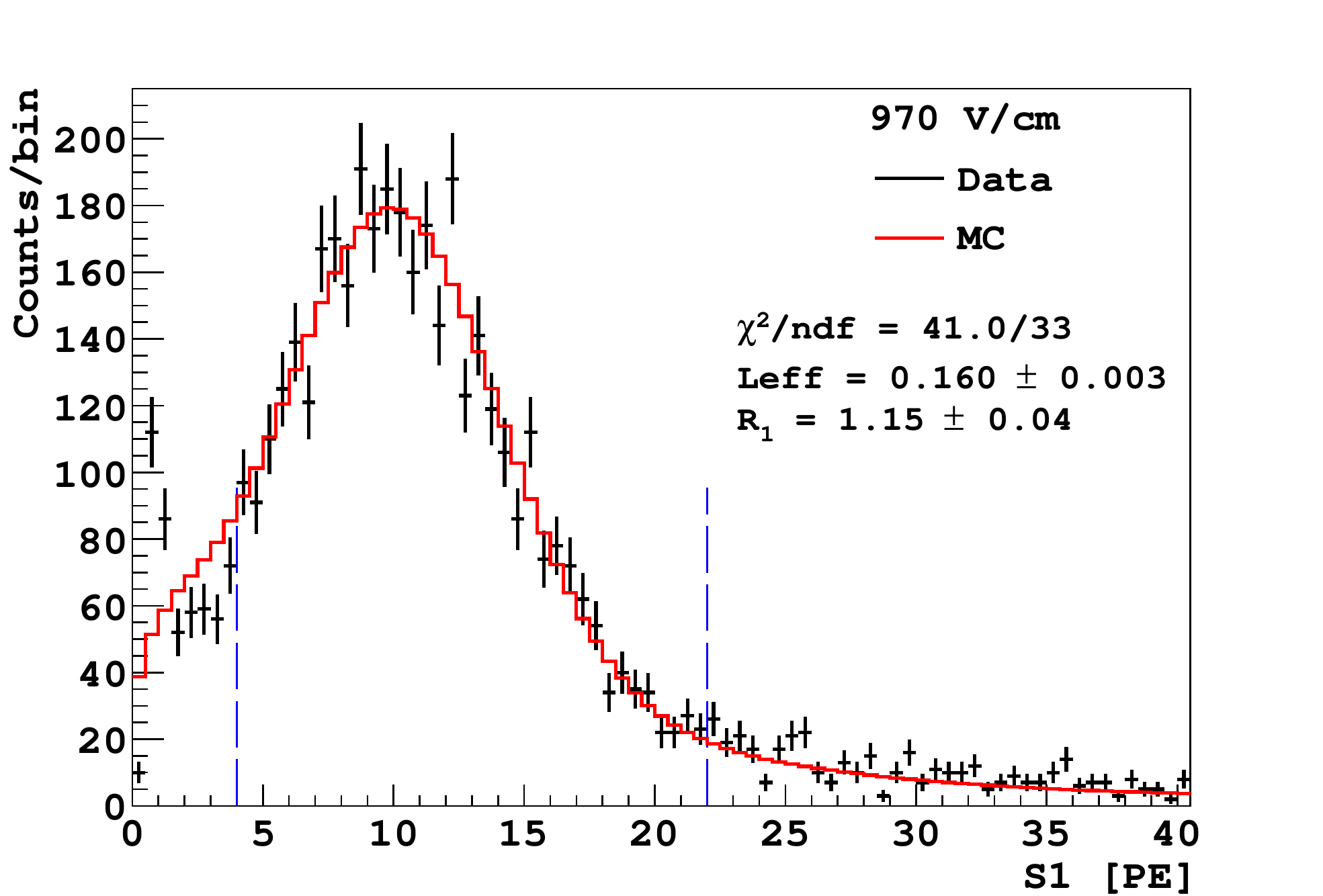}
\caption{\label{fig:leff11}
All panels.  {\bf \color{black} Black:} Experimental data collected for 10.3\,\kevr\ nuclear recoils.  {\bf \color{red} Red:} Monte Carlo fit of the experimental data.  The range used for each fit is indicated by the vertical {\bf \color{blue} blue} dashed lines.
}
\end{figure*}

\begin{figure*}[t!]
\includegraphics[width=0.95\columnwidth]{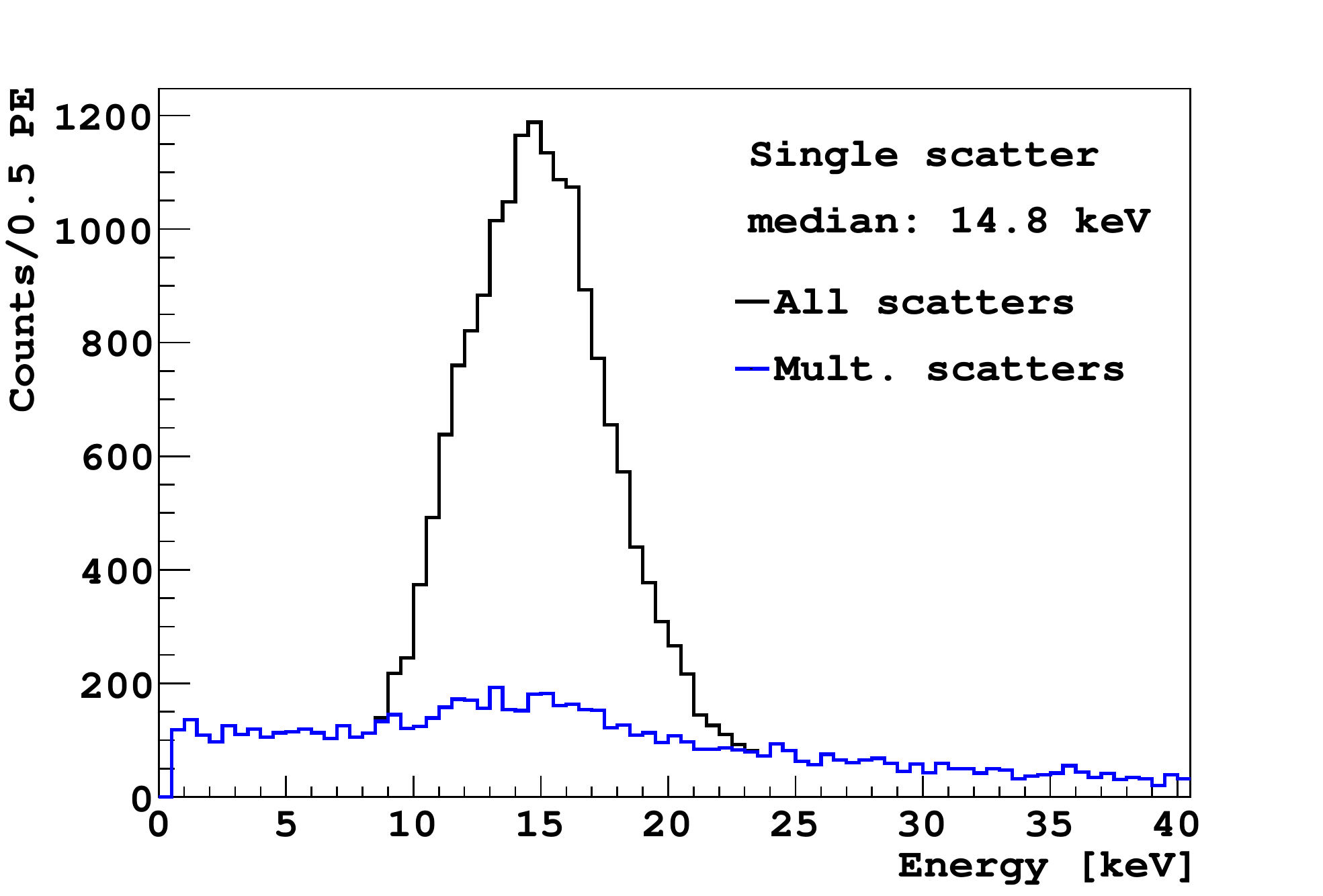}
\includegraphics[width=0.95\columnwidth]{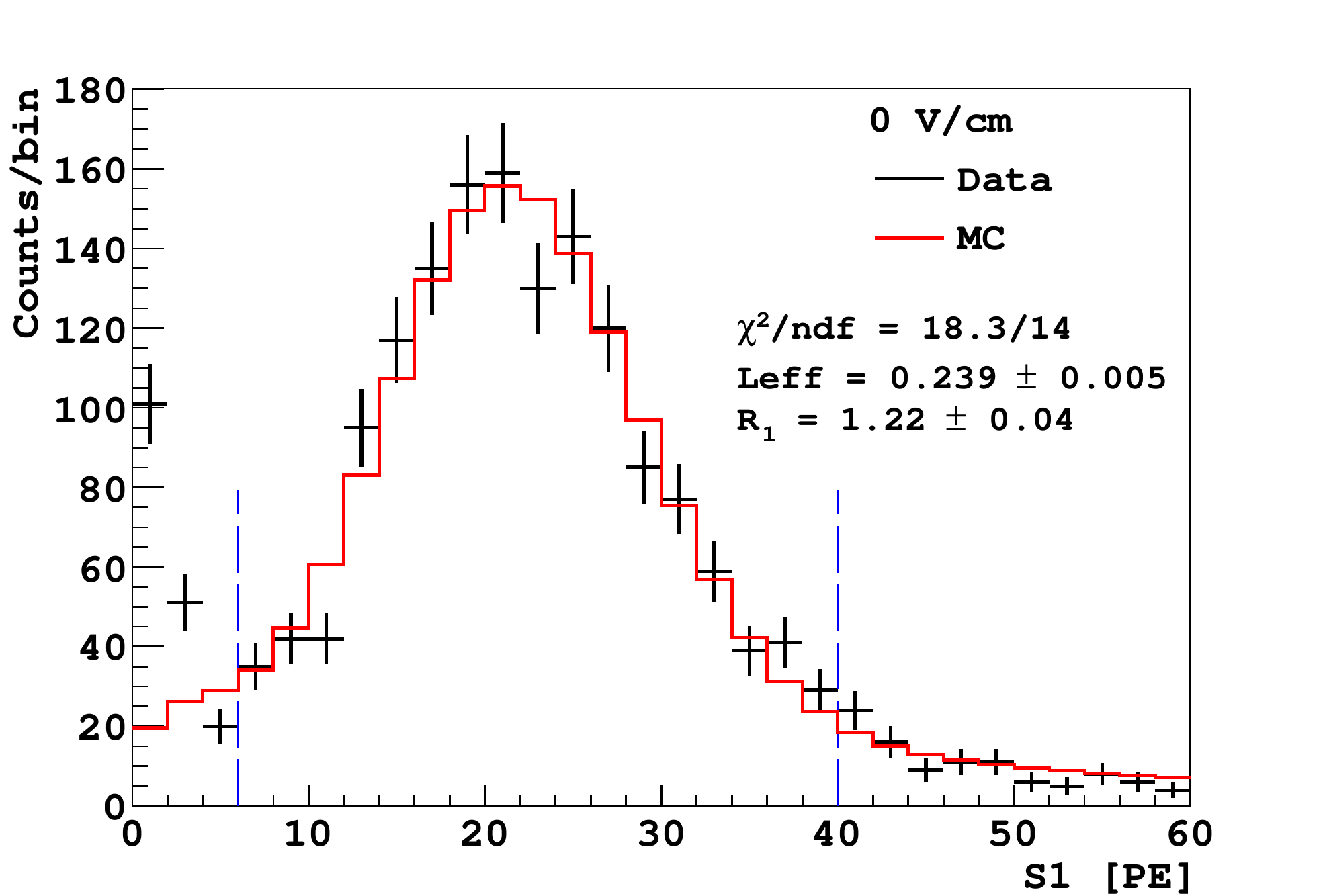}
\includegraphics[width=0.95\columnwidth]{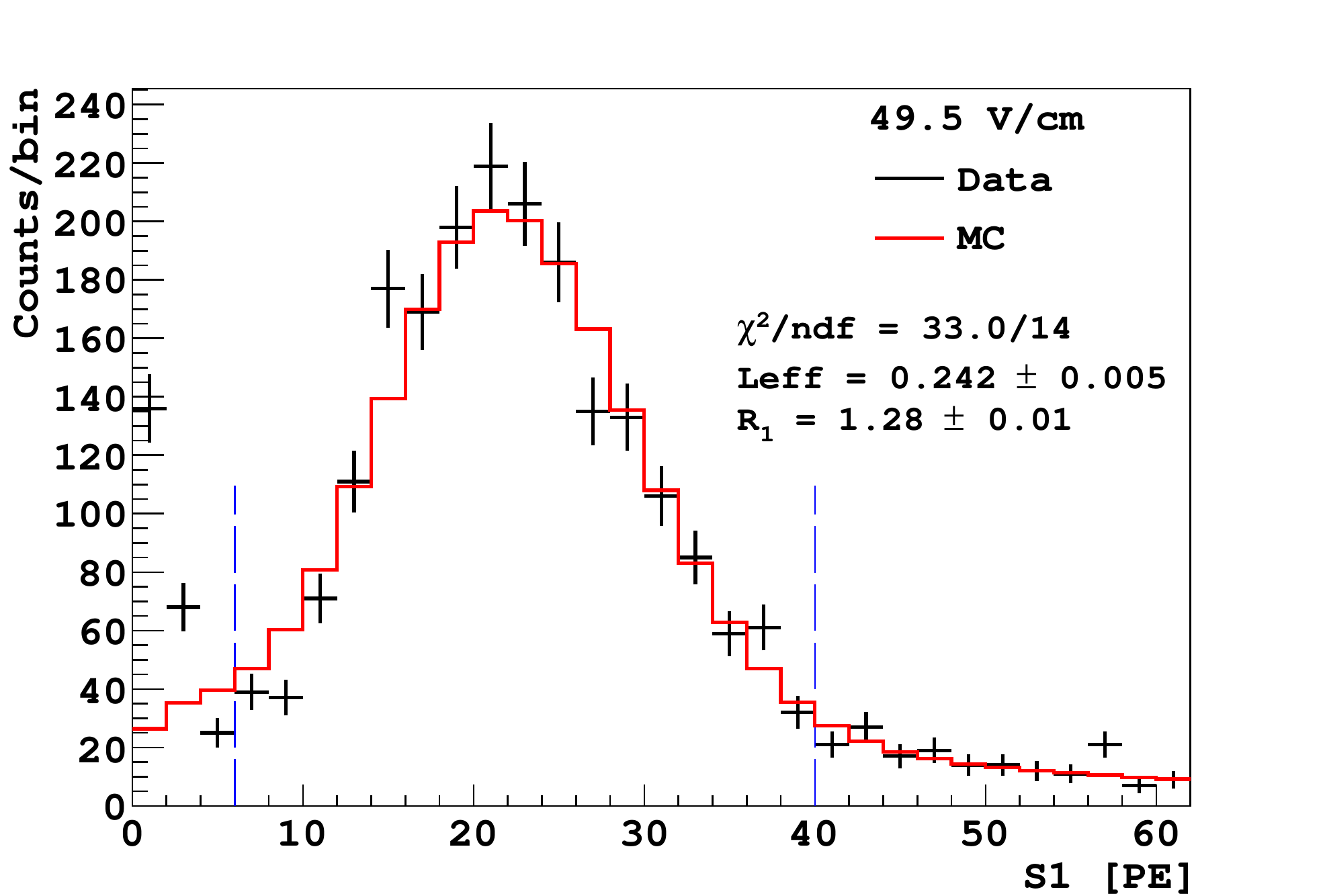}
\includegraphics[width=0.95\columnwidth]{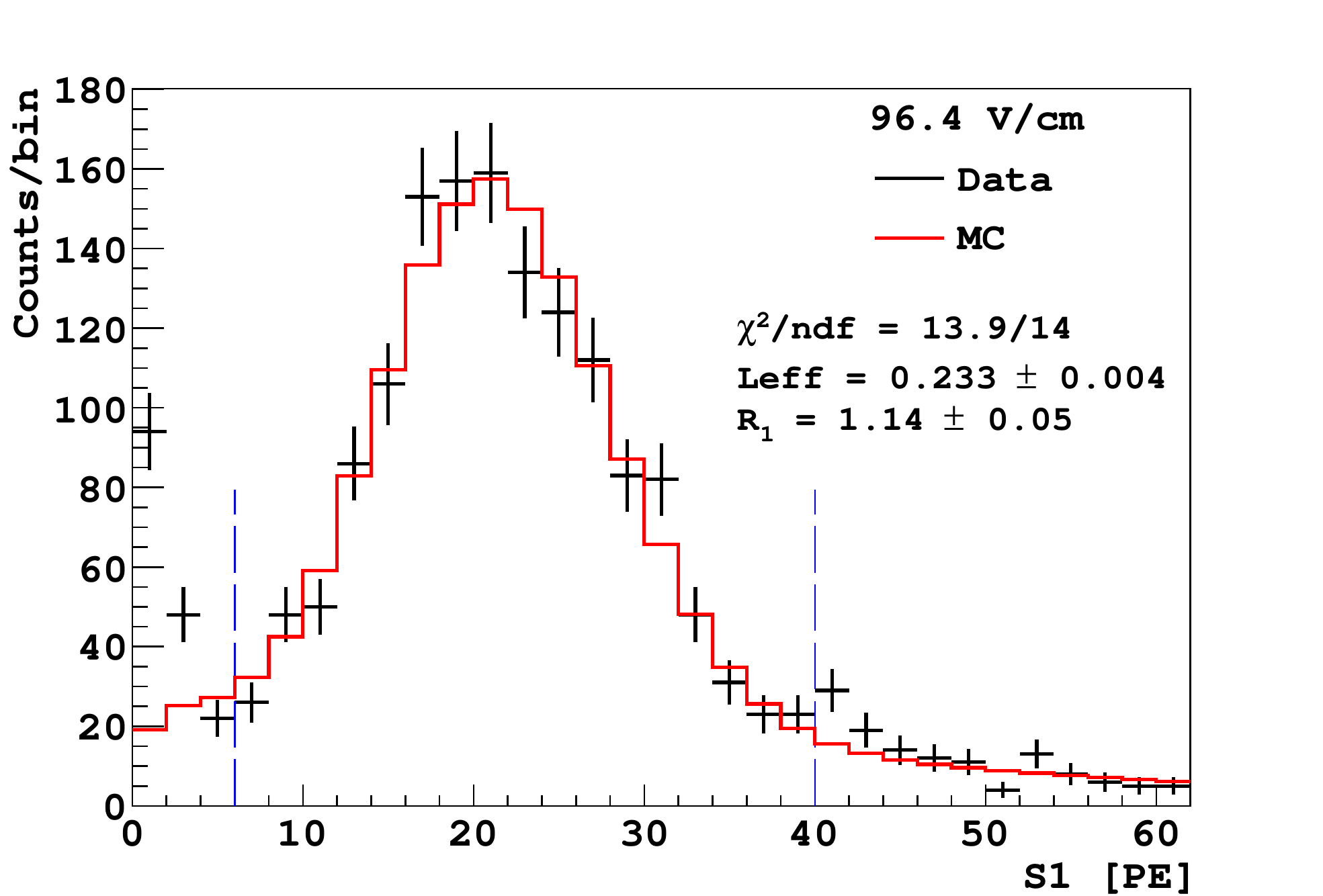}
\includegraphics[width=0.95\columnwidth]{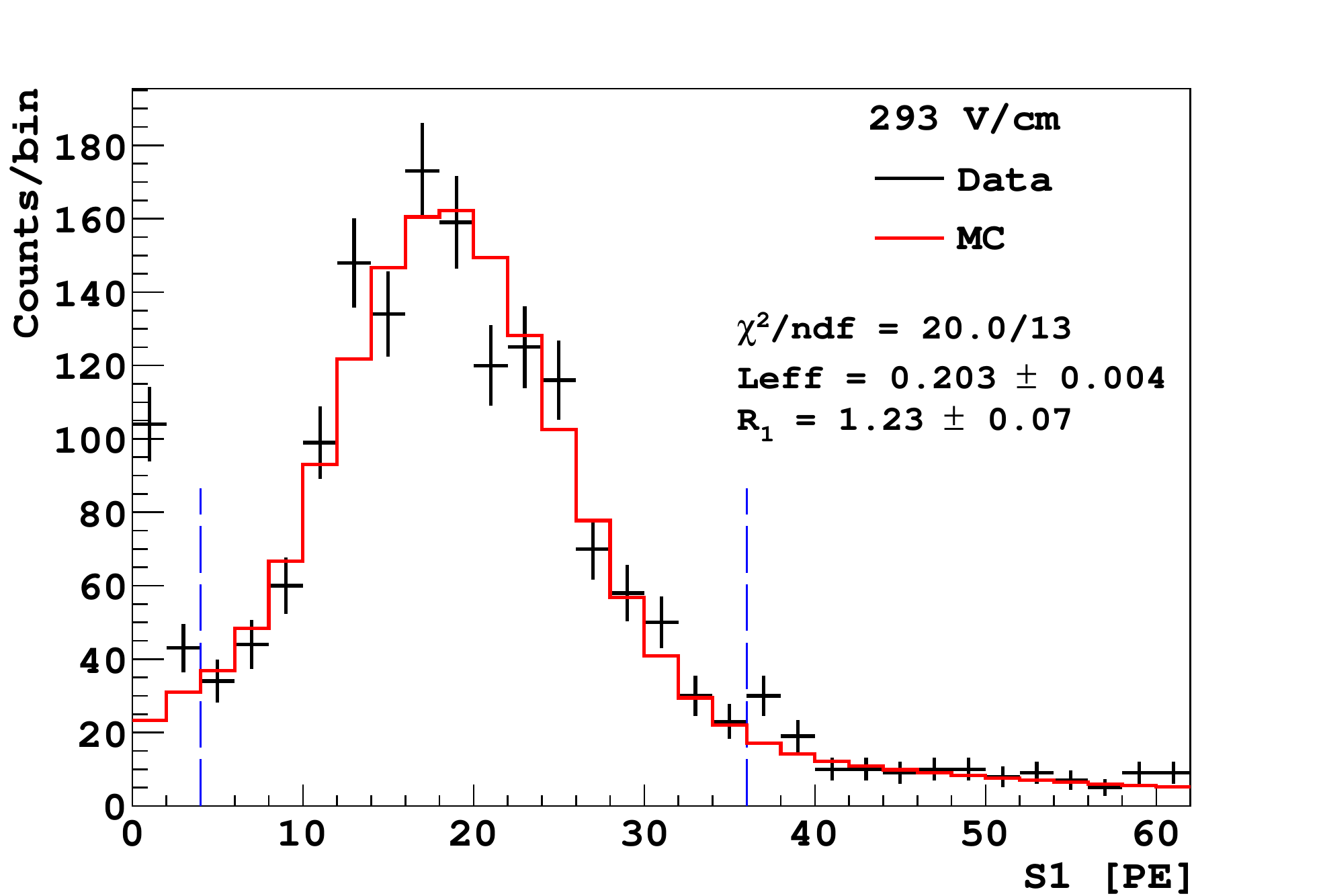}
\includegraphics[width=0.95\columnwidth]{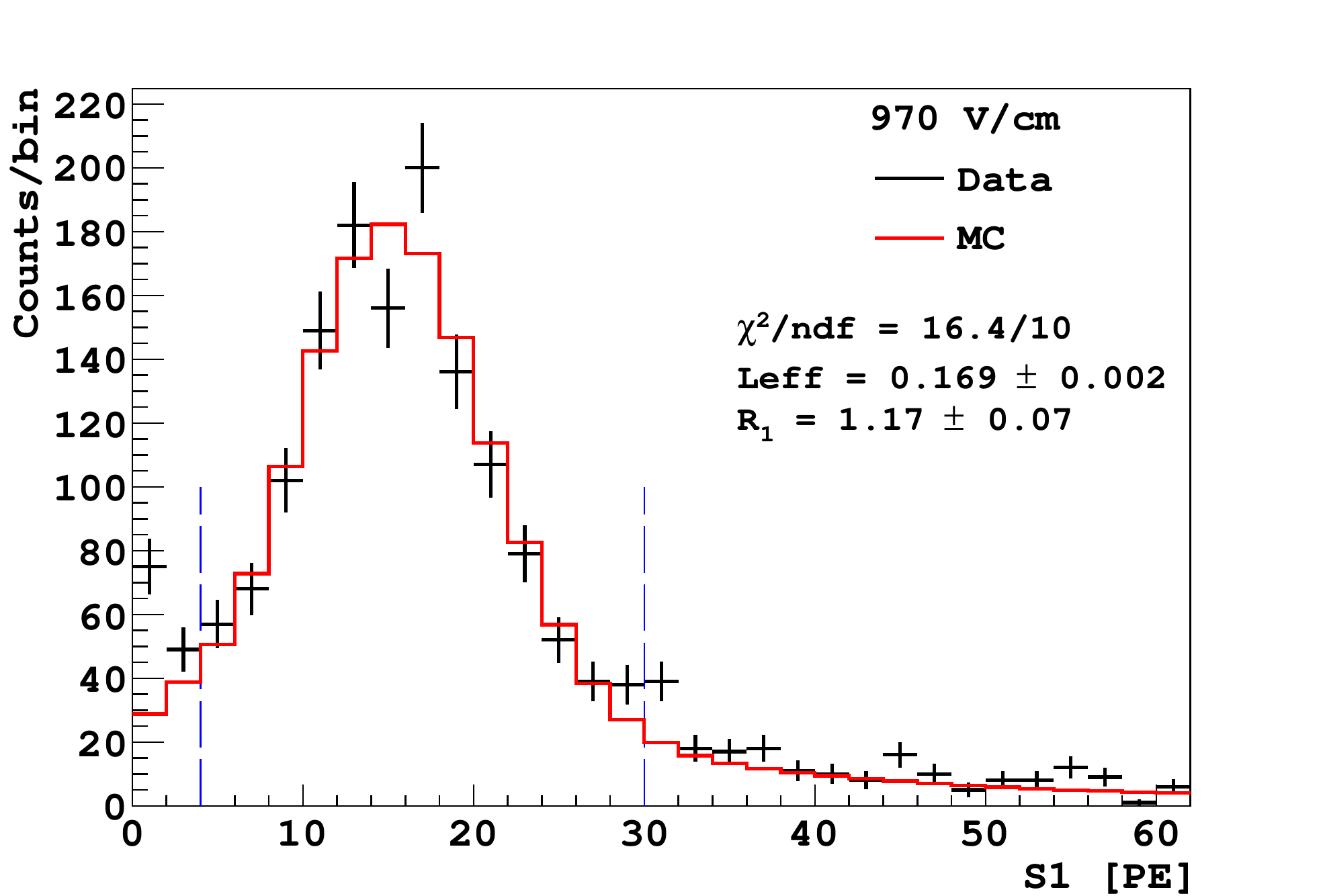}
\caption{\label{fig:leff15}
Top left panel.  {\bf \color{black} Black:} GEANT4-based simulation of the energy deposition in the SCENE detector at the setting devised to produce 14.8\,\kevr\ nuclear recoils.  {\bf \color{blue} Blue:} From neutrons scattered more than once in any part of the entire TPC apparatus before reaching the neutron detector.\\
All other panels.  {\bf \color{black} Black:} Experimental data collected for 14.8\,\kevr\ nuclear recoils.  {\bf \color{red} Red:} Monte Carlo fit of the experimental data.  The range used for each fit is indicated by the vertical {\bf \color{blue} blue} dashed lines.
}
\end{figure*}

\begin{figure*}[t!]
\includegraphics[width=0.95\columnwidth]{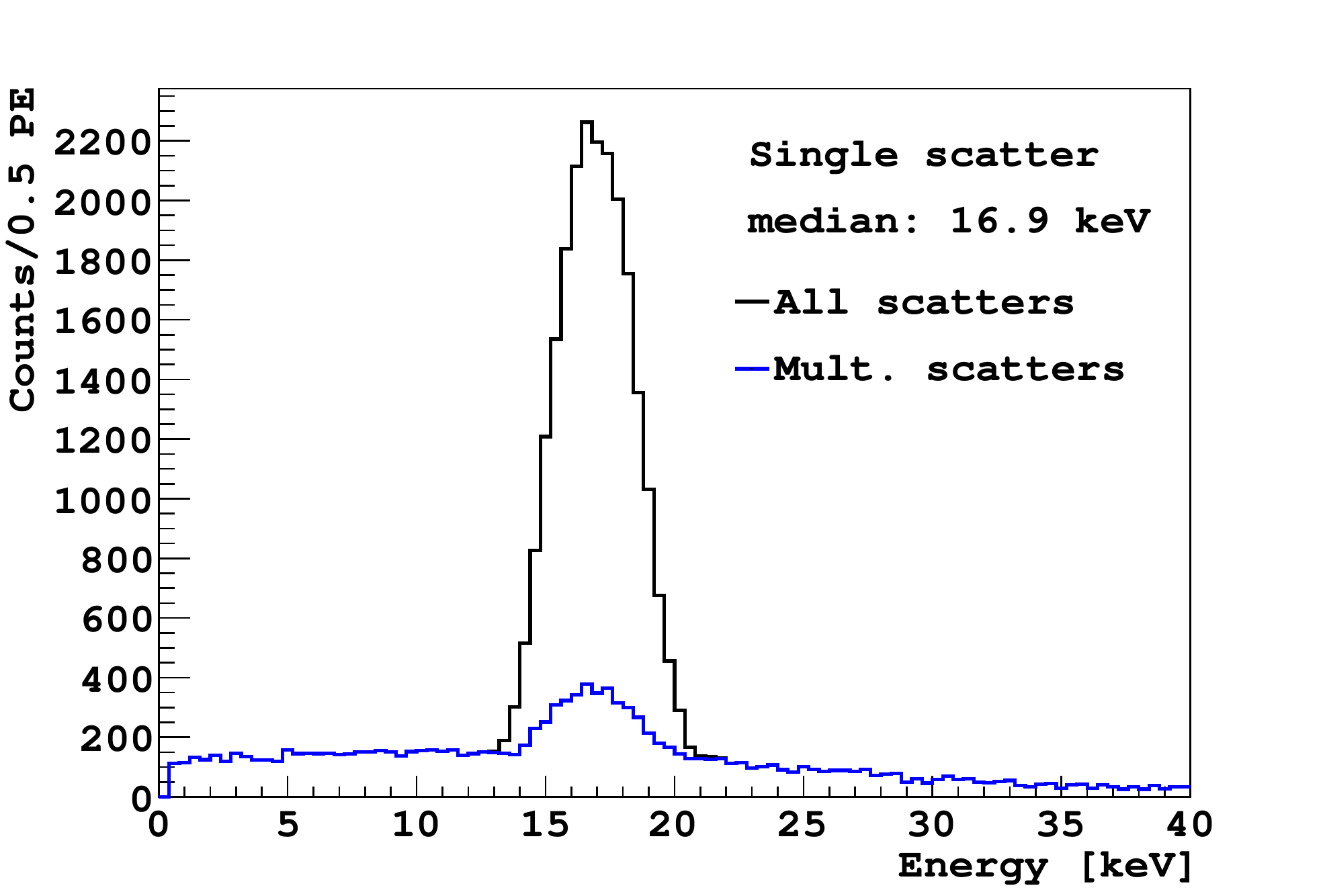}
\includegraphics[width=0.95\columnwidth]{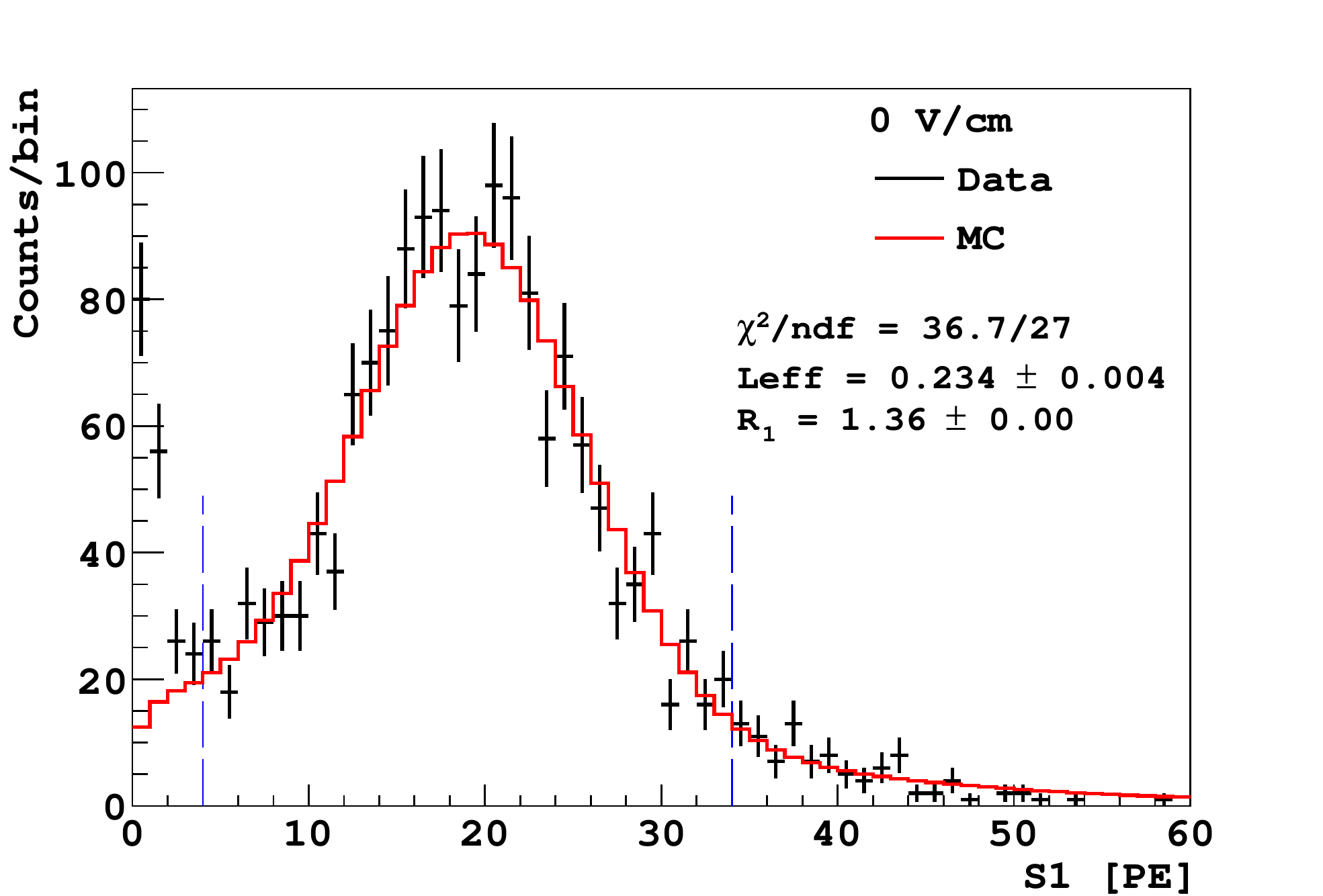}
\includegraphics[width=0.95\columnwidth]{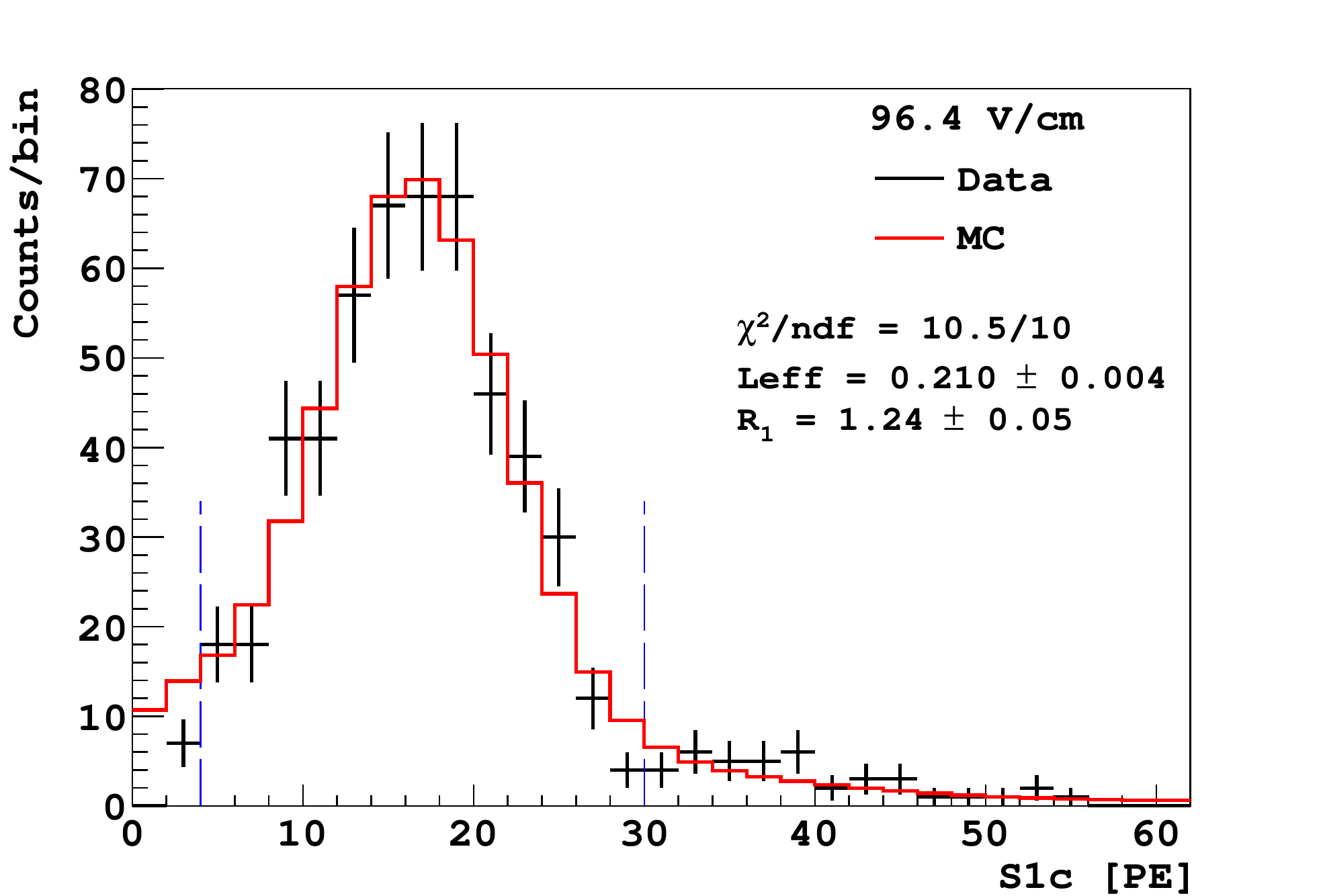}
\includegraphics[width=0.95\columnwidth]{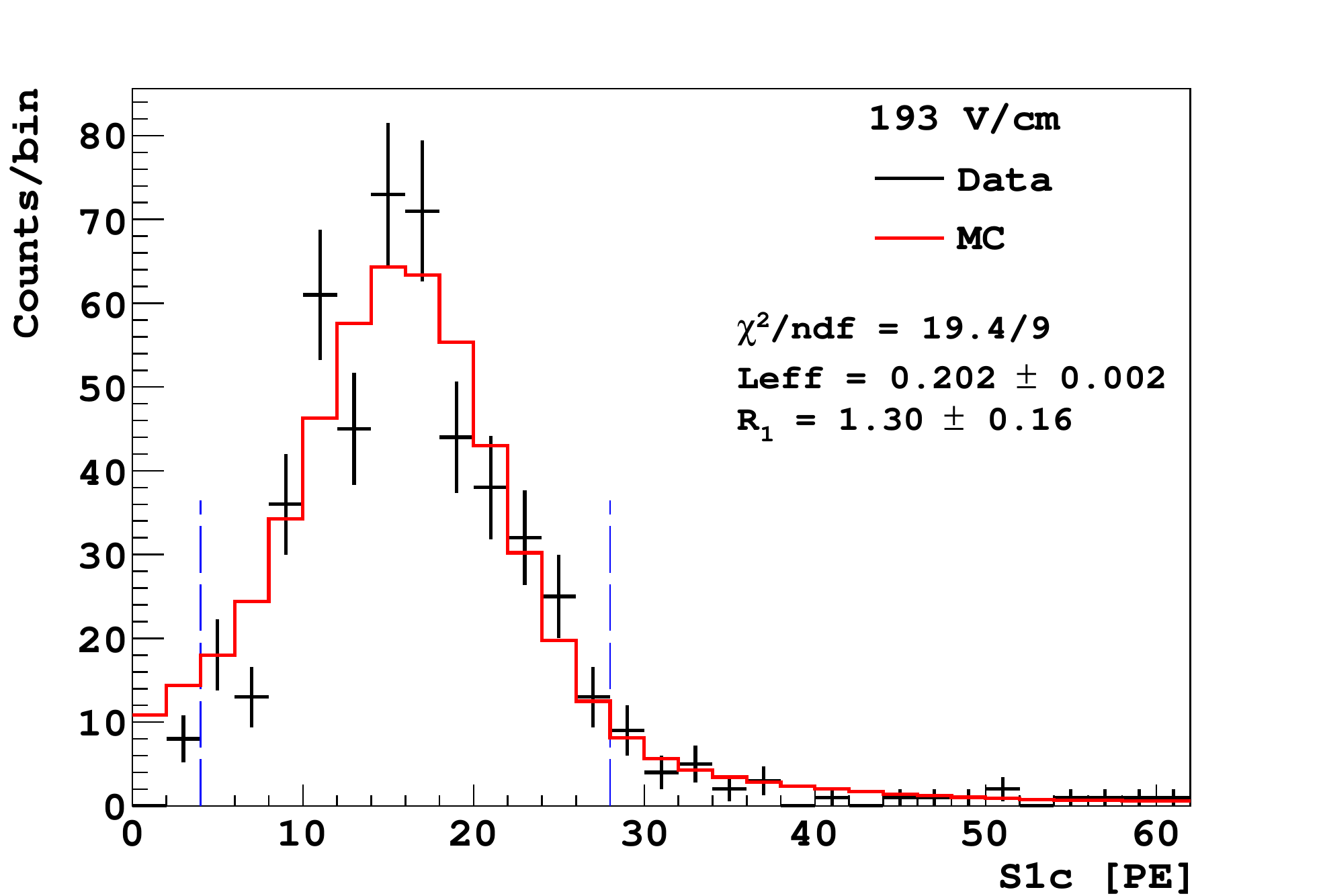}
\includegraphics[width=0.95\columnwidth]{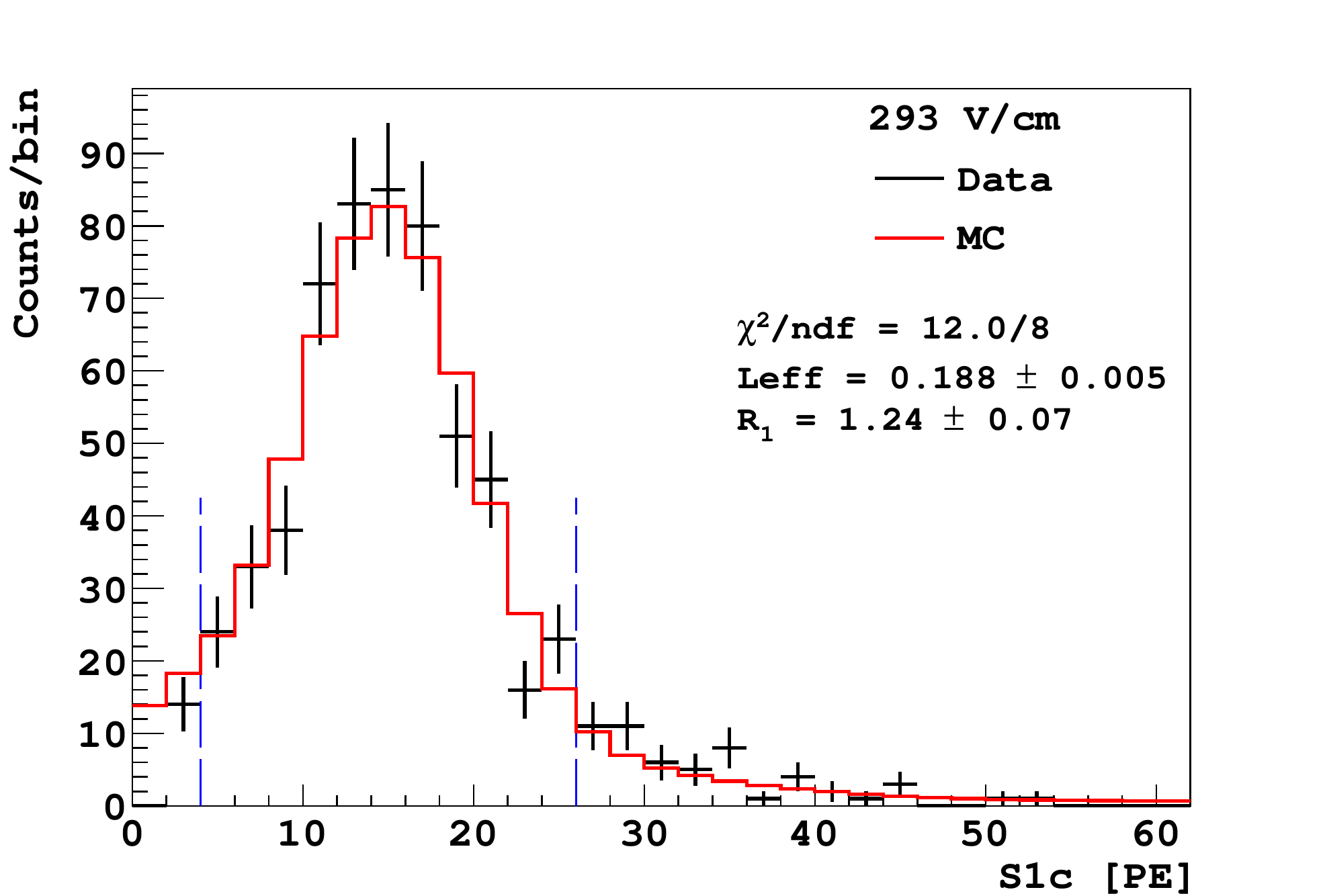}
\includegraphics[width=0.95\columnwidth]{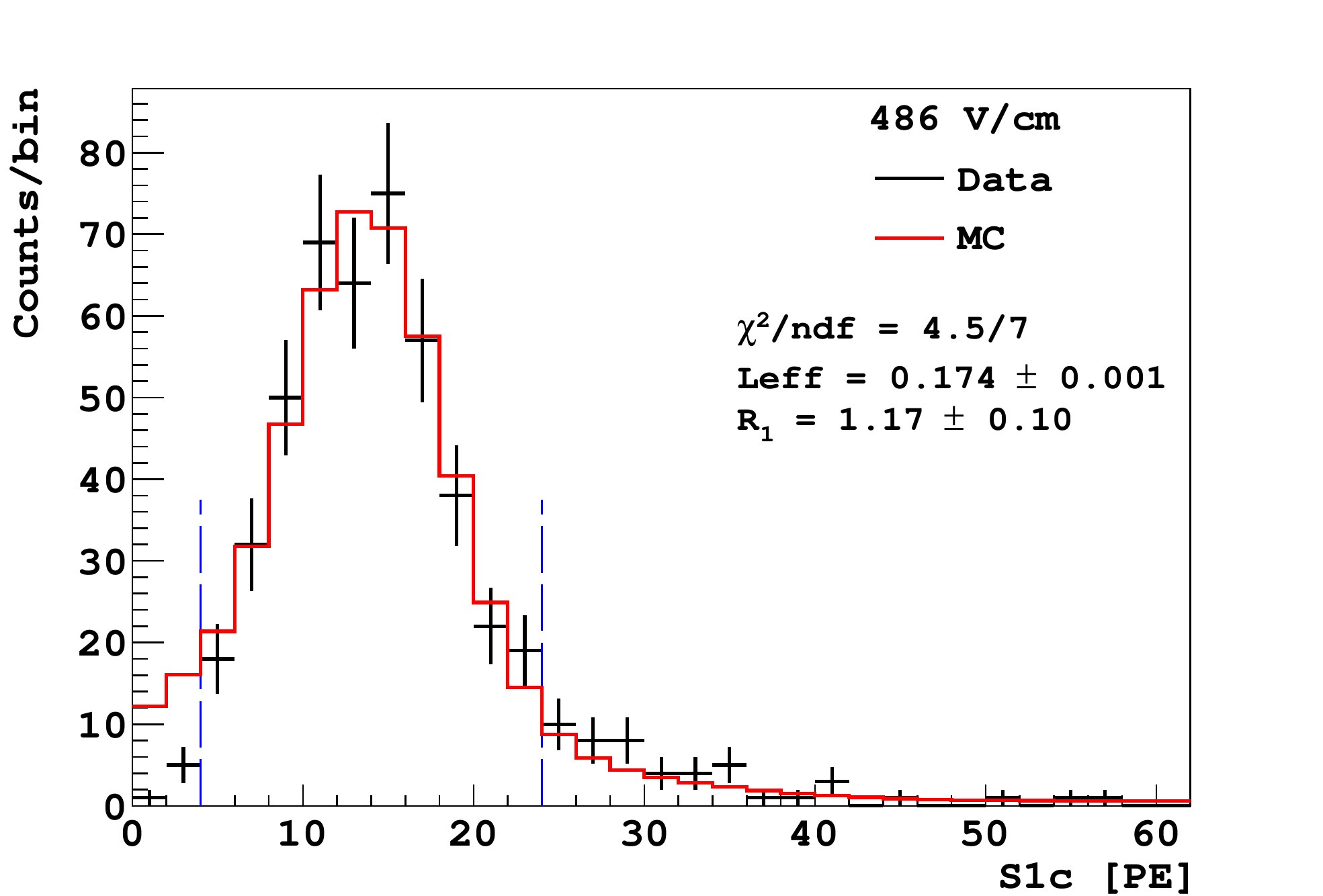}
\caption{\label{fig:leff17}
Top left panel.  {\bf \color{black} Black:} GEANT4-based simulation of the energy deposition in the SCENE detector at the setting devised to produce 16.9\,\kevr\ nuclear recoils.  {\bf \color{blue} Blue:} From neutrons scattered more than once in any part of the entire TPC apparatus before reaching the neutron detector.\\
All other panels.  {\bf \color{black} Black:} Experimental data collected for 16.9\,\kevr\ nuclear recoils.  {\bf \color{red} Red:} Monte Carlo fit of the experimental data.  The range used for each fit is indicated by the vertical {\bf \color{blue} blue} dashed lines.
}
\end{figure*}

\begin{figure*}[t!]
\includegraphics[width=0.95\columnwidth]{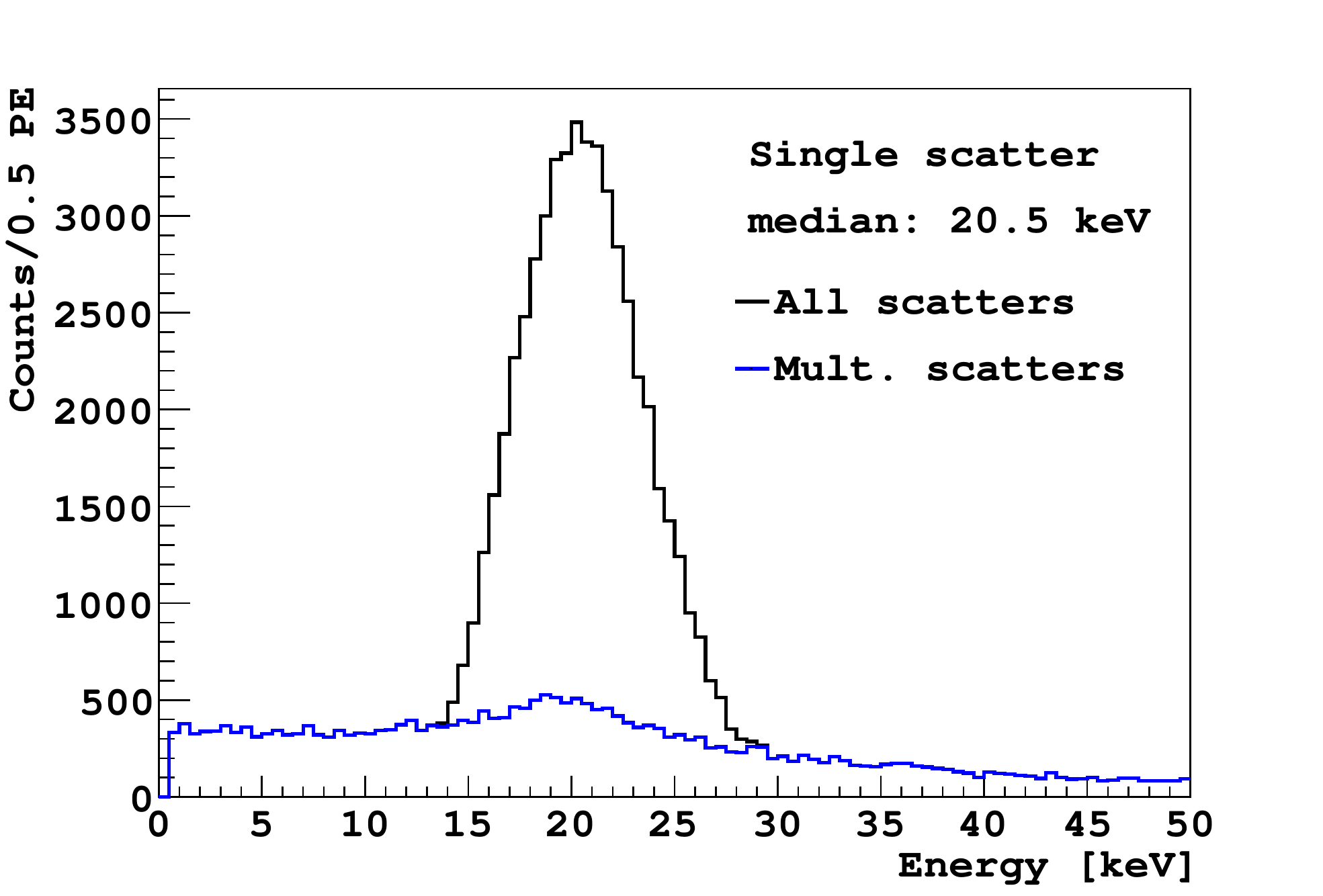}
\includegraphics[width=0.95\columnwidth]{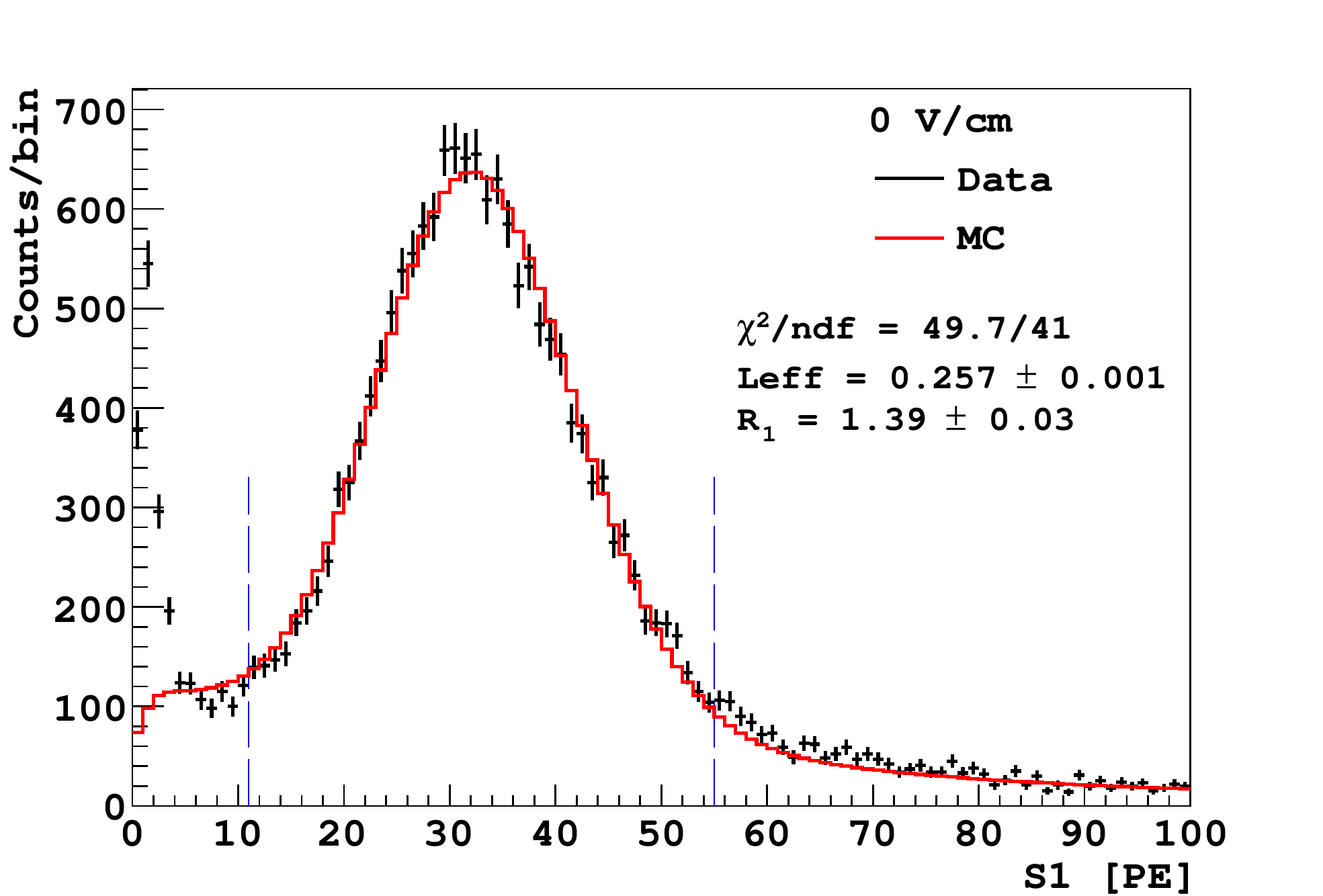}
\includegraphics[width=0.95\columnwidth]{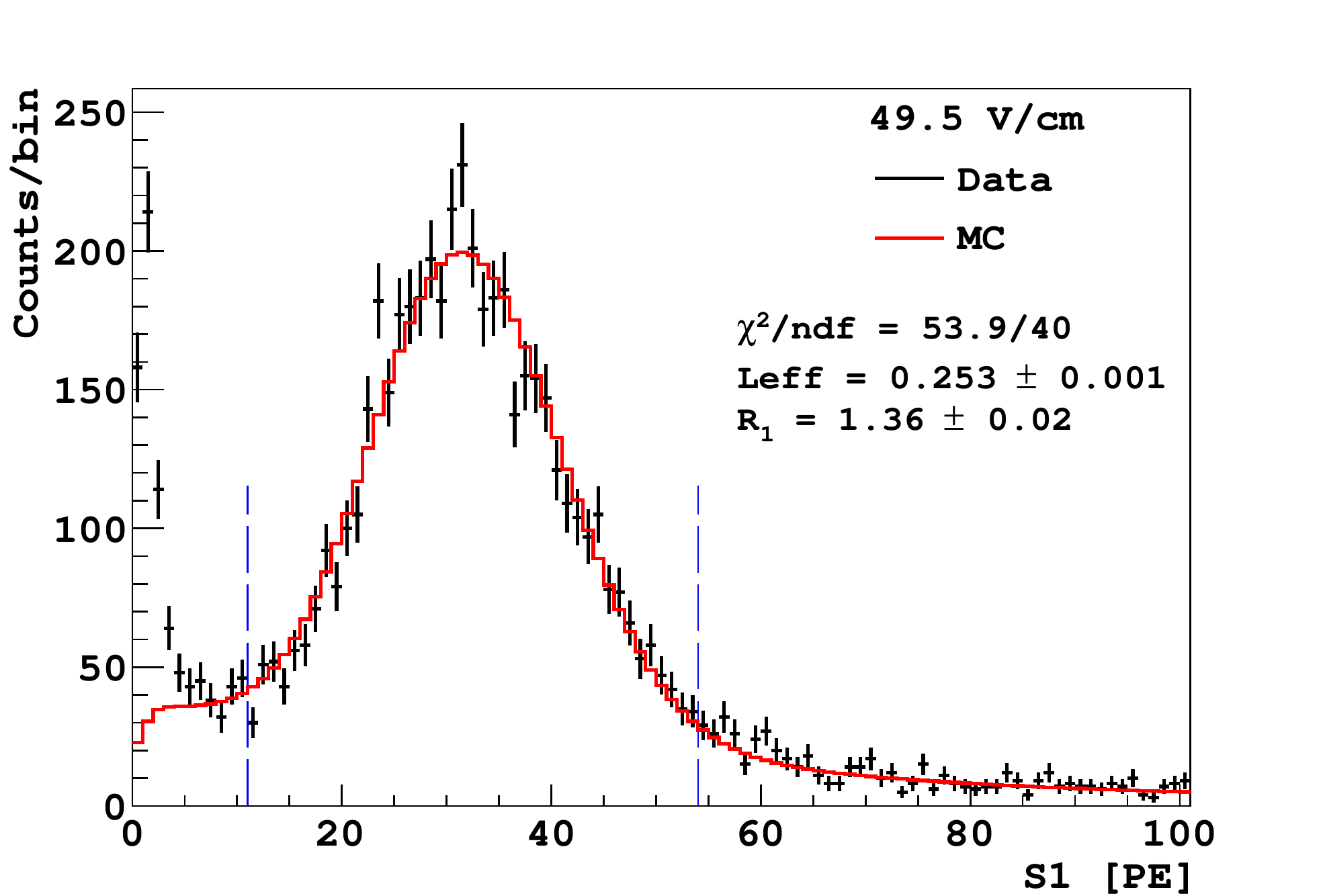}
\includegraphics[width=0.95\columnwidth]{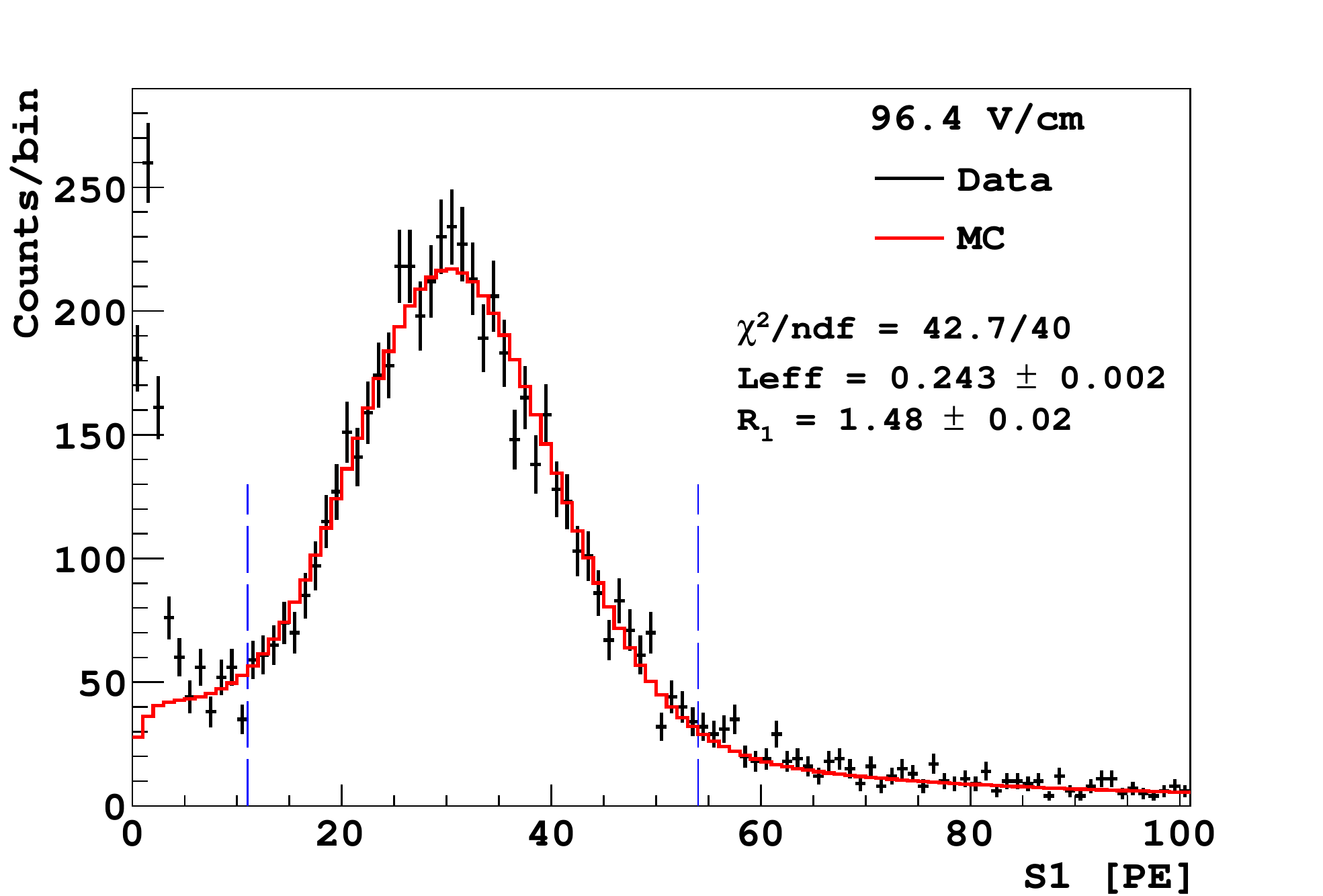}
\includegraphics[width=0.95\columnwidth]{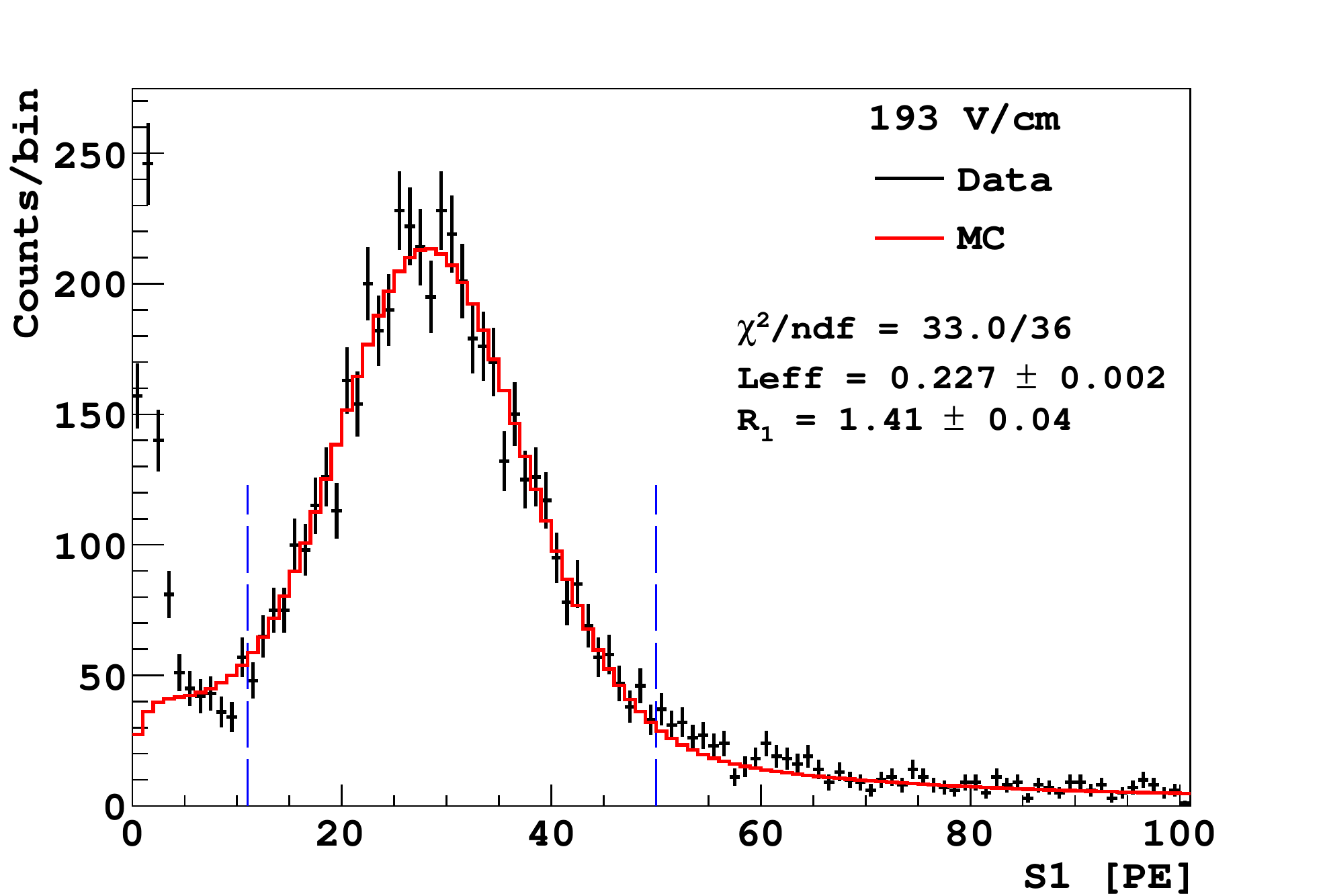}
\includegraphics[width=0.95\columnwidth]{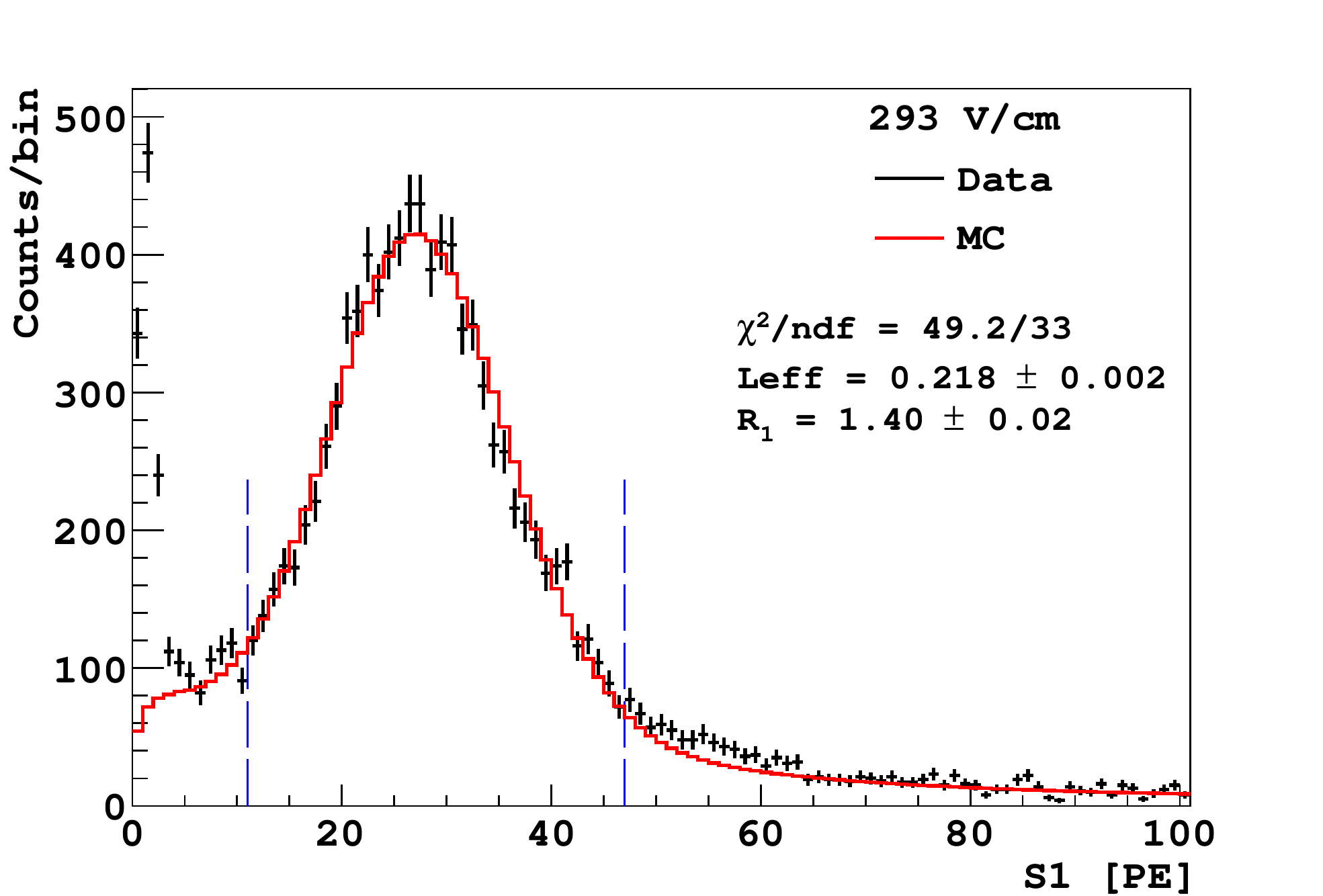}
\includegraphics[width=0.95\columnwidth]{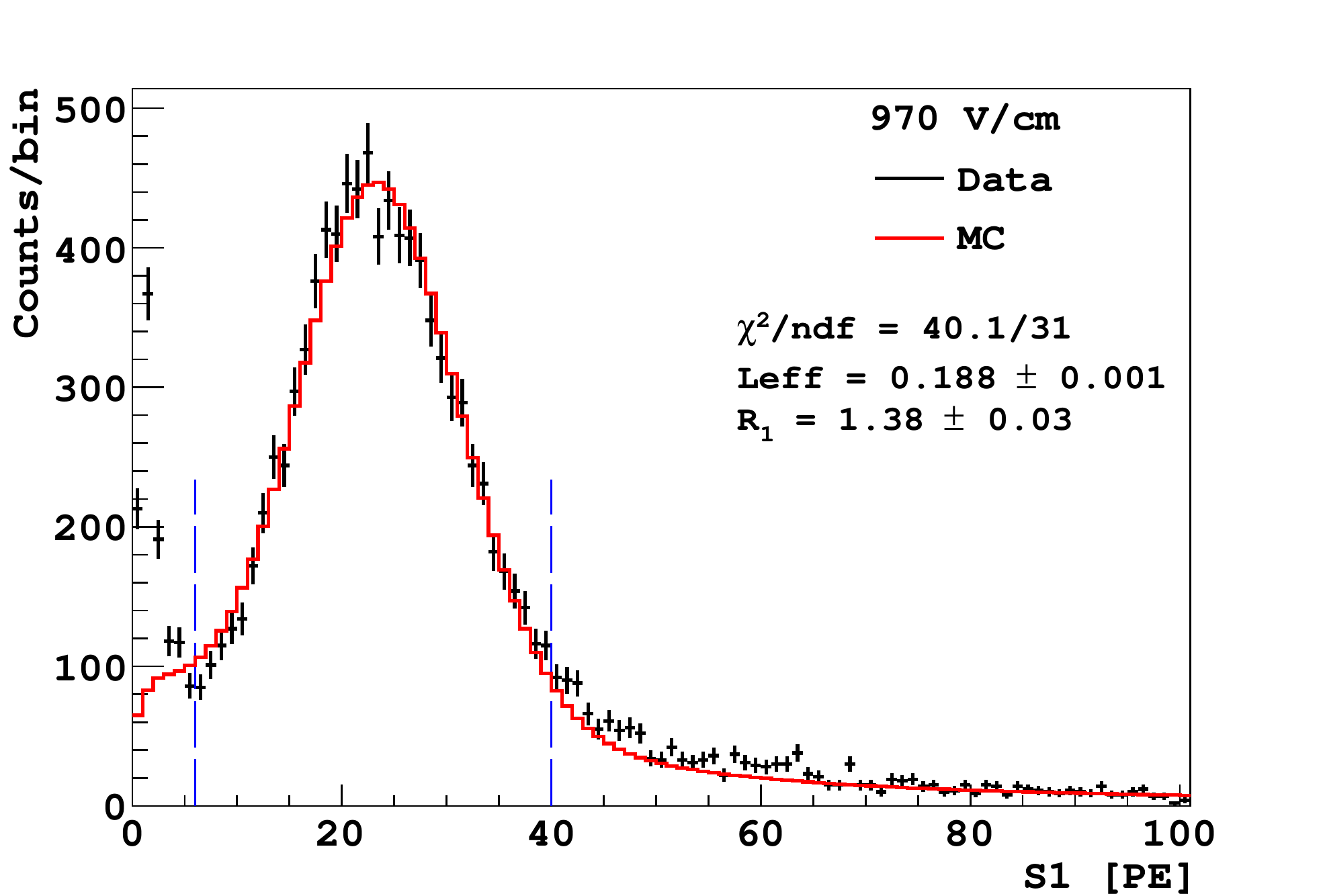}
\caption{\label{fig:leff21}
Top left panel.  {\bf \color{black} Black:} GEANT4-based simulation of the energy deposition in the SCENE detector at the setting devised to produce 20.5\,\kevr\ nuclear recoils. {\bf \color{blue} Blue:} From neutrons scattered more than once in any part of the entire TPC apparatus before reaching the neutron detector.\\
All other panels.  {\bf \color{black} Black:} Experimental data collected for 20.5\,\kevr\ nuclear recoils.  {\bf \color{red} Red:} Monte Carlo fit of the experimental data.  The range used for each fit is indicated by the vertical {\bf \color{blue} blue} dashed lines.
}
\end{figure*}

\begin{figure*}[t!]
\includegraphics[width=0.95\columnwidth]{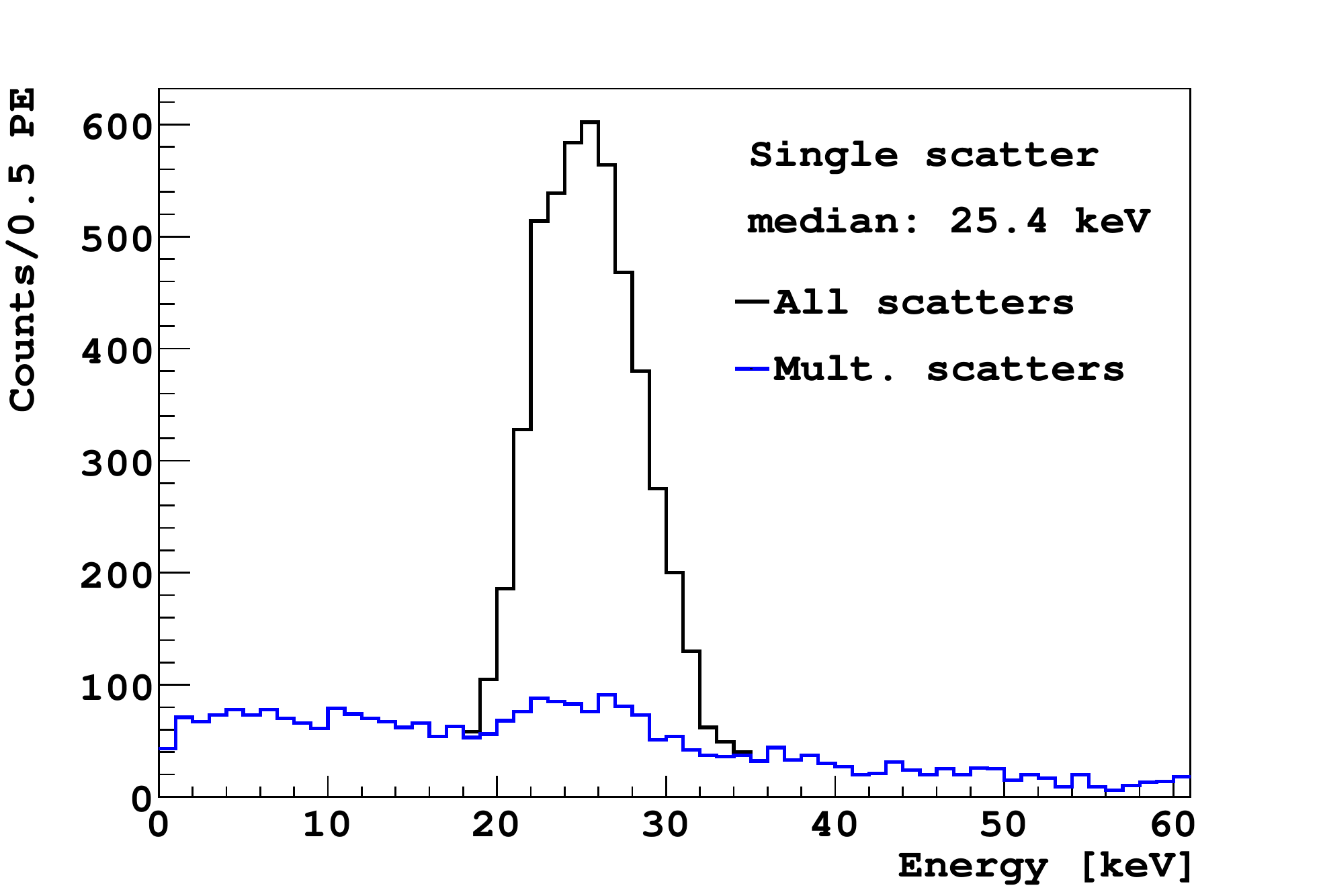}
\includegraphics[width=0.95\columnwidth]{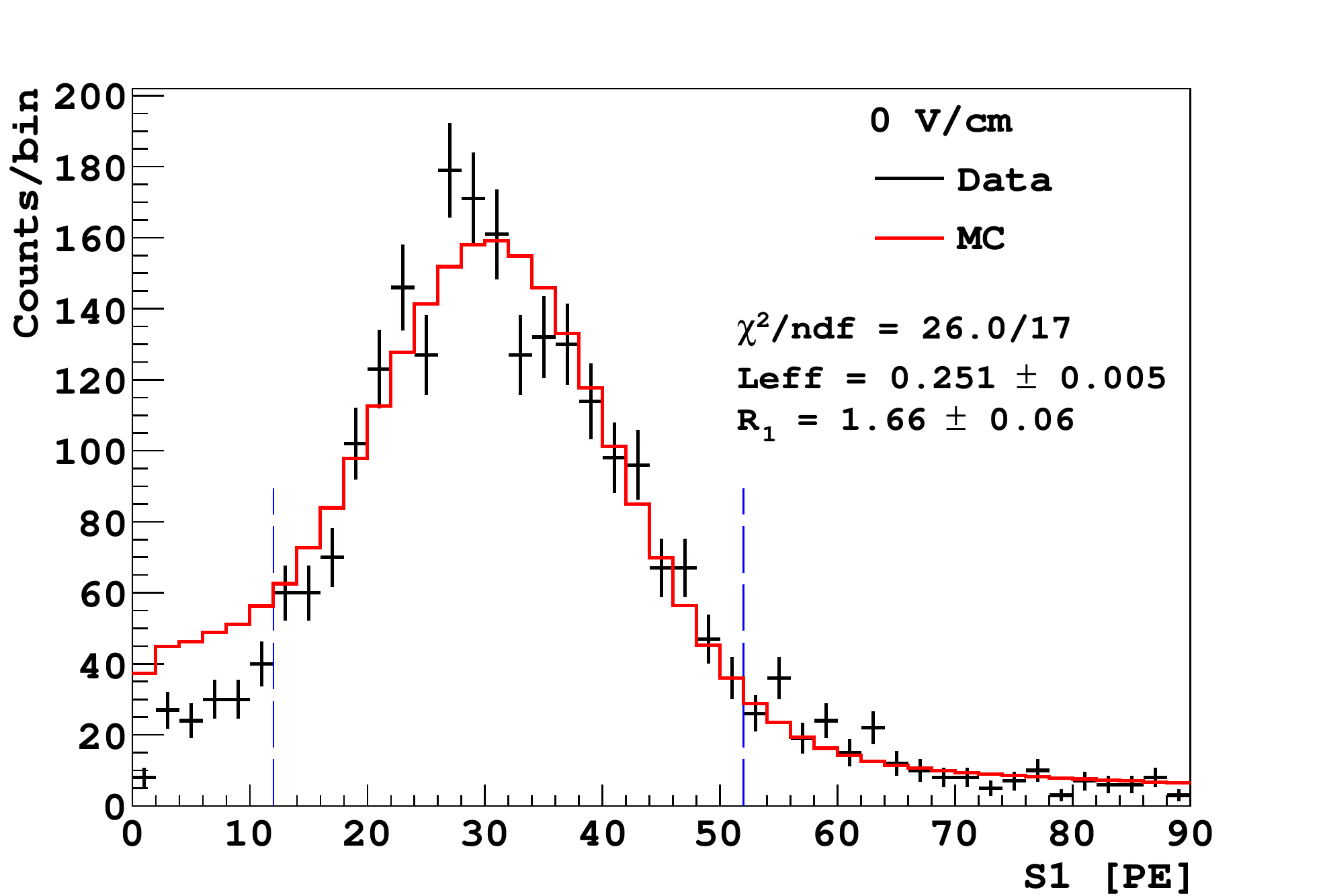}
\includegraphics[width=0.95\columnwidth]{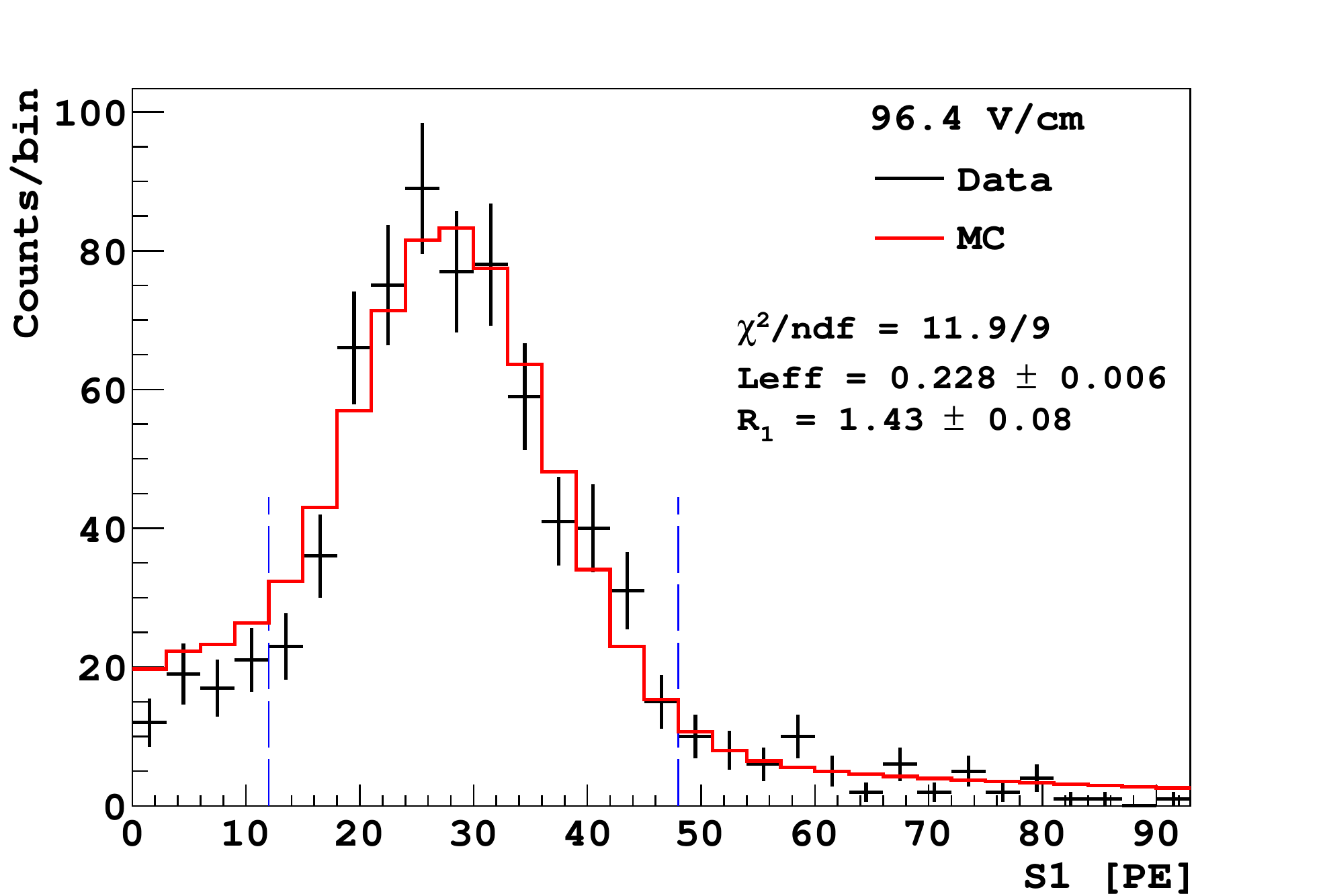}
\includegraphics[width=0.95\columnwidth]{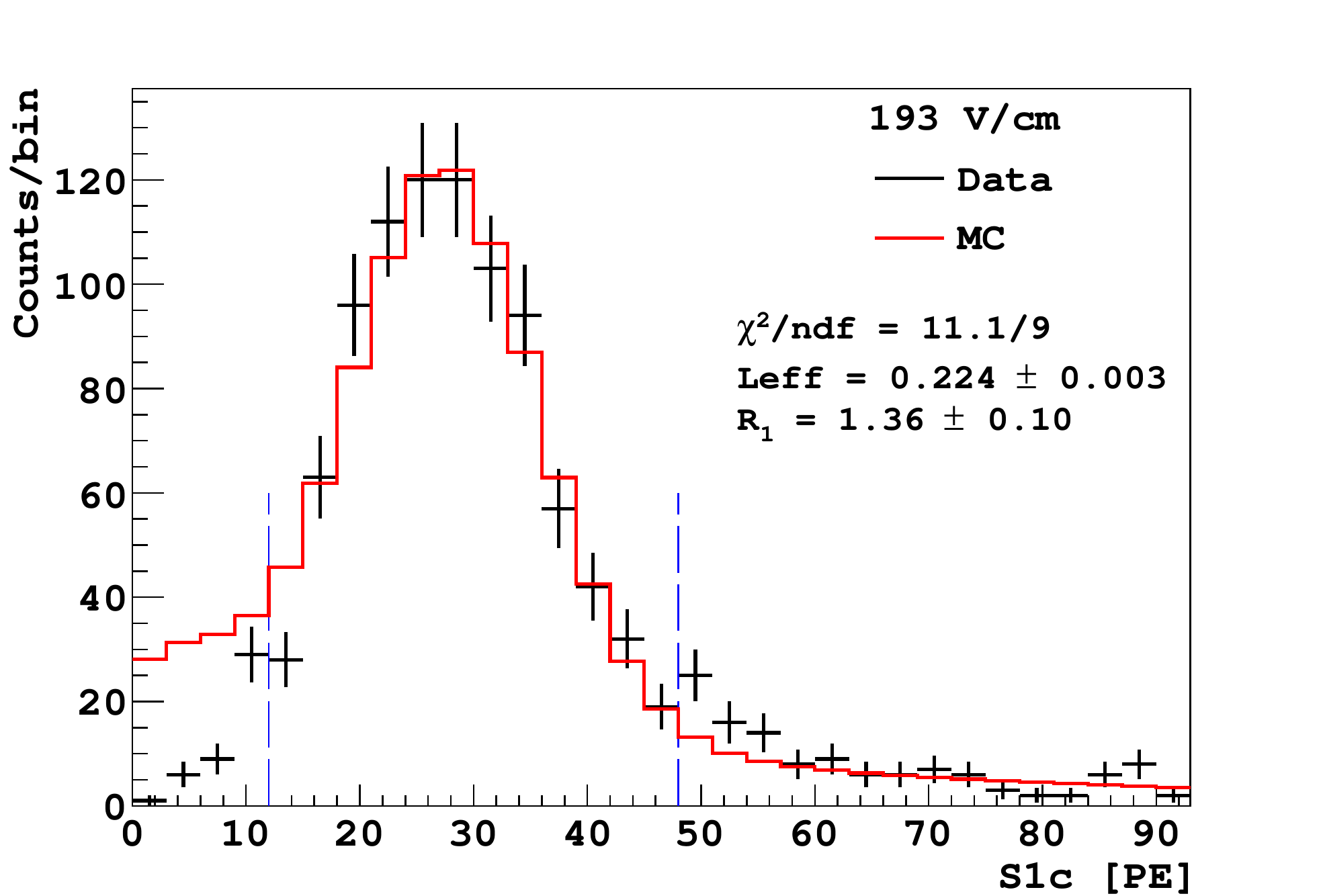}
\includegraphics[width=0.95\columnwidth]{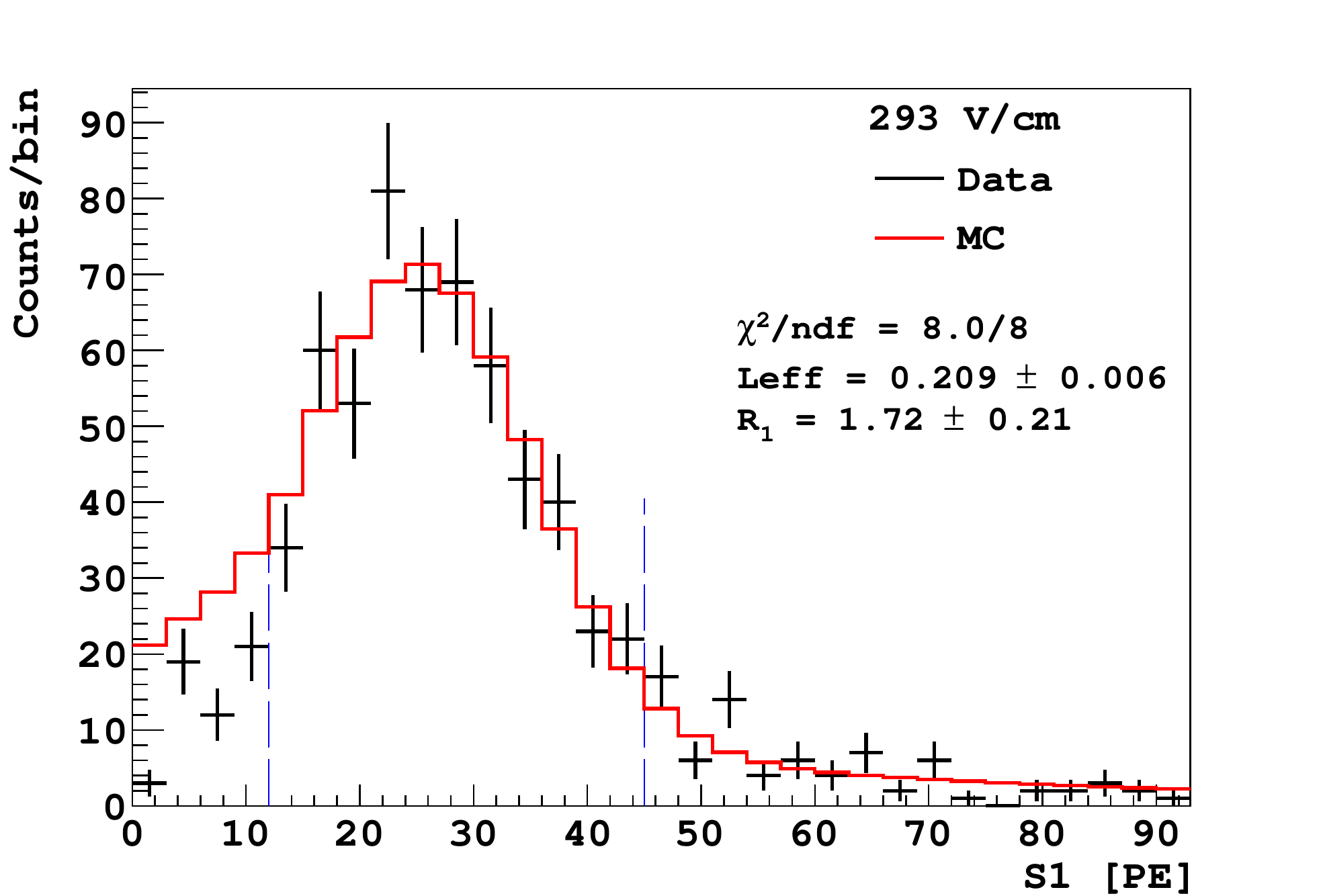}
\includegraphics[width=0.95\columnwidth]{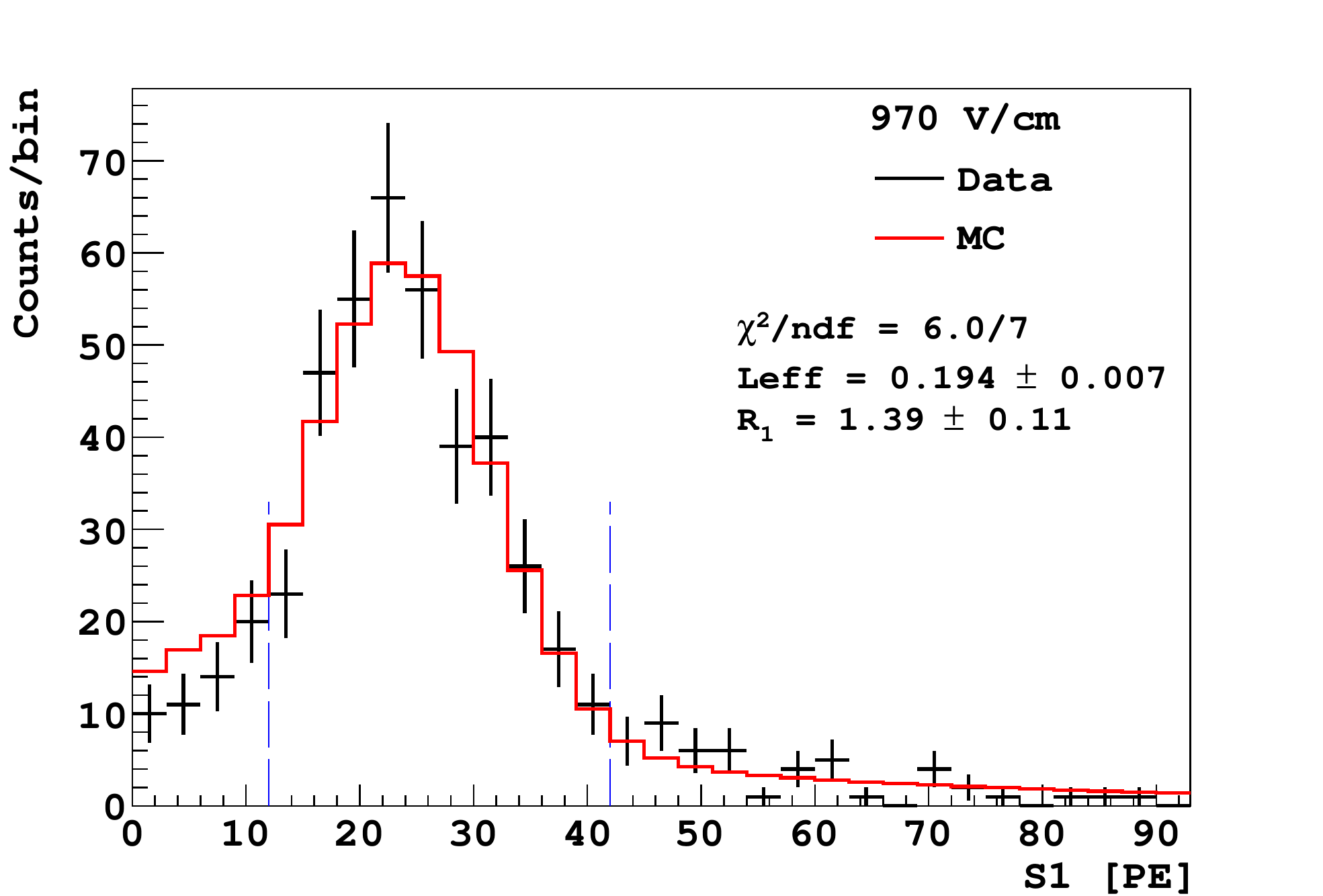}
\caption{\label{fig:leff26}
Top left panel.  {\bf \color{black} Black:} GEANT4-based simulation of the energy deposition in the SCENE detector at the setting devised to produce 25.4\,\kevr\ nuclear recoils.   {\bf \color{blue} Blue:} From neutrons scattered more than once in any part of the entire TPC apparatus before reaching the neutron detector.\\
All other panels.  {\bf \color{black} Black:} Experimental data collected for 25.4\,\kevr\ nuclear recoils.  {\bf \color{red} Red:} Monte Carlo fit of the experimental data.  The range used for each fit is indicated by the vertical {\bf \color{blue} blue} dashed lines.
}
\end{figure*}

\begin{figure*}[t!]
\includegraphics[width=0.95\columnwidth]{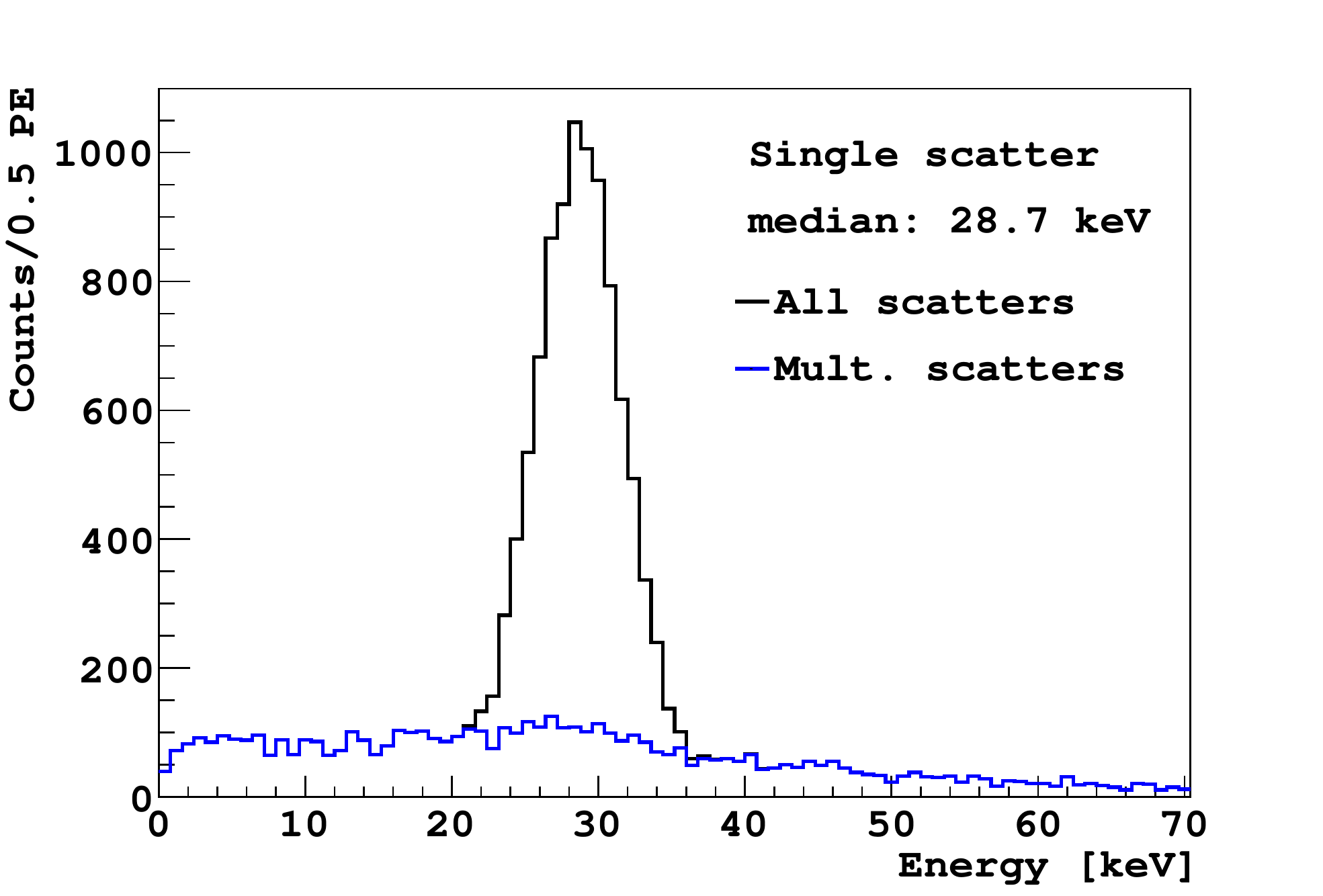}
\includegraphics[width=0.95\columnwidth]{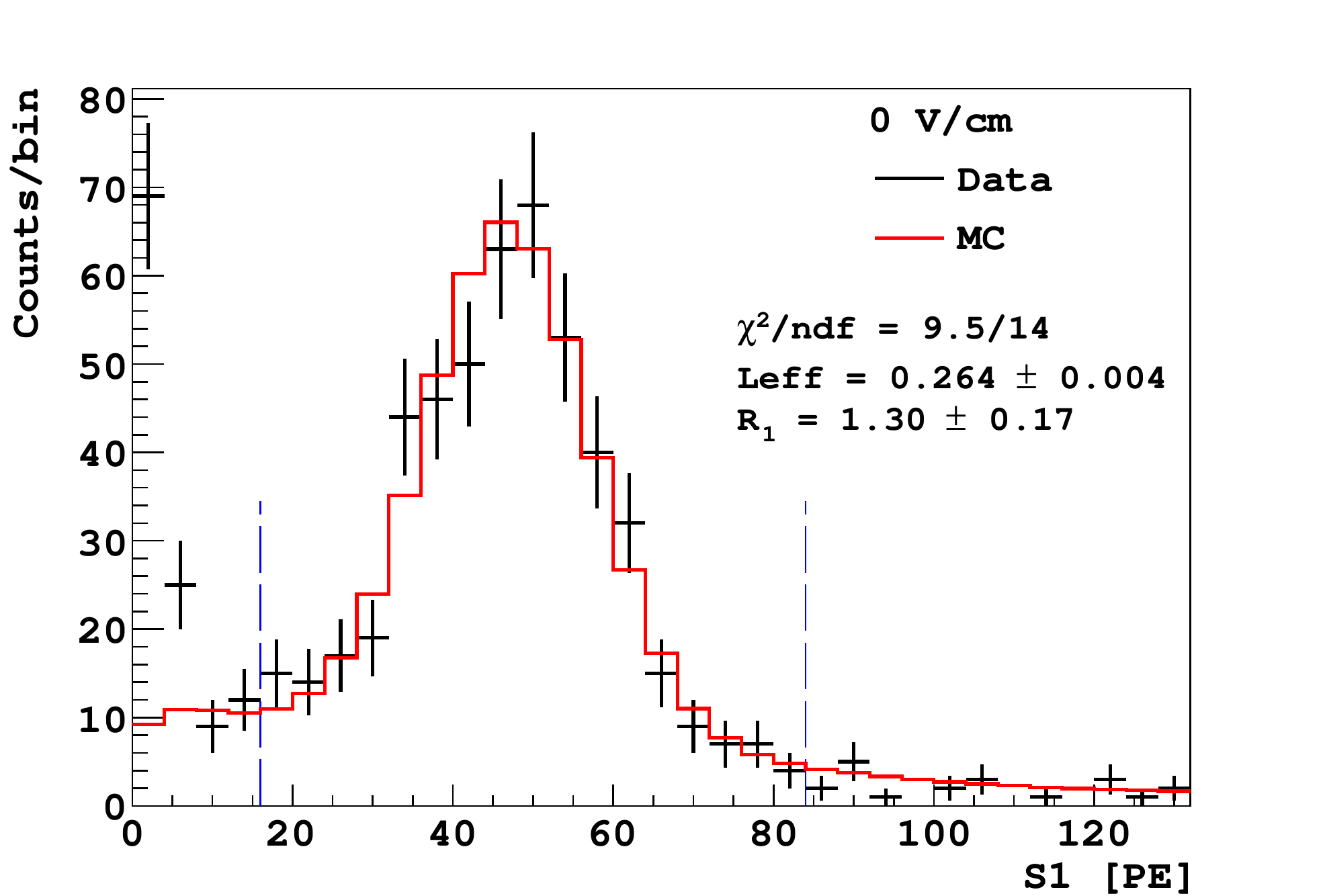}
\includegraphics[width=0.95\columnwidth]{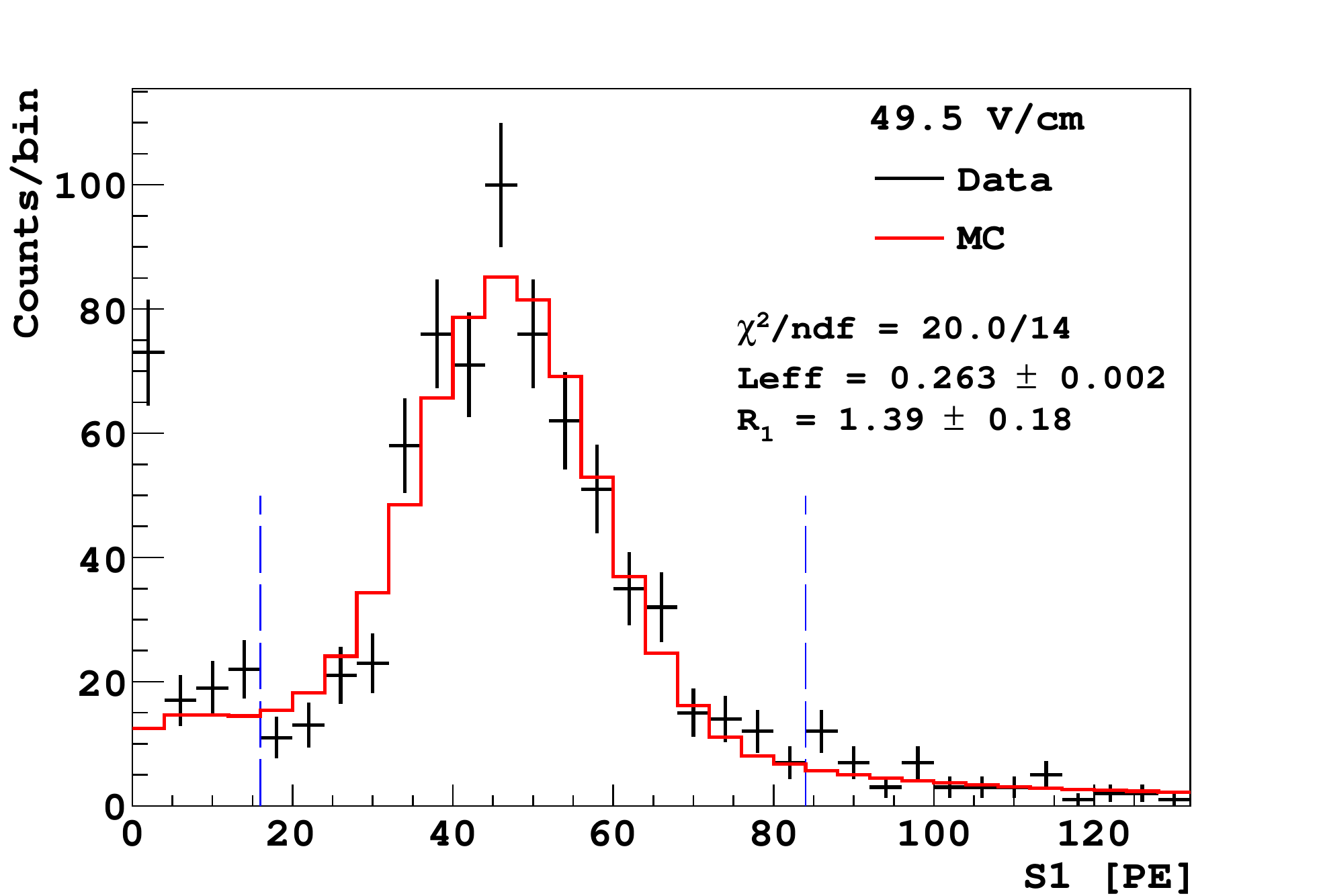}
\includegraphics[width=0.95\columnwidth]{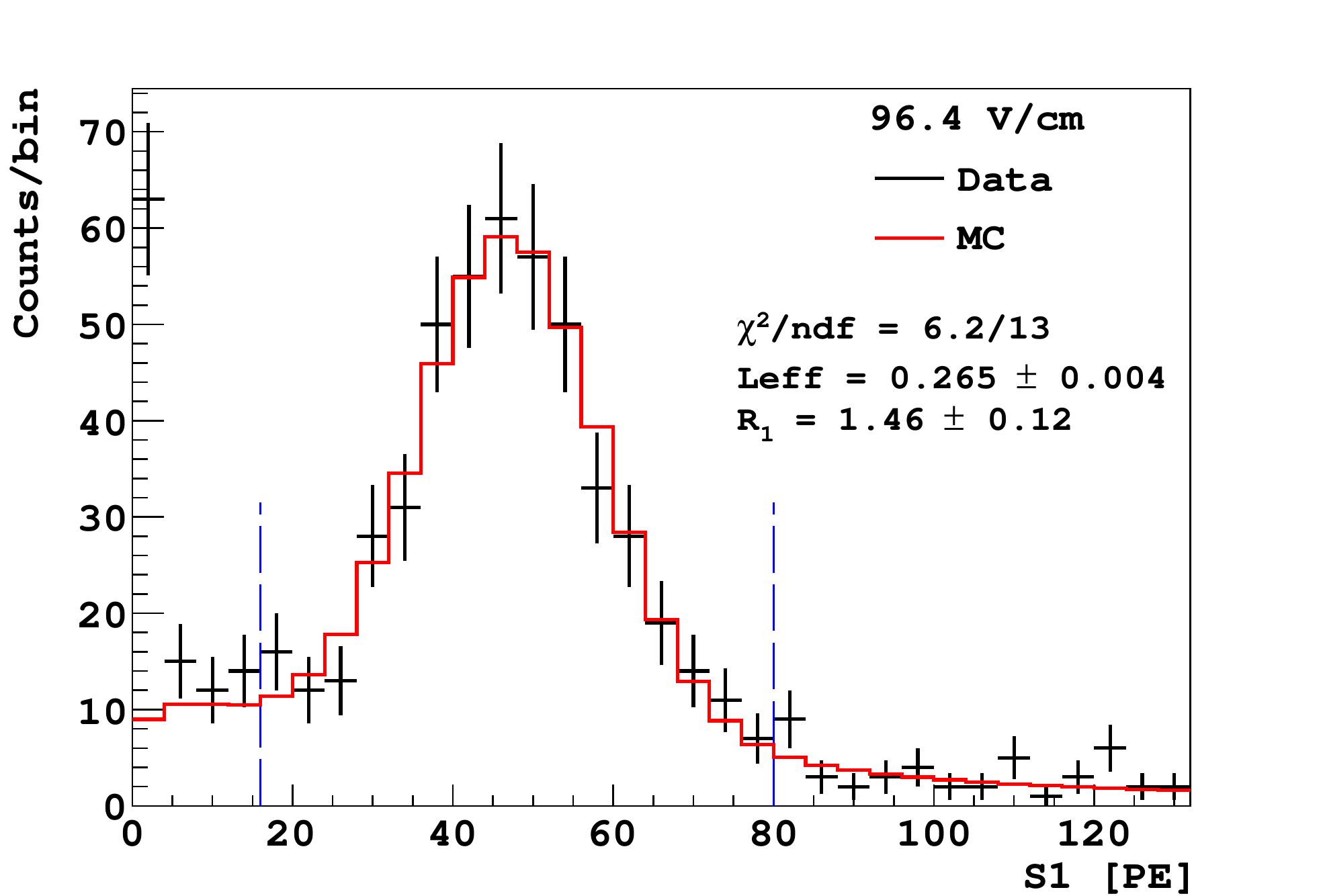}
\includegraphics[width=0.95\columnwidth]{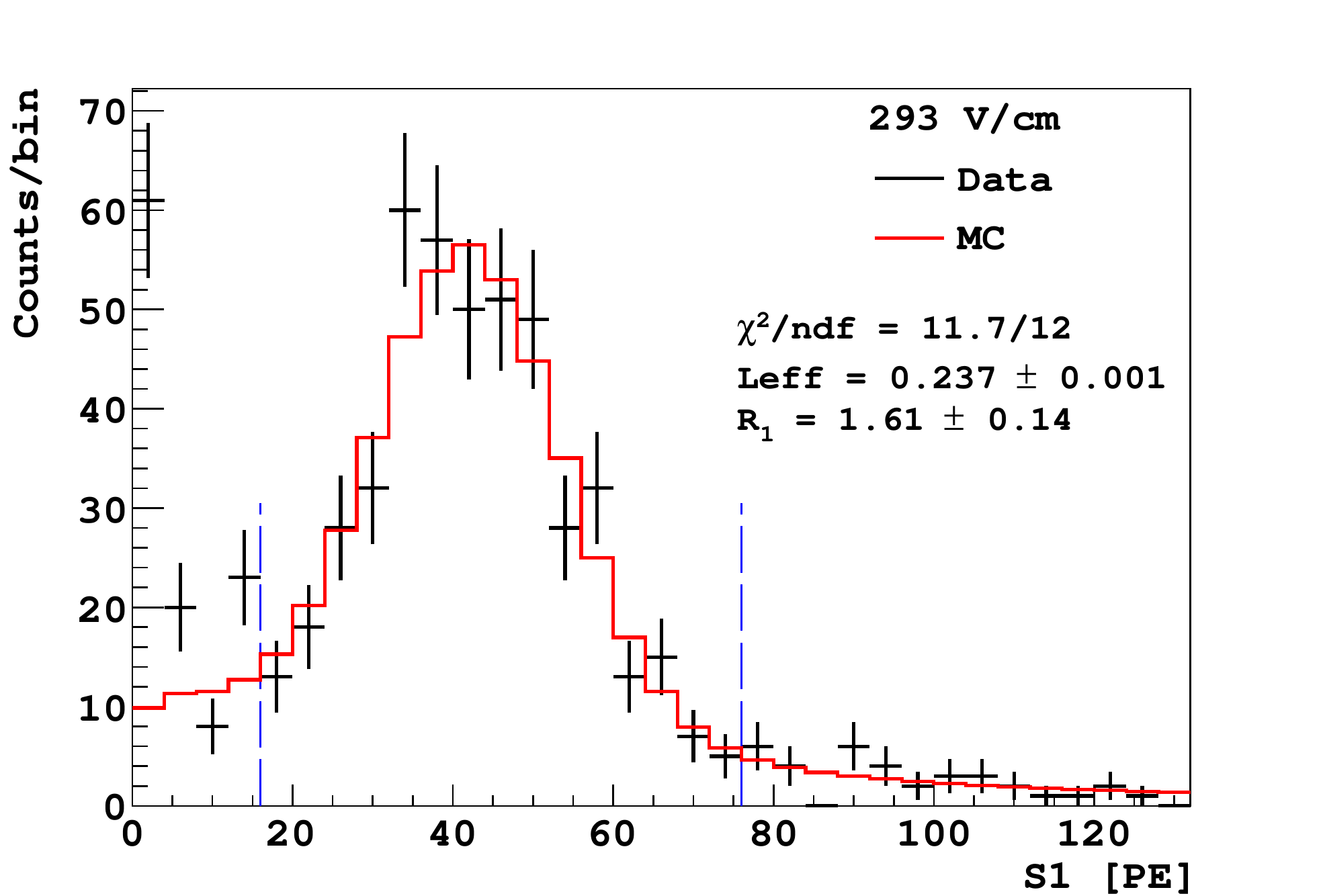}
\includegraphics[width=0.95\columnwidth]{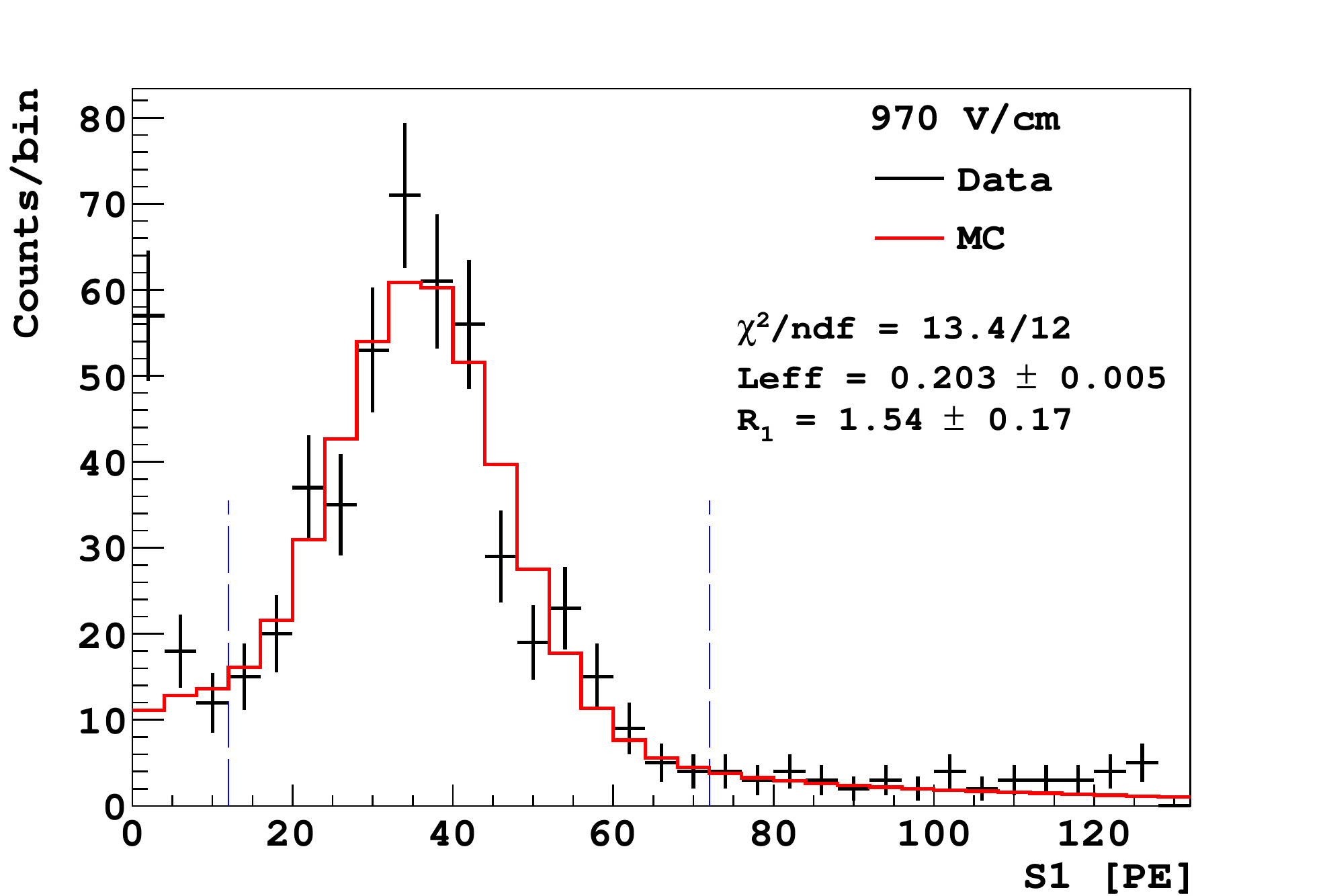}
\caption{\label{fig:leff29}
Top left panel.  {\bf \color{black} Black:} GEANT4-based simulation of the energy deposition in the SCENE detector at the setting devised to produce 28.7\,\kevr\ nuclear recoils.  {\bf \color{blue} Blue:} From neutrons scattered more than once in any part of the entire TPC apparatus before reaching the neutron detector.\\
All other panels.  {\bf \color{black} Black:} Experimental data collected for 28.7\,\kevr\ nuclear recoils.  {\bf \color{red} Red:} Monte Carlo fit of the experimental data.  The range used for each fit is indicated by the vertical {\bf \color{blue} blue} dashed lines.
}
\end{figure*}

\begin{figure*}[t!]
\includegraphics[width=0.95\columnwidth]{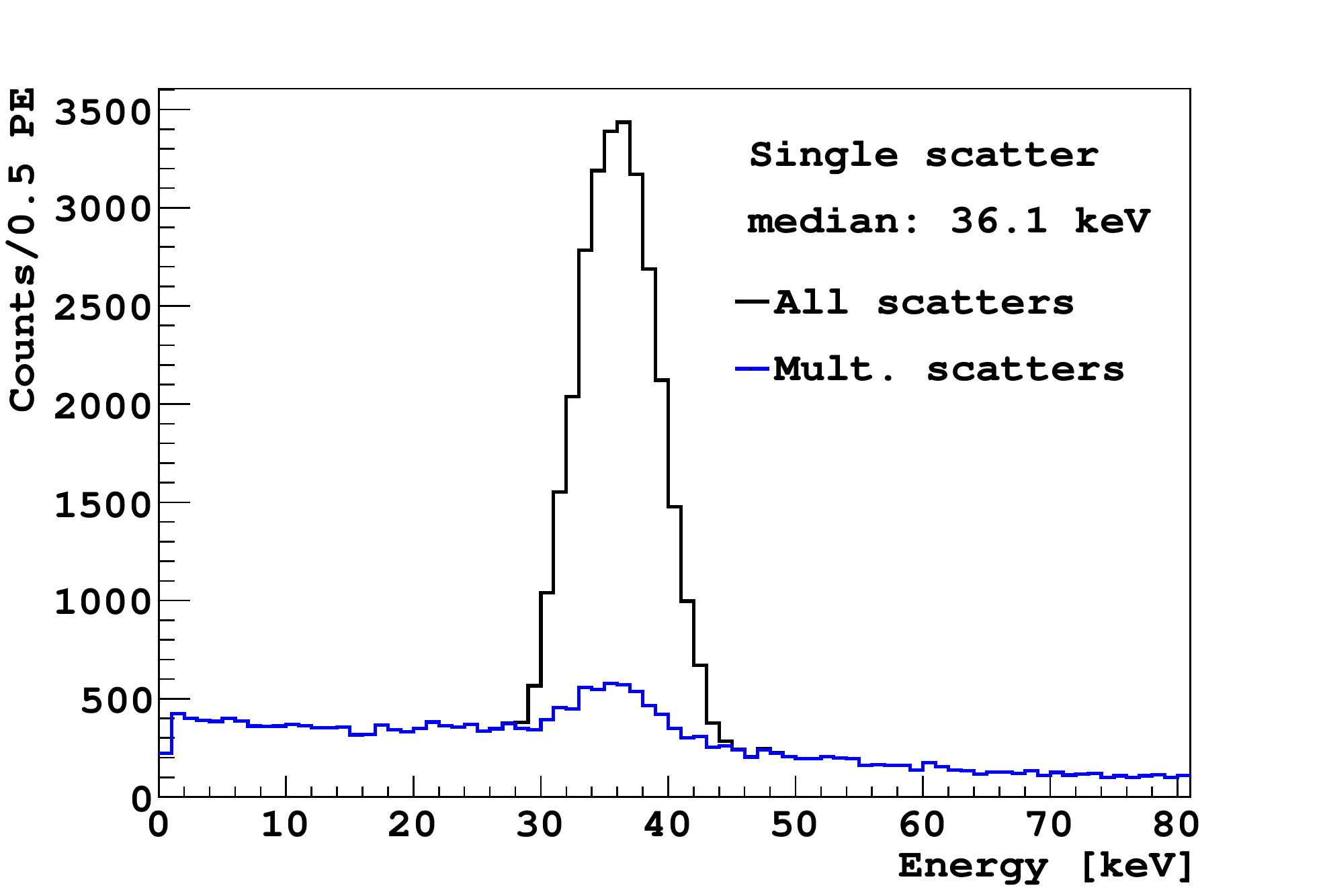}
\includegraphics[width=0.95\columnwidth]{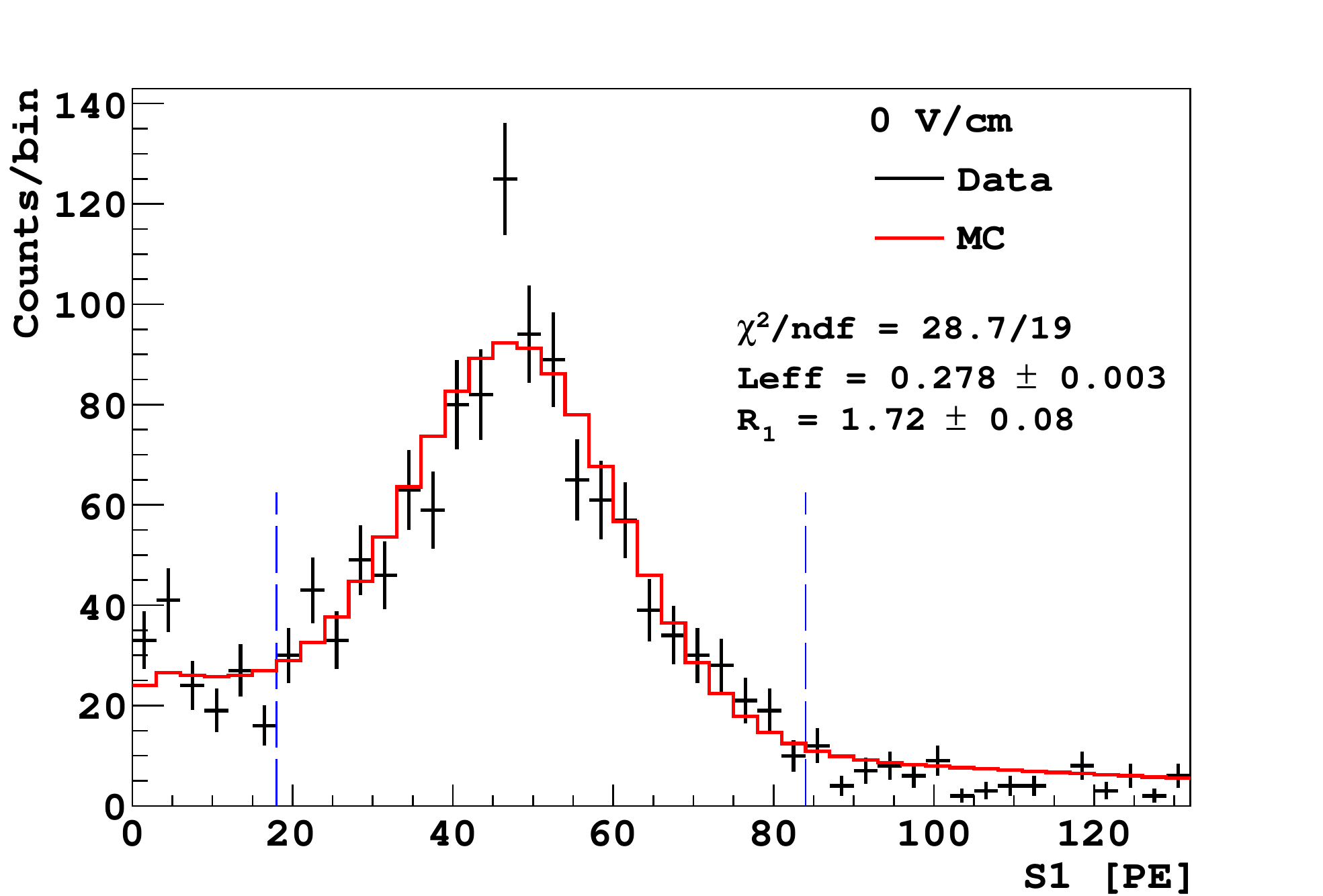}
\includegraphics[width=0.95\columnwidth]{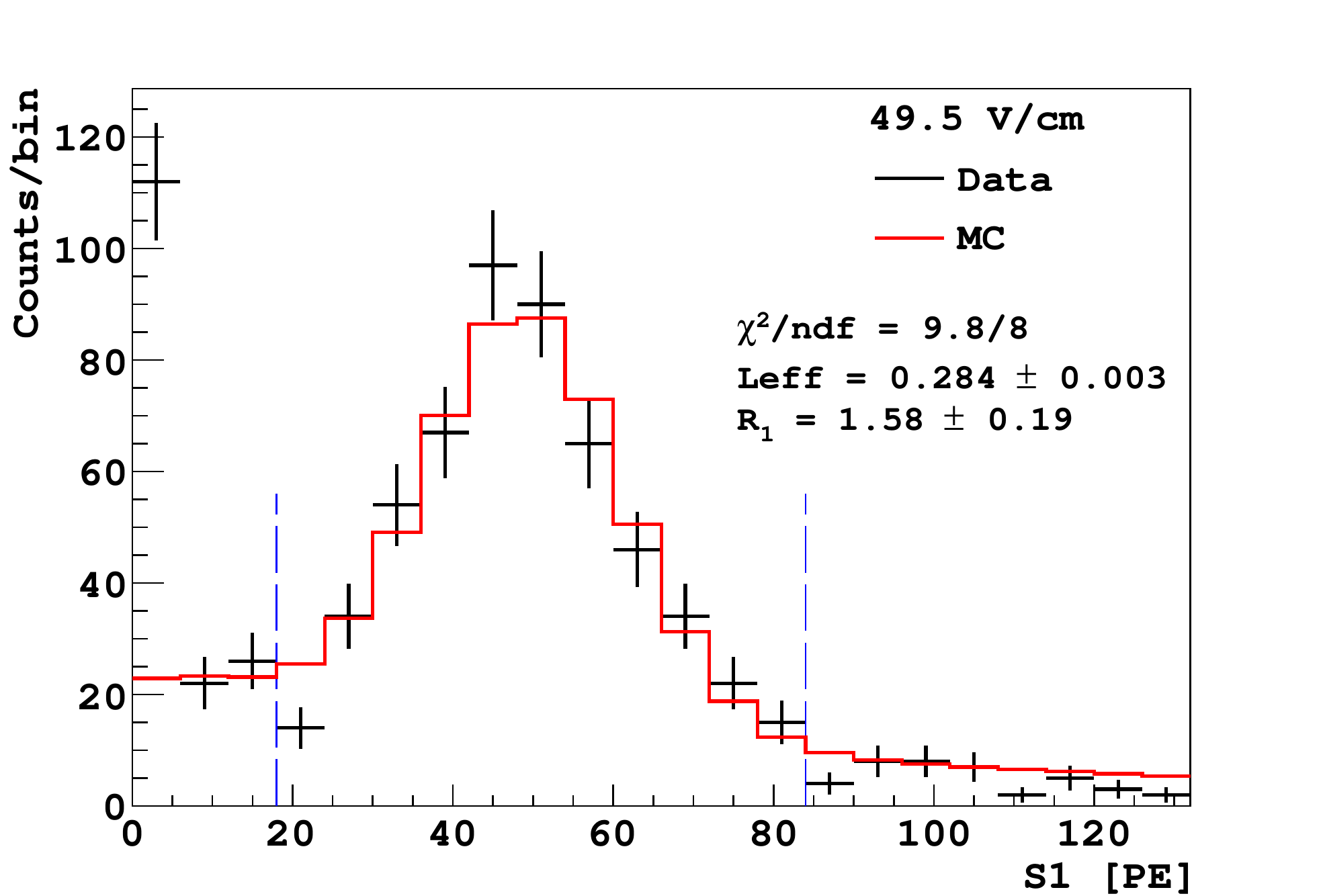}
\includegraphics[width=0.95\columnwidth]{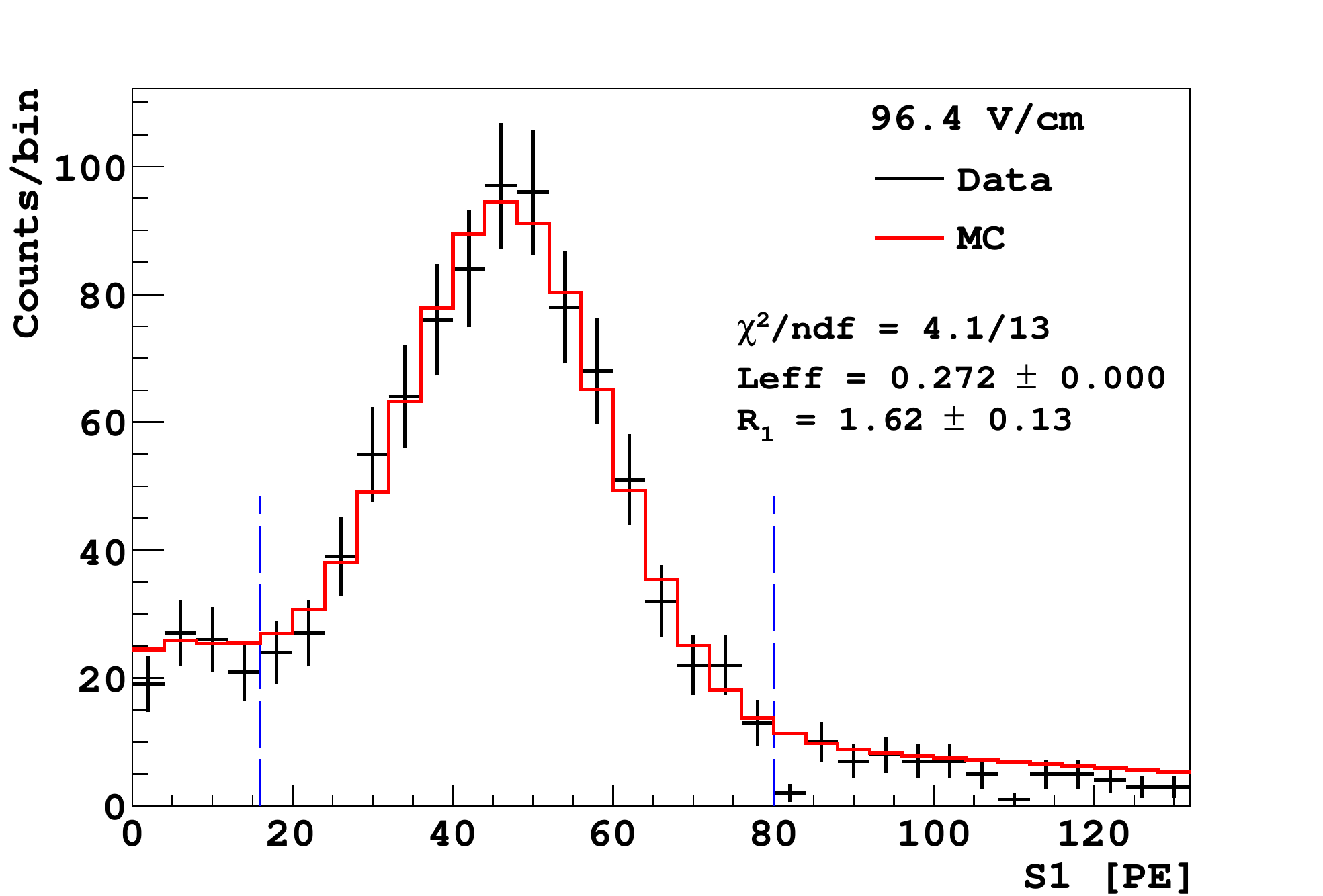}
\includegraphics[width=0.95\columnwidth]{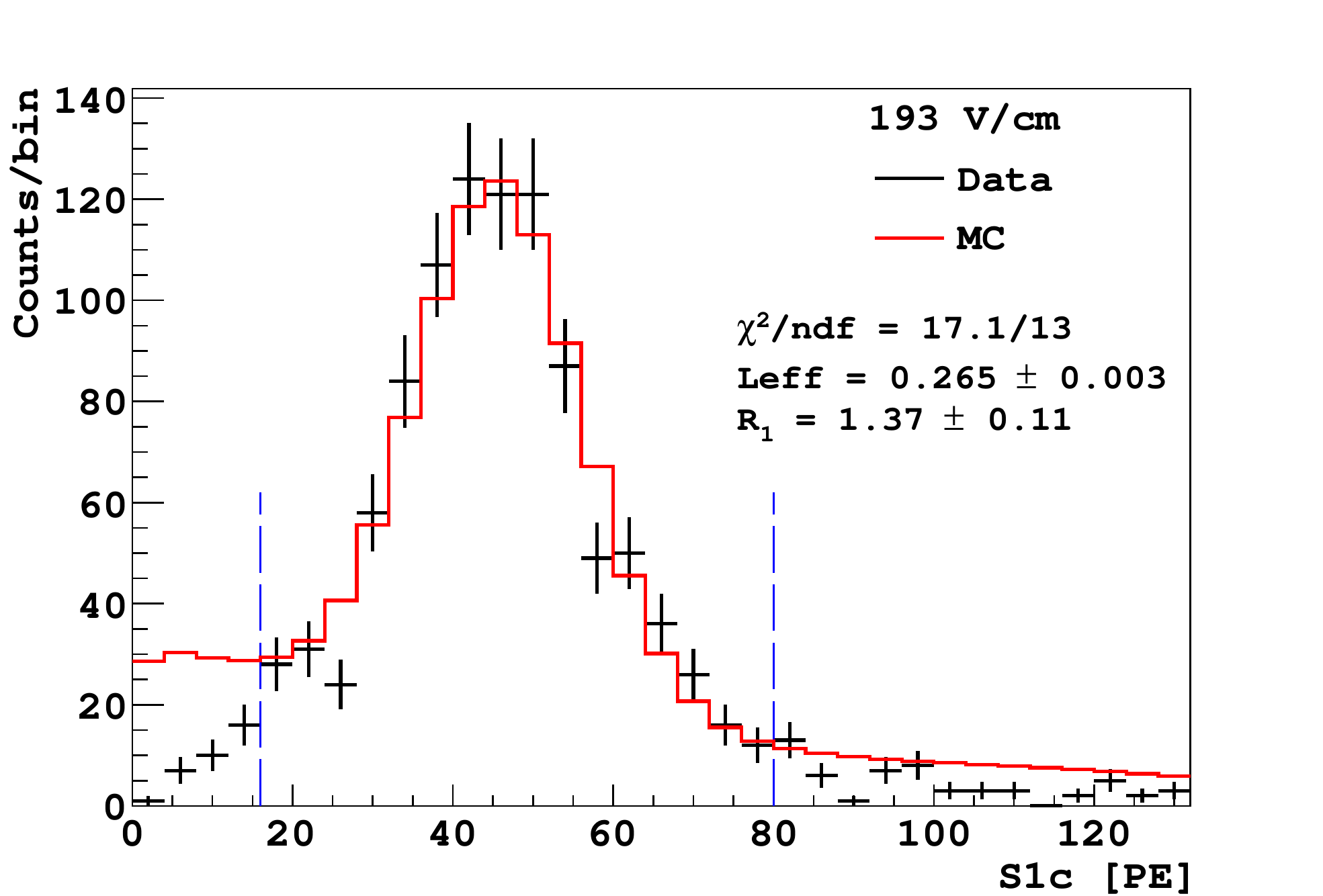}
\includegraphics[width=0.95\columnwidth]{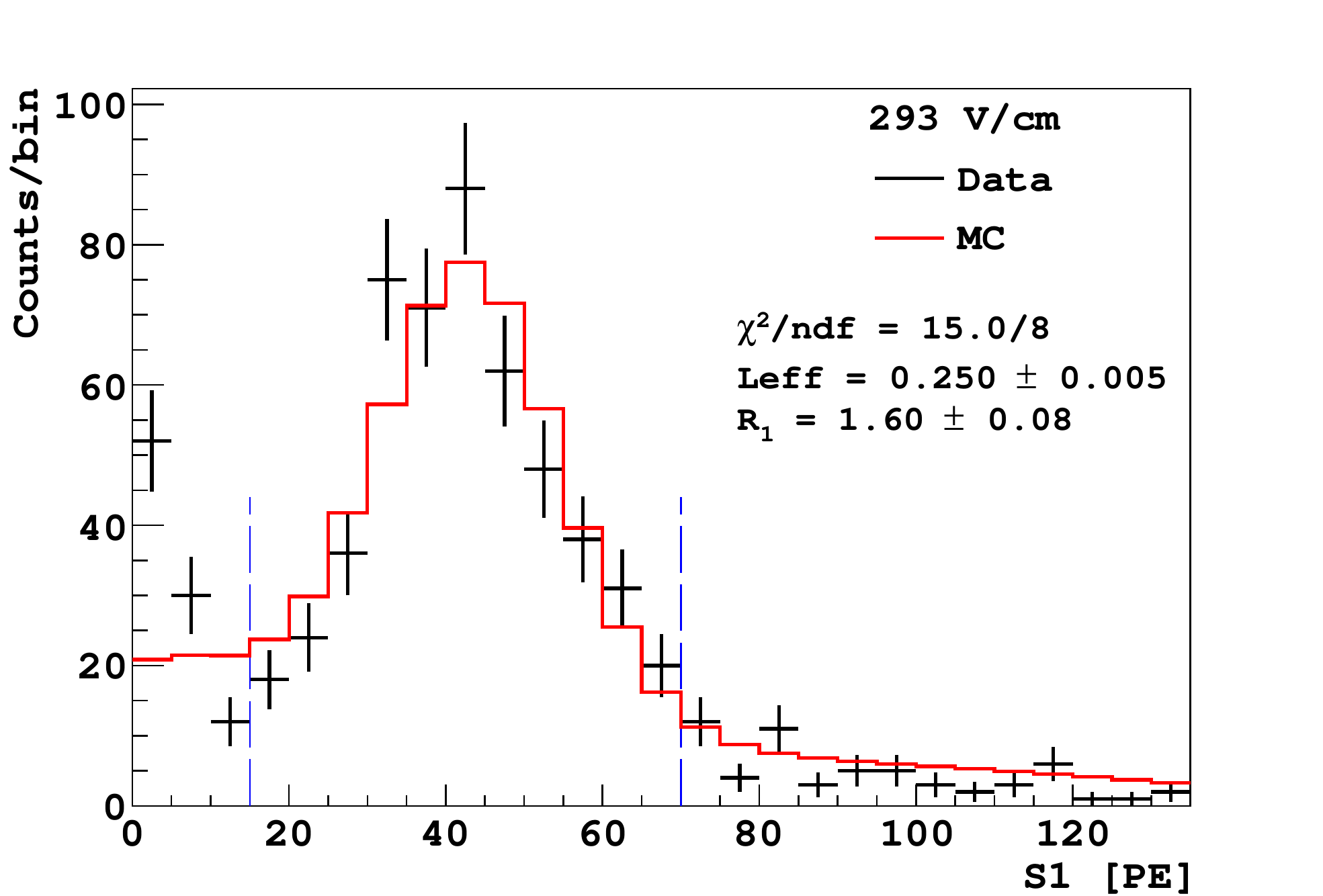}
\includegraphics[width=0.95\columnwidth]{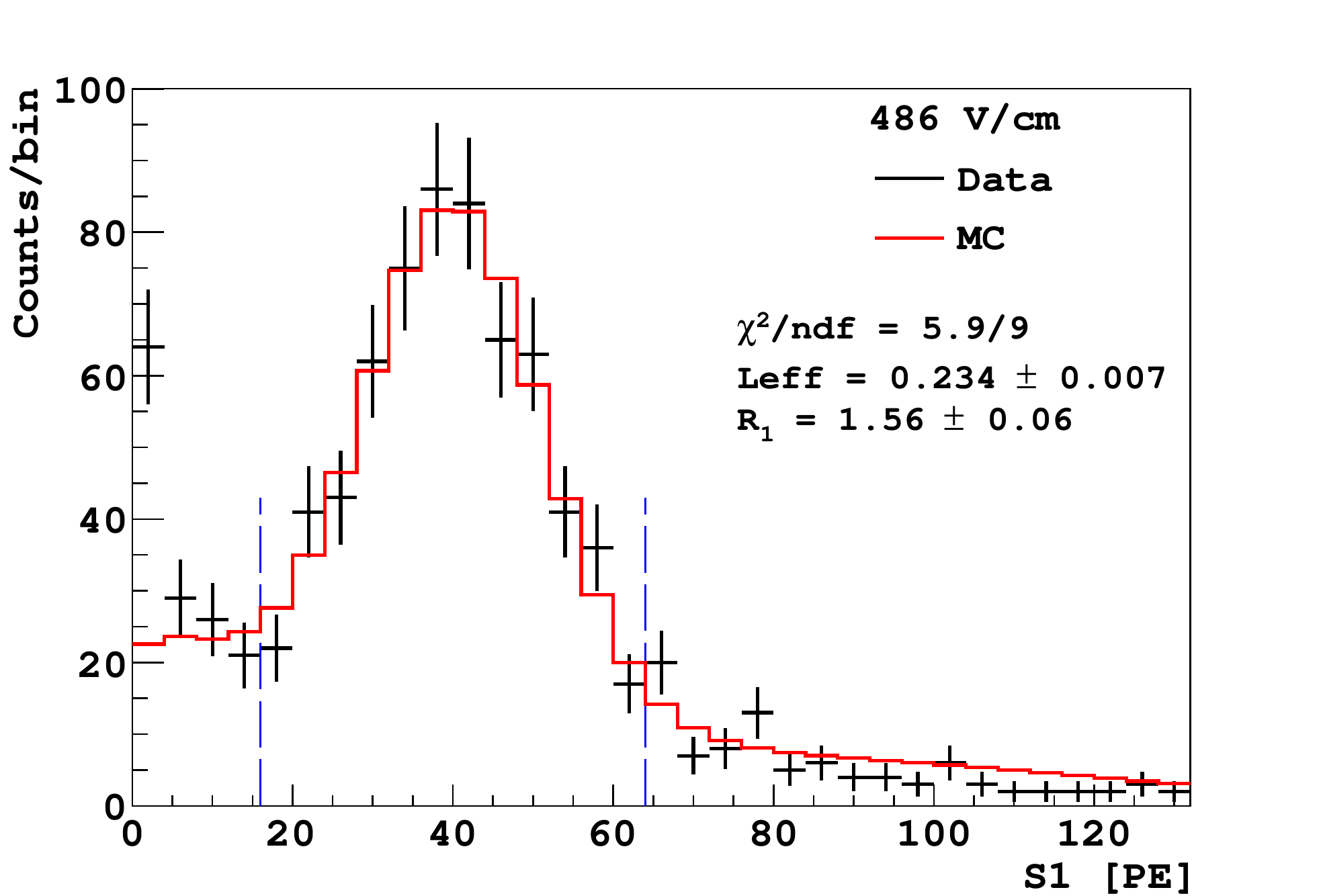}
\includegraphics[width=0.95\columnwidth]{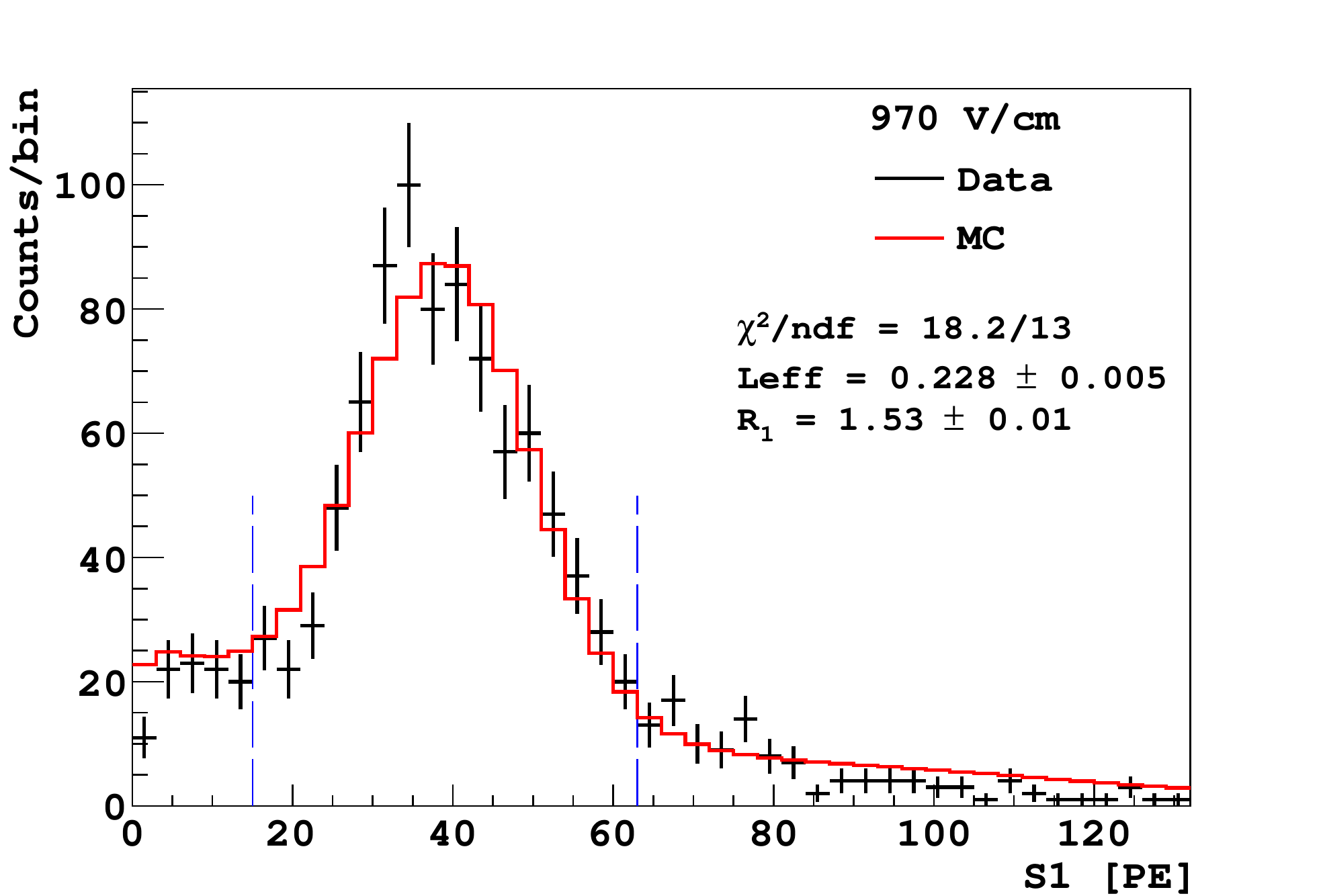}
\caption{\label{fig:leff36}
Top left panel.  {\bf \color{black} Black:} GEANT4-based simulation of the energy deposition in the SCENE detector at the setting devised to produce 36.1\,\kevr\ nuclear recoils.  {\bf \color{blue} Blue:} From neutrons scattered more than once in any part of the entire TPC apparatus before reaching the neutron detector.\\
All other panels.  {\bf \color{black} Black:} Experimental data collected for 36.1\,\kevr\ nuclear recoils.  {\bf \color{red} Red:} Monte Carlo fit of the experimental data.  The range used for each fit is indicated by the vertical {\bf \color{blue} blue} dashed lines.
}
\end{figure*}

\begin{figure*}[t!]
\includegraphics[width=0.95\columnwidth]{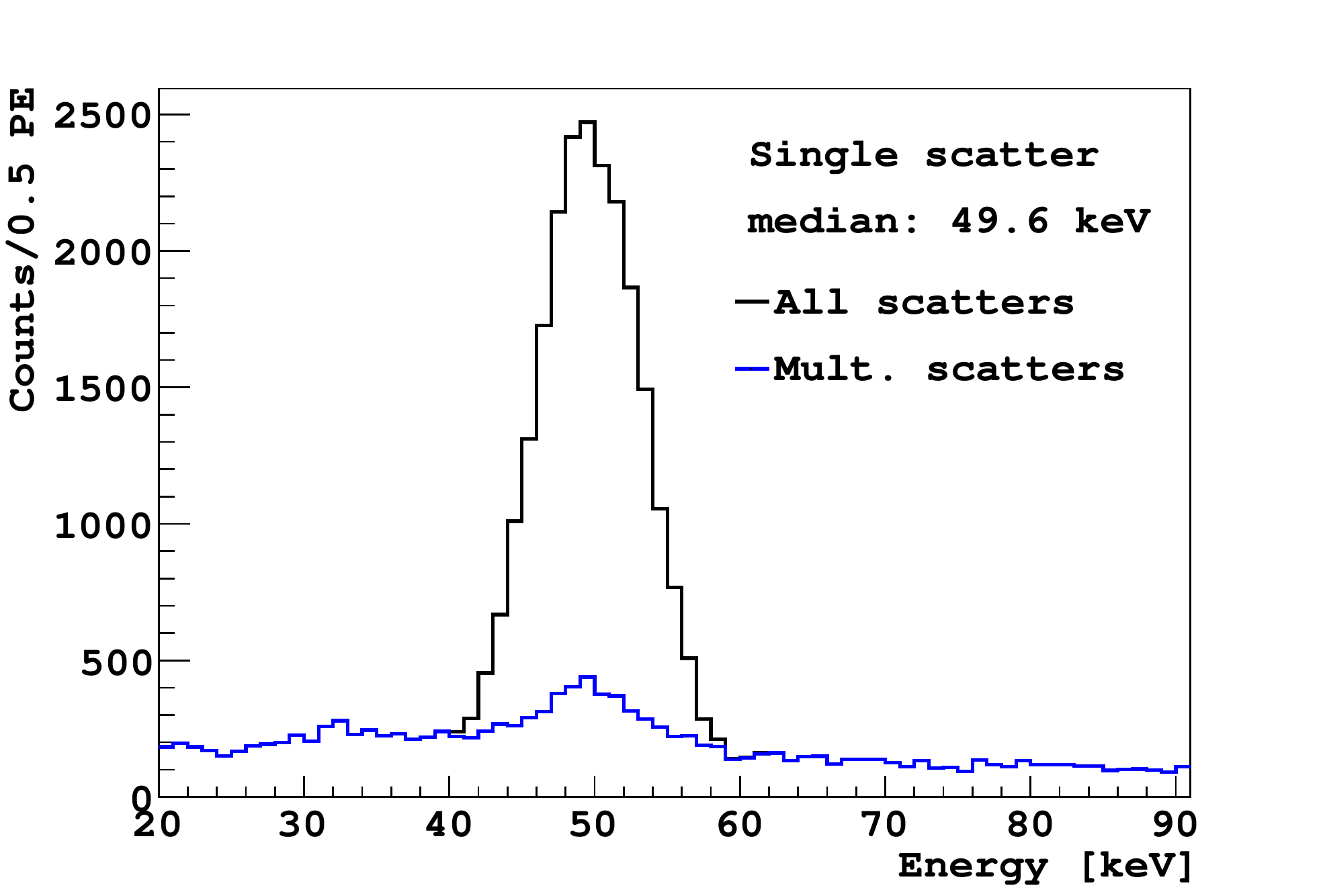}
\includegraphics[width=0.95\columnwidth]{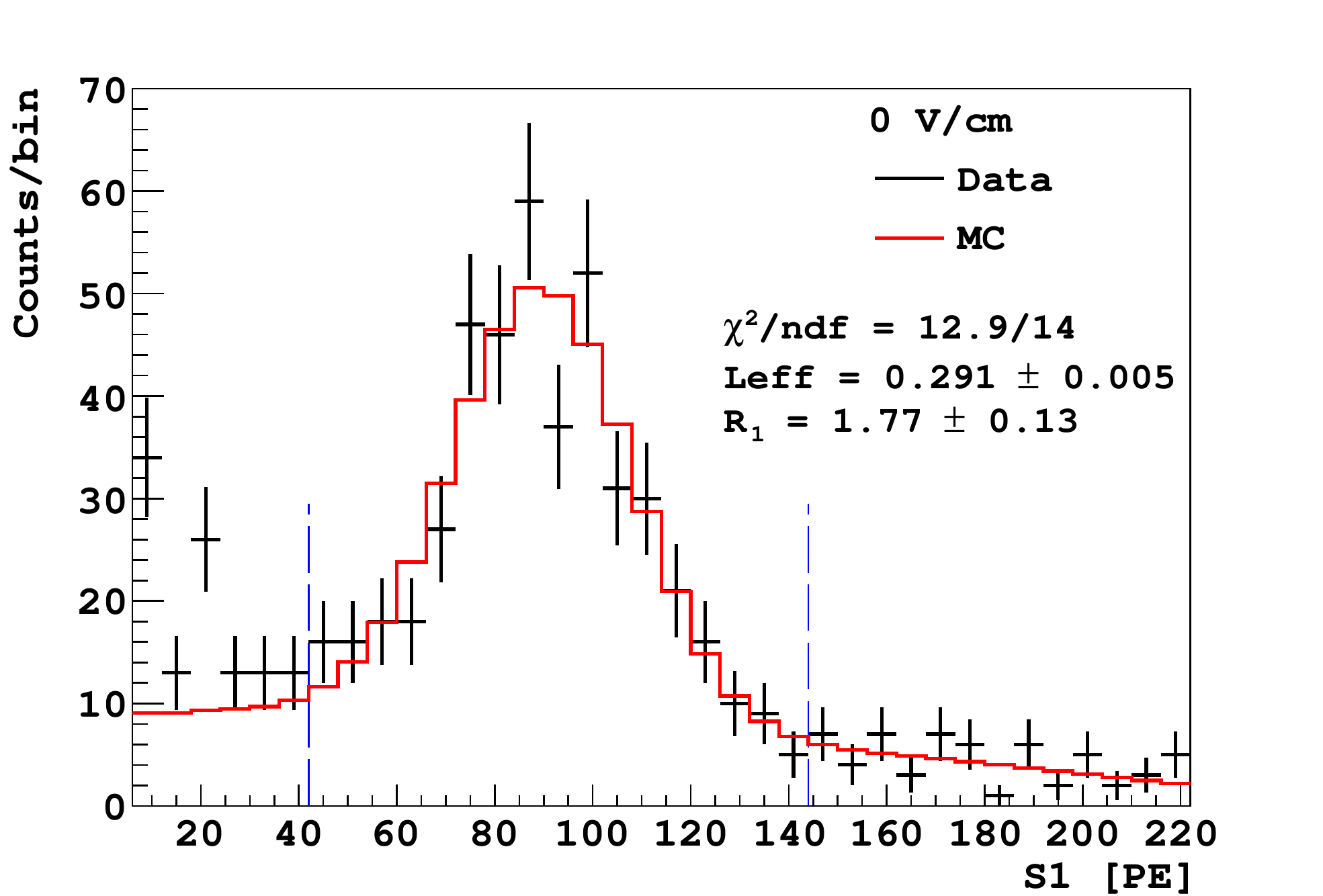}
\includegraphics[width=0.95\columnwidth]{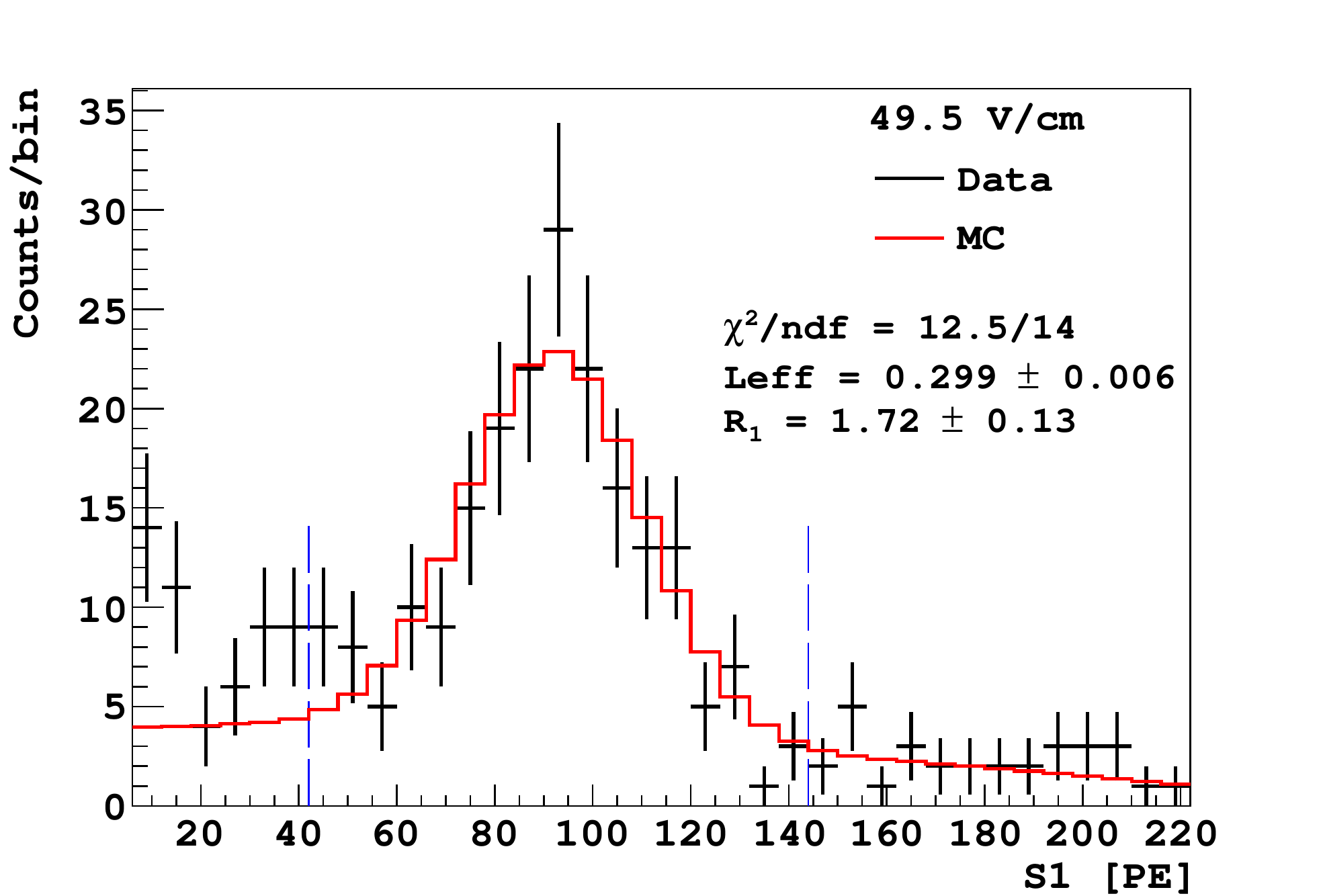}
\includegraphics[width=0.95\columnwidth]{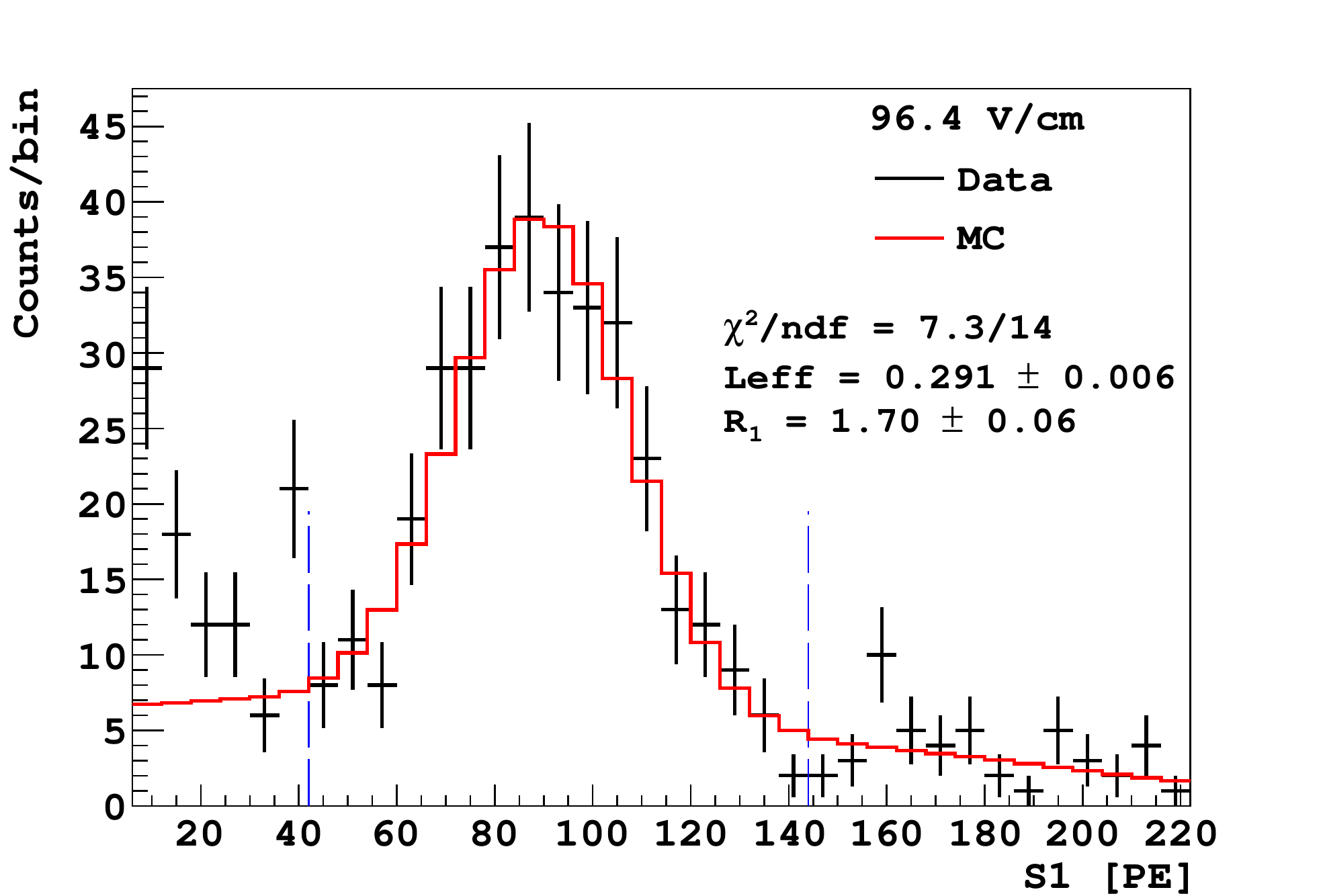}
\includegraphics[width=0.95\columnwidth]{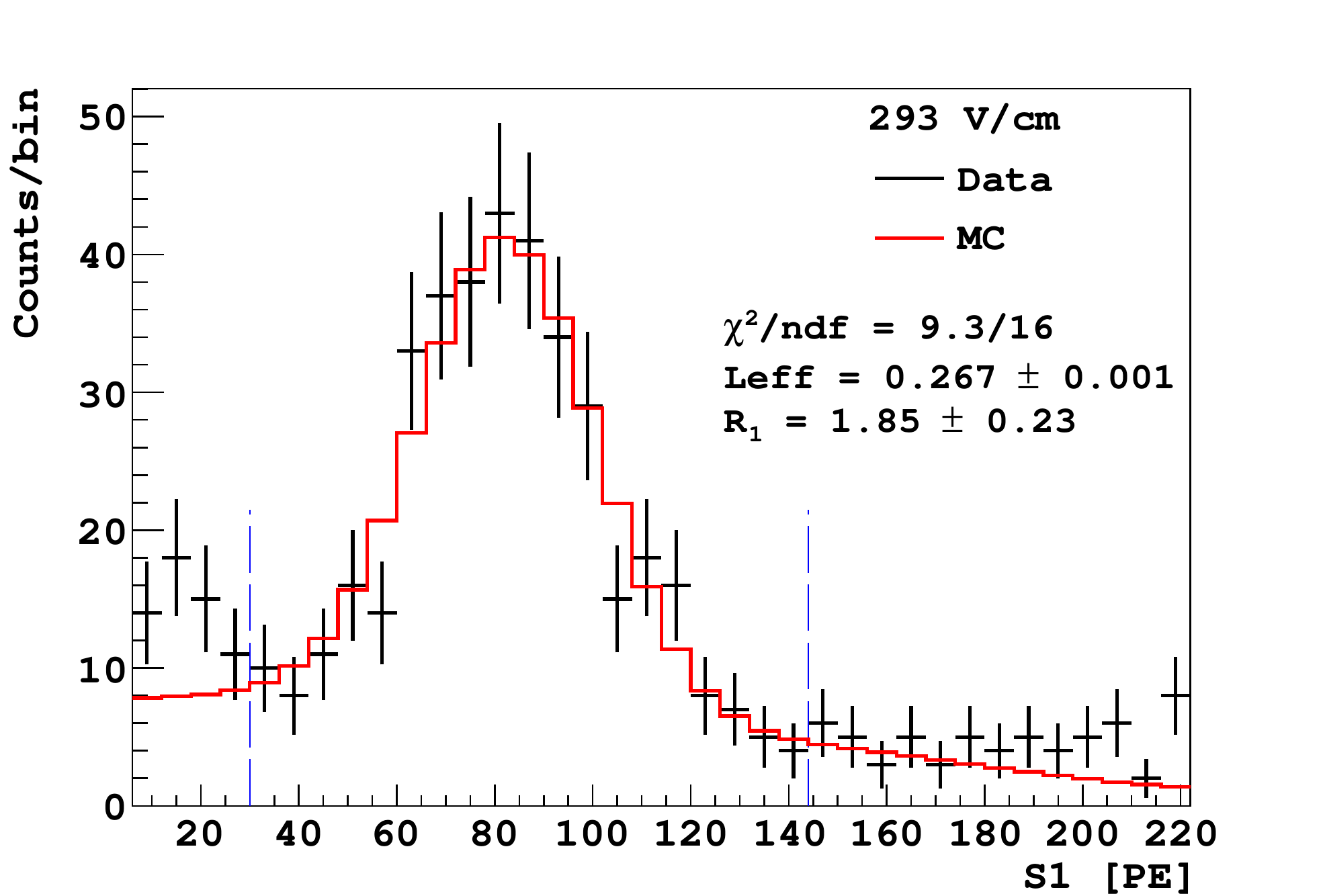}
\includegraphics[width=0.95\columnwidth]{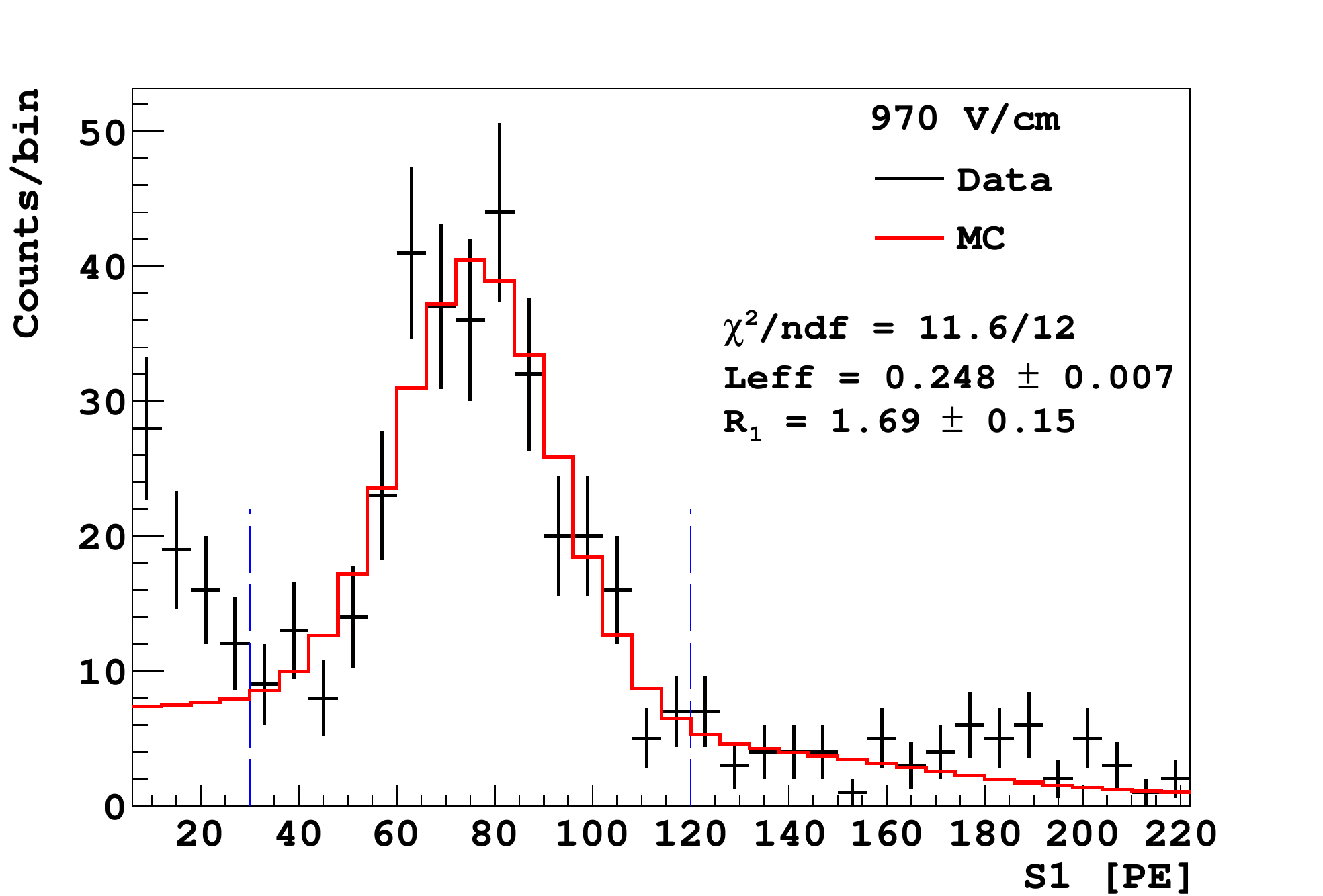}
\caption{\label{fig:leff50}
Top left panel.  {\bf \color{black} Black:} GEANT4-based simulation of the energy deposition in the SCENE detector at the setting devised to produce 49.7\,\kevr\ nuclear recoils.  {\bf \color{blue} Blue:} From neutrons scattered more than once in any part of the entire TPC apparatus before reaching the neutron detector.\\
All other panels.  {\bf \color{black} Black:} Experimental data collected for 49.7\,\kevr\ nuclear recoils.  {\bf \color{red} Red:} Monte Carlo fit of the experimental data.  The range used for each fit is indicated by the vertical {\bf \color{blue} blue} dashed lines.
}
\end{figure*}

\begin{figure*}[t!]
\includegraphics[width=0.95\columnwidth]{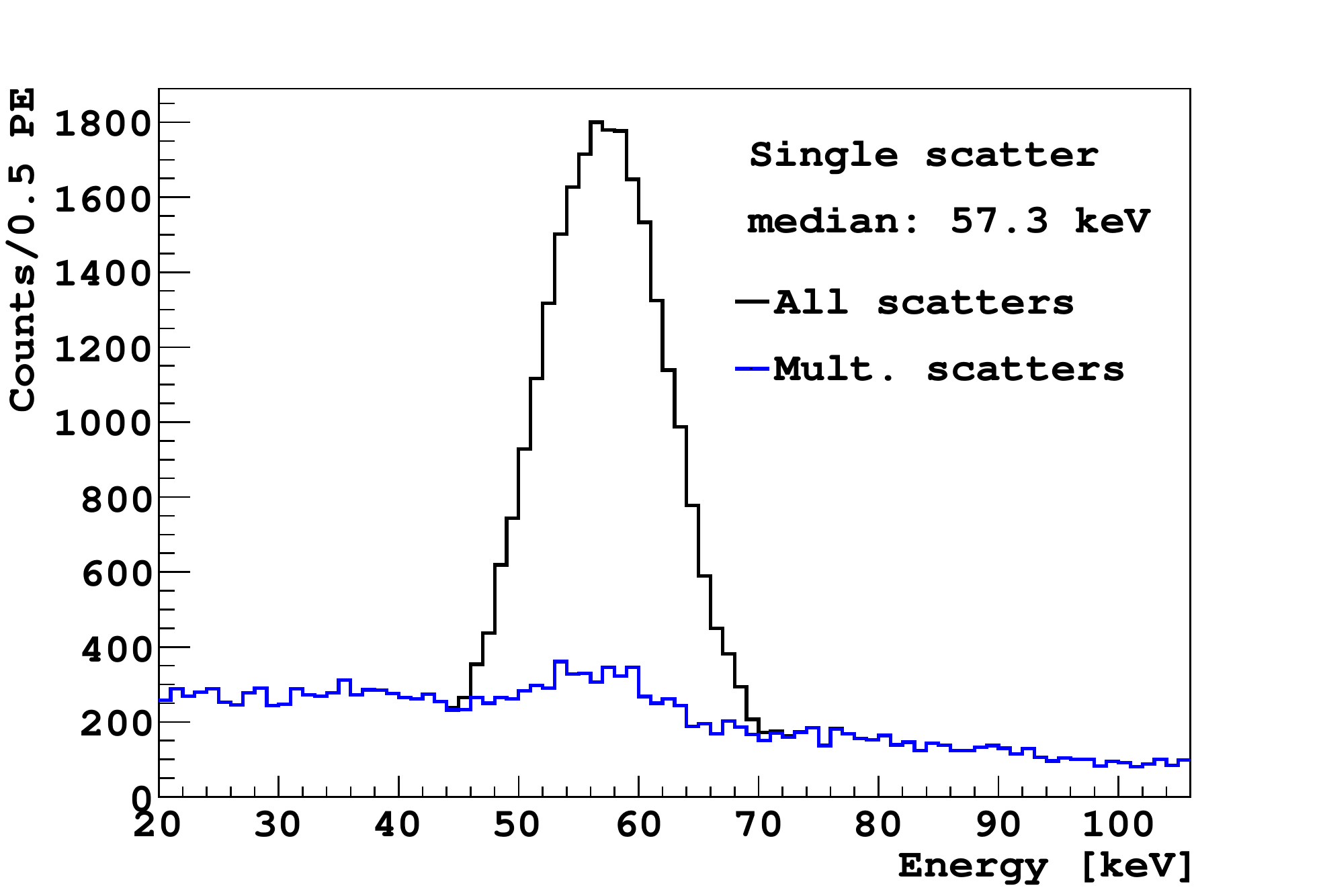}
\includegraphics[width=0.95\columnwidth]{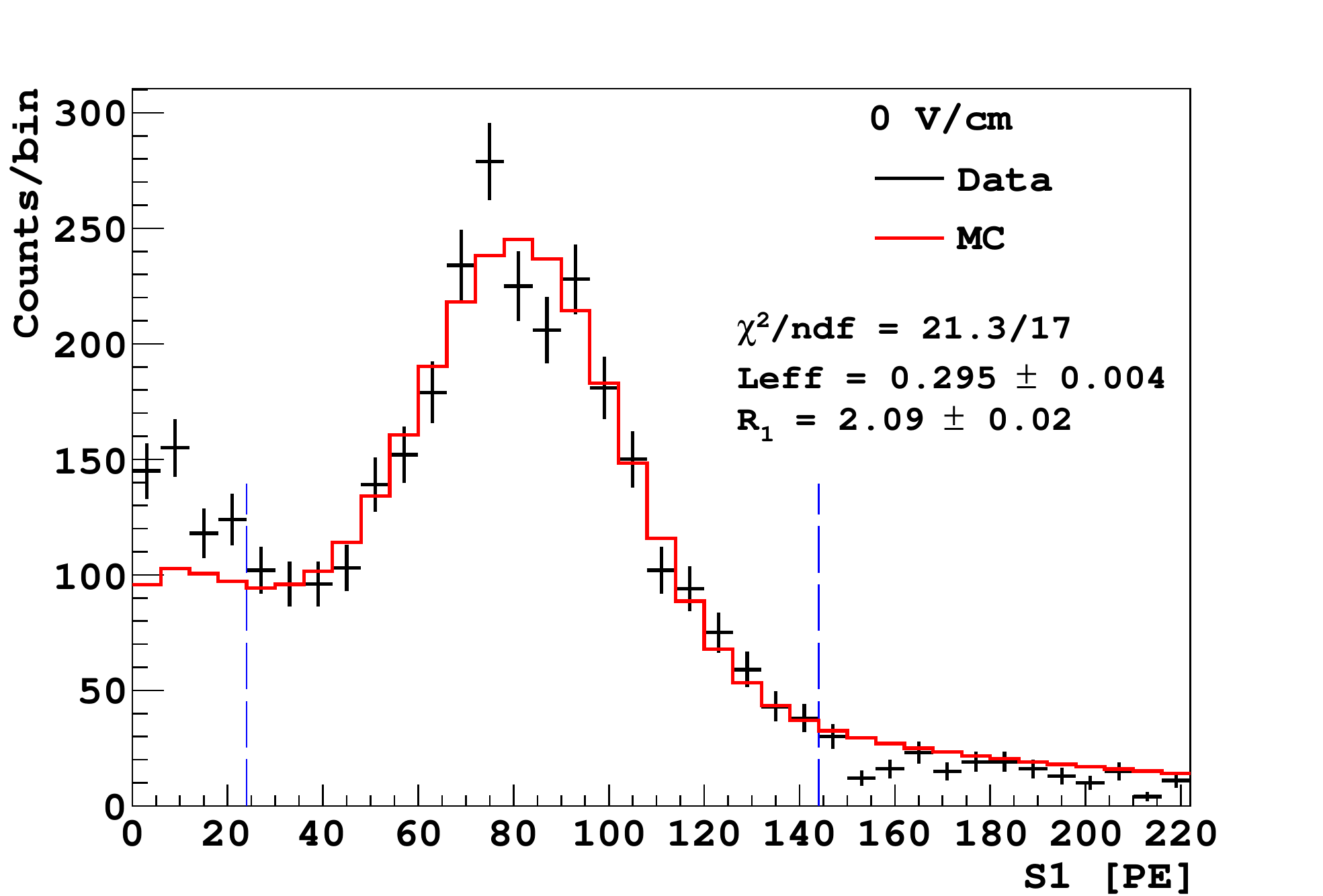}
\includegraphics[width=0.95\columnwidth]{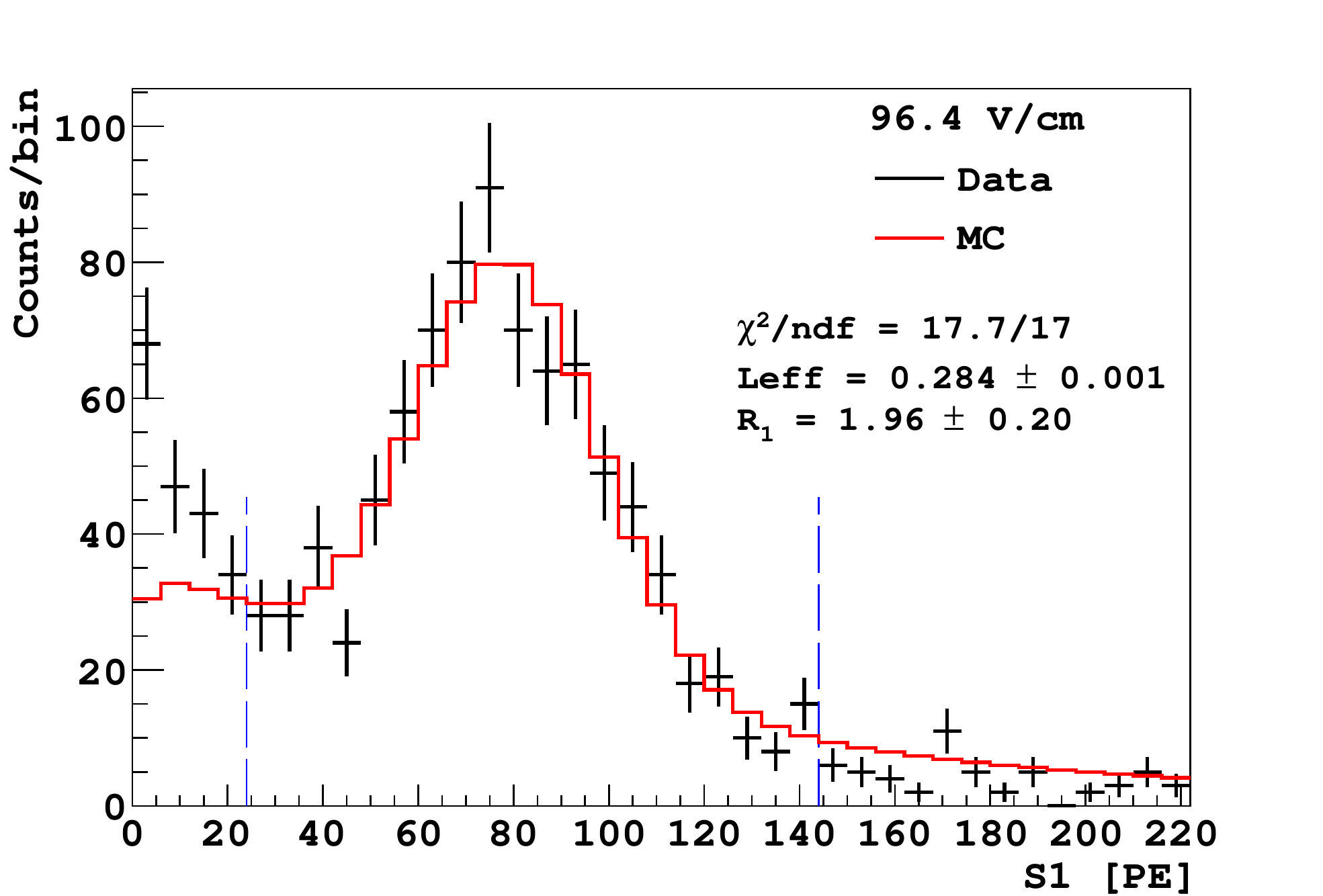}
\includegraphics[width=0.95\columnwidth]{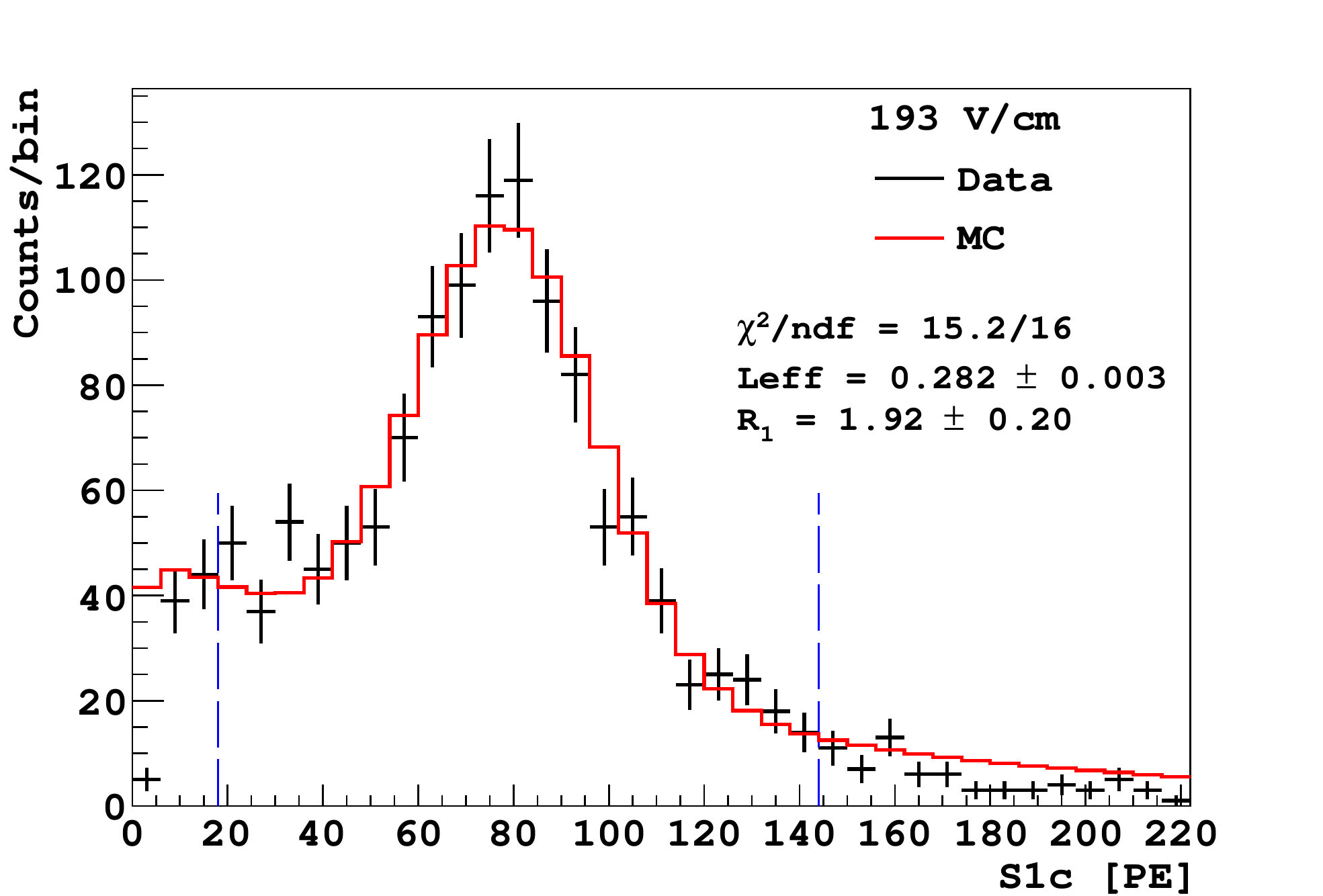}
\includegraphics[width=0.95\columnwidth]{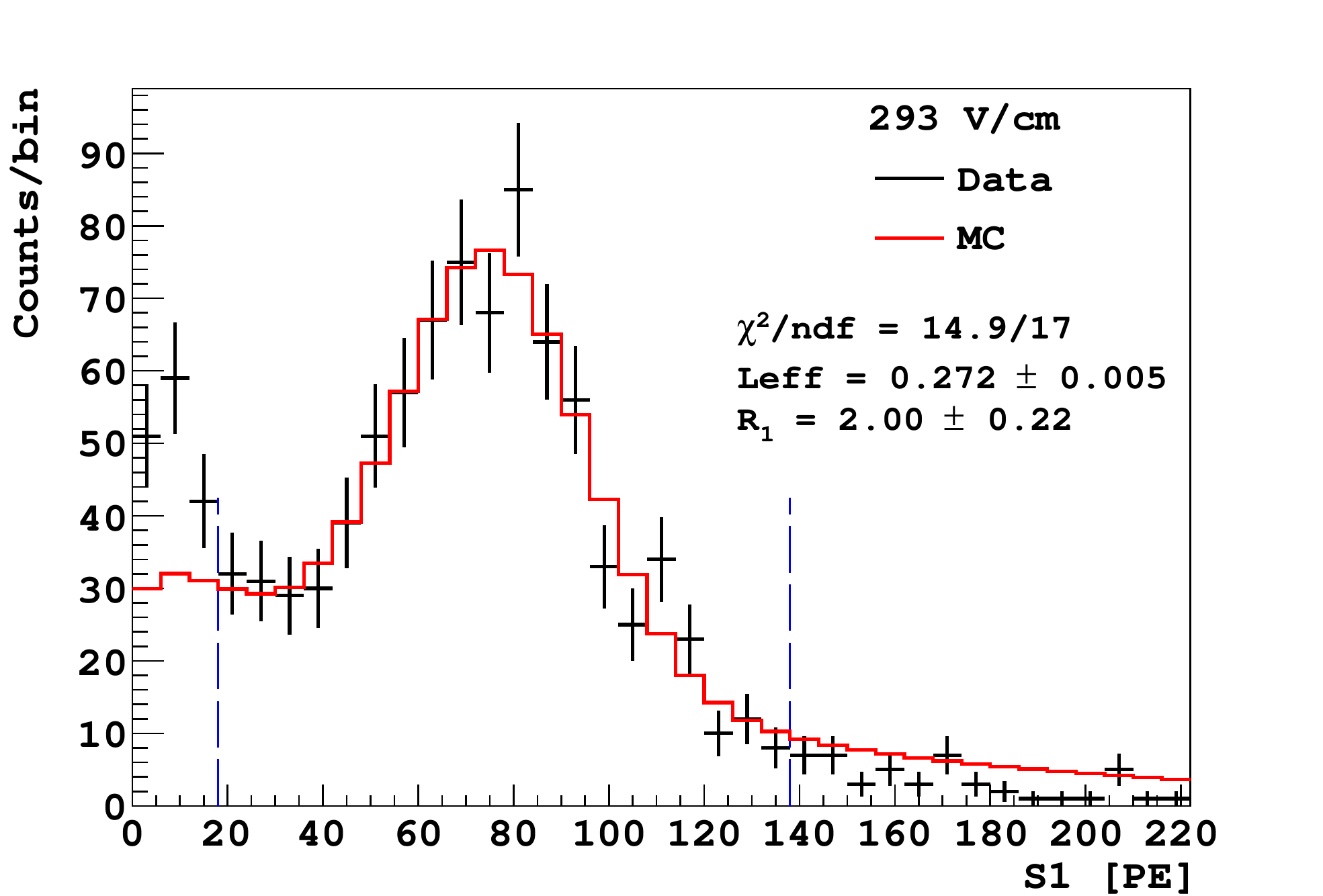}
\includegraphics[width=0.95\columnwidth]{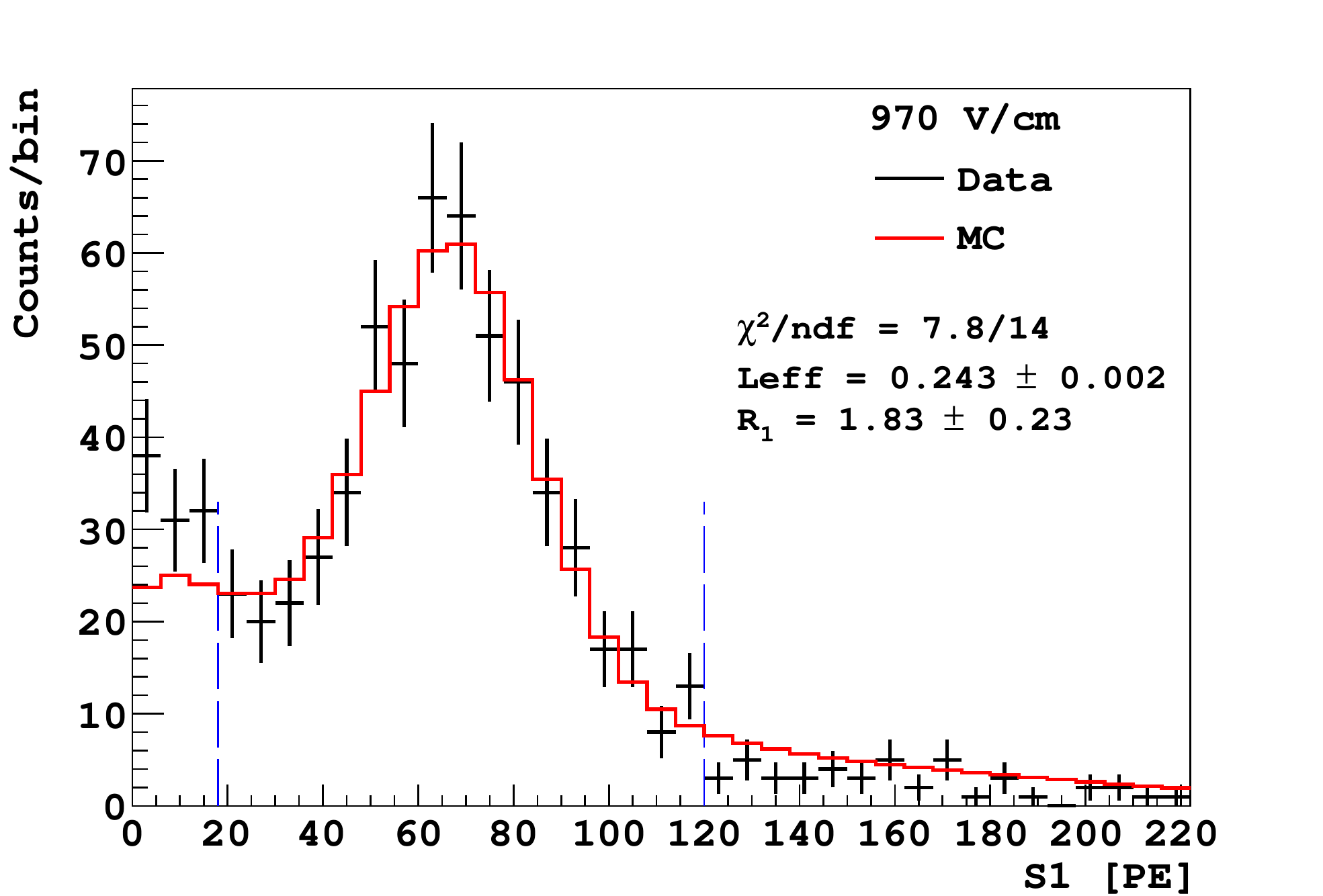}
\caption{\label{fig:leff58}
Top left panel.  {\bf \color{black} Black:} GEANT4-based simulation of the energy deposition in the SCENE detector at the setting devised to produce 57.3\,\kevr\ nuclear recoils.  {\bf \color{blue} Blue:} From neutrons scattered more than once in any part of the entire TPC apparatus before reaching the neutron detector.\\
All other panels.  {\bf \color{black} Black:} Experimental data collected for 57.3\,\kevr\ nuclear recoils.  {\bf \color{red} Red:} Monte Carlo fit of the experimental data.  The range used for each fit is indicated by the vertical {\bf \color{blue} blue} dashed lines.
}
\end{figure*}

The fit results for all ten recoil energies measured $-$ ranging from 10.3 to 57.3\,\kevr\ $-$ and all drift fields investigated $-$ ranging from 0 to 970\,V/cm $-$ are shown in Figs.~\ref{fig:leff11} to \ref{fig:leff58}. 
In each of the figures, the plot in the top left panel shows the simulated energy spectrum for all scatters along with those from multiple scatters (the plot for the 10.3\,\kevr\ nuclear recoils is absent in Fig.~\ref{fig:leff11} since it is already shown in Fig.~\ref{fig:mc}).  All other panels show the experimental data at a given drift field fit with Monte Carlo data. Apart from the low \sone\ region, the agreement between the data and the MC prediction is remarkably good.

\begin{figure}[t!]
\includegraphics[width=\columnwidth]{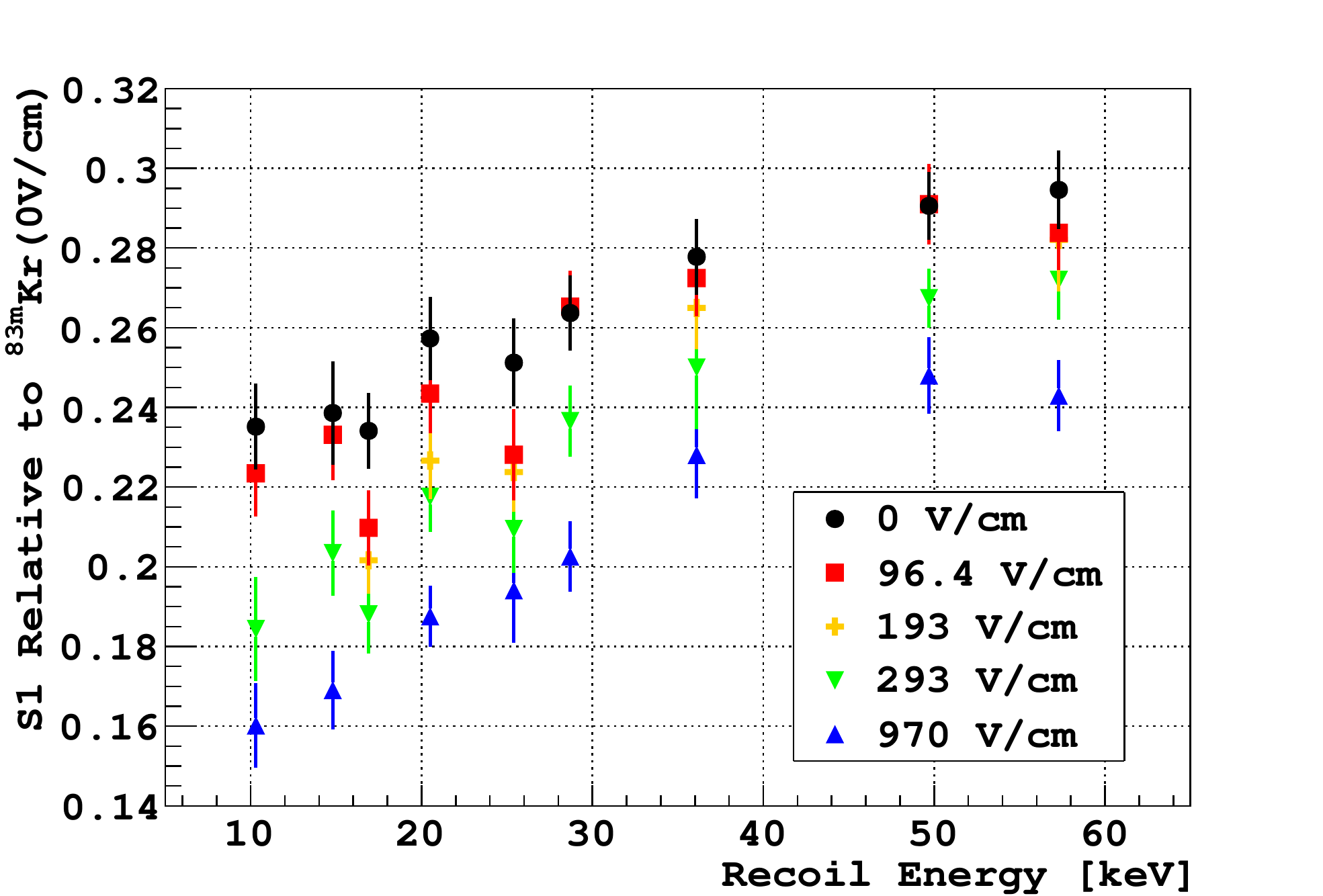}
\caption{\label{fig:leff}S1 yield as a function of nuclear recoil energy measured at five drift fields (0, 96.4, 193, 293 and 970 V/cm) relative to the light yield of \krthree\ at zero field.}
\end{figure}

\begin{figure}[t!]
\includegraphics[width=\columnwidth]{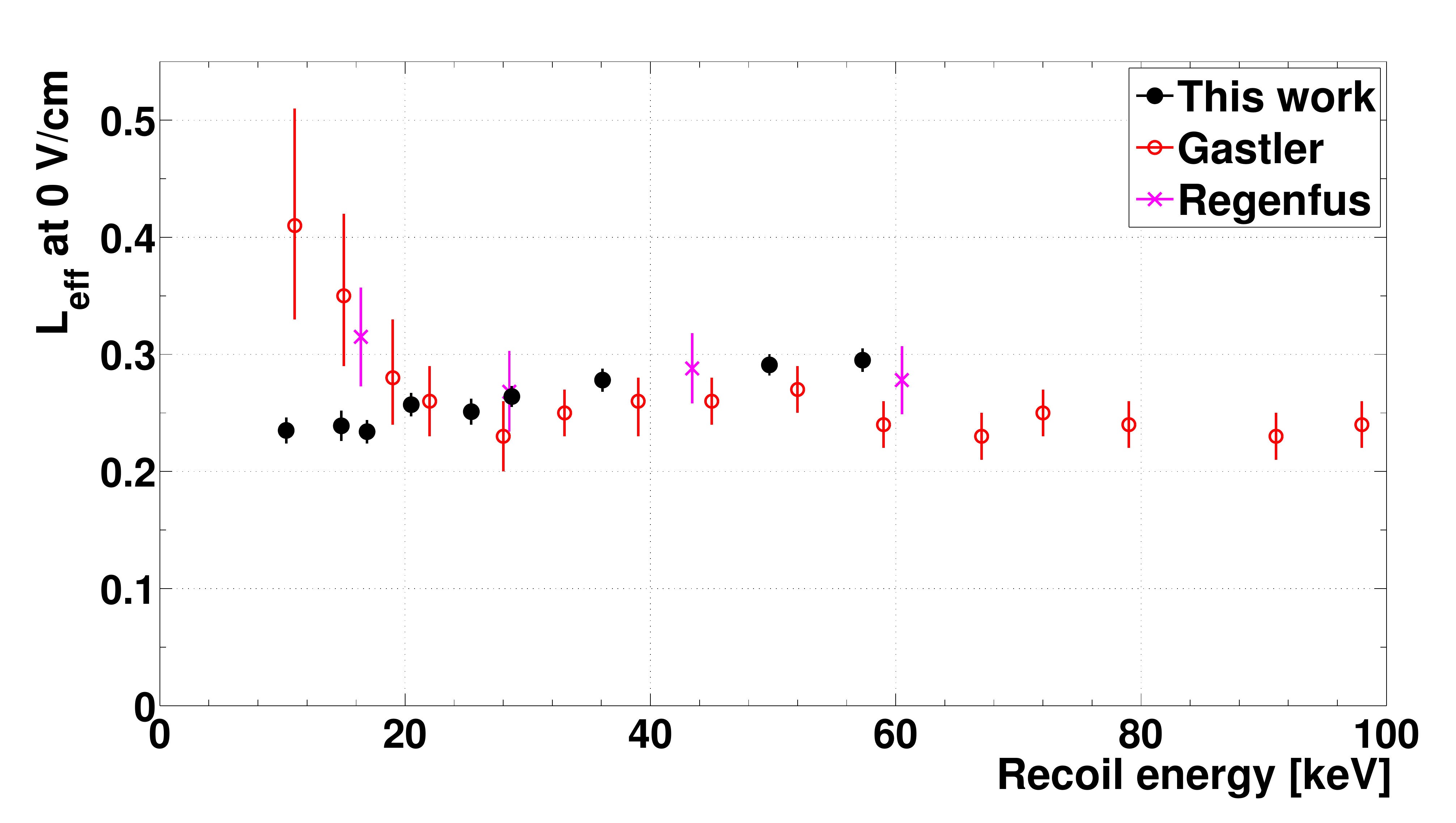}
\caption{\label{fig:leff_comp}S1 yield as a function of nuclear recoil energy measured at zero field relative to the light yield of \krthree\ at zero field, compared to previous measurements\cite{gastler,regenfus}.}
\end{figure}

\begin{table*}[t!]
\begin{center} 
\begin{tabular*}{\textwidth}{@{\extracolsep{\fill} } lrrrrrrrrr} \hline\hline
Recoil energy~[\kevr]	& 10.3 & 14.8 & 16.9 & 20.5 & 25.4 & 28.7 & 36.1 & 49.7 & 57.3 \\ 
\hline 
\leff & 0.235 & 0.239 & 0.234 & 0.257 & 0.251 & 0.264 & 0.278 &  0.291 & 0.295\\
Statistical error	 & 0.003 & 0.005 & 0.004 & 0.001 & 0.005 & 0.004 & 0.003 & 0.005 & 0.004\\
Systematic error source\\
\quad Fit method & 0.001 & 0.000 & 0.004 & 0.004 & 0.002 & 0.001 & 0.003 & 0.001 & 0.002\\
\quad Fit range & 0.000 & 0.002 & 0.000 & 0.001 & 0.002 & 0.000 & 0.001  & 0.000 & 0.000\\
\quad TPCtof cut & 0.002 & 0.003 & 0.003 & 0.001 & 0.002 & 0.001 & 0.001  & 0.001 & 0.001\\
\quad Ntof cut & 0.004 & 0.002 & 0.001 & 0.001 & 0.002 & 0.004 & 0.001  & 0.003 & 0.001\\
\quad f90 cut & 0.004 & 0.004 & 0.003 & 0.001 & 0.000 & 0.001 & 0.000  & 0.000 & 0.000\\
\quad \krthree\ light yield & 0.005 & 0.005 & 0.005 & 0.005 & 0.005 & 0.005 & 0.006  & 0.006 & 0.006\\
\quad Recoil energy\\
\quad \quad TPC position & 0.001 & 0.001 & 0.001 & 0.001 & 0.001 & 0.001 & 0.001  & 0.001 & 0.001\\
\quad \quad EJ301 position & 0.007 & 0.010 & 0.005 & 0.008 & 0.008 & 0.005 &  0.006 & 0.003 & 0.006\\
Combined error total & 0.011 & 0.013 & 0.010 & 0.010 & 0.011 & 0.009 & 0.010 & 0.009 & 0.010\\
\hline
\end{tabular*}
\caption{Summary of error contributions to individual \leff\ measurements at \Edrift\ = 0. Only minor variations in the magnitude of systematic errors were observed across the range of drift field explored. The combined error for each measurement is shown Fig.~\ref{fig:leff}.}
\label{table:leffsys}
\end{center}
\end{table*}

Figure~\ref{fig:leff} shows the resulting values of \leff\ as a function of \Enr\ as measured at five different drift fields (0, 96.4, 193, 293 and 970\,V/cm).  The error bar associated with each \leff\ measurement represents the quadrature combination of the statistical error returned from the fit and the systematic errors due to each of the sources accounted for (see Table~\ref{table:leffsys} for a detailed account of systematic errors at null drift field). Figure~\ref{fig:leff_comp} shows our values of \leff\ at zero field compared to previous measurements~\cite{gastler,regenfus}.  Our results do not show the increase at low energies previously observed. It should be noted that \krthree\ did not provide the electronic recoil energy scale reference in the earlier measurements, but both groups report a linear response to electronic recoils in the relevant energy range~\cite{lippincott_kr,regenfus}.

In order to assess any bias introduced by our Monte Carlo model in the fit, we also fit each of the data sets with a Gaussian function plus a first order polynomial to account for background.  The difference between the results of the two methods is listed in Table~\ref{table:leffsys} in the row ``Fit Method'' for \Edrift\,=\,0.  Across all measured recoil energies and drift electric fields, this systematic error is less than 2\%. The sensitivity of \leff\ to the fit range selection is characterized by comparing the fit results to those obtained with a reduced fit range.  We define the reduced range by raising the lower bound by 10\% of the original fit range and lowering the upper bound by the same amount.   The original fit ranges can be found in Figs.~\ref{fig:leff11} to \ref{fig:leff58}. 

We evaluated the systematic error in \leff\ from the TOF window selection by advancing or delaying the TPCtof cut by 3\,ns while holding the Ntof cut constant, and vice versa, while keeping the same fit function described above and based on the Monte Carlo-generated spectra.  We determined the associated systematic error as the average of the absolute difference in \leff\ obtained by either advancing or delaying the TOF window.

Within the data set from a specific recoil energy and field setting, the \lartpc\ light yield determined with the \krthree\ source fluctuated with a standard deviation of about 1\%.  In addition to such short term fluctuations, changes in the purity of the LAr result in variations of the light yield and the observed \fno\ parameter. Impurities also affect the mean life of the triplet state of the S1 scintillation time profile~\cite{warpO2,warpN2,jones}. In our analysis, we created average S1 waveform traces by summing together the baseline-subtracted waveforms of individual events, aligned by their peak position. We then performed a two parameter fit of the triplet lifetime to a simple exponential function, without a constant baseline term, in the range between 0.75 and 7.5 $\mu$s.  The lifetime is measured to lie in the range from 1.39 to 1.48\,$\mu$s for all data presented here. 

\begin{figure}[t!]
\includegraphics[width=\columnwidth]{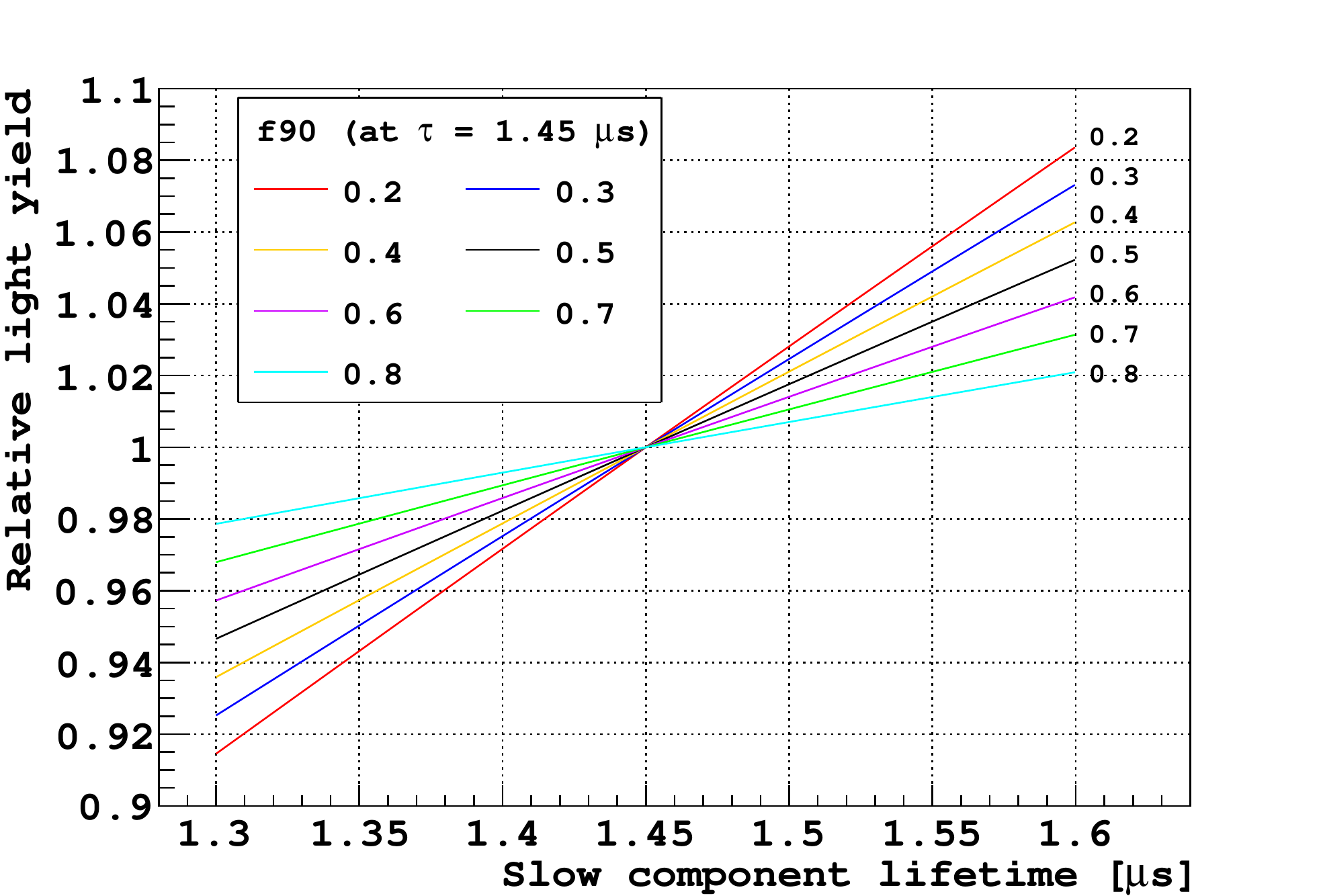}
\caption{\label{fig:f90_sim}Systematic error induced by chemical impurities affecting the mean life of the triplet component of the S1 scintillation spectrum, as a function of mean life in the range of interest. The S1 time profile was simulated with two exponential decay terms. Each line represents the events with a given \fno\ when the slow component lifetime is 1.45 $\mu$s. Note that \fno\ increases slightly with the decrease in the slow component lifetime.}
\end{figure}

The effect of impurities on the light output  and \fno\ both depend on the value of \fno, it being a measure of the relative importance of the slow component in the total light output. Figure~\ref{fig:f90_sim} shows the fractional reduction in light output for different values of the slow component lifetime for different values of \fno, with 1.45~$\mu$s taken to be the nominal value. As an example, for a measured lifetime of 1.38~$\mu$s, the effect on light output is calculated to be a reduction of $3.5\%$ when \fno\ is 0.3 and a reduction of about $1\%$ at \fno\ of 0.7; the effect is larger for low values of \fno\, where the long lifetime component is more important. The effect on \fno\ itself can also be calculated and for the same measured lifetime of 1.38 $\mu$s, the effect is an increase of \fno\ at 0.3 by $4\%$ to 0.311 and of \fno\ at 0.7 by $1.5\%$ to 0.711. These effects are included in the contribution of the \krthree\ light yield to the overall systematic uncertainties on \leff. 

The uncertainty due to the alignment of the TPC and neutron detectors was calculated assuming a $\pm$1\,cm uncertainty in our determination of their absolute positions relative to the production target.

\section{Energy Resolution}

A number of factors, including the width of \ser\ of the PMT's, the position dependence of light collection in the LAr TPC, \pe\ counting statistics, and the intrinsic resolution of LAr scintillation, contributed to the energy resolution of the detector $\sigma_1$=$\sqrt{S1}$\Rone$\left(E,\Edrift\right)$.

\begin{figure*}[t!]
\includegraphics[width=\columnwidth]{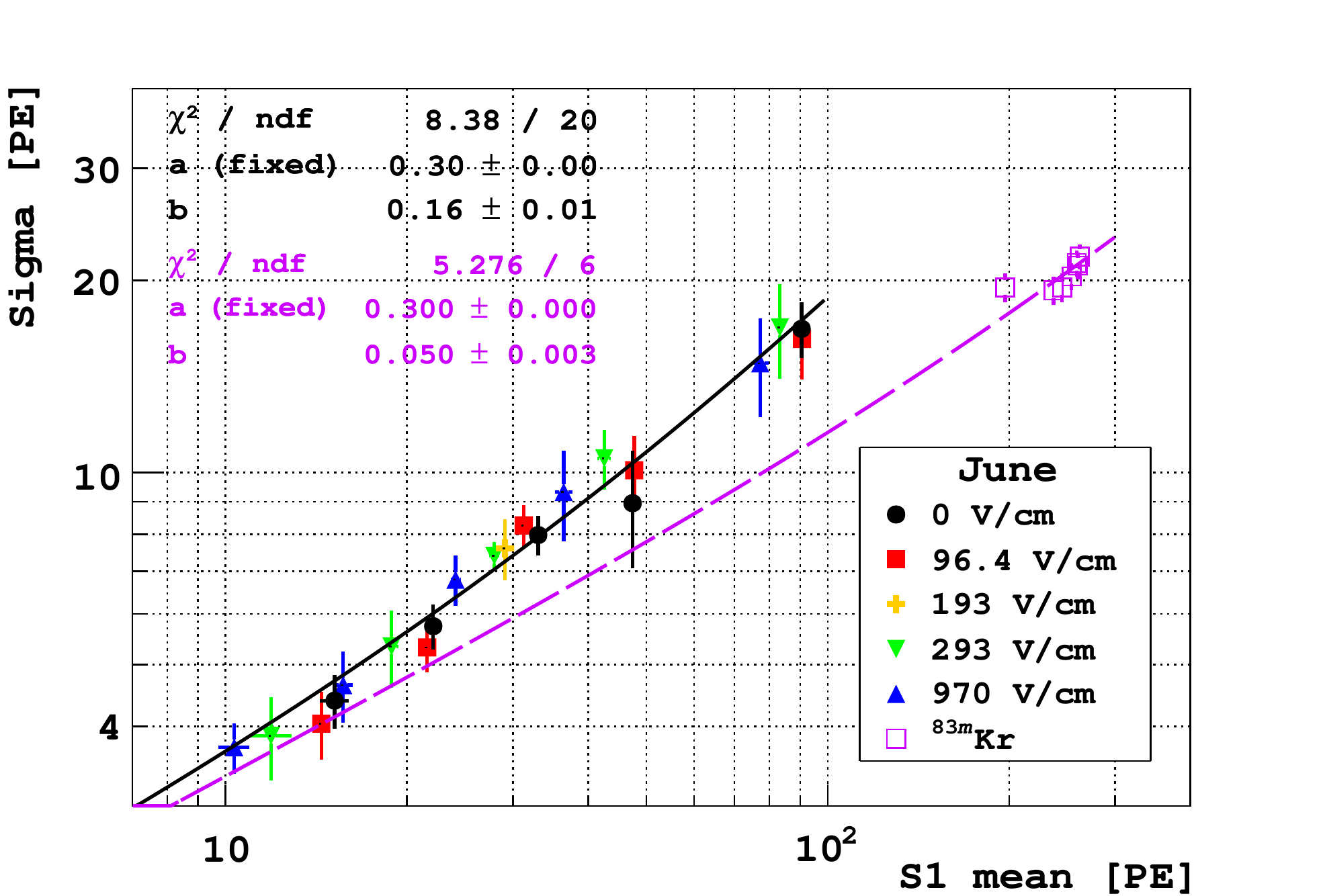}
\includegraphics[width=\columnwidth]{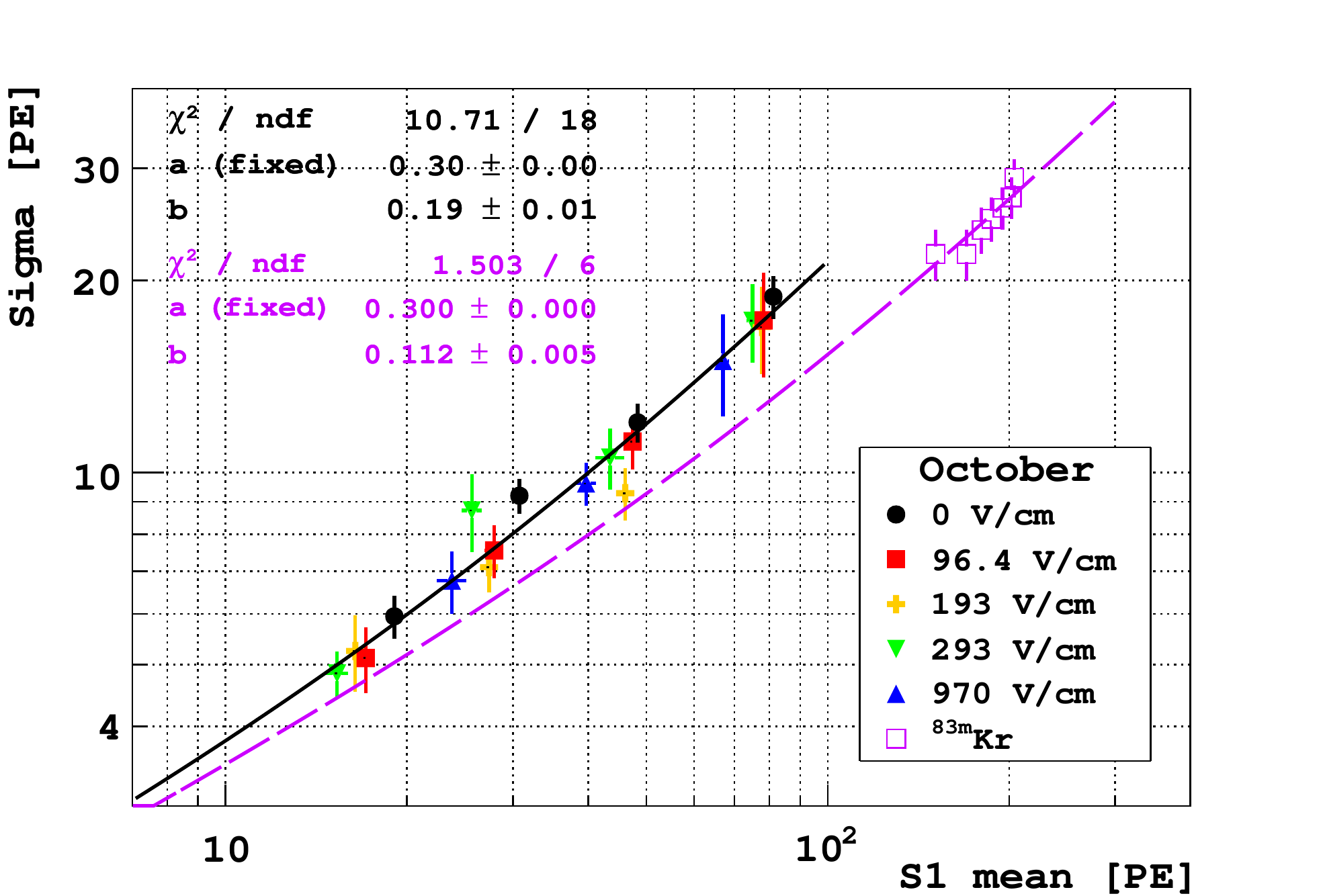}
\caption{\label{fig:resolution}Energy resolution, $\sigma_1$, of the nuclear recoils extracted from the Monte Carlo fit, as a function of recoil energy for all drift field combinations.  We separate the results from the June and October~2013 runs.  The resolution for the nuclear recoils, $\sigma_1$, is fit ({\bf \color{black} black continuous curve}) with the function described in the text and compared with the fit obtained for the \krthree\ ({\bf \color{magenta} purple dashed curve}) with the same function.}
\end{figure*}

We assumed that the contribution from the spread in nuclear recoil energy due to the geometry of the detectors was fully accounted for by the Monte Carlo fit function.  Our fits for \Rone\ as a function of recoil energy and drift field showed a dependence of $\sigma_1$ upon \sone\ deviating from Poisson statistics.  Results from the June and October~2013 runs are plotted separately in Fig.~\ref{fig:resolution}.  The dependence of $\sigma_1$ on \sone\ is in both cases well described by
\begin{equation}
\sigma_1^{2} = \left(1+a^{2}\right)S1 + b^{2}S1^{2},
\label{eq:sigma}
\end{equation}
where $a$ is the ratio of the measured width of the \ser\ to its mean, and $b$ is the combined effect of the intrinsic resolution of \lar\ and the dependence of the TPC light collection upon the position of the event.  We fixed $a$ to the measured value, $a$\,=\,0.3, and extracted $b$ from a fit of $\sigma_1$ versus \sone.  For the \krthree\ data, we calculated the resolution by a simple Gaussian plus first order polynomial fit of the spectrum, then fit the data points with the model described by Eq.~\eqref{eq:sigma}.  

Comparing the resolution fits for nuclear recoils vs~\krthree, there is a substantial difference in the resolution fit parameters between the nuclear recoils and the $\beta$-like events. This could be attributed to the different contributions of recombination light for the two particle types.

The fitted value of the parameter $b$ increased between the June and October runs.  We believe this was related to the observed decrease in the resolution of the \krthree\ peak, also shown in Fig.~\ref{fig:resolution}.  

We attribute the decrease in energy resolution for both nuclear recoils and \krthree\ from the June to the October run to the change in the liquid level.  The level was kept below the mesh during the June run and was raised 1\,mm above the mesh for the October run to ensure the proper production of S2 signals.  The latter configuration was less favorable for light yield and resolution of the TPC as a number of scintillation photons undergoing internal reflection at the liquid-gas boundary passed multiple times through the mesh obstruction. Although the decrease in resolution in the \krthree\ data from June to October appears to be larger than that for nuclear recoils, the fits are consistent under the hypothesis that the liquid level and the recombination physics contribute to $b$ in quadrature. 

\section{Distribution of \fno\ Pulse Shape Parameter}
\begin{figure*}[t!]
\includegraphics[width=\columnwidth]{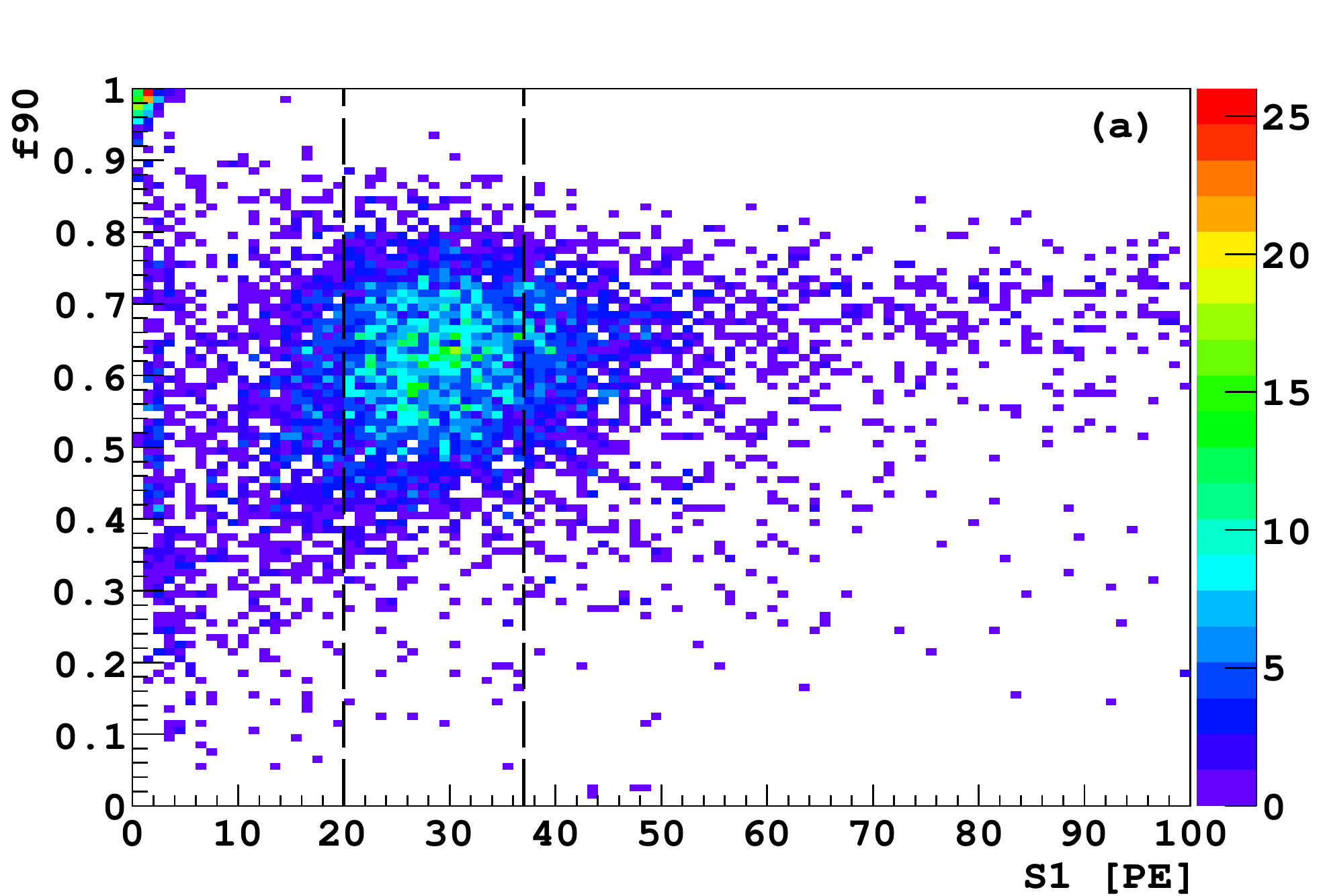}
\includegraphics[width=\columnwidth]{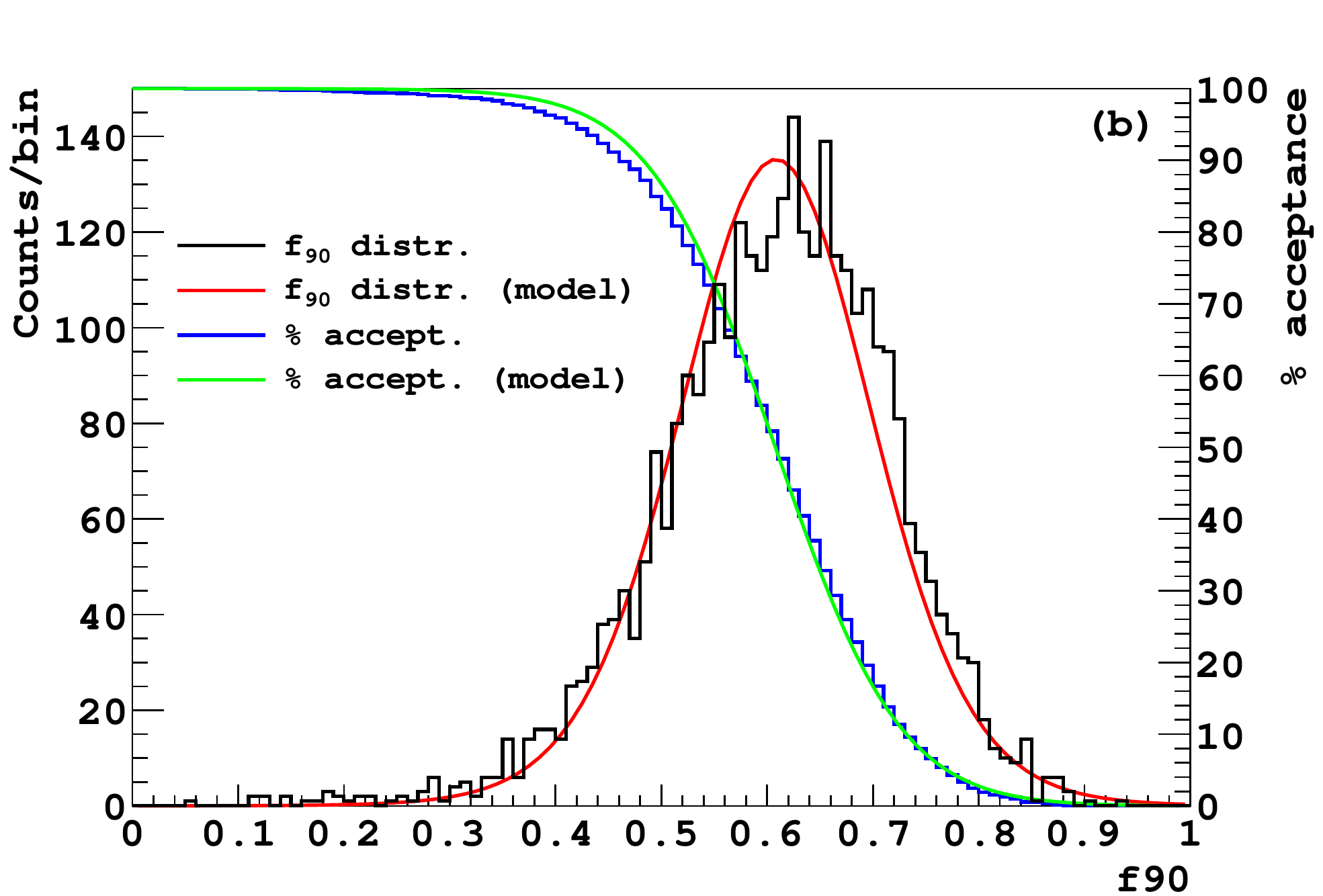}
\includegraphics[width=\columnwidth]{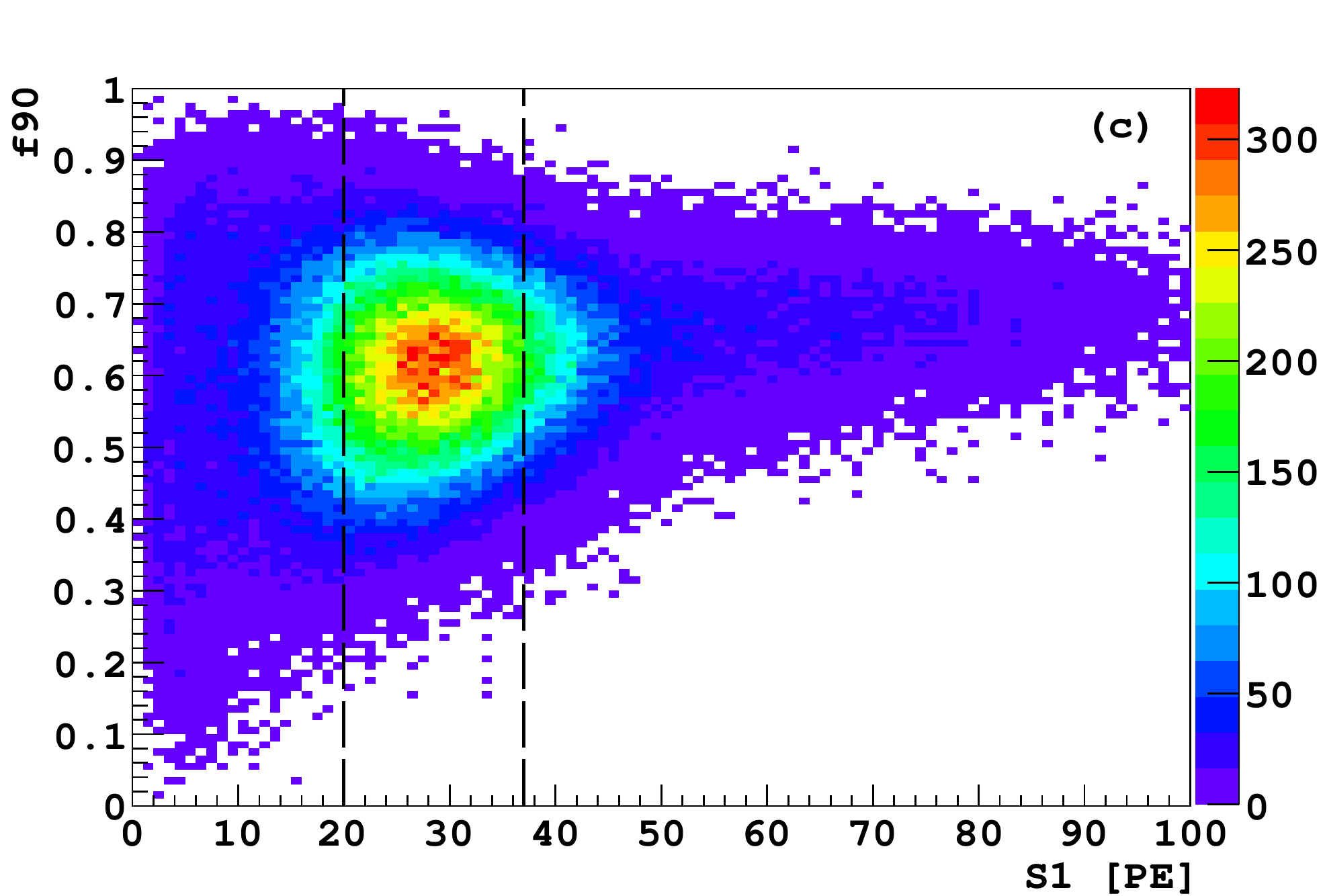}
\includegraphics[width=\columnwidth]{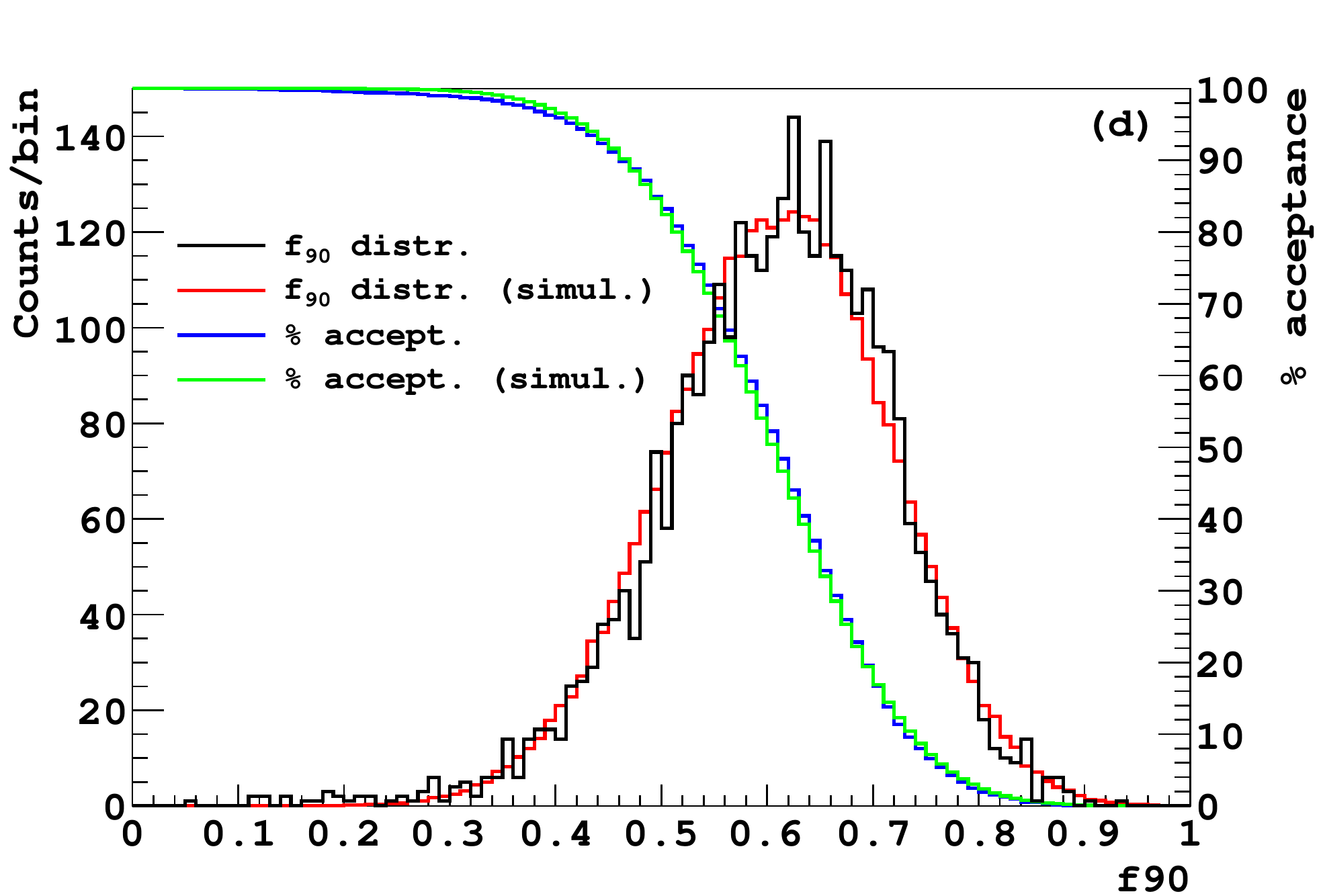}
\caption{\label{fig:f90_21keV}(a) Distribution of \fno\ vs~\sone\ for 20.5\,\kevr\ recoil data taken at \Edrift\,=\,193\,V/cm.  The vertical dashed lines indicate the boundaries of the region where \sone\ is within 1$\sigma$ of the mean of the Gaussian fit $\mu$, as described in the text.  (b) {\color{black} Black:} \fno\ distribution for the 20.5\,\kevr\ nuclear recoil events with S1 falling in the region in [$\mu-\sigma$, $\mu+\sigma$] {\it i.e.}, for the events fall in between the vertical dashed lines in panel (a). {\color{red} Red:} \fno\ distribution model prediction (not a fit, see text). (c) Simulated distribution of \fno\ vs~\sone\ with $\sim$\,30 times the statistics present in the data. (d) {\color{black} Black:} Same in (b). {\color{red} Red:} \fno\ distribution of the simulated events that fall in between the vertical dashed lines in panel (c).}
\end{figure*}

\begin{figure}[htbp!]
\includegraphics[width=\columnwidth]{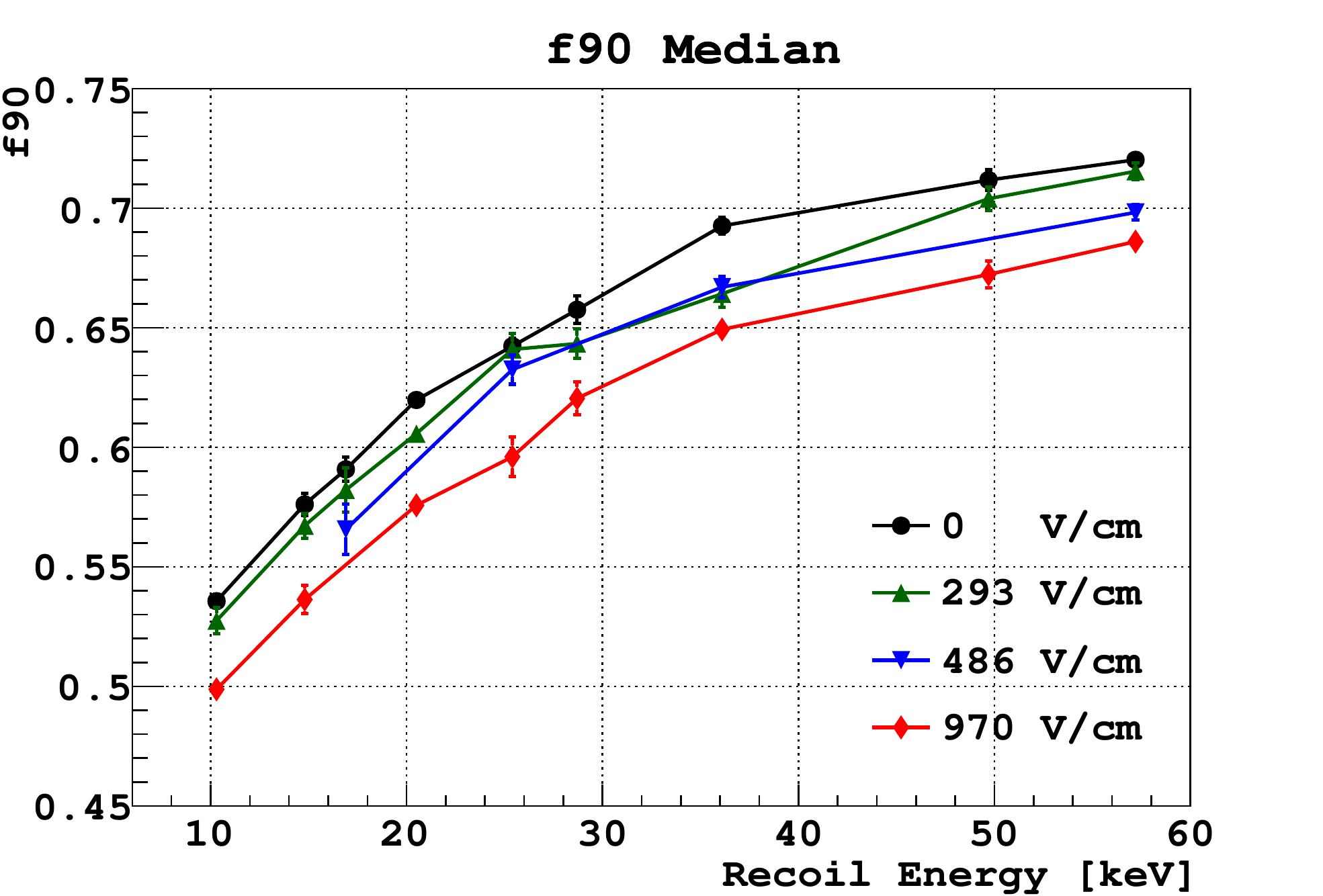}
\caption{\label{fig:psd}Median \fno\ of nuclear recoils as a function of energy at several drift fields. The median values and statistical errors for all energies explored are listed in Table~\ref{tab:f90}. All points have a common systematic error of $0.01$ (see text for discussion). }
\end{figure}

Strong pulse shape discrimination (PSD) against electron recoil background is a key enabling feature of liquid argon weakly interacting massive particle (WIMP) dark matter searches~\cite{warp,ardm,darkside}. The \fno\ parameter, first studied in detail in~\cite{deap,lippincott}, provides a simple and effective method to reject electron recoils on an event-by-event basis. It also serves as a benchmark for the comparisons of more sophisticated PSD techniques~\cite{lippincott}. The samples of nuclear recoil events, which were dominated by single scatters of a given energy, are excellent inputs for studies of the discrimination power and the acceptance levels of nuclear recoils in LAr-TPC. 

We have used the \sone\ data from our experiment for a careful determination of the \fno\ parameter as a function of recoil energy.  For this determination, we first selected events by applying the Ntof, Npsd and TPCtof cuts described above, then by requiring in addition that S1 lay in the range [$\mu-\sigma$, $\mu+\sigma$], where $\mu$ and $\sigma$ are the average value and the standard deviation of S1 as determined with the second fit method (Gaussian plus first order polynomial).  
This additional criterion further reduces the contribution of multiple-scatter events.
As an example of this selection, Fig.~\ref{fig:f90_21keV}(a) shows the 2D distribution of \fno\ vs.~\sone\ for 20.5\,\kevr\ nuclear recoils at \Edrift\,=\,193\,V/cm: in this case, the S1 selection range is the region bounded by the vertical dashed lines.  The resulting \fno\ distribution is shown in Fig.~\ref{fig:f90_21keV}(b). 

\begin{sidewaystable}[htbp]
   \centering
   \begin{tabular}{c|c|c|c|c|c|c|c|} 
      \hline
     Recoil energy  & \multicolumn{7}{c}{Drift field [V/cm]} \\
      $[$keV$]$ & 0 & 49.5 & 96.4 & 193 & 293 & 486 & 970\\
      \hline
     	10.3  &  	0.536 $\pm$ 0.003&                   &				0.541 $\pm$ 0.006&	                &					0.527 $\pm$ 0.006&                           &			0.499 $\pm$ 0.004\\
	14.8  &  	0.576 $\pm$ 0.005&		0.580 $\pm$ 0.004&		0.577 $\pm$ 0.005&                &					0.567 $\pm$ 0.005&	                           &			0.536 $\pm$ 0.006\\
	16.9  &  	0.591 $\pm$ 0.005&                   &				0.601 $\pm$ 0.009&		0.583 $\pm$ 0.009&		0.582 $\pm$ 0.009&		0.566 $\pm$ 0.011&	                     	\\
	20.5  &  	0.620 $\pm$ 0.001&		0.630 $\pm$ 0.002&		0.626 $\pm$ 0.002&		0.616 $\pm$ 0.002&		0.606 $\pm$ 0.002&                           &			0.576 $\pm$ 0.002\\
	25.4  &  	0.642 $\pm$ 0.003&		0.651 $\pm$ 0.006&		0.652 $\pm$ 0.006&		0.642 $\pm$ 0.005&		0.641 $\pm$ 0.007&		0.632 $\pm$ 0.006&		0.596 $\pm$ 0.008\\
	28.7  &  	0.657 $\pm$ 0.006& 	0.662 $\pm$ 0.005& 	0.664 $\pm$ 0.006& 	      &					0.643 $\pm$ 0.006&                            &			0.620 $\pm$ 0.007\\
	36.1  &  	0.693 $\pm$ 0.003& 	0.683 $\pm$ 0.005& 	0.678 $\pm$ 0.004& 	0.672 $\pm$ 0.004& 	0.664 $\pm$ 0.006& 	0.667 $\pm$ 0.004& 	0.649 $\pm$ 0.005\\
	49.7  &  	0.712 $\pm$ 0.004& 	0.718 $\pm$ 0.006& 	0.723 $\pm$ 0.005& 	      &					0.704 $\pm$ 0.005& 			&			0.672 $\pm$ 0.006\\
	57.3  &  	0.720 $\pm$ 0.002& 	0.719 $\pm$ 0.004& 	0.712 $\pm$ 0.003& 	0.720 $\pm$ 0.003& 	0.715 $\pm$ 0.004& 	0.698 $\pm$ 0.003& 	0.686 $\pm$ 0.005\\
      \hline
   \end{tabular}
   \caption{The median vale of \fno\ as a function of recoil energy and drift field, with statistical uncertainties. All points have a systematic uncertainty of $0.01$. }
   \label{tab:f90}
\end{sidewaystable}

Following this procedure, we determine the median \fno\ for nuclear recoils as a function of recoil energy and applied field. The results with statistical uncertainties are plotted in Fig.~\ref{fig:psd} and listed in Table~\ref{tab:f90}. The median value decreases by $\sim$0.01 as the drift field increases from 0 to 293\,V/cm, and by another $\sim$0.03 from 293\,V/cm to 970\,kV/cm. We calculate systematic uncertainties caused by the reconstruction algorithm, data selection cuts, differences in the response of the two PMTs, and impurities, with the two leading contributions from data selection cuts and purity variations in time. To explore the effect of the data selection cuts, we change the size of the S1 acceptance window by $0.25\sigma$ in both directions, finding a  $1\%$ change in \fno. A similar analysis of impurity levels using the observed triplet lifetime to that described previously with regard to \leff\ finds a variation in the measured \fno\ of $1.6\%$. Taken all together, we find a systematic uncertainty on the median \fno\, of 0.01.


We have used two different models to compare with our measured \fno\ distributions: the ``ratio-of-Gaussians" model described in~\cite{deap,lippincott}, and a statistical model that simulates each of the various processes associated with the scintillation signal. The ``ratio-of-Gaussians" model assumes that the number of photoelectrons in the prompt and late time windows, $N_p$ and $N_l$, are normally distributed, independent random variables with means $\mu_p$ and $\mu_l$ and variances $\sigma_p^2$ and $\sigma_l^2$.  By definition \fno\,=\,$N_p/(N_p+N_l)$, therefore the assumption on $N_p$ and $N_l$ turns \fno\ into a ratio of two normally distributed, correlated random variables.  Equation (11) of Ref.~\cite{lippincott} gives a close approximation to the probability density function of \fno.  References~\cite{deap,lippincott} have shown that this model describes the \fno\ distributions of electron recoil to great precision.  The lack of a clean sample of single-scatter nuclear recoils previously prevented the application of this model to nuclear recoils.  

To use this model, we plugged in our measured S1 means and \fno\ means of nuclear recoils to compute $\mu_p$ and $\mu_l$.  For the variances, in addition to Poisson counting statistics, we included the variance due to the width of the SER, $a^2\mu_p$ and $a^2\mu_l$, in $\sigma_p^2$ and $\sigma_l^2$, respectively.  In our experiment, at the recoil energies of interest, the variance due to electronic noise is negligible compared to the contribution from counting statistics.  We superimposed the model output over the measured \fno\ distribution for 20.5\,\kevr\ at \Edrift\,=\,193\,V/cm in Fig.~\ref{fig:f90_21keV}(b).  

In addition to this model, we compared our data to a statistical simulation that uses both the observed mean S1 scintillation yields and \fno\ means of nuclear recoils as a function of energy. Starting from the nuclear recoil energy distribution predicted from the Monte Carlo of the neutron beam and geometrical setup of the detectors (see first panel of Fig.~\ref{fig:leff21}), we simulated the effects of the scintillation of UV photons (which is assumed to follow a Poisson distribution), wavelength shifter, the conversion to photoelectrons, and the final charge distribution (using parameters from the SER calibration). The details of the simulation will be discussed in an upcoming publication. The simulated prompt and late signals are combined to form the S1 and \fno\ variables which are shown in Fig.~\ref{fig:f90_21keV}(c). The simulation contains significantly higher statistics than the data to accurately depict the shape of the tails. To obtain the final simulated \fno\ distribution [the red curve in Fig.~\ref{fig:f90_21keV}(d)], we selected only those events whose S1 signal falls in the same range as the one used for the data, and normalized the distribution to match the data. 

Although we do not expect an exact match between each of the models and the measured distribution because of the multiple-scattered neutron background and electron/neutron recoil background from random coincidence, the level of agreement suggests that both models are suitable for application on the \fno\ distribution of nuclear recoils as well as electron recoils. The good agreement between the simulation and the data indicates that the measured \fno\ distributions are consistent with the expected statistical distributions of the various physical processes involved. 

With either model, one can use the \leff\ and \fno\ medians reported in this paper to deduce \fno\ acceptance in any LAr dark matter detector, whether single or double phase, either in absence or as a function of the drift field value, as a function of the light yield, electronics, and noise specific for the dark matter detector of interest.  The starting elements are the \leff\ and \fno\ values reported here as a function of the applied drift field.  With the additional input of the light yield of the LAr dark matter detector at null field, one can calculate precisely the correspondence between the nuclear recoil energy scale and the \pe\ scale, and assign a mean of the \fno\ distribution as a function of the detected number of \pe.  At this point, with the final input coming from the contributions of the electronics and noise specific to the LAr detector under consideration, 
fluctuations of the \fno\ distribution and acceptance curves as a function of the detected number of \pe\ can be calculated using the ``ratio-of-Gaussians" model introduced and elaborated in Refs.~\cite{deap,lippincott} or using a simulation which individually accounts for all the major physical processes related to the scintillation signal.


\section{Analysis of the S2 Spectra and Determination of \qy}

We define \qy\ as the ionization yield of nuclear recoils.  Earlier measurements have shown that the ionization yield from electrons, relativistic heavy ions, $\alpha$ particles, fission fragments~\cite{kubota78,doke85,hitachi87} and 6.7\,keV nuclear recoils~\cite{llnl} can be enhanced by stronger drift electric fields (\Edrift).  Our data confirmed this for internal conversion electrons from \krthree, and nuclear recoils in the energy range of $16.9 - 57.3$ keV.  Ideally, \qy\ should be expressed in detector-independent units of extracted electrons per unit of recoil energy (such as \e/keV), but in practice conversion to these units is susceptible to significant systematic uncertainties due to the requirement of single electron calibration for \stwo. With an extraction field of 3.0\,kV/cm, a multiplication field of 4.5\,kV/cm and a gas region of 6\,mm in height, we did not observe resolved single-electron S2 signals by applying the technique described in Ref.~\cite{xesingle}.  We will show in the next section an indirect method of determining the single electron S2 gain in a \lartpc\ by taking advantage of the simultaneous measurements of scintillation and ionization. The single electron S2 gain of our data was estimated to be $3.1\pm0.3$\,PE/\e\ by this method. 

We also report \qy\ in detector-dependent units of PE/keV along with the ionization yield of \krthree. A measurement of \qy\ relative to the ionization yield of \krthree, like \leff, permits direct computation of the nuclear recoil ionization yield from the measured ionization yield of \krthree\ in any liquid argon \lartpc.

\begin{figure}[t!]
\includegraphics[width=\columnwidth]{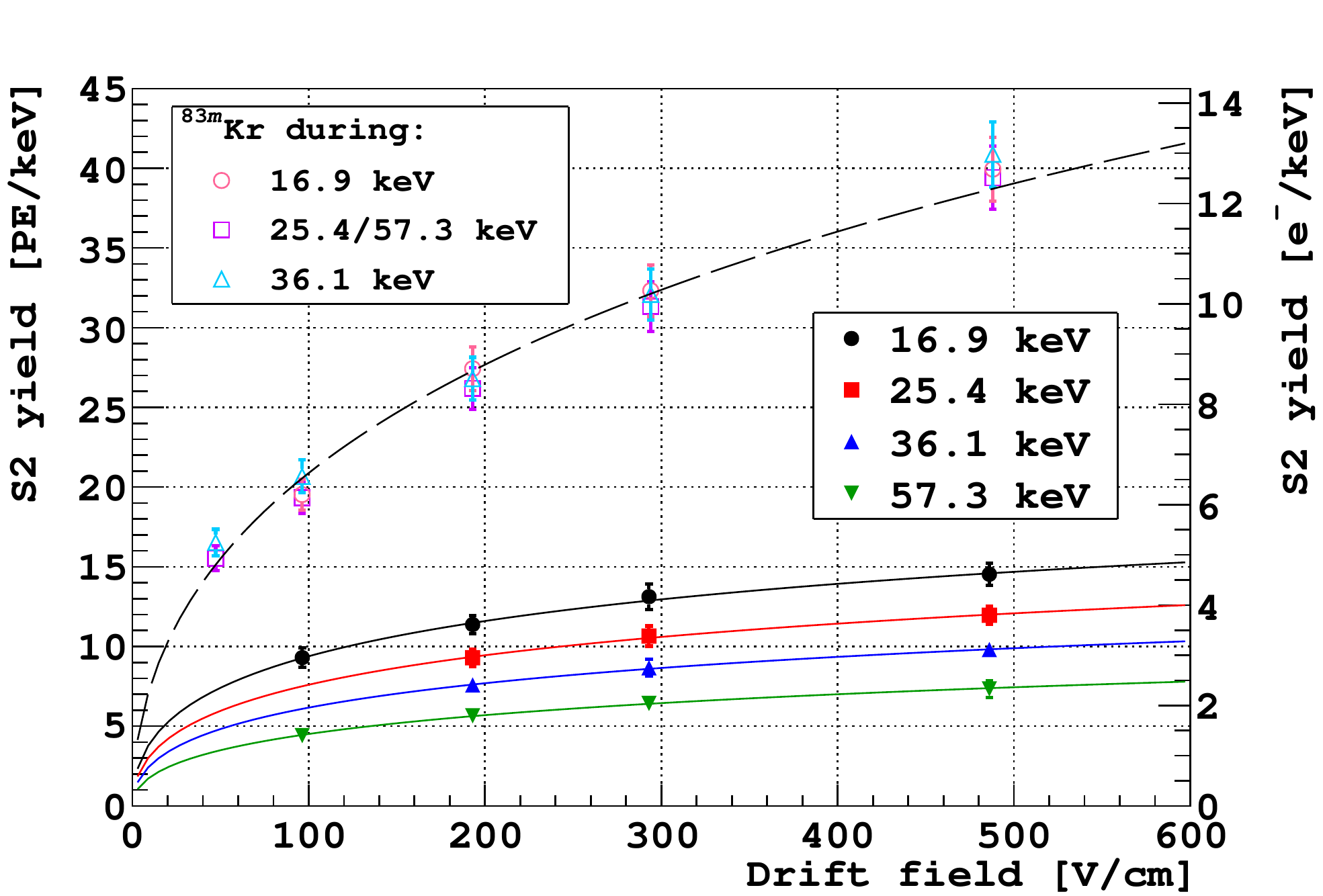}
\caption{\label{fig:s2}Measured S2 yield as a function of \Edrift\ at four recoil energies. Extraction field is fixed at 3.0\,kV/cm and multiplication field at 4.5\,kV/cm. To quote S2 yield in [\e/keV], an additional 10\% systematic uncertainty must be combined with each error bar shown, to take into account the uncertainty in the single-electron calibration. The dashed curve shows the best fit  of the modified Thomas-Imel model (see text) to \krthree\ data.  The solid curves show the best fits of the same model (see text) to the nuclear recoil data.}
\end{figure}

We determined \qy\ in a manner similar to \leff, {\it i.e.}~by fitting experimental data with Monte Carlo-generated spectra that took into account the complete geometry of the experiment.  Instead of extracting \qy\ independently for each \Enr\ and \Edrift, we assumed that S2 at a given drift field can be modeled by a logarithmic function of recoil energy:
\begin{equation}
\mathrm{S2_{nr}} = a \ln({bE_{\mathrm{nr}}} )+ c,
\label{eq:log}
\end{equation}
 and fit all S2 spectra acquired at the same \Edrift\ with the same function.  All coefficients of Eq.~\ref{eq:log} were treated as free parameters.  This procedure improved the goodness of the fit between data and Monte Carlo, particularly on the left (low PE) side of the peaks, as \qy\ depends more strongly on recoil energy than \leff\ does (in our \sone\ fits, we assumed \leff\ is constant in the fit region).  Also similar to what was done before for the S1 study, the resolution in S2 was taken as a free parameter in the fit.  The resolution in S2 was parametrized as $\sigma_2\,=\,\sqrt{\left(1+a^2\right) S2 + R_{2}^{2} S2^{2}}$, where the ratio of the width of the SER to its mean, $a$, was fixed to 0.3. 

By varying \Rtwo\ and the coefficients of Eq.~\ref{eq:log}, the fit procedure minimizes the \chisq\ defined as: 
\begin{equation}
\chi^{2}(\mathcal{E}_{\rm{d}}) = \sum\limits_{j=1}^m\sum\limits_{i=1}^{n_{j}}\frac{\left(O_{j,i}-S_{j,i}\right)^2}{S_{j,i}} ,
\end{equation}
where $m$ is the number of recoil spectra acquired with the same \Edrift, $n_j$ is the total number of bins in the chosen fit region for the $j$-th spectrum, $O_{j,i}$ is the number of events observed in bin $i$ for the $j$-th spectrum, and $S_{j,i}$ is the number of events in bin $i$ of the $j$-th spectrum generated by the Monte Carlo simulations. Each Monte Carlo-generated spectrum was normalized so that within the fit range the total number of events was equal to that in the corresponding experimental spectrum.

\begin{figure}[t!]
\includegraphics[width=\columnwidth]{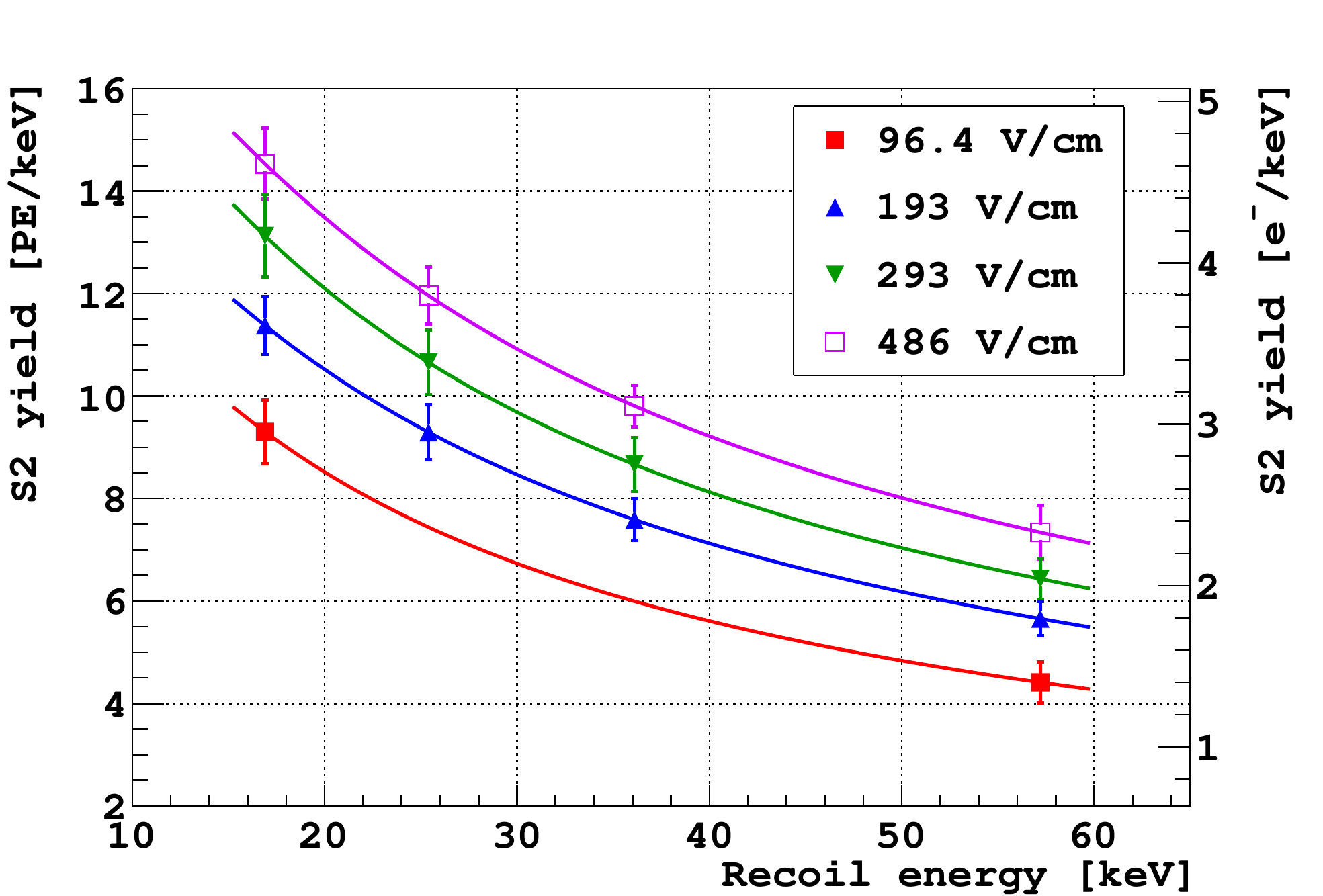}
\caption{\label{fig:s2qy}Best fit S2 yield as a function of recoil energy at four different drift fields (96.4, 193, 293 and 486 V/cm), with a fixed extraction field of 3.0\,kV/cm and multiplication field of 4.5\,kV/cm. To quote S2 yield in [\e/keV], an additional 10\% systematic uncertainty must be combined with each error bar shown, to take into account the uncertainty in the single-electron calibration.}
\end{figure}

\begin{table}[t!]
\begin{center} 
\begin{tabular*}{0.45\textwidth}{@{\extracolsep{\fill} } lrrrr} \hline\hline
Recoil energy~[\kevr] & 16.9 & 25.4 & 36.1 & 57.3 \\ 
\hline 
\qy~[PE/keV]  & 11.4 & 9.3 & 7.6 & 5.7\\
 Statistical error  & 0.2 & 0.2 & 0.1 & 0.1\\
 Systematic errors & & & & \\
\quad Fit model  & 0.2 & 0.2 & 0.2 & 0.1\\
\quad Fit method  & 0.1 & 0.2 & 0.0 & 0.2\\
\quad Fit range  & 0.1 & 0.1 & 0.1 & 0.0\\
\quad TPC tof  & 0.1 & 0.1 & 0.1 & 0.1\\
\quad N tof  & 0.2 & 0.2 & 0.1 & 0.0\\
\quad f90  & 0.2 & 0.2 & 0.1 & 0.0\\
\quad Kr LY  & 0.2 & 0.2 & 0.2 & 0.1\\
\quad Recoil energy\\
\quad \quad TPC pos  & 0.1 & 0.0 & 0.0 & 0.0\\
\quad \quad EJ pos  & 0.2 & 0.3 & 0.2 & 0.1\\
 Combined error  & 0.6 & 0.5 & 0.4 & 0.3\\ 
\hline
\end{tabular*}
\caption{Summary of error contributions to individual \qy\ measurements at \Edrift\,=\,193\,V/cm. Only minor variations in the magnitude of systematic errors are observed across the range of drift field explored. The combined error for each measurement is shown Fig.~\ref{fig:s2qy}.}
\label{table:qysys}
\end{center}
\end{table}

\begin{figure}[t!]
\includegraphics[width=\columnwidth]{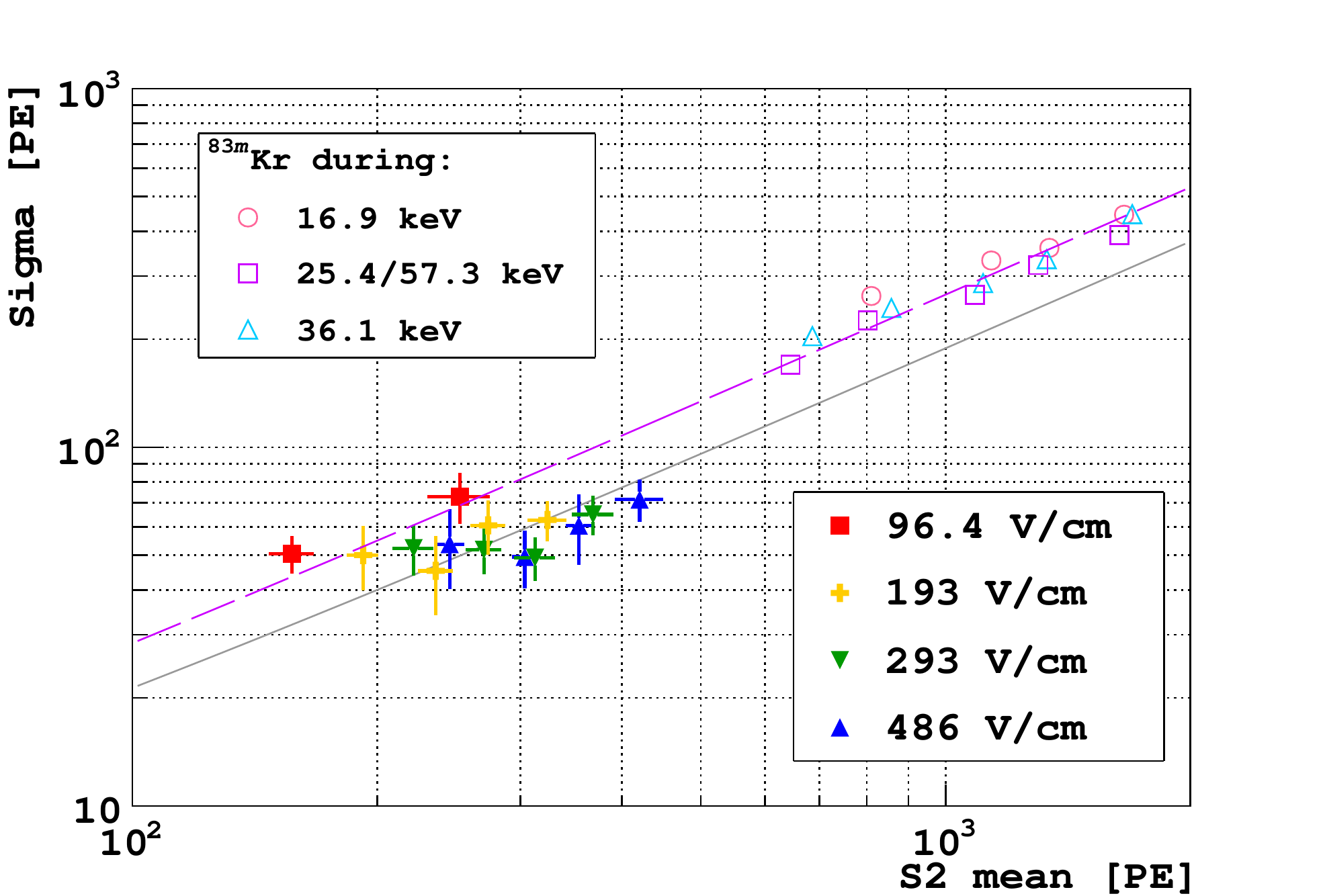}
\caption{\label{fig:s2sigma}Resolution vs. S2 in \pe\ at each recoil energy and drift field. The resolution is determined through the Monte Carlo fit. The resolutions of \krthree\ are shown in the same plot. The best overall fits of $R_2$ (indicated by the fit curves) are $0.19 \pm 0.01$ for nuclear recoils and $0.26 \pm 0.02$ for \krthree.}
\end{figure}

\begin{figure*}[t!]
\includegraphics[width=0.9\columnwidth]{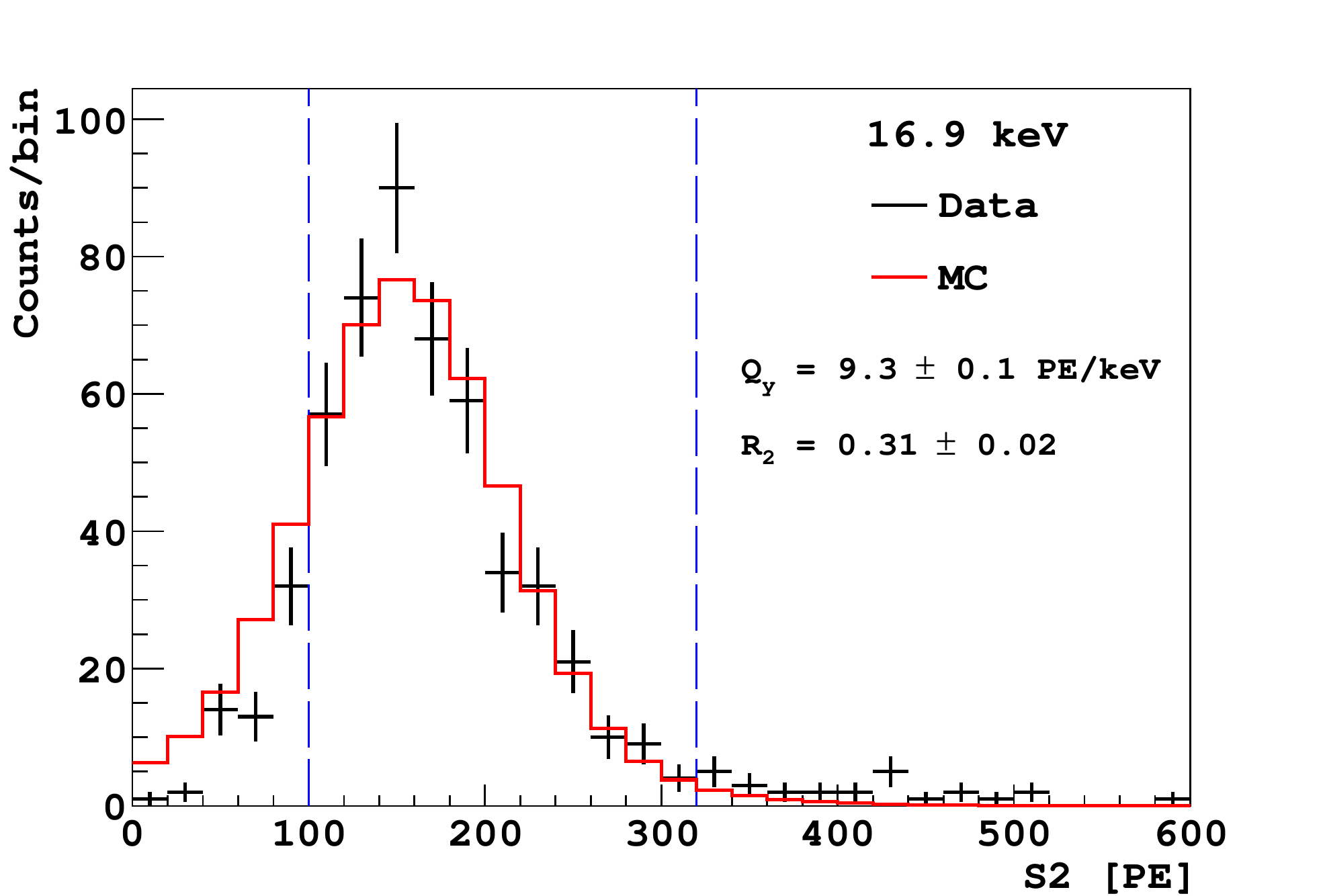}
\includegraphics[width=0.9\columnwidth]{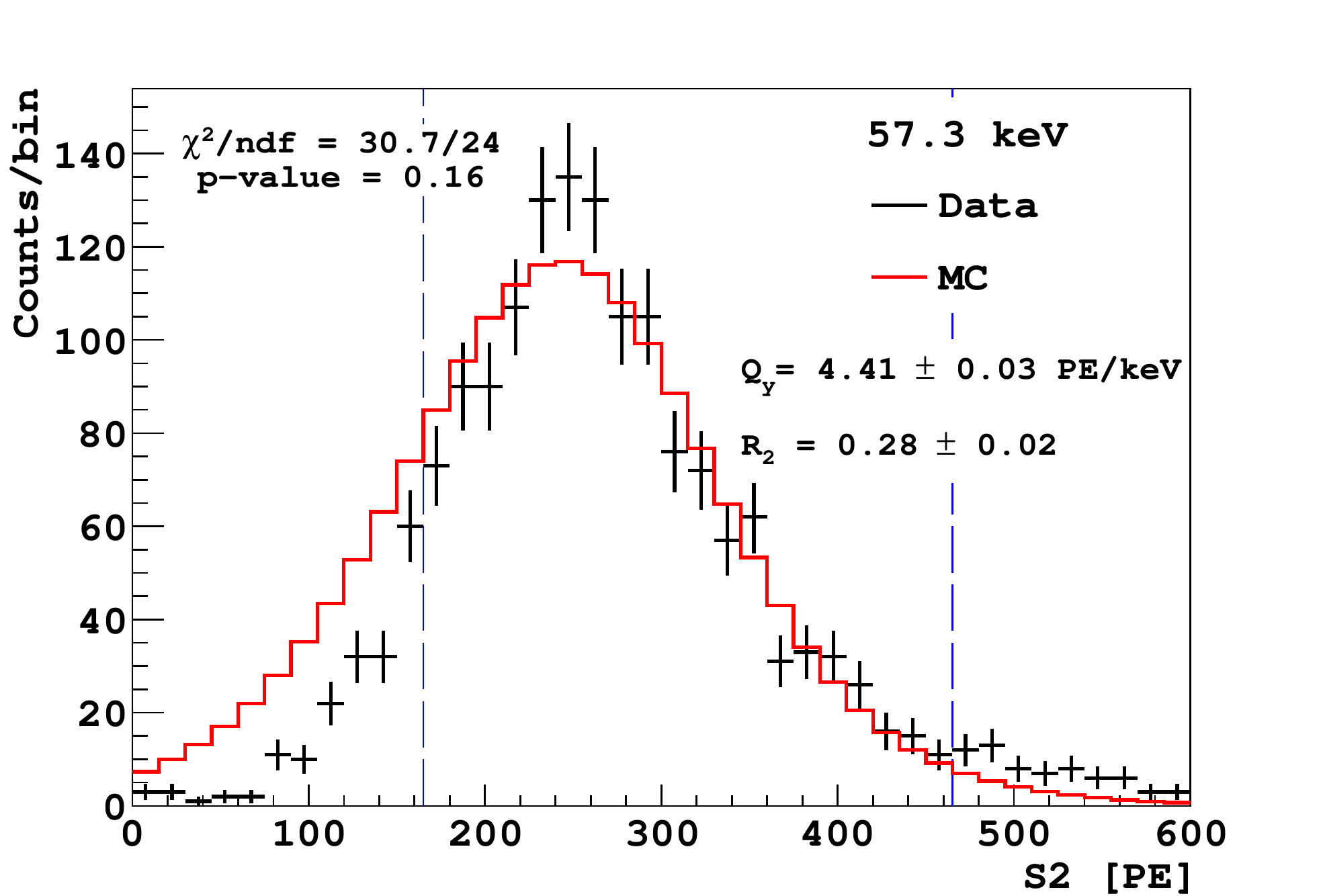}
\caption{\label{fig:qy100}
{\bf \color{black} Black:} Experimental data collected with \Edrift\ = 96.5\,V/cm.  {\bf \color{red} Red:} Monte Carlo fit of the experimental data.  The range used for each fit is indicated by the vertical {\bf \color{blue} blue} dashed lines. The $\chi^{2}$ (sum across all spectra as defined in the text) and the total number of degrees of freedom are shown in the last panel. 
}
\end{figure*}

\begin{figure*}[t!]
\includegraphics[width=0.9\columnwidth]{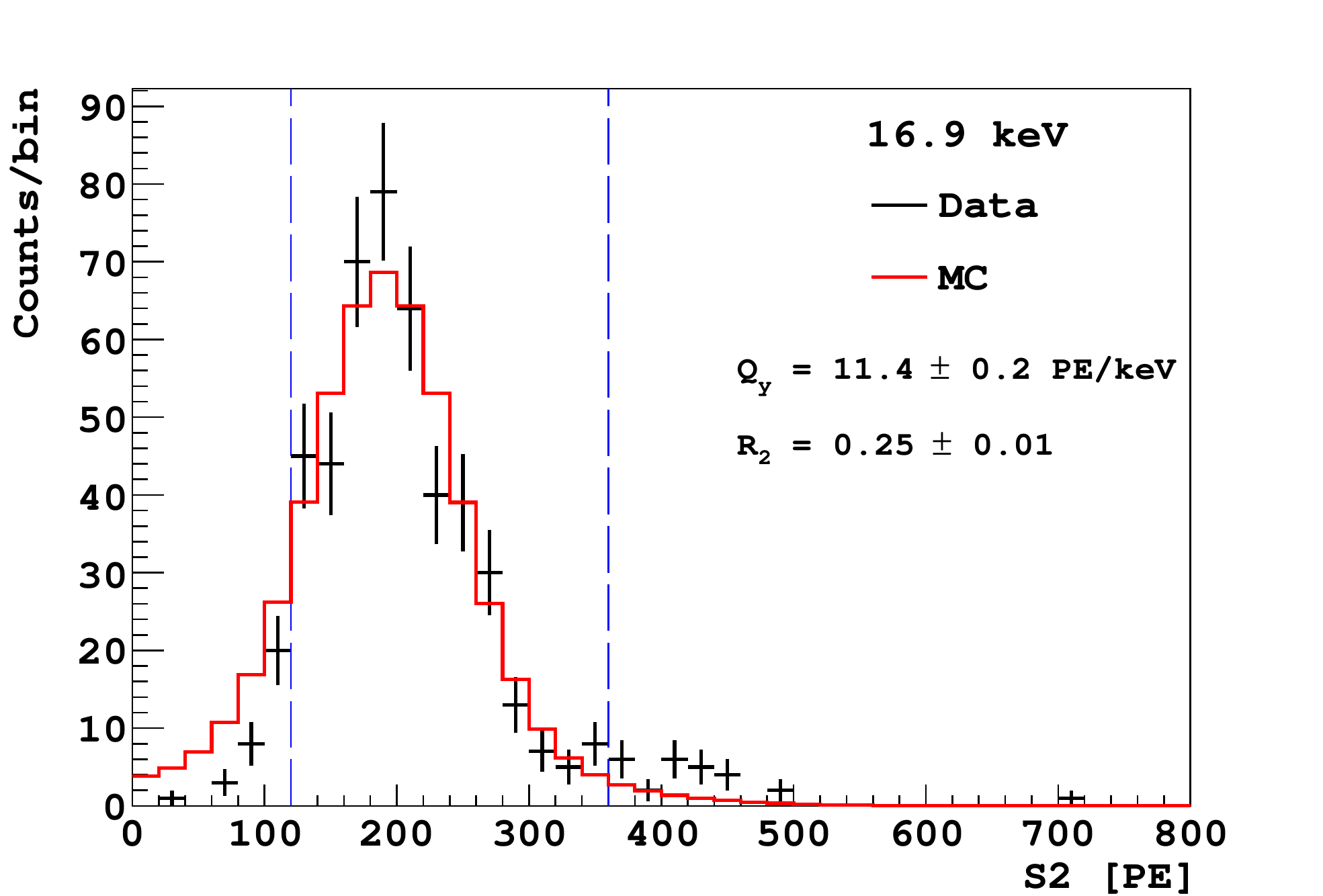}
\includegraphics[width=0.9\columnwidth]{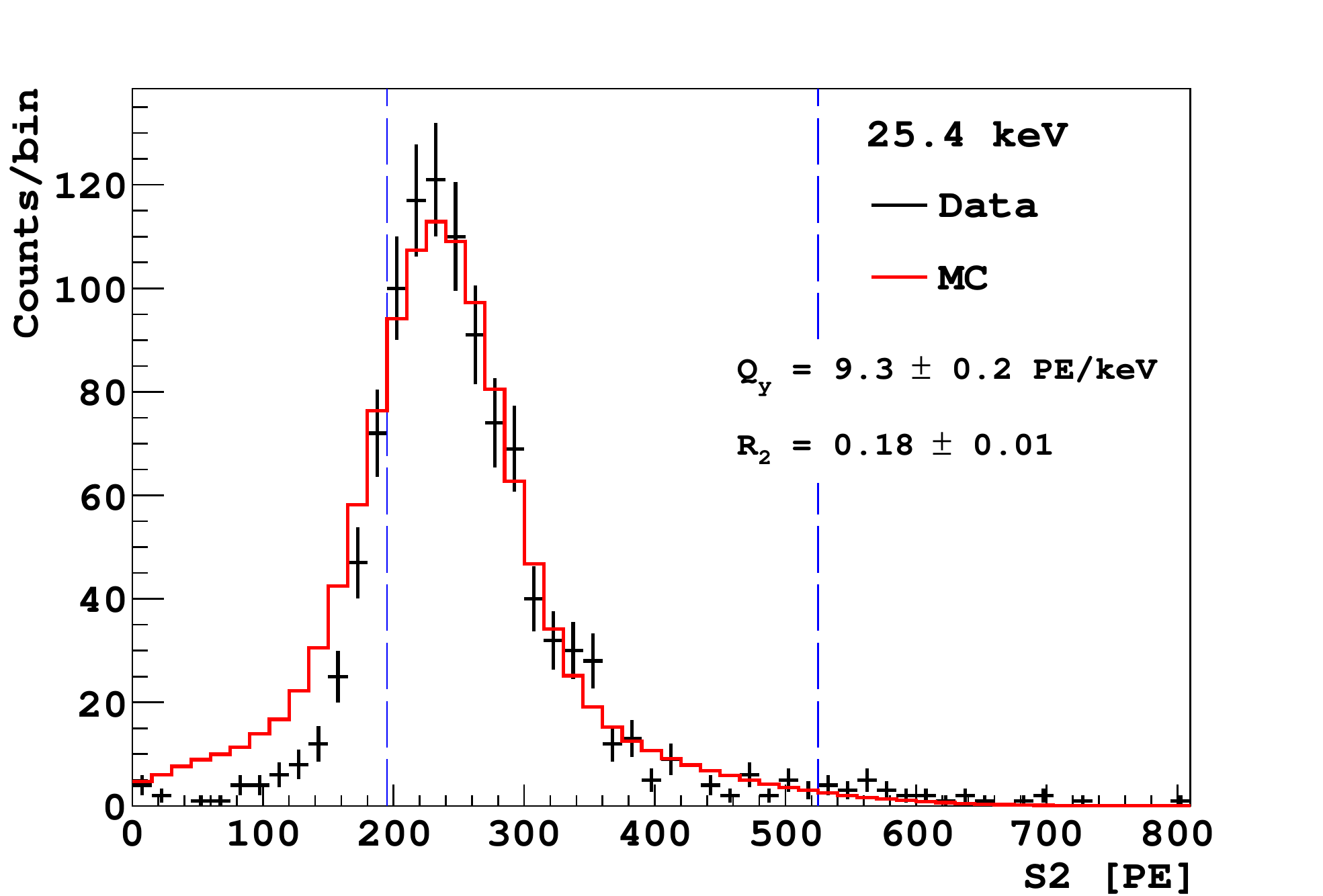}
\includegraphics[width=0.9\columnwidth]{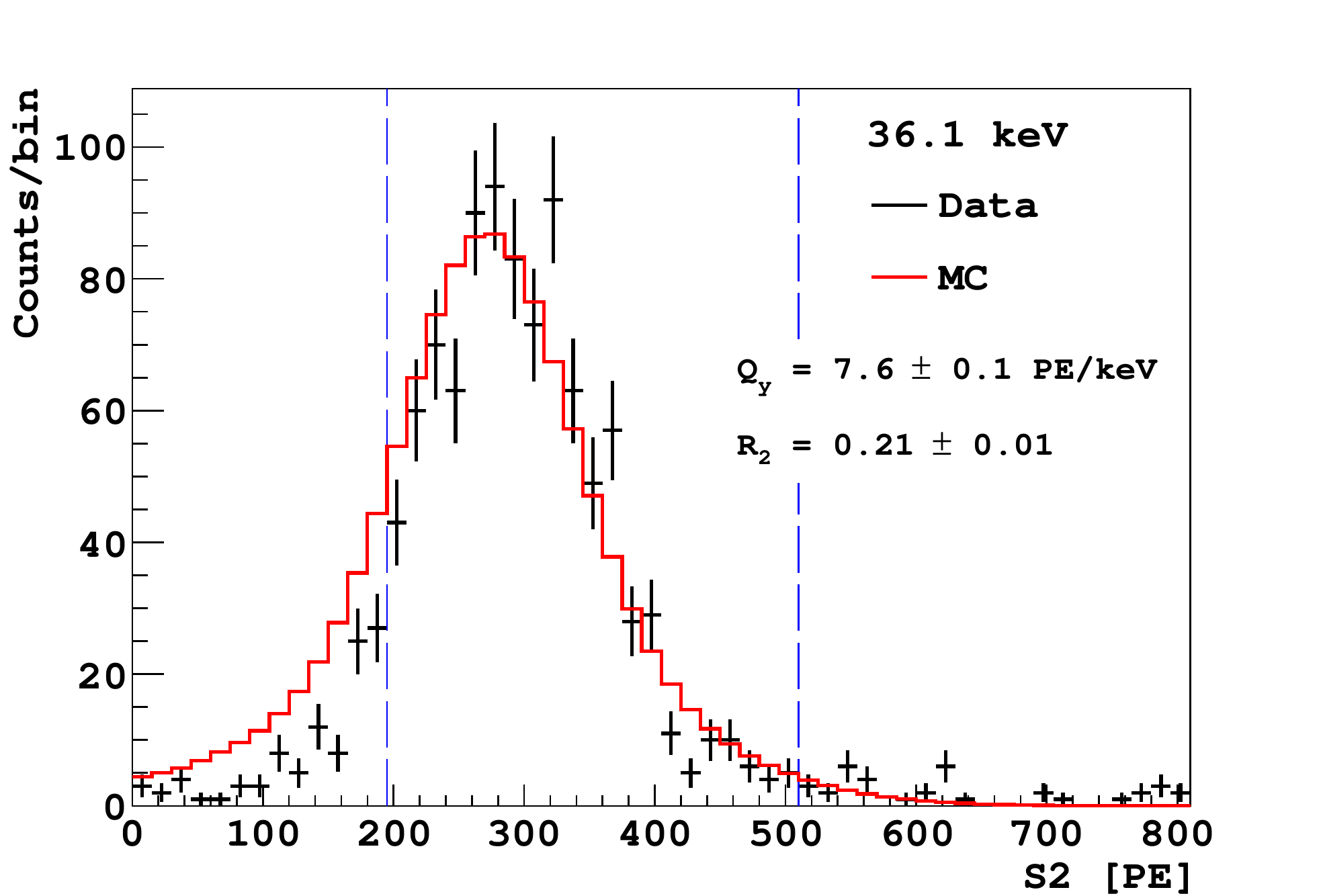}
\includegraphics[width=0.9\columnwidth]{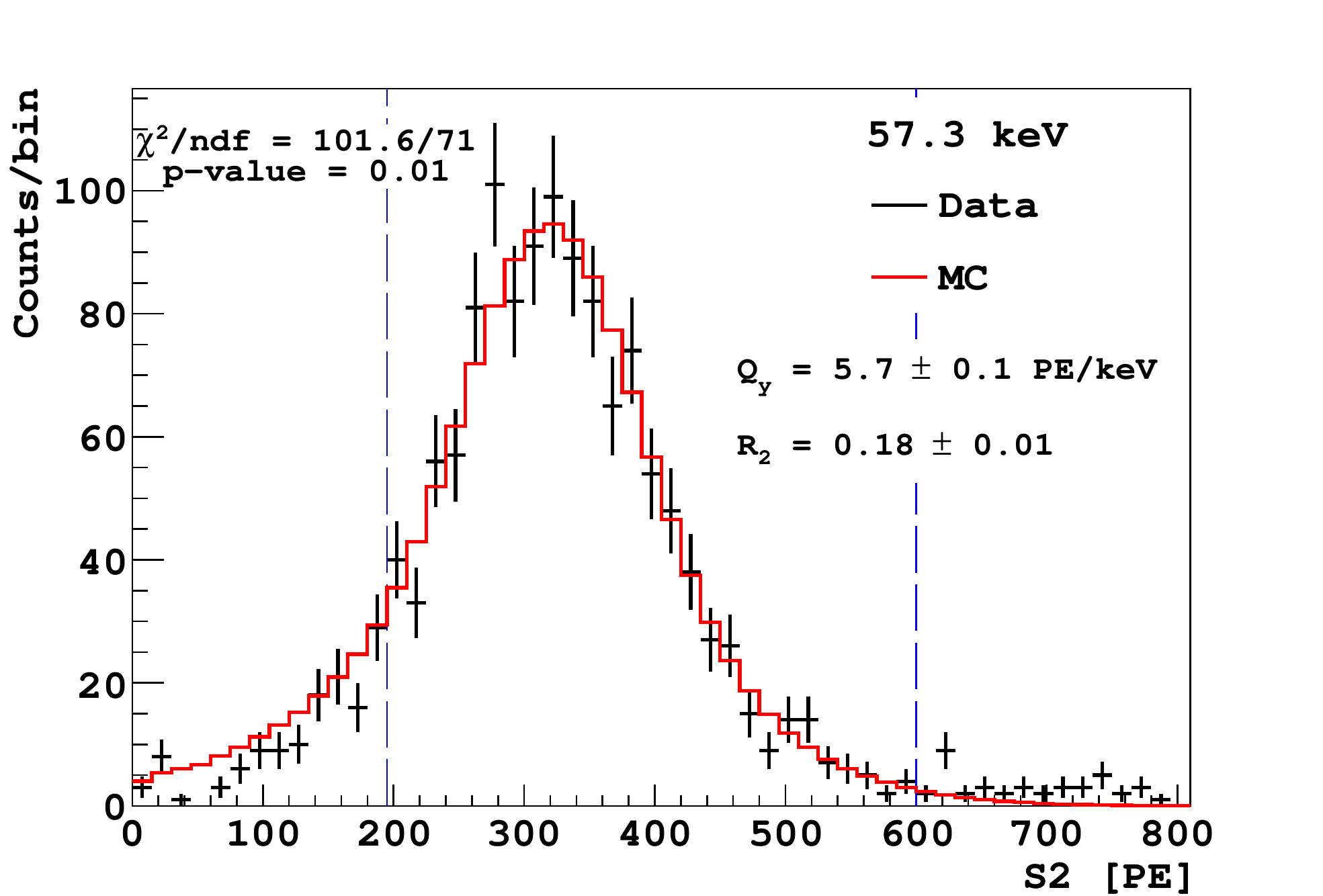}
\caption{\label{fig:qy200}
{\bf \color{black} Black:} Experimental data collected with \Edrift\ = 193\,V/cm.  {\bf \color{red} Red:} Monte Carlo fit of the experimental data.  The range used for each fit is indicated by the vertical {\bf \color{blue} blue} dashed lines. The $\chi^{2}$ (sum across all spectra as defined in the text) and the total number of degrees of freedom are shown in the last panel. 
}
\end{figure*}

\begin{figure*}[t!]
\includegraphics[width=0.9\columnwidth]{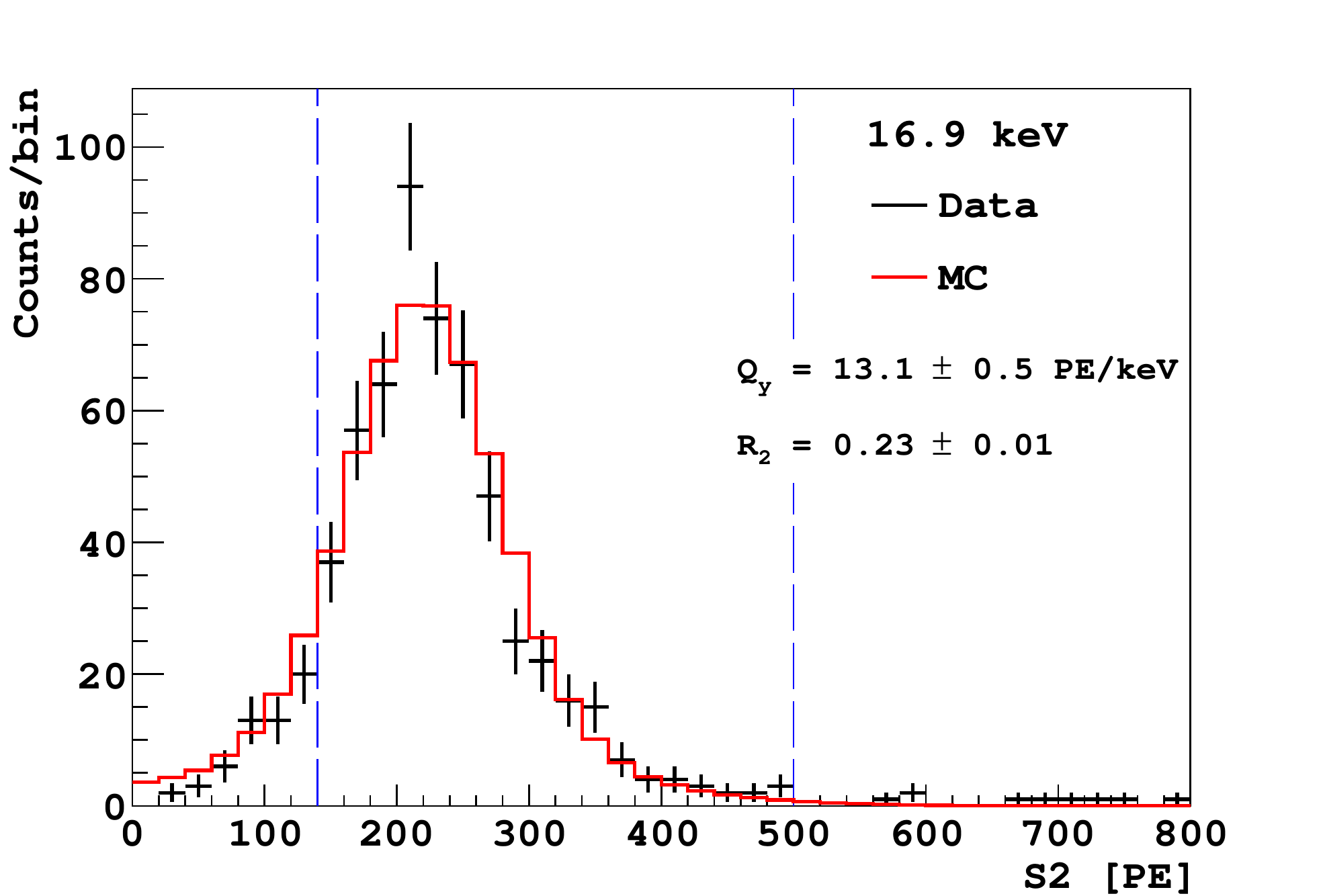}
\includegraphics[width=0.9\columnwidth]{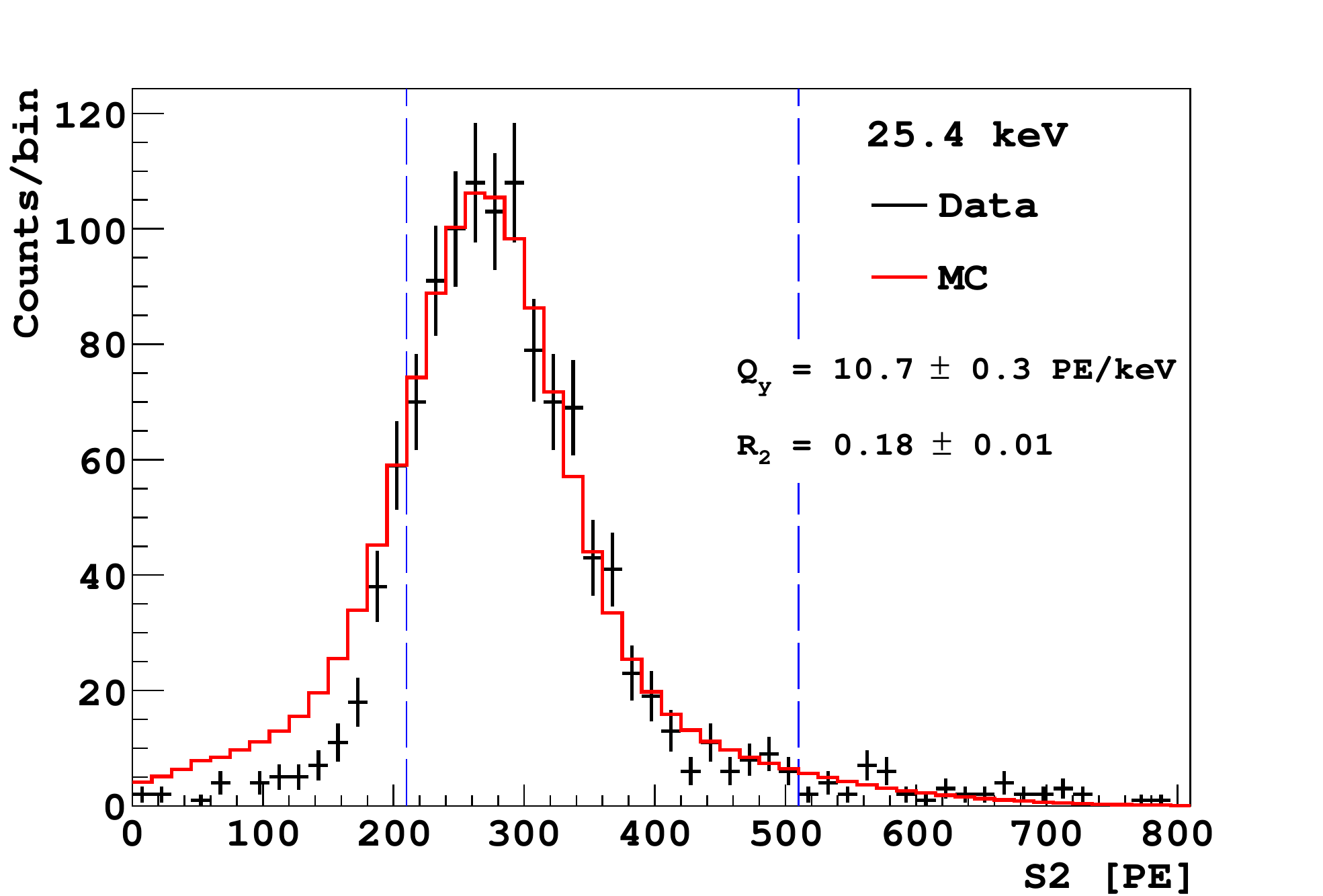}
\includegraphics[width=0.9\columnwidth]{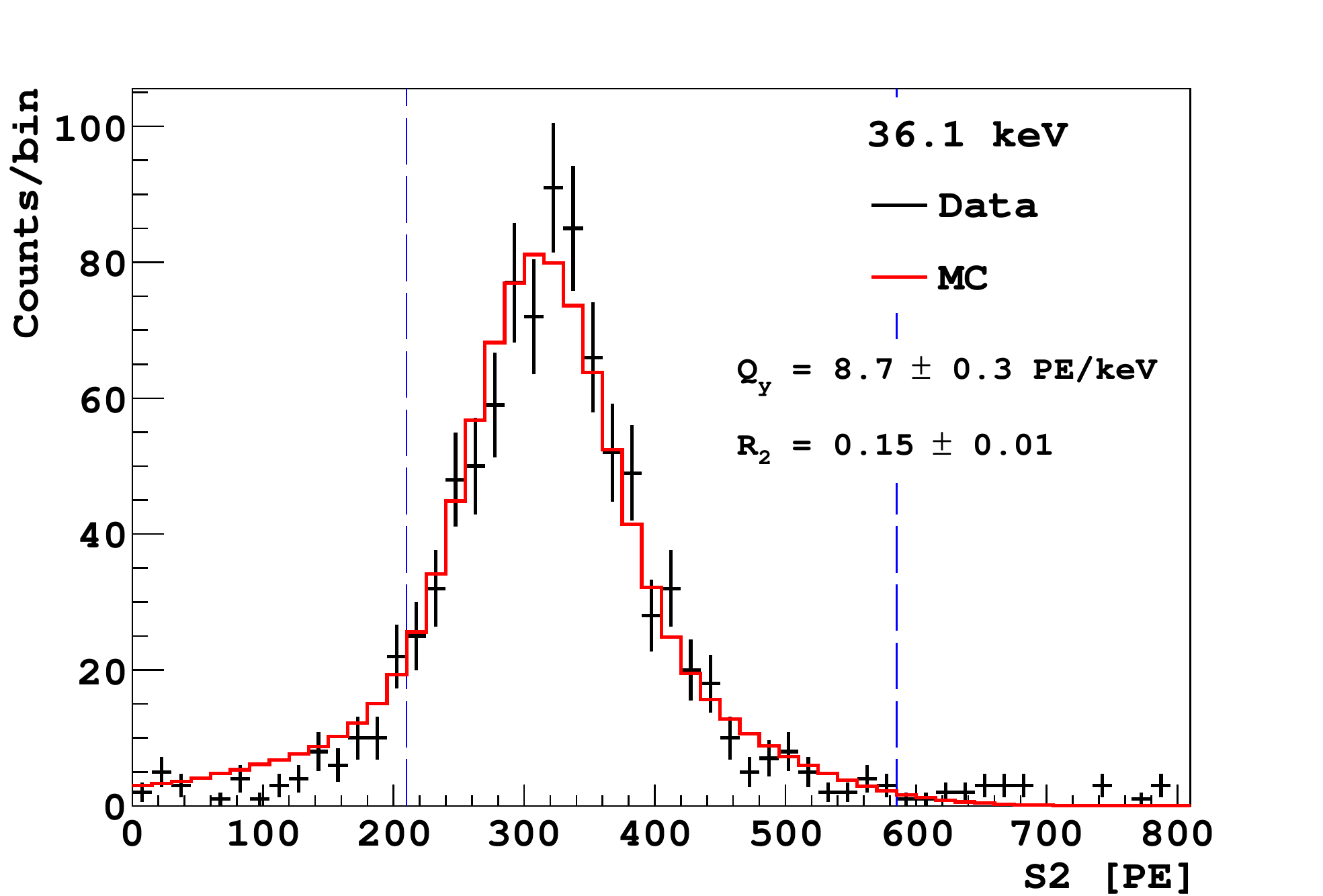}
\includegraphics[width=0.9\columnwidth]{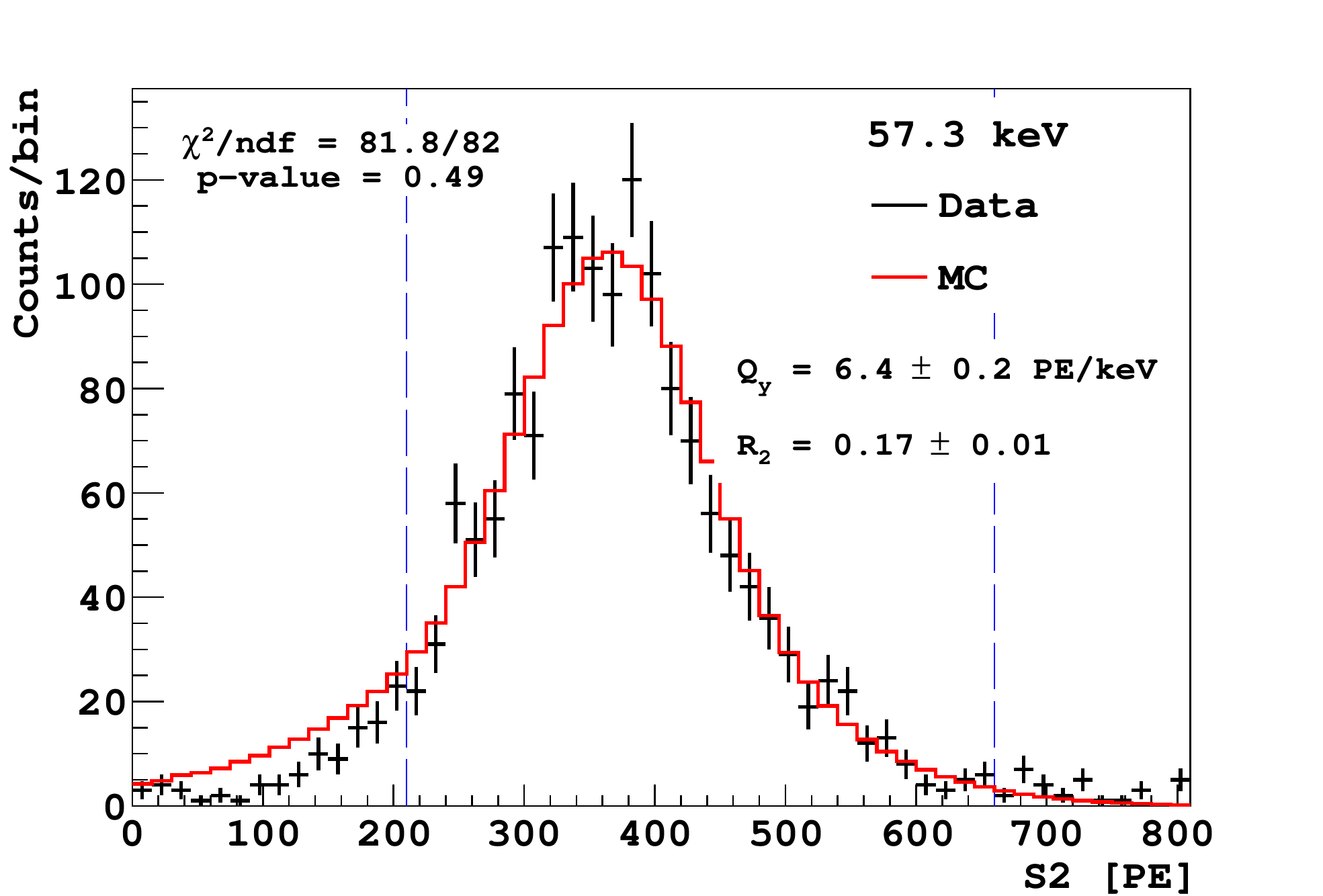}
\caption{\label{fig:qy300}
{\bf \color{black} Black:} Experimental data collected with \Edrift\ = 293\,V/cm.  {\bf \color{red} Red:} Monte Carlo fit of the experimental data.  The range used for each fit is indicated by the vertical {\bf \color{blue} blue} dashed lines. The $\chi^{2}$ (sum across all spectra as defined in the text) and the total number of degrees of freedom are shown in the last panel. 
}
\end{figure*}

\begin{figure*}[t!]
\includegraphics[width=0.9\columnwidth]{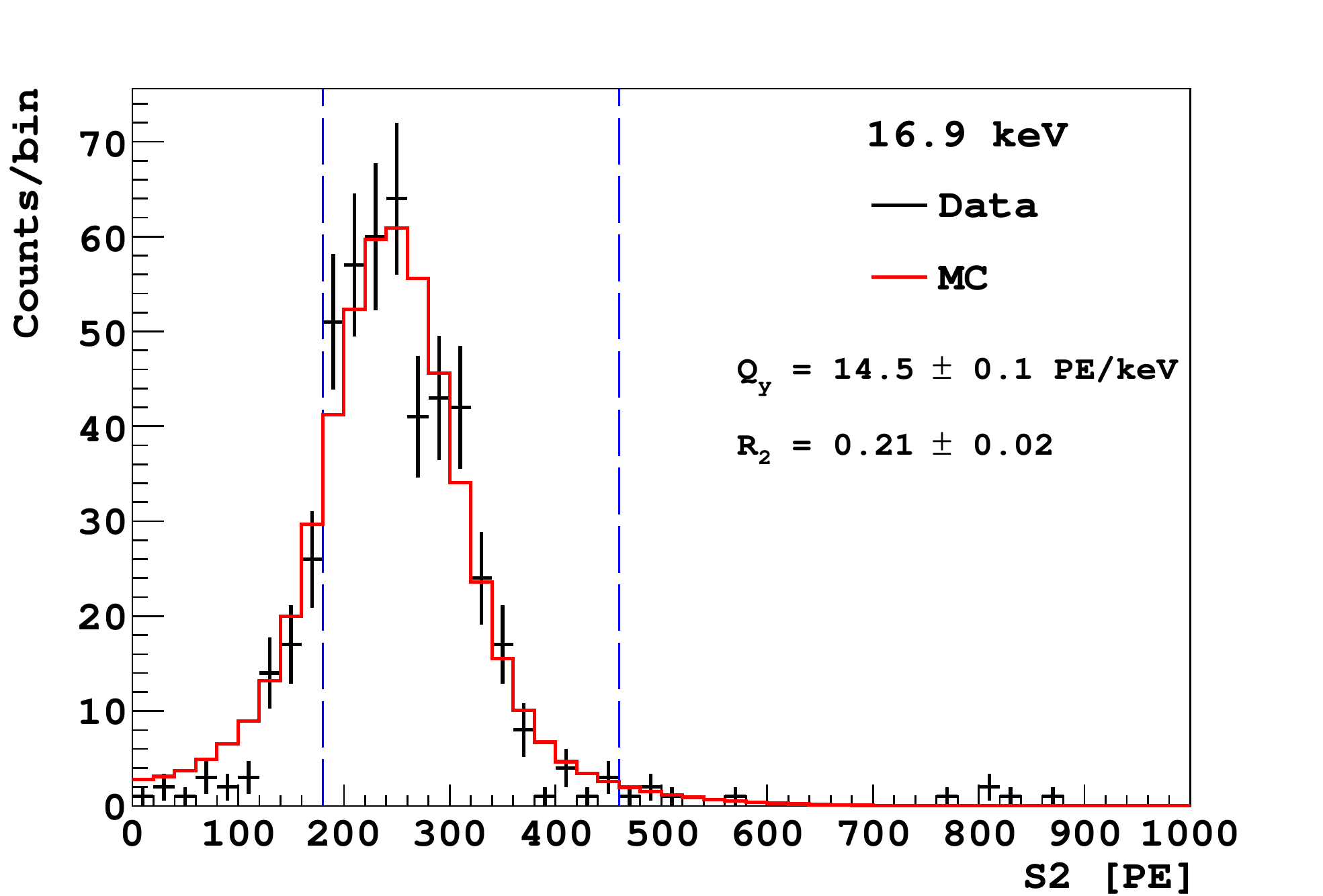}
\includegraphics[width=0.9\columnwidth]{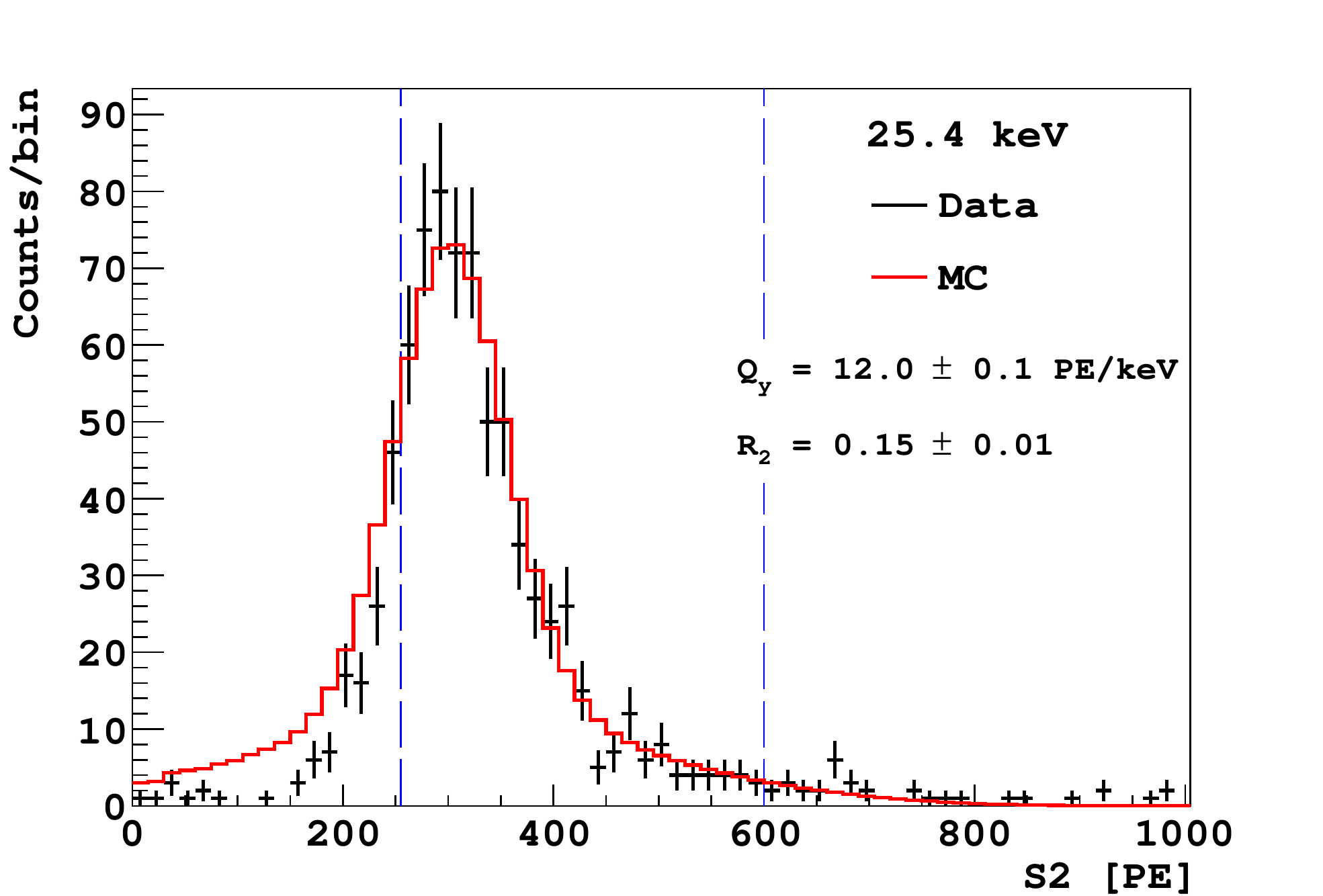}
\includegraphics[width=0.9\columnwidth]{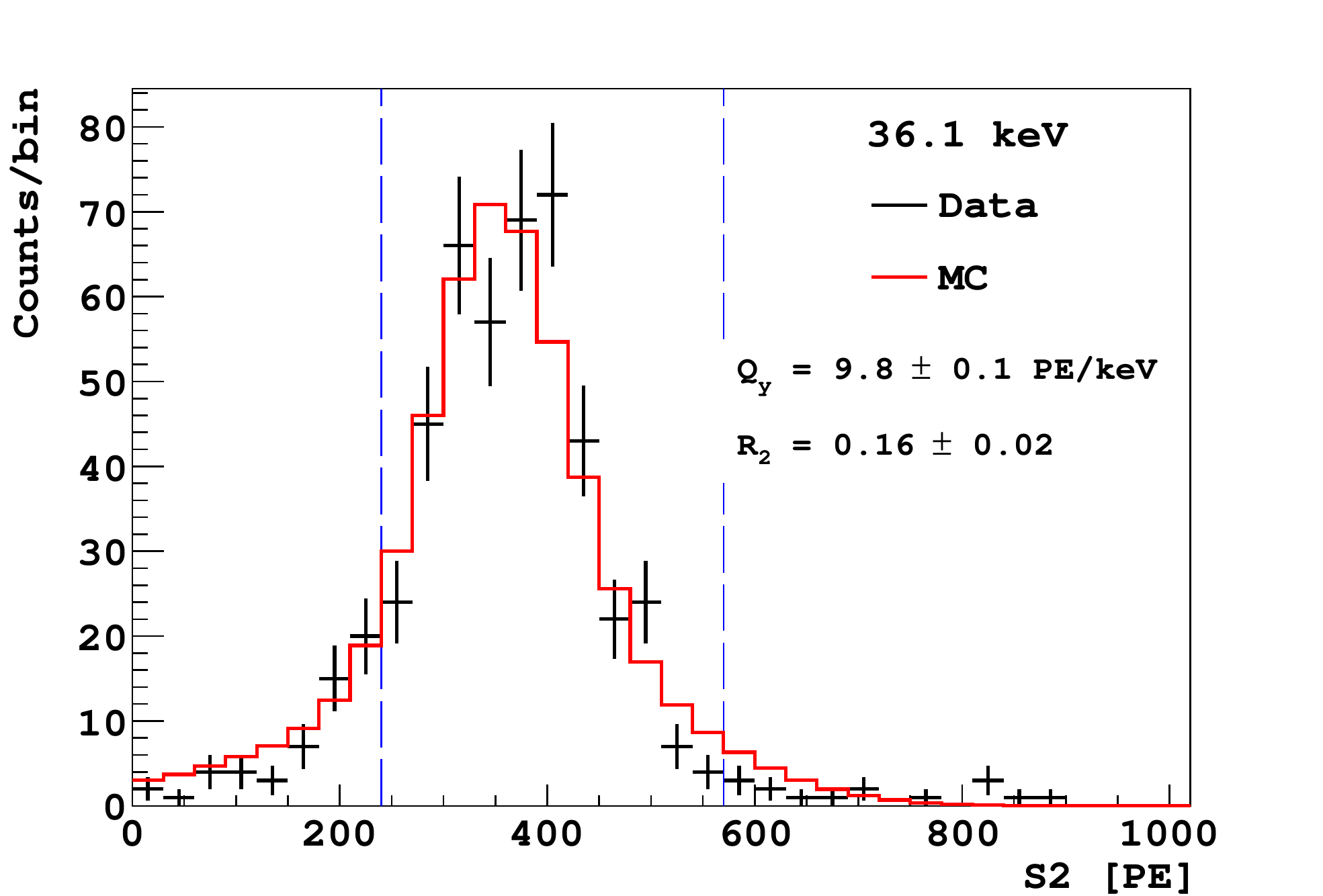}
\includegraphics[width=0.9\columnwidth]{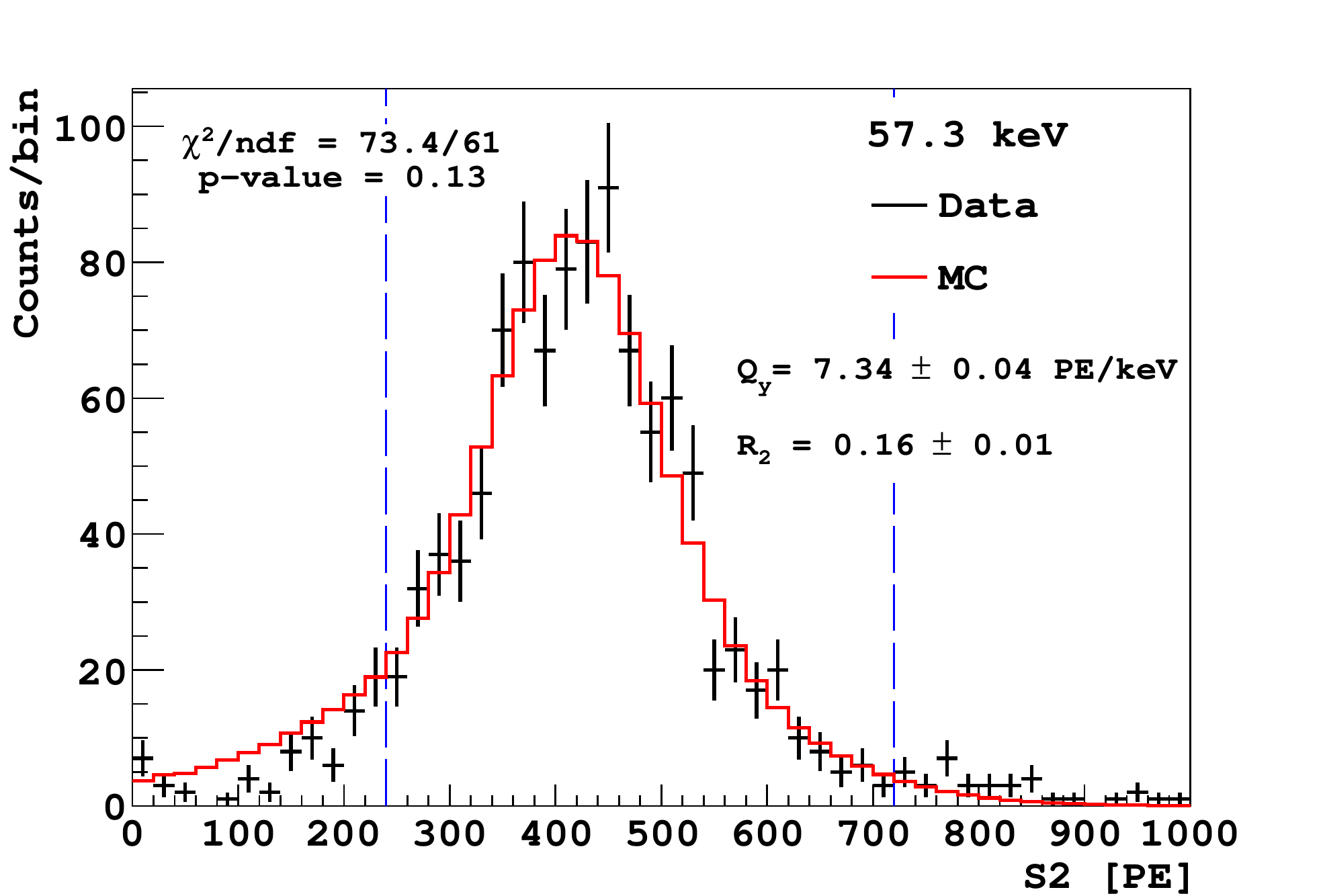}
\caption{\label{fig:qy500}
{\bf \color{black} Black:} Experimental data collected with \Edrift\ = 486\,V/cm.  {\bf \color{red} Red:} Monte Carlo fit of the experimental data.  The range used for each fit is indicated by the vertical {\bf \color{blue} blue} dashed lines. The $\chi^{2}$ (sum across all spectra as defined in the text) and the total number of degrees of freedom are shown in the last panel. 
}
\end{figure*}

The fit results for all five drift fields investigated - ranging from 96.4 to 486\,V/cm - and all four recoil energies under consideration - ranging from 16.9 to 57.3\,\kevr\ - are shown in Figs.~\ref{fig:qy100}, \ref{fig:qy200}, \ref{fig:qy300}, and \ref{fig:qy500}.  In each of the figures, the panels show the experimental data at a given recoil energy fit with Monte Carlo data.  The \chisq\ and the total number of degrees of freedom (ndf) are shown in the last (57.3\,keV) panel of each figure.  The agreement between the data and the MC is adequate, although the data is systematically lower than MC on the left tail below the fit bound. This deficit is likely due to decreased trigger efficiency for small signals (see Sec.~\ref{sec:trigger}).  Because we chose not to fit the S2 spectra below the fit bounds listed in Table~\ref{table:S2trigger}, we did not resolve nuclear recoil peaks for the 25.4 and 36.1 keV data taken at 96.4\,V/cm and did not resolve any nuclear recoil peaks for the 49.5\,V/cm data.  We therefore do not report S2 results for those combinations.

\begin{figure*}[t!]
\includegraphics[width=\columnwidth]{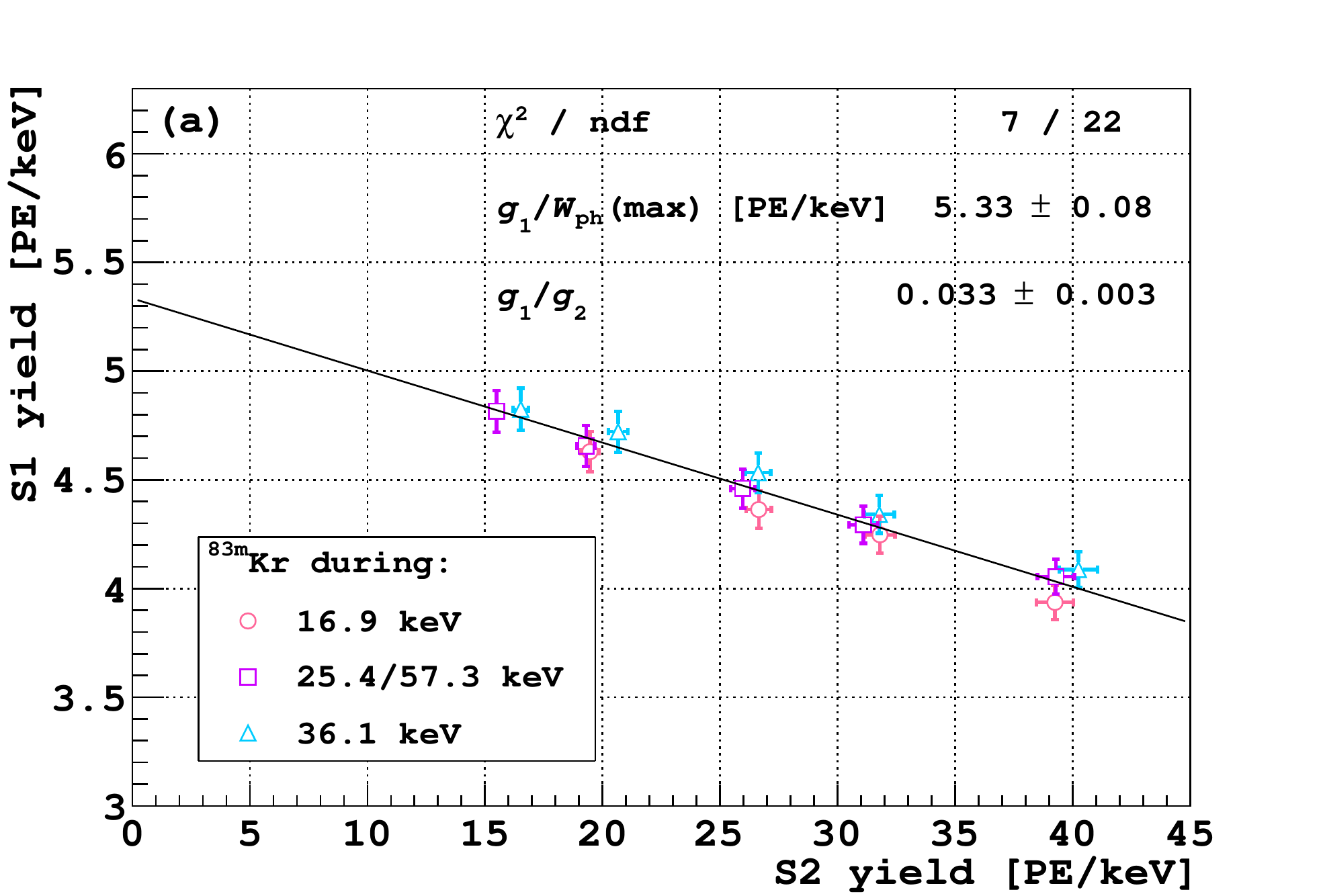}
\includegraphics[width=\columnwidth]{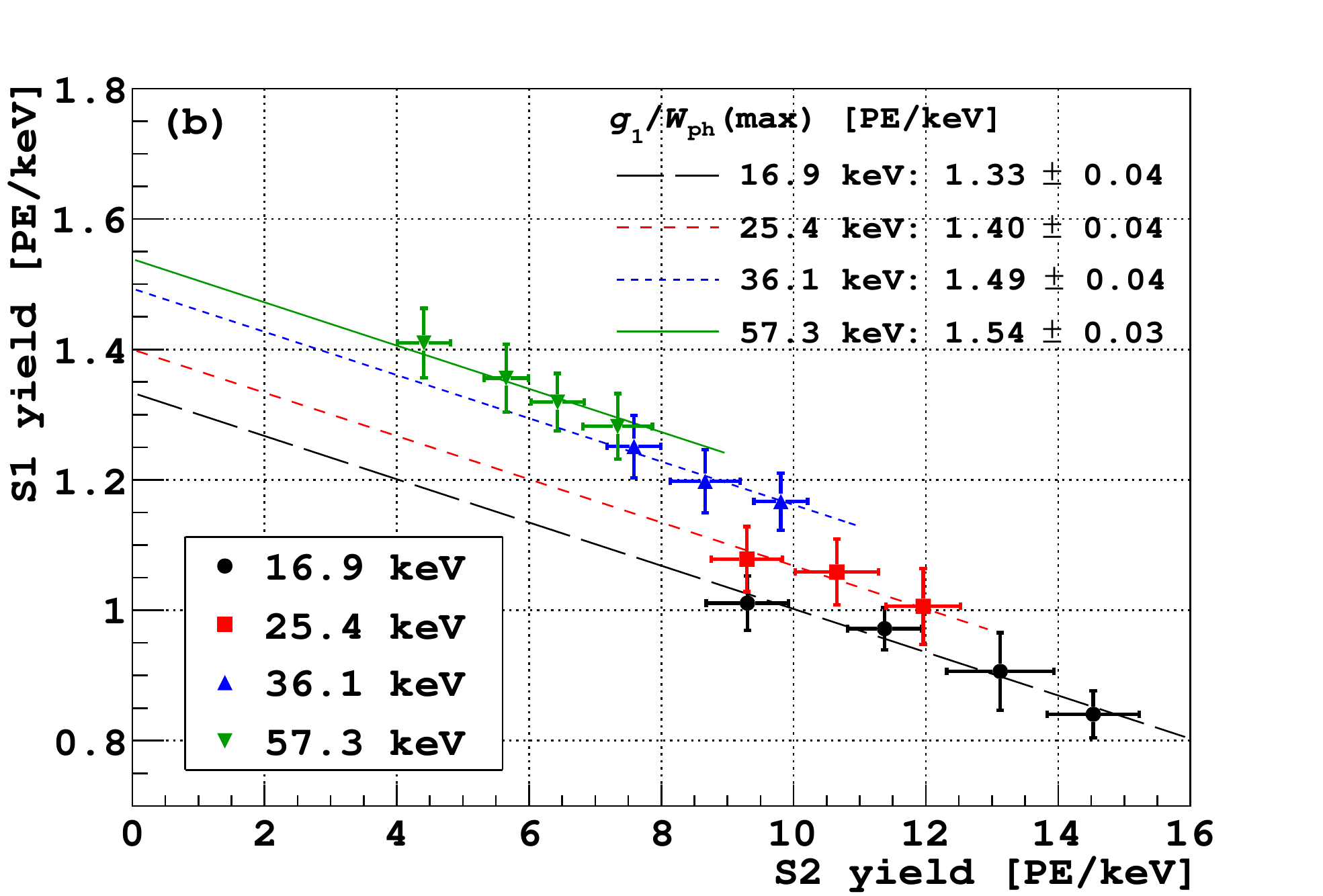}
\caption{\label{fig:s1_vs_s2}(a) S1 yield vs. S2 yield for \krthree. The best fit results for Eq.~\eqref{eq:anticorr} are shown. (b) S1 yield vs. S2 yield for nuclear recoils. The data in (a) and (b) are fit simultaneously with the intercepts free and the slopes taken as a common parameter. }
\end{figure*}

Figure~\ref{fig:s2} shows the fitted S2 yield at each recoil energy as a function of \Edrift. The charge yield of \krthree\ is plotted in the same figure for comparison.  Figure~\ref{fig:s2qy} shows S2 yield as a function of \Enr\ measured at the four different drift fields for which the S2 peak was resolved (96.4, 193, 293 and 486\,V/cm).  The error bars represent the quadrature combination of the statistical error returned from the fit and the systematic errors due to each of the sources examined.
We evaluated the systematic uncertainties of \qy\ following the same procedures described in the section of \leff\ analysis.  In Table~\ref{table:qysys}, we show, as an example, the statistical, systematic and combined errors for \qy\ at \Edrift = 193\,V/cm. Table~\ref{table:S2results} shows fit results for each drift field and recoil energy combination with total combined uncertainties. 

\begin{table}[htdp]
\begin{center}
\begin{tabular}{c|cccc}
\hline
\hline
Drift field & \multicolumn{4}{c}{Recoil energy} \\
$[$V$/$cm$]$ & \multicolumn{4}{c}{[keV]} \\
      & 16.9 & 25.4 & 36.1 & 57.3\\
\hline 
96.4 & 9.3 $\pm$ 0.6 &  &  & 4.4 $\pm$ 0.4\\ 
193 & 11.4 $\pm$ 0.6 & 9.3 $\pm$ 0.5 & 7.6 $\pm$ 0.4 & 5.7 $\pm$ 0.3\\ 
293 & 13.1 $\pm$ 0.8 & 10.7 $\pm$ 0.6 & 8.7 $\pm$ 0.5 & 6.4 $\pm$ 0.4\\ 
486 & 14.5 $\pm$ 0.7 & 12.0 $\pm$ 0.6 & 9.8 $\pm$ 0.4 & 7.3 $\pm$ 0.5\\ 
\hline
\end{tabular}
\end{center}
\caption{$Q_y$ values in units of [PE/keV] with total combined errors.}
\label{table:S2results}
\end{table}

Fits of the resolution of \stwo\ to $\sigma_2\,=\,\sqrt{\left(1+a^2\right) S2 + R_{2}^{2} S2^{2}}$, shown in Fig.~\ref{fig:s2sigma}, indicate a better resolution of nuclear recoils than $\beta$-like events. This is opposite to the \sone\ case. Differences in how the recombination ratio fluctuates could again play a role in determining this result.

\section{Anticorrelation between S1 and S2}

Figure~\ref{fig:s1_vs_s2} shows our simultaneous measurement of \sone\ and \stwo\ yields for both \krthree\ and nuclear recoils up to \Edrift=\,486\,V/cm. We found in both cases the decrease of S1 yield with drift field was accompanied by an increase in S2 yield.  Such anticorrelation was previously observed and reported for electrons, relativistic heavy ions, $\alpha$ particles and fission fragments~\cite{kubota78,doke85,hitachi87}. In liquid xenon, \sone-\stwo\ anticorrelation has also been observed for $\beta$-like events (see Ref.~\cite{kubota78} for \bios\ response in LXe TPCs and Ref.~\cite{Aprile:2007qd} for studies on $\gamma$-ray response).  Our observation is the first reported for nuclear recoils.  In the case of $\beta$-like events, the decrease of S1 and increase of S2 is linked to the partial inhibition of the recombination of electron-ion pairs caused by the drift fields~\cite{kubota78}.  It is not surprising that the anticorrelation is observed for nuclear recoils in argon, given the dependence of the S1 yield on the drift field~\cite{scene1}.

The \sone-\stwo\ anticorrelation allows the S1 and S2 measurement gains to be determined, if we consider the recombination model a good approximation in this drift field and ionization density regime. In this model, the origin of scintillation produced by ionizing radiation in a liquefied noble gas is attributed to the ions R$^+$ and excitons R$^*$ created along the particle track.  Each R$^+$ (after recombination with \e) and each R$^*$ quickly form an excited dimer R$_{2}^{*}$, and deexcitation of this dimer to the ground state, ${\rm R}_{2}^{*} \rightarrow 2 {\rm R} + h\nu$, is assumed to emit a single UV photon due to the transition between the lowest excited molecular level and the ground level~\cite{doke02}. 

Reference~\cite{dahl} provided a detailed description of the recombination model that illustrates the relationship between the number of excitons, \Nex\ and electron-ion pairs, \Ni\ produced by ionizing radiation, and the \sone\ and \stwo\ signals in a liquid noble gas TPC, which we summarize below. The total number of scintillation photons can be written as,
\begin{equation}
N_{\rm ph} = \eta_{\rm ex} N_{\rm ex}+ \eta_{\rm i} r N_{\rm i},
\end{equation}
where $r$ is the fraction of ions that recombine, and $\eta_{\rm ex}$ and $\eta_{\rm i}$ are the efficiencies with which direct excitons and recombined ions produce scintillation photons respectively.  In the absence of 
nonradiative relaxation processes affecting isolated excitons or recombined ions, we expect $\eta_{\rm ex}$ and $\eta_{\rm i}$ to both be unity. Here, we explicitly do not include Penning or Hitachi quenching processes in the definition of $\eta_{\rm ex}$ and $\eta_{\rm i}$, instead assuming these processes affect \Nex\ and \Ni. We define the \sone\ and \stwo\ measurement gains $g_1$ and $g_2$ such that
\begin{equation}
S1 = g_1 N_{\rm ph}\text{, and }
S2 = g_2 \left( 1 - r \right) N_{\rm i},
\end{equation}
where $S1$ and $S2$ are the scintillation and ionization signals respectively in units of PE (both are corrected for z-dependence). We believe $g_1$ and $g_2$ are detector properties, hence they remain constant from electron recoils to nuclear recoils. Following~\cite{doke02}, the average energy required for the production of a single photon in the limit $r$\,$\rightarrow$\,1, \wph, can be written as
\begin{equation}
W_{\rm ph}{\rm (max)} = \frac{E}{N_{\rm ex} + N_{\rm i}} = \frac{W}{1+N_{\rm ex}/N_{\rm i}}.
\end{equation}
Here the average energy required for an electron-ion pair production, the so-called $W$-value, ($W = E/N_{\rm i}$, where $E$ is the energy of the recoil) is determined to be 23.6$\pm$0.3\,eV in \lar\ using internal conversion electrons emitted from \bios~\cite{doke74}. The inherent \sone-\stwo\ anticorrelation in the recombination model can now be expressed as
\begin{equation}
\frac{S1}{E} = \frac{g_1}{W_{\rm ph}{\rm (max)}} - \frac{g_1}{g_2} \frac{S2}{E}.
\label{eq:anticorr}
\end{equation}

\wph\ was measured as $19.5\pm 1.0$ eV by Doke \emph{et al.}~\cite{doke02} for \bios\ conversion electrons.  We fit the data shown in Figs.~\ref{fig:s1_vs_s2}(a) and~\ref{fig:s1_vs_s2}(b) simultaneously with all y-intercepts free [$g_1/$\wph], and the slope ($g_1/g_2$) as a common parameter. We assume that the measured value of \wph\, by Doke also holds for \krthree\, to extract $g_1= 0.104\pm0.006$\,PE/photon, and $g_2= 3.1\pm0.3$\,PE/\e. We used this $g_2$ value to convert our measured \qy\ into units of \e/keV in the previous section. 

Because $g_1$ is assumed to be a detector constant, the increasing y-intercepts of the nuclear recoil data in Fig.~\ref{fig:s1_vs_s2}(b) imply that $W_{\rm ph}$(max) decreases with increasing nuclear recoil energy.  

We also fit the measured S2 yields shown in Fig.~\ref{fig:s2} as a function of drift field simultaneously for both \krthree\, and nuclear recoils with an empirical modification~\cite{llnl} of the Thomas-Imel box model~\cite{thomas-imel}, 
\begin{equation}
Q_{\rm y} = g_2\frac{N_{\rm i}}{E \xi} {\rm ln}\left( 1 + \xi \right) \mbox{, } \xi = \frac{N_{\rm i} C }{\mathcal{E}_{\rm d}^{B}},
\label{eq:ti}
\end{equation}
where $B$ and $C$ are constants. In this modified model, $\xi$\,$\propto$\,$\mathcal{E}_{\rm d}^{-B}$ instead of $\mathcal{E}_{\rm d}^{-1}$ as originally assumed by Thomas and Imel. Following our earlier assumption that the $W$-value of \krthree\ is the same as \bios, we fix \Ni\,=\,1.76$\times 10^3$\,\e\ for \krthree. In the combined fit, $B$ is treated as a common parameter for both \krthree\, and nuclear recoil data, while $C$ and $N_{i,\mathrm{nr}}$ are allowed to vary for each data set. The best fit results yield $C_{^{83m}\rm Kr}$\,=\,0.18\,$\pm$\,0.03\,(V/cm)$^{B}$/\e and $B$\,=\,0.61\,$\pm$\,0.03. The best fit values and errors for $C$ and \Ni\ of the nuclear recoils as a function of energy are listed in Table~\ref{table:s2_ti}. 

\begin{table*}[t!]
\begin{center} 
\begin{tabular*}{\textwidth}{@{\extracolsep{\fill} } cccccc} \hline\hline
Recoil energy~[\kevr] & $C$ [(V/cm)$^{B}$/\e] & \Ni & \Nex+\Ni & \Nex/\Ni & $\mathcal{L}$ \\ 
\hline 
16.9 & 0.58$\pm$0.17 & 139$\pm$32 & 217$\pm$12 & 0.6$\pm$0.4 & 0.250$\pm$0.005 \\
25.4 & 0.50$\pm$0.23 & 179$\pm$63 & 342$\pm$19 & 0.9$\pm$0.7 & 0.262$\pm$0.006 \\
36.1 & 0.45$\pm$0.19 & 214$\pm$71 & 518$\pm$28 & 1.4$\pm$0.8 & 0.280$\pm$0.005 \\
57.3 & 0.42$\pm$0.16 & 276$\pm$105 & 848$\pm$46 & 2.1$\pm$1.2 & 0.288$\pm$0.005 \\
\hline
\end{tabular*}
\caption{Columns $C$ and \Ni\ are the fit results of the \krthree\, and nuclear recoil data in Fig.~\ref{fig:s2} to the modified Thomas-Imel box model with $N_i$ for \krthree\, fixed to the value obtained by Doke (see text). Columns \Nex\,+\,\Ni\ and $\mathcal{L}$ are the computed values based on Fig.~\ref{fig:s1_vs_s2} (see text). We note that the $\mathcal{L}$-factor is referenced to electronic recoils from \krthree, not electronic recoils of the same energy as the nuclear recoils under investigation.  Column \Nex/\Ni\ is computed with columns \Ni\ and \Nex\,+\,\Ni.}
\label{table:s2_ti}
\end{center}
\end{table*} 

We finally calculate \Nex\,+\,\Ni\  using the ratio of y-intercepts in Fig.~\ref{fig:s1_vs_s2}(b) to the y-intercept of Fig.~\ref{fig:s1_vs_s2}(a) and the value of $W_{\rm ph}$(max) from~\cite{doke02}. We note again that \Nex\, and \Ni\, as defined here are the number of excitons and ions remaining after any track-dependent quenching processes (such as Penning or Hitachi quenching) have completed. We use the \Ni\ and \Nex\,+\,\Ni\ columns in Table~\ref{table:s2_ti} to compute \Nex/\Ni. The last column in the table, $\mathcal{L}$, is the overall quenching factor of nuclear recoils, which we define as the ratio of (\Nex\,+\,\Ni)/$E$ for nuclear recoils to that for electrons from \krthree. We note that this differs somewhat from the canonical definition of the Lindhard factor in referring to specific energy electronic recoils. It is equal to the ratio of each y-intercept in Fig.~\ref{fig:s1_vs_s2}(b) to the y-intercept of Fig.~\ref{fig:s1_vs_s2}(a). 

\begin{figure}[t!]
\includegraphics[width=\columnwidth]{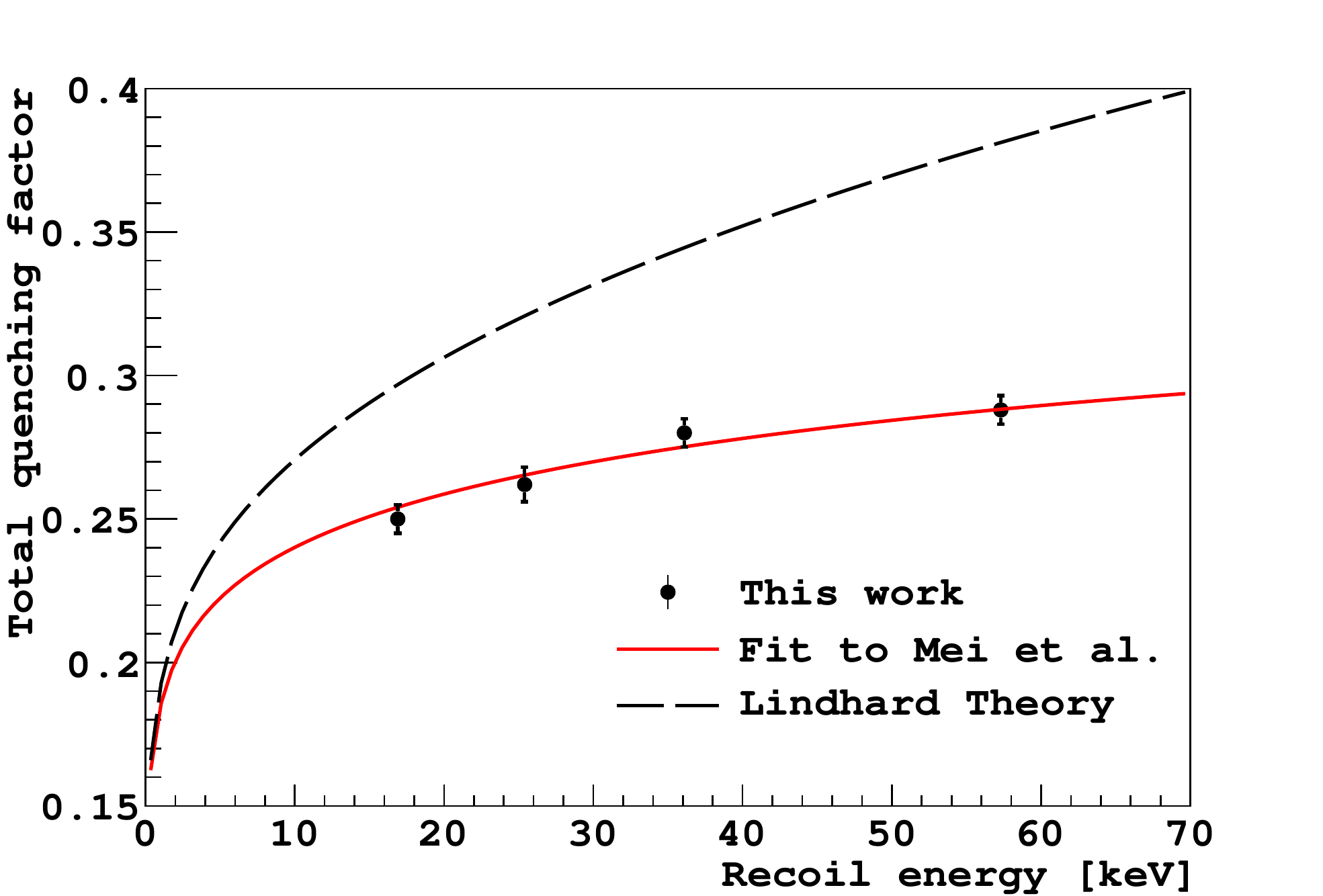}
\caption{\label{fig:mei}Total quenching factor $\mathcal{L}$ (relative to \krthree\,events as defined in the text) compared to Lindhard's theory and the Lindhard-Birk combined model proposed by Mei \emph{et al.} The best fit curve to Mei's model with Birk's constant $kB$ as free parameter yields $kB=5.0\pm0.2\times10^{-4}$\,MeV$^{-1}$g\,cm$^{-2}$.}
\end{figure}

Mei \emph{et al.} attributed the reduction of scintillation efficiency to two major mechanisms: (1) energy loss due to nuclear collisions, and (2) scintillation quenching due to high ionization and excitation density induced by nuclear recoils~\cite{mei}. Lindhard's theory~\cite{lindhard} describes the first mechanism (\fn), and Birk's saturation law~\cite{birk} models the latter (\fl). They argued that since these two effects are independent of each other, one could combine the two directly (\fn$\cdot$\fl) to explain the observed reduction of scintillation yield for nuclear recoils in noble liquids. We compared their prediction to our data. Instead of interpreting \fn$\cdot$\fl\ as the reduction in scintillation alone, we consider it equal to the total reduction factor of scintillation and ionization combined, {\it i.e.}, $\mathcal{L}$\,=\,\fn$\cdot$\fl.  The best fit curve to Mei's model with Birk's constant $kB$ as free parameter yields $kB=5.0\pm0.2\times10^{-4}$\,MeV$^{-1}$g\,cm$^{-2}$ and is shown in Fig.~\ref{fig:mei}. Using the experimental results of~\cite{laverne}, Mei \emph{et al.} found a value of $kB=7.4\times10^{-4}$\,MeV$^{-1}$g\,cm$^{-2}$. If one uses the quenching factor of $^{36}$Ar ion at \Edrift\,=\,3.2\,kV/cm instead of at null field with Mei's approach~\cite{mei}, the $kB$ value will be within 1$\sigma$ of our fit result. 

\begin{figure}[t!]
\includegraphics[width=\columnwidth]{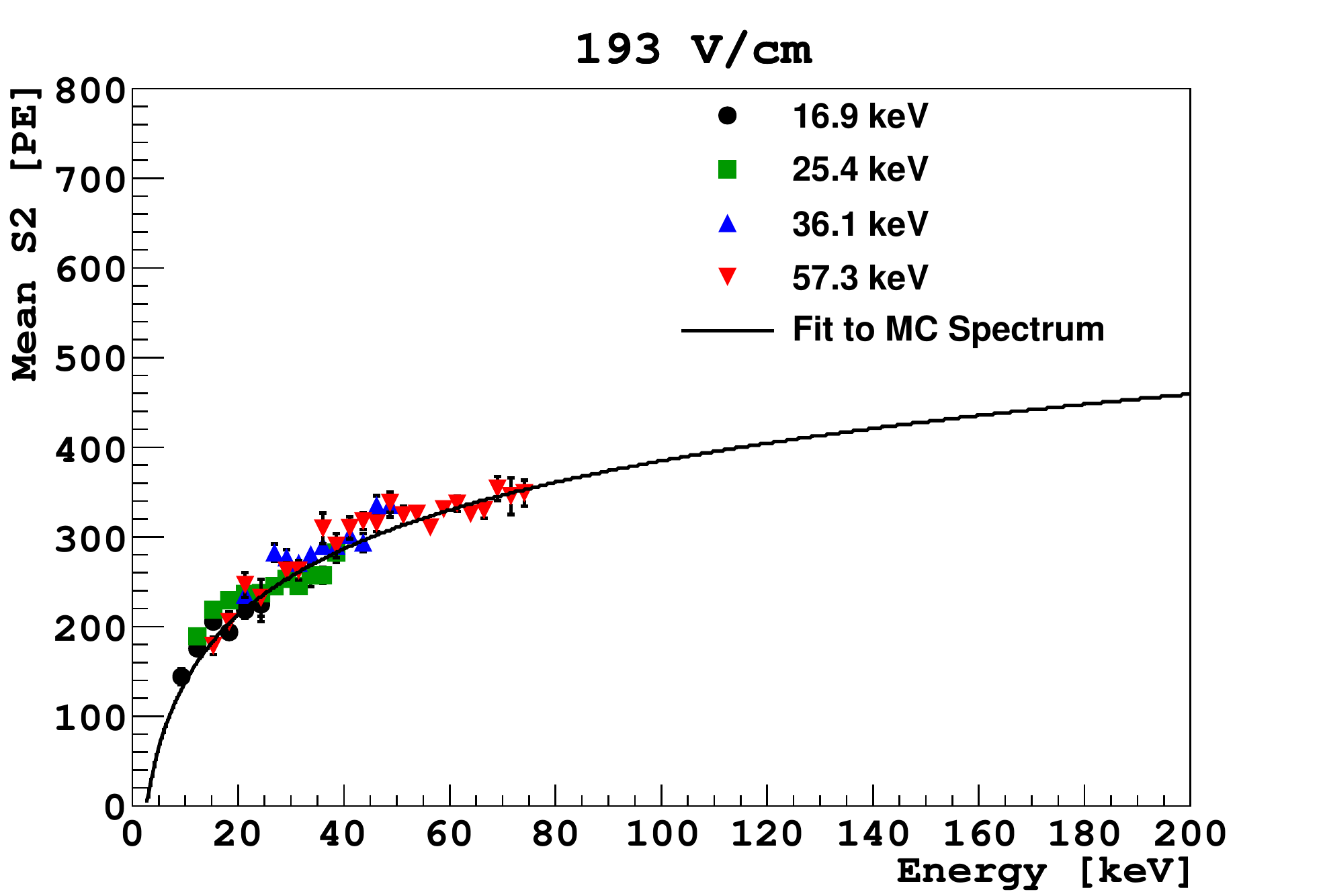}
\caption{\label{fig:S2vEnergy}The data points show S2 vs energy for \Edrift\,=\,193\,V/cm, where the energy axis is evaluated using S1 and our measured \leff\,values. The line shows the logarithmic relationship [Eq.~\ref{eq:log}] between S2 and energy as obtained from fitting S2 to the MC. }
\end{figure}

As a final consistency check of our analysis, Fig.~\ref{fig:S2vEnergy} shows Eq.~\ref{eq:log} with the best fit parameter values obtained by fitting the S2 data at \Edrift\,=\,193\,V/cm to the MC energy spectra. The data points shown were obtained from S2 vs. S1 distributions [such as Fig.~\ref{fig:s1s2}(d)] at the same drift field, where the energy is evaluated using our \leff\,measurement. The two independent estimations of \leff\,and $Q_y$ show general consistency. 

\section{Comparison of Scintillation and Ionization from Recoils Parallel and Perpendicular to the Drift Field}

\begin{figure*}[t!]
\includegraphics[width=\columnwidth]{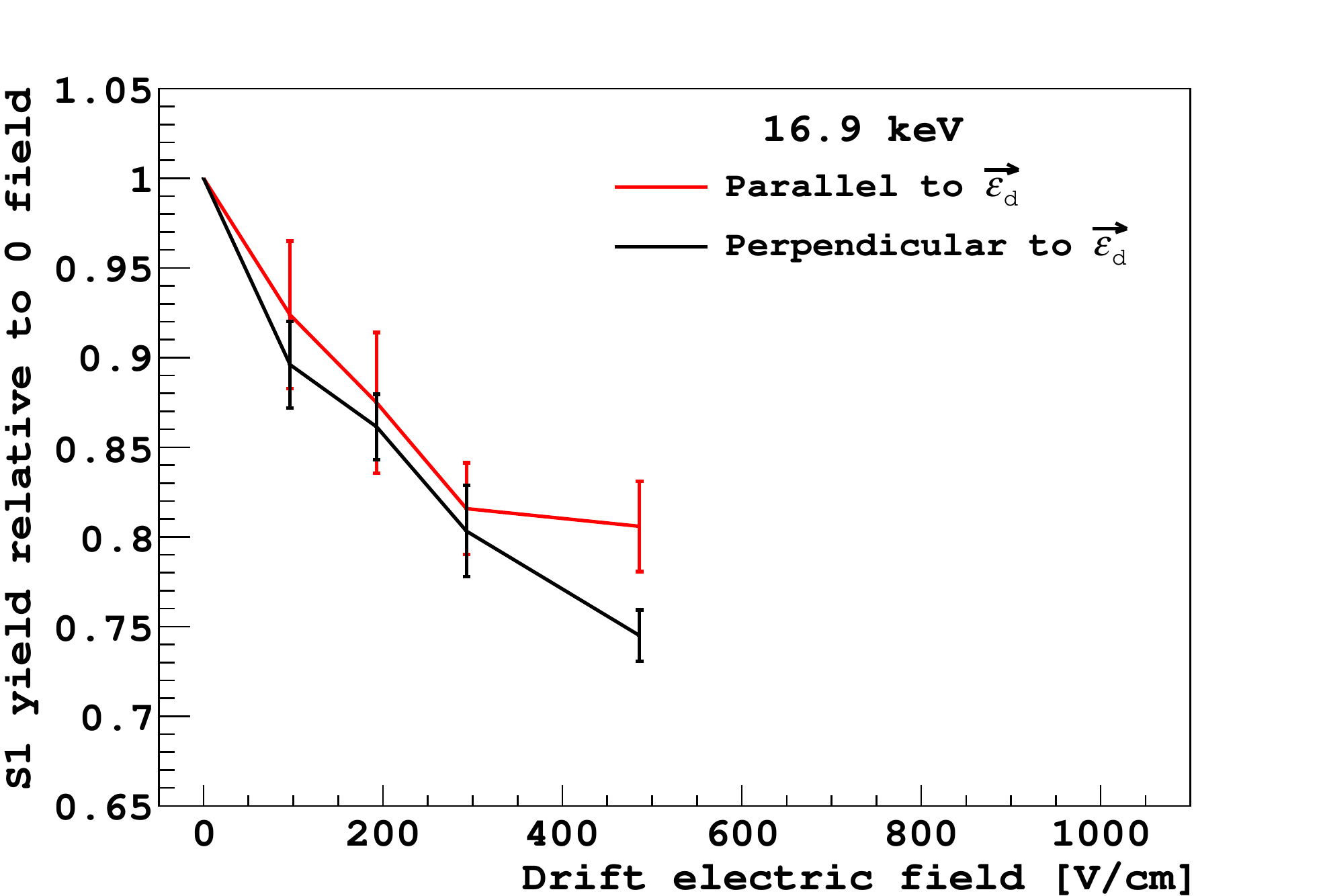}
\includegraphics[width=\columnwidth]{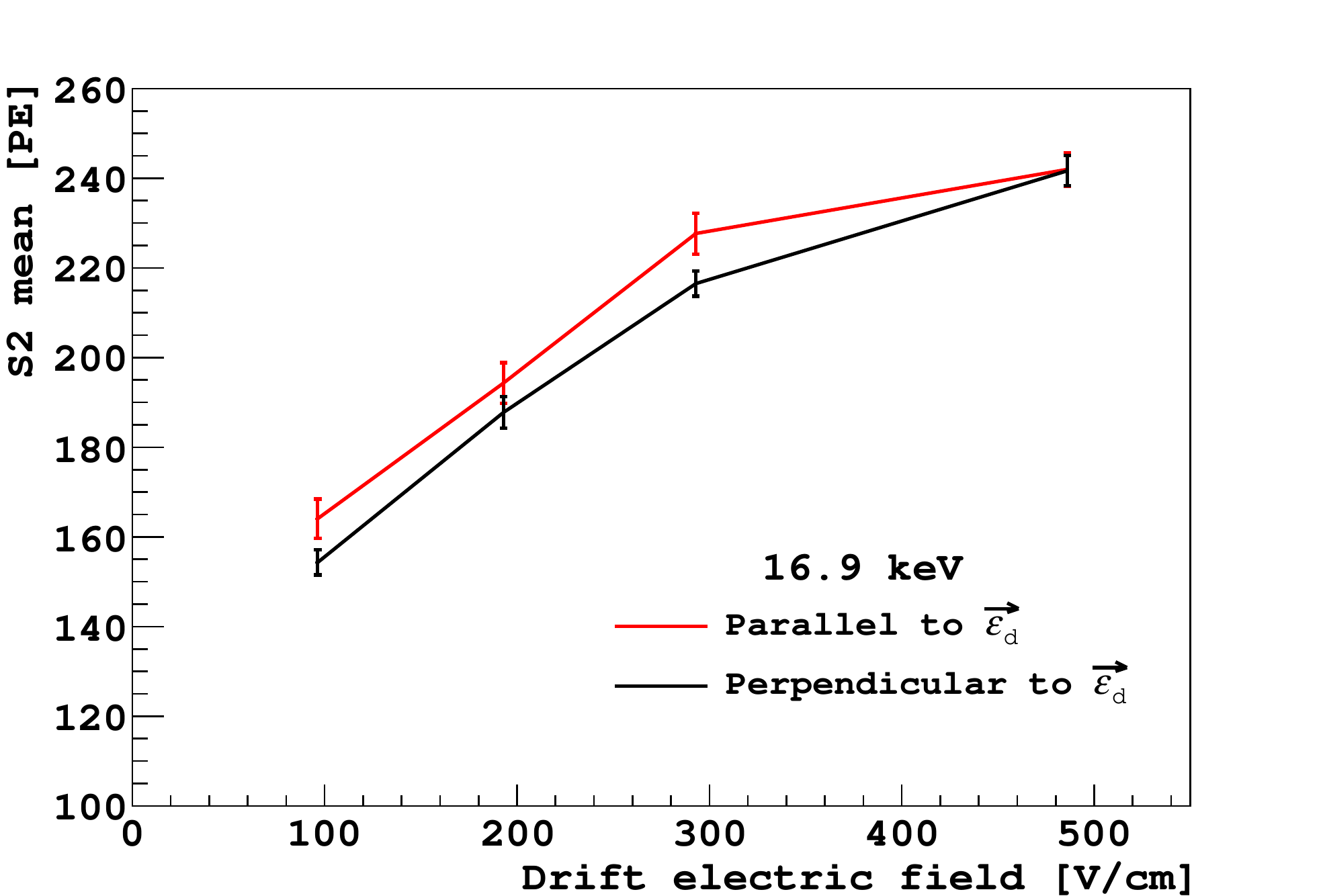}
\includegraphics[width=\columnwidth]{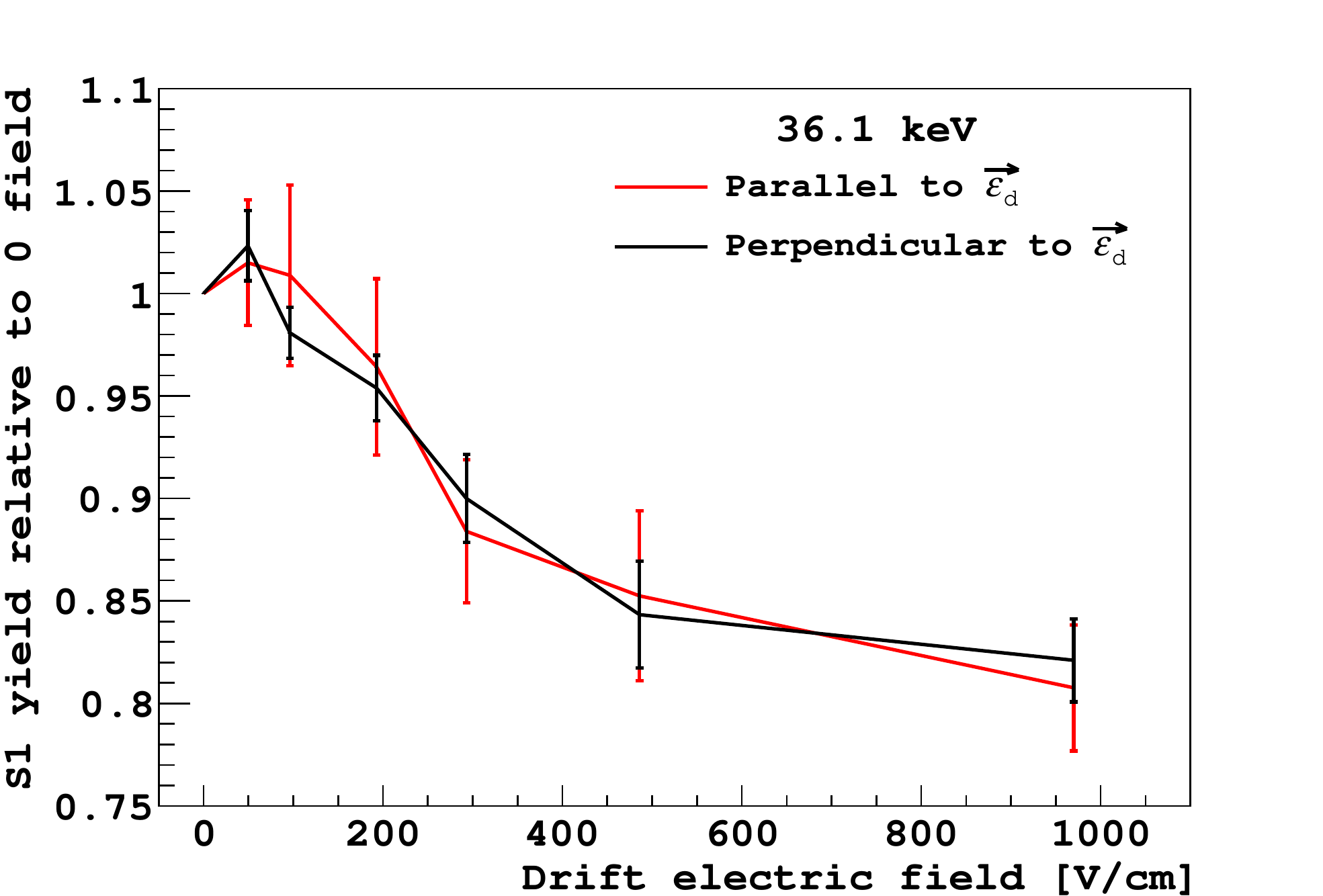}
\includegraphics[width=\columnwidth]{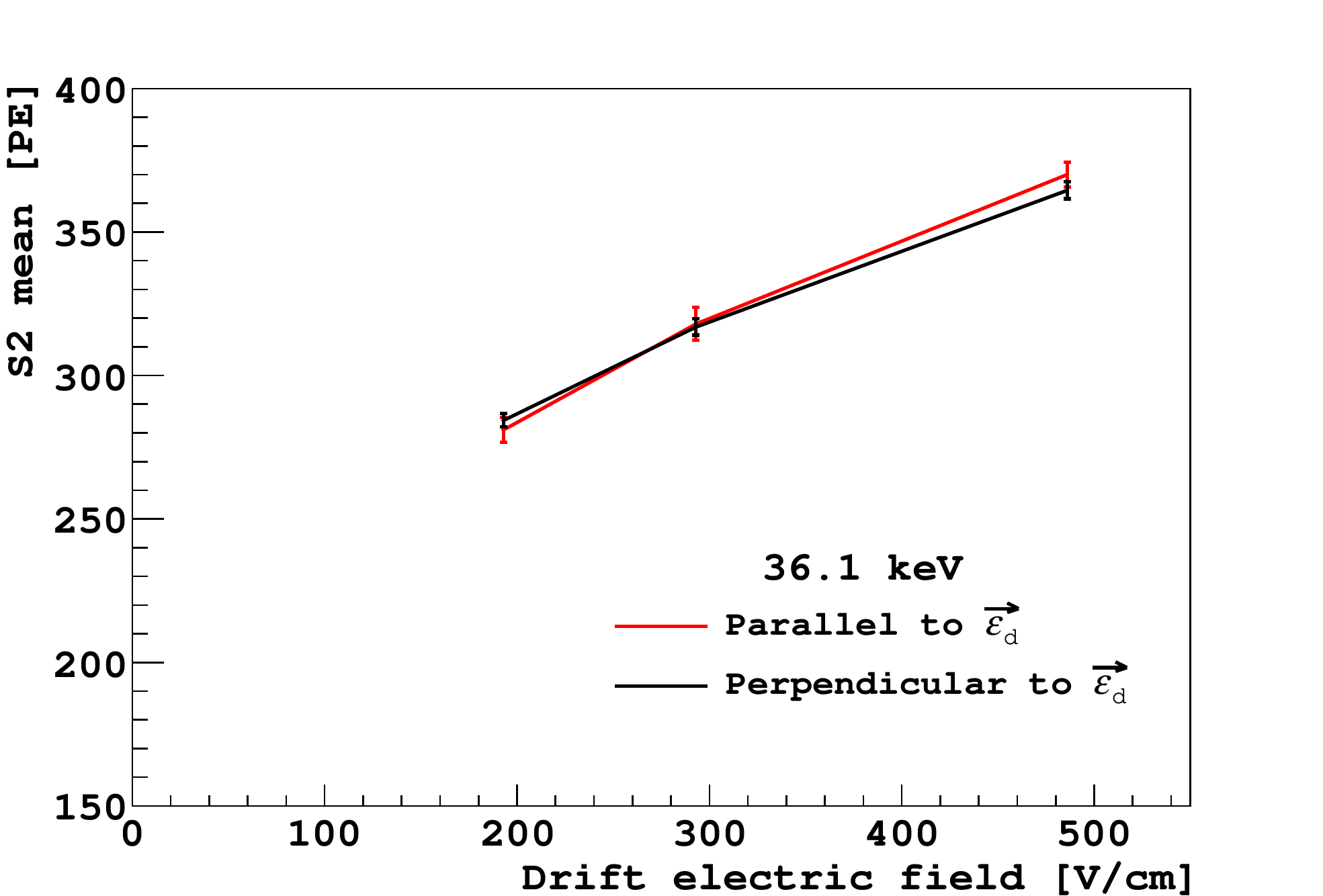}
\includegraphics[width=\columnwidth]{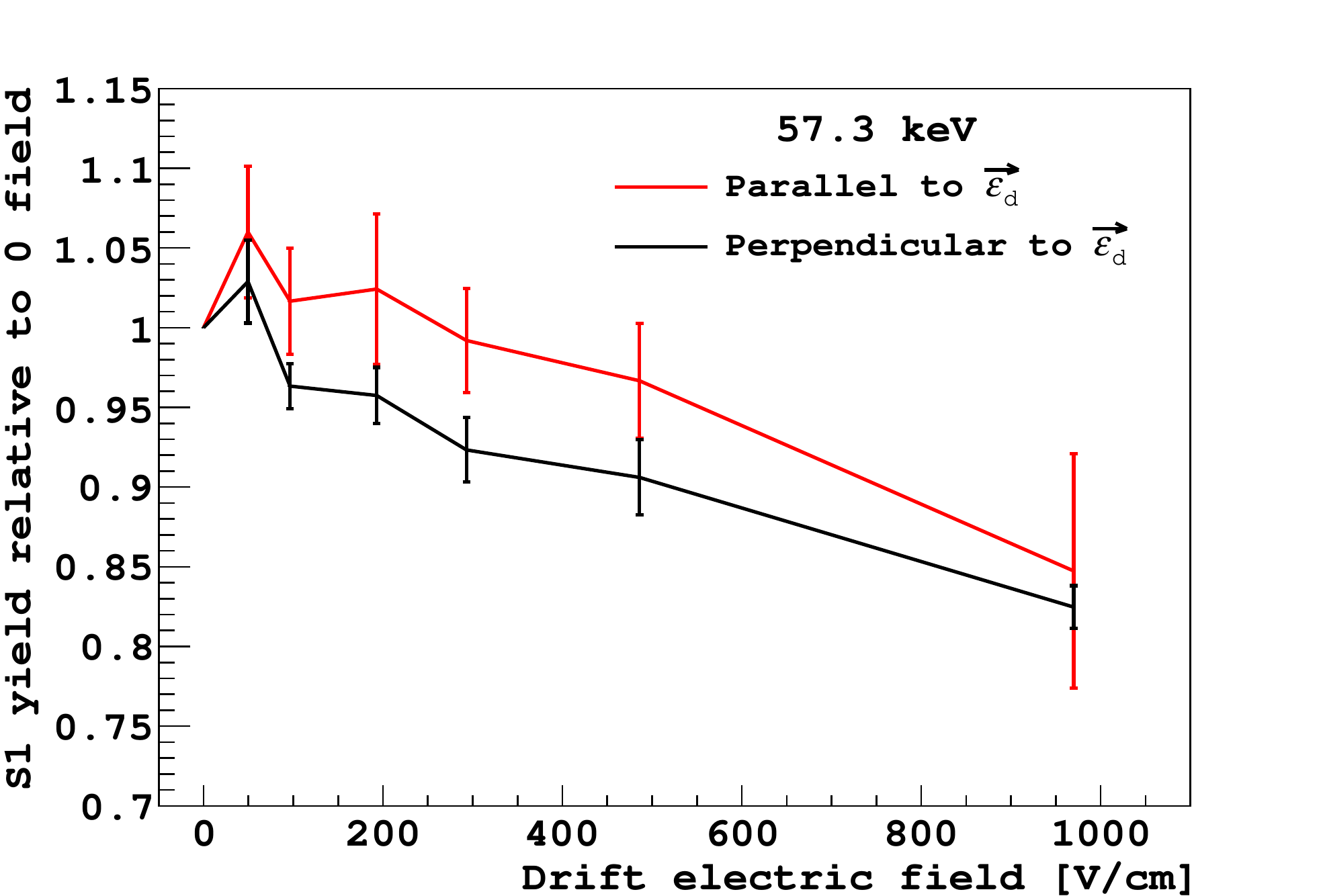}
\includegraphics[width=\columnwidth]{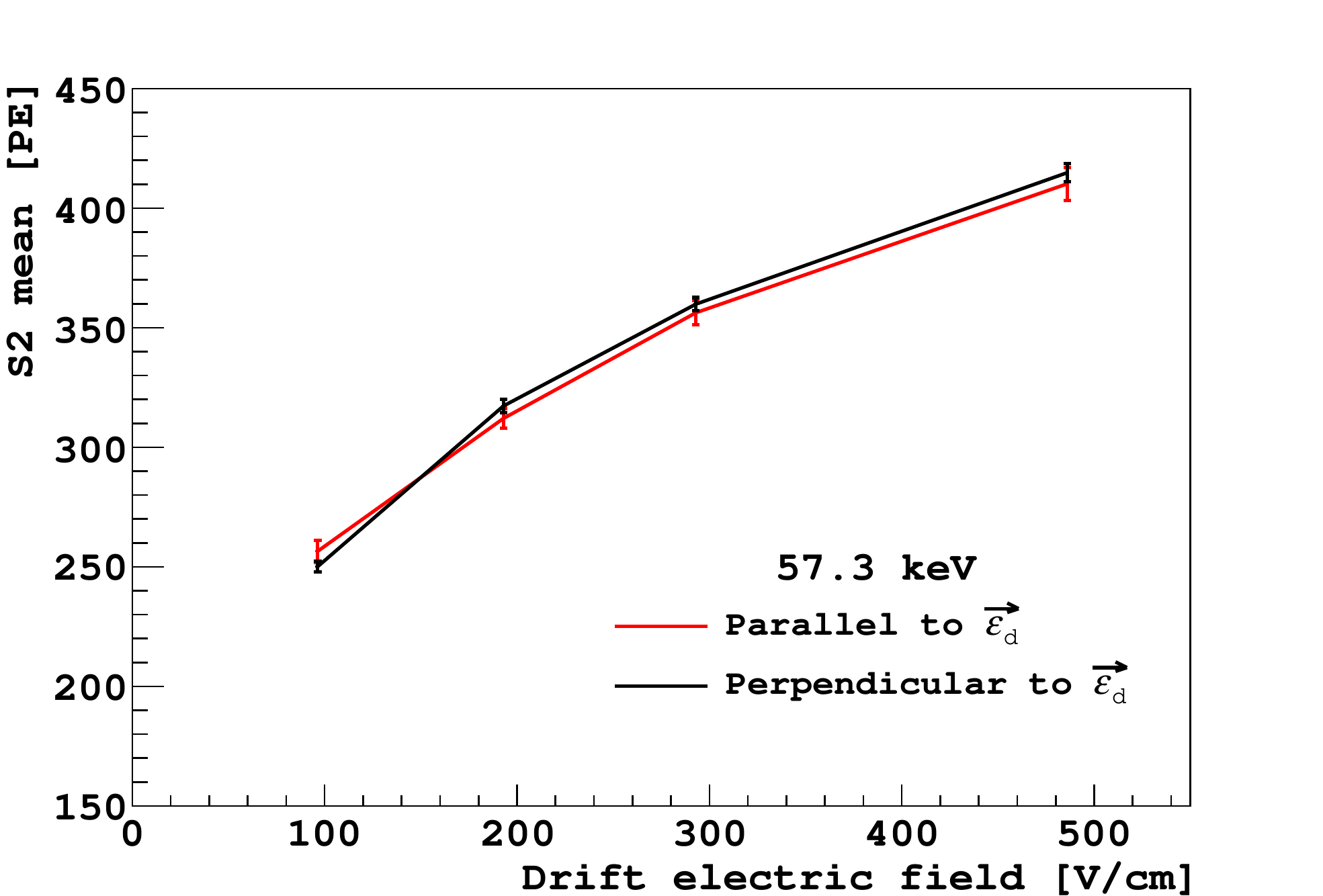}
\caption{\label{fig:directional}Scintillation yield relative to null field (left panels) and ionization yield with nonzero drift field (right panels) of nuclear recoils at 16.9, 36.1 and 57.3\,\kevr. {\bf \color{black} Black:} Momentum of nuclear recoil is perpendicular to \boldsymbol{\Edrift}.  {\bf \color{red} Red:} Momentum of nuclear recoil is parallel to \boldsymbol{\Edrift}. Sources of systematic uncertainties common to both field orientations are not included in the error bars.}
\end{figure*}

Sensitivity to the direction of detected WIMP recoils would give a powerful signature for identifying a signal observed in a direct-detection dark matter experiment with the galactic dark matter~\cite{Spergel}.  The main velocity component of an earthbound laboratory with respect to the Galactic center of mass is due to the revolution of the solar system about the Galactic center.  This rotational velocity is nearly equal to the virial velocity of an isothermal dark matter WIMP halo~\cite{smith}.  In this situation, the kinematics of WIMP-nucleus scattering will result in recoil nuclei from WIMP scattering which are predominantly directed into the hemisphere antiparallel to the rotational velocity.  However, this direction has a fixed location in celestial coordinates of right ascension and declination.  In the laboratory frame, the Earth's rotation makes this direction rotate around the polar axis with a period of one sidereal day.  The strength of the correlation varies for different WIMP halo models and different detector characteristics and has been extensively studied theoretically~\cite{green}. 

These studies show that in practically any WIMP model, even modest direction sensitivity for a limited number of detected events gives a powerful discriminant for identifying a signal with the galactic halo, as opposed to any isotropic or fixed-location source in the laboratory.  Direction sensitivity is therefore a highly desirable characteristic for a direct detection experiment, and has been actively sought after in many signal modalities for many years~(see Ref.~\cite{ahlen} and references cited therein).

Applied electric fields are known to modify the recombination of electron-ion pairs in ionizing radiation tracks.  Columnar recombination~\cite{jaffe} models suggest that the magnitude of these effects should in some circumstances vary with the angle between the field and the track direction, and these effects have been discussed as possible ways to achieve direction sensitive WIMP recoil detection in LAr or GXe targets~\cite{pu,nygren}.  Such directional effects have been reported from experiments using tracks from $\alpha$ particles~\cite{swan} and protons~\cite{argoneut} in liquid argon.
 
As we have shown, electron-ion recombination for nuclear recoil tracks in liquid argon also depends strongly on the applied electric field.  If a directional effect on recombination is present in LAr, we would expect to measure different scintillation and/or ionization responses for nuclear recoils of the same energy but with different track orientations.  We therefore configured our neutron beam and neutron detector placement so as to allow us to simultaneously record nuclear recoil events with tagged initial momentum in directions parallel and perpendicular to the drift field applied to the liquid argon TPC.  The neutron beam direction was selected at a downward angle with respect to the horizontal, dictated by the kinematics at the neutron energy in use. The two-angle goniometric mount then allowed the neutron detectors to be placed at positions corresponding to a single scattering angle but at different azimuthal angles corresponding to recoil nucleus directions parallel or perpendicular to the (vertical) drift field.  The results reported in the preceding sections of this paper combined the data from the two neutron counters with the initial recoil direction perpendicular or near perpendicular to the drift field.

In order to produce a direction-sensitive response, the recoil nucleus must have enough energy (range) to form a track with a definite direction.  Following the arguments of~\cite{nygren}, one might expect such a response to start for recoils above the energy where the length of the track exceeds the Onsager radius, $r_{\rm O}$\,=\,$e^2/4 \pi \epsilon K$.  This is the distance between a positive ion and a free electron for which the potential energy of the electrostatic field, $e^2$/$4\pi\epsilon r_{\rm O}$, is equal to the kinetic energy of a thermal electron, $K$\,=\,$3kT/2$.  In liquid argon ($T$\,=\,87\,K, $\epsilon$\,=\,1.5) $r_{\rm O}$\,$\simeq$\,80\,nm.  The range of argon recoils in liquid argon~\cite{srim} is about 90\,nm at 36.1\,\kevr.  It increases to 135\,nm at 57.3\,keV, substantially exceeding the $r_{\rm O}$.  A similar value for the energy at which directional effects might start is obtained in the line-charge model of Ref.~\cite{hitachi92}.

Figure~\ref{fig:directional} shows the comparison of average S1 and S2 responses for the two track orientations, with the scintillation and ionization yields plotted as a function of the applied electric field.  Any differences for parallel and perpendicular tracks for both signals are seen to be very small compared to the statistical errors and the overall trend of field dependence for the 16.9 and 36.1\,\kevr\ energies.  The S1 response at 57.3\,keV does exhibit an orientation difference, but with marginal statistical significance.  Further investigation with more precise measurements at higher recoil energies is planned.

\section{Conclusion}

In a previous paper, we presented results showing for the first time an electric field dependence in the \sone\ scintillation efficiency of nuclear recoils in liquid argon.  The current paper presents the following new results for argon recoils in liquid argon in the energy range 10.3 to 57.3 \kevr\ and the drift field range $0-970$ V/cm ($16.9-57.3$ \kevr\ and $96.5-486$ V/cm for ionization):
\begin{compactitem}
\item values for the nuclear recoil scintillation yield relative to that of \krthree\ (\leff) and the associated uncertainties
\item detailed information on the distributions of the pulse shape discrimination parameter \fno
\item values and uncertainties for \qy, the apparatus-independent absolute yield of extracted ionization electrons per keV kinetic energy for both nuclear recoils and \krthree\ at an extraction field of 3.0\,kV/cm
\item a method and results of a search for sensitivity of the LAr-TPC response to the initial direction of nuclear recoils with respect to the applied electric field.
\end{compactitem}

These data were intended for use in calibration and parameter optimization for the DarkSide series of LAr-TPC's for dark matter searches.  The results show that the real effects of electric field on the responses of LAr-TPC's are substantially more complicated than the small, energy-independent changes that have generally been assumed up to now.  The present results should be valuable in connection with the design and calibration of any detector using scintillation and ionization in liquid argon to detect nuclear recoils.  The results also suggest a line of further investigation of a direction-sensitive effect in the response of LAr-TPC's. Direction sensitivity would be of great interest in unambiguously associating any WIMP-like signal in such a device with the apparent motion of the galactic halo. The technique described here can also be used for calibrating other dark matter targets. 

\section{Acknowledgments}

We particularly thank the technical staff at Fermilab, and A.~Nelson of Princeton University and E.~Kaczanowicz of Temple University for their contributions to the construction of the SCENE apparatus.  We thank W.~McClain of Princeton University for the preparation of the phenylphosphonic acid subtrate on ITO. We thank Dr.~G. Korga and Dr.~A. Razeto for providing the low-noise amplifiers used on the TPC PMT signals.  We thank Professor~D.~N.~McKinsey, Dr.~S. Cahn, and K.~Charbonneau of Yale University for the preparation of the \krthree\ source.  We thank Eric V\'azquez-J\'auregui of SNOLAB for assistance with MC simulations.  Finally, we thank the staff at the Institute for Structure \& Nuclear Physics and the operators of the Tandem accelerator of the University of Notre Dame for their hospitality and for the smooth operation of the beam. 

The SCENE program is supported by NSF (U.S., Grants No. PHY-1314507, No. PHY-1242611, No. PHY-1001454, No. PHY-1068192, No. PHY-1242625, No. PHY-1211308, and associated collaborative grants), the University of Chicago and DOE (U.S., under section H.44 of DOE Contract No. DE-AC02-07CH11359 awarded to Fermi Research Alliance, LLC), the Istituto Nazionale di Fisica Nucleare (Italy ASPERA 1st common call, DARWIN project), and Lawrence Livermore National Laboratory (Contract No. DE-AC52-07NA27344).



\end{document}